\documentclass[prd,aps,showpacs,nofootinbib,preprint,eqsecnum]{revtex4}
\usepackage{amsmath}
\usepackage{amssymb}
\usepackage{amsfonts}
\usepackage{graphicx,bm}
\usepackage{dcolumn}
\usepackage{color,amsxtra}
\usepackage{epsf}
\usepackage{enumerate}
\usepackage{hhline}
\usepackage{array}
\usepackage{tabularx}
\newcommand{\be}{\begin{equation}}
\newcommand{\ee}{\end{equation}}
\newcommand{\bea}{\begin{eqnarray}}
\newcommand{\eea}{\end{eqnarray}}
\newcommand{\beaa}{\begin{eqnarray*}}
\newcommand{\eeaa}{\end{eqnarray*}}

\newcommand{\e}{\mathrm{e}}

\newcommand{\Eqn}[1]{&\hspace{-0.2em}#1\hspace{-0.2em}&}
\newcommand{\abs}[1]{\vert{#1}\vert}
\def\Vec#1{\mbox{\boldmath $#1$}}

\def\be{\begin{equation}}
\def\ee{\end{equation}}
\def\bea{\begin{eqnarray}}
\def\eea{\end{eqnarray}}

\def\e{\mathrm{e}}

\begin{document}


\title{Dark energy cosmology: the equivalent description via different 
theoretical models and cosmography tests
}

\author{Kazuharu Bamba$^{1,}$\footnote{
E-mail address: bamba@kmi.nagoya-u.ac.jp},
Salvatore Capozziello$^{2, 3,}$\footnote{E-mail address: capozziello@na.infn.it},
Shin'ichi Nojiri$^{1, 4,}$\footnote{E-mail address:
nojiri@phys.nagoya-u.ac.jp} 
and 
Sergei D. Odintsov$^{5, 6, 7, 8,}$\footnote{
E-mail address: odintsov@ieec.uab.es}
}
\affiliation{
$^1$Kobayashi-Maskawa Institute for the Origin of Particles and the
Universe,
Nagoya University, Nagoya 464-8602, Japan\\
$^2$Dipartimento di Scienze Fisiche,
Universit\`{a} di Napoli {}``Federico II'' and 
$^3$INFN Sez. di Napoli, Compl. Univ. di Monte S. Angelo, Edificio G, Via Cinthia, I-80126, Napoli, Italy \\
$^4$Department of Physics, Nagoya University, Nagoya 464-8602, Japan\\
$^5$Instituci\`{o} Catalana de Recerca i Estudis Avan\c{c}ats (ICREA), 
Barcelona, Spain\\ 
$^6$Instituto de Ciencias del Espacio (CSIC) and Institut d Estudis Espacials 
de Catalunya (IEEC-CSIC), Campus UAB, Facultat de Ciencies, Torre C5,
08193 Bellaterra (Barcelona), Spain\\ 
$^7$Tomsk State Pedagogical University, Tomsk, Russia\\
$^8$Eurasian International Center for Theoretical Physics and Department of General \& Theoretical Physics, Eurasian National University, Astana 010008, Kazakhstan}


\begin{abstract}

We review different dark energy cosmologies. In particular, we present the 
$\Lambda$CDM cosmology, Little Rip and Pseudo-Rip universes, the phantom and 
quintessence cosmologies with Type I, II, III and IV finite-time future 
singularities and non-singular dark energy universes. 
In the first part, we explain the $\Lambda$CDM model and well-established 
observational tests which constrain the current cosmic acceleration. 
After that, we investigate the dark fluid universe where a fluid has quite general equation of state (EoS) [including inhomogeneous or imperfect EoS]. 
All the above dark energy cosmologies for different 
fluids are explicitly realized, and their properties are also explored. 
It is shown that all the above dark energy universes may mimic the $\Lambda$CDM model currently, consistent with the recent observational data. 
Furthermore, special attention is paid to the equivalence of different dark 
energy models. We consider single and multiple scalar field theories, tachyon scalar theory and holographic dark energy as models for 
current acceleration with the features of quintessence/phantom cosmology, 
and demonstrate their equivalence to the corresponding fluid descriptions. 
In the second part, we study another equivalent class of dark energy 
models which includes $F(R)$ gravity as well as $F(R)$ Ho\v{r}ava-Lifshitz 
gravity and the teleparallel $f(T)$ gravity. The cosmology of such models 
representing the $\Lambda$CDM-like universe or the accelerating expansion with 
the quintessence/phantom nature is described. 
Finally, we approach the problem of testing dark energy and alternative 
gravity models to general relativity by cosmography. We show that degeneration 
among parameters can be removed by accurate data analysis of large data 
samples and also present the examples. 

\end{abstract}

\pacs{
04.50.Kd, 95.36.+x, 
98.80.-k
}

\maketitle

\tableofcontents \clearpage

\section{Introduction}

Cosmic observations from Supernovae Ia (SNe Ia)~\cite{SN1}, 
cosmic microwave background (CMB) radiation~\cite{WMAP, Komatsu:2008hk, 
Komatsu:2010fb}, 
large scale structure (LSS)~\cite{LSS}, 
baryon acoustic oscillations (BAO)~\cite{Eisenstein:2005su}, 
and weak lensing~\cite{Jain:2003tba} 
have implied that the expansion of the universe is 
accelerating at the present stage.  
Approaches to account for the late time cosmic acceleration 
fall into two representative categories: 
One is to introduce ``dark energy'' in 
the right-hand side of the Einstein equation 
in the framework of general relativity 
(for recent reviews on dark energy, see~\cite{Caldwell:2009ix, Book-Amendola-Tsujikawa, Li:2011sd, Kunz:2012aw}). 
The other is to modify the left-hand side of the Einstein equation, called as 
a modified gravitational theory, 
e.g., $F(R)$ gravity (for recent reviews, see~\cite{Review-Nojiri-Odintsov, Book-Capozziello-Faraoni, Clifton:2011jh, Capozziello:2011et, Harko:2012ar, Capozziello:2012hm}). 

The various cosmological observational data supports 
the $\Lambda$ cold dark matter ($\Lambda$CDM) model, 
in which the cosmological constant $\Lambda$ plays a role of 
dark energy in general relativity. At the current stage, 
the $\Lambda$CDM model is considered to be a standard cosmological model. 
However, the theoretical origin of the cosmological constant $\Lambda$ has not been understood yet~\cite{Weinberg:1988cp}. 
A number of models for dark energy to explain the late-time cosmic 
acceleration without the cosmological constant 
has been proposed. For example, 
a canonical scalar field, so-called quintessence~\cite{quintessence-DE}, 
a non-canonical scalar field such as phantom~\cite{Caldwell:1999ew}, 
tachyon scalar field motivated by string theories~\cite{Padmanabhan:2002cp}, 
and a fluid with a special equation of state (EoS) called 
as Chaplygin gas~\cite{Kamenshchik:2001cp, Bento:2002ps, BTV-DM}. 
There also exists a proposal of holographic dark energy~\cite{Li:2004rb, Elizalde:2005ju, Nojiri:2005pu}.

One of the most important quantity to describe the features of dark energy 
models is the equation of state (EoS) $w_{\mathrm{DE}}$, which is the ratio of 
the pressure $P$ to the energy density $\rho_{\mathrm{DE}}$ of dark energy, 
defined as $w_{\mathrm{DE}} \equiv P_{\mathrm{DE}}/\rho_{\mathrm{DE}}$. 
We suppose that in the background level, the universe is homogeneous and isotropic and hence assume the Friedmann-Lema\^{i}tre-Robertson-Walker (FLRW) 
space-time. 
There are two ways to describe dark energy models. 
One is a fluid description~\cite{Nojiri:2005sx, Nojiri:2005sr, Stefancic:2004kb} and the other is to describe the action of a scalar field theory. 
In the former fluid description, we express the pressure as a function of 
$\rho$ (in more general, and other background quantities such as the Hubble parameter $H$). On the other hand, in the latter scalar field theory we derive 
the expressions of the energy density and pressure of the scalar field from 
the action. In both descriptions, 
we can write the gravitational field equations, so that 
we can describe various cosmologies, e.g., 
the $\Lambda$CDM model, in which $w_{\mathrm{DE}}$ is a constant and exactly 
equal to $-1$, 
quintessence model, where $w_{\mathrm{DE}}$ is a dynamical quantity 
and $-1< w_{\mathrm{DE}} < -1/3$, 
and phantom model, where $w_{\mathrm{DE}}$ also varies in time and 
$w_{\mathrm{DE}} < -1$. 
This means that one cosmology may be described equivalently 
by different model descriptions.  

In this review, we explicitly show that one cosmology can be described by 
not only a fluid description, but also by the description of a scalar field 
theory. 
In other words, the main subject of this work is to demonstrate that 
one dark energy model may be expressed as the other dark energy models, 
so that such a resultant unified picture of dark energy models could be 
applied to any specific cosmology. 

This review consists of two parts. 
In the first part, various dark energy models in the 
framework of general relativity are presented. 
First, we introduce the $\Lambda$CDM model and the recent cosmological 
observations. At the current stage, the $\Lambda$CDM model is consistent with 
the observational data. 
We then explain a fluid description of dark energy 
and the action representing a scalar field theory. 
In both descriptions of a fluid and a scalar field theory, 
we reconstruct representative cosmologies such as the $\Lambda$CDM, 
quintessence and phantom models. 
Through these procedures, we show the equivalence between 
a fluid description and a scalar field theory. 
We also consider a tachyon scalar field theory. 
Furthermore, we extend the investigations to multiple scalar field theories. 
In addition, we explore holographic dark energy scenarios. 
On the other hand, in the second part, modified gravity models, 
in particular, $F(R)$ gravity as well as $F(R)$ Ho\v{r}ava-Lifshitz gravity 
and $f(T)$ gravity with $T$ being a torsion scalar~\cite{Teleparallelism, Bengochea:2008gz, Linder:2010py}, 
i.e., pictures of geometrical dark energy, are given. 
It is illustrated that by making a conformal transformation, 
an $F(R)$ theory in the Jordan frame can be moved to 
a corresponding scalar field theory in the Einstein frame. 
It is also important to remark that 
as another modified gravitational theory to account for dark energy and 
the late-time cosmic acceleration, $F(R, \mathcal{T})$ theory has been 
proposed in Ref.~\cite{Harko:2011kv}, where $\mathcal{T}$ is the trace of 
the stress-energy tensor. 
We use units of $k_\mathrm{B} = c = \hbar = 1$ and denote the
gravitational constant $8 \pi G$ by
${\kappa}^2 \equiv 8\pi/{M_{\mathrm{Pl}}}^2$
with the Planck mass of $M_{\mathrm{Pl}} = G^{-1/2} = 1.2 \times 10^{19}$GeV. 
Throughout this paper, 
the subscriptions ``DE'', ``m'', and ``r'' represent the quantities of 
dark energy, non-relativistic matter (i.e., cold dark matter and baryons), and 
relativistic matter (e.g., radiation and neutrinos) respectively. 

The review is organized as follows. 
In the first part, 
in Sec.\ II we explain the $\Lambda$CDM model. 
We also present the recent cosmological observational data, 
in particular, in terms of SNe Ia, BAO and CMB radiation, 
by defining the related cosmological quantities. 
These data are consistent with the $\Lambda$CDM model. 

In Sec.\ III, 
we investigate a description of dark fluid universe. 
We represent basic formulations for the EoS of dark energy. 
We also introduce the four types of the finite-time future singularities as well as the energy conditions and give examples of fluid descriptions for 
the $\Lambda$CDM model, 
the GCG model and a model of coupled dark energy with dark matter. 
Next, we explore various phantom cosmologies such as 
a coupled phantom scenario, Little Rip scenario and Pseudo-Rip model. 
Furthermore, we show that the fluid description of the EoS of dark energy 
can yield all the four types of the finite-time future singularities. 
A fluid description with realizing asymptotically de Sitter phantom universe 
is also examined. 
In addition, we investigate the inhomogeneous (imperfect) dark fluid universe. 
We study the inhomogeneous EoS of dark energy and its cosmological effects on 
the structure of the finite-time future singularities. 
Moreover, its generalization of the implicit inhomogeneous EoS is presented. 

In Sec.\ IV, 
we explore scalar field theories in general relativity. 
We explicitly demonstrate the equivalence of fluid descriptions to 
scalar field theories. 
In particular, 
we concretely reconstruct scalar field theories describing 
the $\Lambda$CDM model, the quintessence cosmology, the phantom cosmology 
and a unified scenario of inflation and late-time cosmic 
acceleration. 
We also consider scalar field models with realizing the crossing the phantom divide and its stability problem. 

In Sec.\ V, 
we examine a tachyon scalar field theory. 
We explain the origin, the model action and its stability conditions.

In Sec.\ VI, 
we describe multiple scalar field theories. 
First, we examine two scalar field theories. 
We investigate the standard type of two scalar field theories and 
the stability of the system. 
We then introduce a new type of two scalar field theories, in 
which the crossing of the phantom divide can happen, 
and also explore its stability conditions. 
Next, we extend the considerations for two scalar field theories 
to multiple scalar field theories which consist of more scalars. 
and clearly illustrate those equivalence to fluid descriptions. 

In Sec.\ VII, 
we study holographic dark energy. 
We explain a model of holographic dark energy as well as its generalized 
scenario. In addition, we examine the Hubble entropy in the holographic 
principle. 

In the second part, 
in Sec.\ VIII, 
we consider accelerating cosmology in $F(R)$ gravity. 
First, by using a conformal transformation, we investigate the relations between a scalar field theory in the Einstein frame and an $F(R)$ theory in the Jordan frame. 
Next, we explore the reconstruction method of $F(R)$ gravity. 
We explicitly reconstruct the forms of $F(R)$ with realizing 
the $\Lambda$CDM, quintessence and phantom cosmologies. 
In addition, we study dark energy cosmology in $F(R)$ Ho\v{r}ava-Lifshitz 
gravity. We first present the model action and then reconstruct the $F(R)$ 
forms with performing the $\Lambda$CDM model and the phantom cosmology. 
Furthermore, we explain $F(R, \mathcal{T})$ gravity.

In Sec.\ IX, 
we describe $f(T)$ gravity. 
To begin with, we give fundamental formalism and basic equations. 
We reconstruct a form of $f(T)$ in which the finite-time future singularities 
can occur. We also discuss the removal way of those singularities. 
Furthermore, we represent the reconstructed $f(T)$ models in which  
inflation in the early universe, the $\Lambda$CDM model, the Little Rip scenario and the Pseudo-Rip cosmology are realized. 
In addition, as one of the most important theoretical touch stones 
to examine whether $f(T)$ gravity can be an alternative gravitational theory 
to general relativity, we explore thermodynamics in $f(T)$ gravity. 
We show the first law of thermodynamics and then discuss the second law of 
thermodynamics, and derive the condition for the second law to be satisfied.


Next, in the following sections, we develop the observational investigations 
on dark energy and modified gravity. 
In Sec.\ X, we discuss the basic ideas and concepts of cosmography in order to 
compare concurring models with the Hubble series expansion coming from 
the scale factor. In particular, we derive cosmographic parameters without 
choosing any cosmological model a priori. 

In Sec.\ XI, we examine how it is possible to connect $F(R)$ gravity by 
cosmography and it is possible to reproduce by it the most popular dark 
energy models as so called Chevallier-Polarski-Linder 
(CPL)~\cite{CPL, Linder03}, which is a parameterization of the EoS for dark 
energy, or the $\Lambda$CDM. 

In Sec.\ XII, we show, as examples, how it is possible to constrain $F(R)$ models theoretically. However, the approach works for any dark energy or alternative gravity model. 

In Sec.\ XIII, we discuss constraints coming from observational data. It is 
clear that the quality and the richness of data play a fundamental role in 
this context.


Finally, conclusions with the summary of this review are presented in 
Sec.\ XIV. 

\section{The $\Lambda$ cold dark matter ($\Lambda$CDM) model}

The action of the $\Lambda$CDM model 
in general relativity is described as 
\begin{equation} 
S = 
\int d^4 x \sqrt{-g} 
\frac{1}{2\kappa^2} 
\left( R - 2\Lambda \right) + 
\int d^4 x 
{\mathcal{L}}_{\mathrm{M}} 
\left( g_{\mu\nu}, {\Psi}_{\mathrm{M}} \right)\,,
\label{eq:Add-2-0-1}
\end{equation}
where $R$ is the Ricci scalar, 
$g$ is the determinant of the metric tensor $g_{\mu\nu}$, 
$\Lambda$ is the cosmological constant, and 
${\mathcal{L}}_{\mathrm{M}}$ with a matter field ${\Psi}_{\mathrm{M}}$ 
is the matter Lagrangian. 
The EoS of the cosmological constant, which is the 
ratio of the pressure $P_{\Lambda}$ to the energy density $\rho_{\Lambda}$ 
of the cosmological constant, is given by 
\begin{equation} 
w_{\Lambda} \equiv \frac{P_{\Lambda}}{\rho_{\Lambda}} = 
-1\,. 
\label{eq:II.01}
\end{equation}

We assume the 4-dimensional FLRW metric 
which describes 
the homogeneous and isotropic univese 
\begin{equation}  
ds^2 = - dt^2 + a^2(t) \left[ 
\frac{dr^2}{1-Kr^2} + r^2 d \Omega^2 
\right]\,, 
\label{eq:2.2}
\end{equation}
where $a(t)$ is the scale factor, $K$ is the cosmic curvature 
($K=+1, 0, -1$ denotes closed, flat, and open universe, respectively), and 
$d \Omega^2$ is the metric of 2-dimensional sphere with unit radius. 
The redshift $z$ is defined as $z \equiv a_0/a -1$ with $a_0 = 1$ being 
the current value of the scale factor. 

In the FLRW background~(\ref{eq:2.2}), the Einstein equations are given by
\begin{eqnarray}
H^2 \Eqn{=} 
\frac{\kappa^2}{3} \rho_{\mathrm{M}} + \frac{\Lambda}{3} 
- \frac{K}{a^2}\,,
\label{eq:II.02} \\ 
\dot{H} \Eqn{=} -\frac{\kappa^2}{2} \left( 
\rho_{\mathrm{M}} + P_{\mathrm{M}} \right) + \frac{K}{a^2}\,, 
\label{eq:II.03}
\end{eqnarray}
where 
$H=\dot{a}/a$ is the Hubble parameter and 
the dot denotes the time derivative of $\partial/\partial t$. 
The fractional densities of dark energy, non-relativistic matter, 
radiation and the density parameter of the curvature are defined as 
\begin{equation}
\Omega_{\mathrm{DE}} 
\equiv \frac{\rho_{\mathrm{DE}}}{\rho_{\mathrm{crit}}^{(0)}}\,, 
\quad 
\Omega_{\mathrm{m}} \equiv \frac{\rho_{\mathrm{m}}}{\rho_{\mathrm{crit}}^{(0)}
}\,, 
\quad 
\Omega_{\mathrm{r}} \equiv 
\frac{\rho_{\mathrm{r}}}{\rho_{\mathrm{crit}}^{(0)}}\,, 
\quad 
\Omega_{K} \equiv 
- \frac{K}{\left( aH \right)^2}\,, 
\label{eq:II.04}
\end{equation}
where 
$\rho_{\mathrm{crit}}^{(0)} = 3H_0^2/\kappa^2 
$ is the critical 
density with $H_0$ being the current Hubble parameter. 
By combinning Eq.~(\ref{eq:II.02}) with the quantities in (\ref{eq:II.04}), 
we find 
\begin{equation}
\Omega_{\mathrm{DE}} + 
\Omega_{\mathrm{m}} + 
\Omega_{\mathrm{r}} + 
\Omega_{K} = 1\,, 
\label{eq:II.05}
\end{equation}
%

If there only exists cosmological constant $\Lambda$, i.e., 
$\rho_{\mathrm{M}} = 0$ and $P_{\mathrm{M}} = 0$, from 
Eq.~(\ref{eq:II.03}) we have $H = H_\mathrm{c} = \mathrm{constant}$, so that 
de Sitter expansion can be realized. 
We also note that by comparing Eq.~(\ref{eq:II.02}) with 
Eq.~(\ref{eq:II.01}), we obtain 
$\rho_{\Lambda} = \Lambda/\kappa^2 = - P_{\Lambda}$. 

In addition, the scale factor is expressed as 
\begin{equation}  
a = a_\mathrm{c} \e^{H_\mathrm{c} t}\,, 
\label{eq:II.06}
\end{equation}
where $a_\mathrm{c} (> 0)$ is a positive constant. 

In this section, we present the observational data of SNe Ia, BAO  
and CMB radiation, which supports the $\Lambda$CDM model
(for the way of an analysis of observational data, 
see, e.g.,~\cite{Li:2009jx}).

\subsection{Type Ia Supernovae (SNe Ia)}

With SNe Ia observations, we find the luminosity distance $d_{L}$ 
as a function of the redshift $z$. 
We define the theoretical distance modulus as 
%
$
\mu_{\mathrm{th}}(z_{i})\equiv5\log_{10}D_{L}(z_{i})+\mu_{0}
$. 
%
Here, $D_{L} \equiv H_0 d_{L}$ is the Hubble-free luminosity 
distance and 
$d_{L} \equiv \sqrt{ L_s/\left(4\pi \mathcal{F} \right)}$, where 
$L_s$ and $\mathcal{F}$ are the absolute luminosity of a source 
and an observed flux, respectively, is the luminosity distance. 
Moreover, 
$\mu_{0}\equiv42.38-5\log_{10}h$, where 
$h \equiv H_{0}/100/[\mathrm{km} \, \mathrm{sec}^{-1} \, 
\mathrm{Mpc}^{-1}]$~\cite{Kolb and Turner}. 
We express $D_{L}$ as 
%
\begin{eqnarray}
D_{L}(z) \Eqn{\equiv} H_0 d_{L} 
=\left(1+z\right) f_K (\mathcal{Y})\,,
\label{eq:A.2} \\ 
f_K (\mathcal{Y}) \Eqn{\equiv} \frac{1}{\sqrt{\Omega_{K}^{(0)}}} 
\sinh \left( \sqrt{\Omega_{K}^{(0)}} \mathcal{Y} \right)\,, 
\label{eq:A.2-adding-02} \\ 
\mathcal{Y} \Eqn{\equiv} \int_{0}^{z}\frac{dz'}{E(z')}\,, 
\label{eq:A.2-adding-03} 
\end{eqnarray}
%
%
with
\begin{eqnarray}  
\hspace{-10mm}
&&
E(z) \equiv \frac{H(z)}{H_{0}} 
\label{eq:ED1-19-IIA-Add-1} \\
\hspace{-10mm}
&&
=
\sqrt{
\Omega_{\mathrm{m}}^{(0)}\left(1+z\right)^{3}
+\Omega_{\mathrm{r}}^{(0)}\left(1+z\right)^{4}
+\Omega_{\mathrm{DE}}^{(0)}
\exp \left[ 
\int_{0}^{z}\frac{3\left(1+w_{\mathrm{DE}}\right)}{1+z'}dz' \right]
+ \Omega_{K}^{(0)}\left(1+z\right)^{2}
}\,, 
\label{eq:A.3} 
\end{eqnarray}
%
where 
$\Omega_{\mathrm{r}}^{(0)}=\Omega_{\gamma}^{(0)}
\left(1+0.2271N_{\mathrm{eff}}\right)$ 
with $\Omega_{\gamma}^{(0)}$ being the present fractional photon energy 
density and $N_{\mathrm{eff}}=3.04$ the effective number of neutrino 
species~\cite{Komatsu:2010fb}. 
In what follows, the superscription ``(0)'' represents the values 
at the present time. 
Moreover, in deriving Eq.~(\ref{eq:A.3}) we have used 
the continuity equations 
\begin{equation} 
\dot{\rho_{j}}+ 3H \left( \rho_{j} + P_{j} \right) = 0\,,
\label{eq:A-add-01}
\end{equation}
where $j = $ ``DE'', ``m'' and ``r'', and 
$P_\mathrm{m} = 0$. 
Furthermore, 
$f_K (\mathcal{Y})$ in Eq.~(\ref{eq:A.2-adding-02}) is described by 
\begin{eqnarray}
f_K (\mathcal{Y}) = \begin{cases} 
                    \sin \mathcal{Y}  &\mbox{for $K=+1$\,,} \\
                    \mathcal{Y}       &\mbox{for $K=0$\,,} \\ 
                    \sinh \mathcal{Y} &\mbox{for $K=-1$\,.} 
                    \end{cases}
\label{eq:A-add-02} 
\end{eqnarray}
By using Eqs.~(\ref{eq:A.2}), (\ref{eq:A.3}) and (\ref{eq:A-add-02}), 
for the flat universe, we have 
%
$
D_{L}(z) = \left(1+z\right) \int_{0}^{z} dz'/E(z')
$, 
where 
$E(z)$ in Eq.~(\ref{eq:A.3}) with $\Omega_{K}^{(0)} = 0$. 
%
%
{}From this relation, we find 
%
$
H(z) = \left\{  
\left(d/dz\right) \left[d_L (z)/\left(1+z\right)\right] 
\right\}^{-1}
$. 
Accordingly, 
for $z < \mathcal{O}(1)$,  
the cosmic expansion history can be obtained 
through the measurement of the luminosity distance $d_L (z)$. 
We expand Eq.~(\ref{eq:A.2}) around $z = 0$ as 
\begin{equation}
D_{L}(z) \equiv H_0 d_{L} 
= 
z + \left(1 - \frac{1}{2} \frac{dE(z=0)}{dz} \right)z^2 + \mathcal{O}(z^3) 
= 
z + \frac{1}{4} \left(1-3w_{\mathrm{DE}} \Omega_{\mathrm{DE}}^{(0)} 
+ \Omega_K^{(0)} \right)z^2 + \mathcal{O}(z^3)\,, 
\label{eq:A-add-06} 
\end{equation} 
where in deriving the second equality we have used Eq.~(\ref{eq:A.3}). 
It is clearly seen from Eq.~(\ref{eq:A-add-06}) that 
suppose there exists dark energy, 
$\Omega_{\mathrm{DE}}^{(0)} > 0$ and $w_{\mathrm{DE}} <0$, 
and hence the luminosity distance becomes large. 
We note that 
since the universe is very close to flat as 
$-0.0179 < \Omega_K^{(0)} < 0.0081$ 
(95\% confidence level (CL))~\cite{Komatsu:2008hk}, 
even though the universe is open ($\Omega_K^{(0)} > 0$), 
the change of the luminosity distance is small. 

By applying 
the Seven-Year Wilkinson Microwave Anisotropy Probe (WMAP) 
Observations data~\cite{Komatsu:2010fb}, 
the latest distance measurements from the 
BAO in the distribution of galaxies, 
and the Hubble constant measurement, 
for a flat universe, the current value of a constant EoS 
for dark energy has been estimated as 
$w_{\mathrm{DE}} = -1.10 \pm 0.14 \, 
(68 \% \, \mathrm{CL})$ in Ref.~\cite{Komatsu:2010fb}. 
Moreover, as an example of a time-dependent EoS for dark energy, 
for a linear form $w_{\mathrm{DE}}(a) = w_{\mathrm{DE}\,0} + 
w_{\mathrm{DE}\,a} \left( 1-a \right)$~\cite{CPL, Linder03}, 
where $w_{\mathrm{DE}\,0}$ and $w_{\mathrm{DE}\,a}$ 
are the current value of $w_{\mathrm{DE}}$ and 
its derivative, respectively, 
by using the WMAP data, the BAO data and 
the Hubble constant measurement and the high-redshift 
SNe Ia data, 
$w_{\mathrm{DE}\,0}$ and $w_{\mathrm{DE}\,a}$ 
have been analyzed as 
$w_{\mathrm{DE}\,0} = -0.93 \pm 0.13$ and 
$w_{\mathrm{DE}\,a} = -0.41^{+0.72}_{-0.71} \, (68 \% \, \mathrm{CL})$. 
This form is called as the CPL
model~\cite{CPL, Linder03}. 
Consequently, for the flat universe, 
the various recent observational data are consistent with 
the cosmological constant, i.e., $w_{\mathrm{DE}} = -1$. 

We also mention the way of analyzing the $\chi^{2}$ of the SNe Ia data, 
given by 
%
$
\chi_{\mathrm{SN}}^{2}=\sum_{i}
\left[\mu_{\mathrm{obs}}(z_{i})-
\mu_{\mathrm{th}}(z_{i})\right]^{2}/\sigma_{i}^{2}
$ 
%
with $\mu_{\mathrm{obs}}$ being the observed distance modulus. 
In the following, the subscripts ``th" and ``obs" mean 
the theoretical and observational values, respectively. 
We expand 
$\chi_{\mathrm{SN}}^{2}$ as~\cite{CS-SN} 
%
\begin{eqnarray}
\chi_{\mathrm{SN}}^{2} \Eqn{=} 
A_{\mathrm{SN}}-2\mu_{0}B_{\mathrm{SN}}
+\mu_{0}^{2}C_{\mathrm{SN}}\,,
\label{eq:A.5} \\
%
%
A_{\mathrm{SN}} 
\Eqn{=}  
\sum_{i}\frac{\left[\mu_{\mathrm{obs}}(z_{i})-\mu_{\mathrm{th}}(z_{i};\mu_{0}=0)\right]^{2}}{\sigma_{i}^{2}}\,, 
\label{eq:ED1-19-IIA-Add-2} \\
B_{\mathrm{SN}} 
\Eqn{=} 
\sum_{i}\frac{\mu_{\mathrm{obs}}(z_{i})-\mu_{\mathrm{th}}(z_{i};\mu_{0}=0)}
{\sigma_{i}^{2}}\,, 
\label{eq:ED1-19-IIA-Add-3} \\
C_{\mathrm{SN}} 
\Eqn{=} 
\sum_{i}\frac{1}{\sigma_{i}^{2}}\,.
\label{eq:A.6} 
\end{eqnarray} 
Since we do not know the absolute magnitude of SNe Ia, 
we should minimize 
$\chi_{\mathrm{SN}}^{2}$ with respect to 
$\mu_{0}$ related to the absolute magnitude. 
We describe 
the minimum of $\chi_{\mathrm{SN}}^{2}$ with respect to $\mu_{0}$ as 
$
\tilde{\chi}_{\mathrm{SN}}^{2}=
A_{\mathrm{SN}} - B_{\mathrm{SN}}^{2}/C_{\mathrm{SN}}
$ by using, for example, 
the Supernova Cosmology Project (SCP) Union2 compilation with 
557 supernovae, whose redshift range is $0.015 \leq z \leq 1.4$~\cite{Amanullah:2010vv}.

\subsection{Baryon Acoustic Oscillations (BAO)}

Baryons couple to photons strongly until the decoupling era, 
and therefore we can detect the oscillation of sound waves 
in baryon perturbations. The BAO is a special pattern in the large-scale correlation function of Sloan Digital Sky Survey (SDSS) luminous red galaxies. 
Hence, we can use the BAO data to explore the features of dark energy. 

We measure the distance ratio 
$d_{z} \equiv r_{s}(z_{\mathrm{d}})/D_{V}(z)$
through the observations of the BAO. 
The volume-averaged distance $D_{V}(z)$~\cite{Eisenstein:2005su} 
and the proper angular diameter distance $D_{A}(z)$ 
for the flat universe are defined by 
\begin{eqnarray}
D_{V}(z) \Eqn{\equiv} \left[\left(1+z\right)^{2}
D_{A}^{2}(z)\frac{z}{H(z)}\right]^{1/3}\,,
\label{eq:A.8} \\ 
D_{A}(z) \Eqn{\equiv}
\frac{1}{1+z}\int_{0}^{z}\frac{dz'}{H(z')}\,. 
\label{eq:A.9} 
\end{eqnarray} 
The comoving sound horizon $r_{s}(z)$ is expressed by 
\begin{equation}
r_{s}(z)=\frac{1}{\sqrt{3}}\int_{0}^{1/\left(1+z\right)}\frac{-dz^{\prime}}{H(z^{\prime}) \sqrt{1+\left(3\Omega_{b}^{(0)}/4\Omega_{\gamma}^{(0)}
\right)/\left(1+z^{\prime}\right)}}\,,
\label{eq:A.10} 
\end{equation} 
with $\Omega_{b}^{(0)} = 2.2765 \times10^{-2} h^{-2}$ and 
$\Omega_{\gamma}^{(0)} = 2.469\times10^{-5}h^{-2}$ being the current 
values of baryon and photon density parameters, 
respectively~\cite{Komatsu:2010fb}. 
The fitting formula of the redshift $z_{\mathrm{d}}$ 
at the drag epoch~\cite{Percival:2009xn}, 
when the sound horizon determines 
the location of the BAO because  
baryons are free from the Compton drag of photons, 
is represented as~\cite{Eisenstein:1997ik} 
\begin{eqnarray}
z_{\mathrm{d}} \Eqn{=} 
\frac{1291(\Omega_{\mathrm{m}}^{(0)}h^{2})^{0.251}}
{1+0.659(\Omega_{\mathrm{m}}^{(0)}h^{2})^{0.828}}\left[1+b_{1}
\left(\Omega_{b}^{(0)}h^{2}\right)^{b_{2}}\right]\,,
\label{eq:A.11} \\ 
b_{1} \Eqn{=} 
0.313(\Omega_{\mathrm{m}}^{(0)}h^{2})^{-0.419}\left[
1+0.607\left(\Omega_{\mathrm{m}}^{(0)}h^{2}\right)^{0.674}\right]\,, 
\label{eq:ED1-19-IIA-Add-4} \\
b_{2} \Eqn{=} 
0.238\left(\Omega_{\mathrm{m}}^{(0)}h^{2}\right)^{0.223}\,.
\label{eq:A.12} 
\end{eqnarray} 
For $\Omega_{\mathrm{m}}^{(0)}=0.276$ and $h=0.705$, we have 
$z_{\mathrm{d}} \approx 1021$.

By using the BAO data from 
the Two-Degree Field Galaxy Redshift Survey (2dFGRS)
and the Sloan Digital Sky Survey Data Release 7 
(SDSS DR7)~\cite{Percival:2009xn}, 
we measure 
the distance ratio $d_{z}$ at two redshifts $z=0.2$ and 
$z=0.35$ 
as $d_{z=0.2}^{\mathrm{obs}}=0.1905\pm0.0061$ and 
$d_{z=0.35}^{\mathrm{obs}}=0.1097\pm0.0036$
with the inverse covariance matrix, defined by 
%
\begin{equation}
{\mathcal{M}}_{\mathrm{BAO}}^{-1} \equiv 
\left(
\begin{array}{cc}
30124 & -17227\\
-17227 & 86977
\end{array}\right)\,.
\label{eq:A.13} 
\end{equation} 
%
We express the $\chi^{2}$ of the BAO data as 
$
\chi_{\mathrm{BAO}}^{2}=
\left(x_{i,\mathrm{BAO}}^{\mathrm{th}}-x_{i,\mathrm{BAO}}^{\mathrm{obs}}\right)
\left({\mathcal{M}}_{\mathrm{BAO}}^{-1}\right)_{ij}
\left(x_{j,\mathrm{BAO}}^{\mathrm{th}}-x_{j,\mathrm{BAO}}^{\mathrm{obs}}
\right)
$ 
with $x_{i,\mathrm{BAO}} \equiv \left(d_{0.2},d_{0.35}\right)$.

\subsection{Cosmic Microwave Background (CMB) radiation}

By using the CMB data, we can derive 
the distance to the decoupling epoch $z_{*} 
(\simeq 1090)$~\cite{Komatsu:2008hk}, 
and hence we constrain the model describing the high-$z$ epoch.  
Since the expansion history of the universe from 
the decoupling era to the present time influences on the positions of acoustic 
peaks in the CMB anisotropies, those are shifted provided that there exists dark energy. 

We define 
the angle for the location of the CMB acoustic peaks as 
\begin{equation} 
\theta_A \equiv \frac{r_s (z_{*})}{d_A^{(\mathrm{c})} (z_{*})}\,,
\label{eq:IIC-add-01} 
\end{equation} 
with $d_A^{(\mathrm{c})} = d_L/\left(1+z\right)$ being the comoving angular 
diameter distance. 
The acoustic scale $l_{A}$~\cite{Hu:1994uz, Hu:1995en} representing with the CMB multipole corresponding to $\theta_A$ and 
the shift parameter $\mathcal{R}$~\cite{Bond:1997wr, Efstathiou:1998xx} 
are defined as 
\begin{eqnarray}
l_{A}(z_{*})
\Eqn{\equiv} \frac{\pi}{\theta_A} 
= 
\left(1+z_{*}\right)\frac{\pi D_{A}(z_{*})}{r_{s}(z_{*})}
\label{eq:A.15} \\ 
\Eqn{=} 
\frac{3\pi}{4} \sqrt{\frac{\Omega_{b}^{(0)} h^2}{\Omega_{\gamma}^{(0)} h^2}} 
\left[ 
\ln \left( 
\frac{\sqrt{R_s (a_{\mathrm{*}}) + R_s (a_{\mathrm{eq}})} 
+ \sqrt{1+R_s (a_{\mathrm{*}})}}{1+\sqrt{R_s (a_{\mathrm{eq}})}}
\right) \right]^{-1} \mathcal{R}\,,
\label{eq:A.15-02} \\ 
R_s (a) \Eqn{\equiv} \frac{3 \rho_b}{4 \rho_{\gamma}}
= \left( \frac{3 \Omega_{b}^{(0)}h^{2}}{4 \Omega_{\gamma}^{(0)}h^{2}} \right) a\,, 
\label{eq:IIC-add-02} \\
\mathcal{R}(z_{*})
\Eqn{\equiv}
\sqrt{\frac{\Omega_{\mathrm{m}}^{(0)}}{\Omega_K^{(0)}}} H_{0} 
\sinh \left[ \sqrt{\Omega_K^{(0)}} 
\left(1+z_{*}\right)D_{A}(z_{*}) \right]\,, 
\label{eq:A.16} 
\end{eqnarray}
with $z_{*}$ being the redshift of the decoupling epoch~\cite{Hu:1995en}, 
given by 
\begin{eqnarray}
z_{*} \Eqn{=} 
1048\left[1+0.00124\left(\Omega_{b}^{(0)}h^{2}\right)^{-0.738}\right]\left[1+g_{1}\left(\Omega_{\mathrm{m}}^{(0)}h^{2}\right)^{g_{2}}\right]\,,
\label{eq:A.17} \\
g_{1} \Eqn{=} 
\frac{0.0783\left(\Omega_{b}^{(0)}h^{2}\right)^{-0.238}}{1+39.5\left(\Omega_{b}^{(0)}h^{2}\right)^{0.763}}\,,
\label{eq:ED1-19-IIA-Add-5} \\
g_{2} \Eqn{=} 
\frac{0.560}{1+21.1\left(\Omega_{b}^{(0)}h^{2}\right)^{1.81}}\,. 
\label{eq:A.18} 
\end{eqnarray} 
In Eq.~(\ref{eq:A.15-02}), 
$\rho_b$ and $\rho_{\gamma}$ are the energy density of baryons and 
photons, respectively. 
Furthermore, 
$a_{\mathrm{*}}$ and $a_{\mathrm{eq}}$ are the scale factors at the decoupling epoch and the radiation-matter equality time. 

It is seen from Eq.~(\ref{eq:A.15-02}) that since 
the change of the cosmic expansion history from the decoupling era to the present time influences on the CMB shift parameter, 
the multipole $l_A$ is shifted. 
The relation between all peaks and 
troughs of the observed CMB anisotropies is represented by~\cite{Doran:2001yw} 
\begin{equation}  
l_n = l_A \left( n - \phi_n \right)\,, 
\label{eq:IIC-add-03} 
\end{equation} 
where $n$ denotes peak numbers and $\phi_n$ is the shift of multipoles. 
For instance, for the first peak $n=1$ and for the first trough $n=1.5$. 
Moreover, 
the WMAP 5-year results~\cite{Komatsu:2008hk} present 
the limit on the shift parameter of 
$\mathcal{R} = 1.710 \pm 0.019$ (68 \% CL). 
In the flat universe, we have 
$\mathcal{R}(z_{*}) = 
\sqrt{\Omega_{\mathrm{m}}^{(0)}}H_{0}
\left(1+z_{*}\right)D_{A}(z_{*})$. 
Hence, the smaller $\Omega_{\mathrm{m}}^{(0)}$ is, namely, 
the larger $\Omega_{\mathrm{DE}}^{(0)}$ is, 
the smaller $\mathcal{R}$ is. 
For an estimation executed by the WMAP 5-year data analysis~\cite{Komatsu:2008hk} of $\mathcal{R} = 1.710$, 
$\Omega_{b}^{(0)} h^2 = 0.02265$, 
$\Omega_{\mathrm{m}}^{(0)} h^2 = 0.1369$ and 
$\Omega_{\gamma}^{(0)} h^2 = 2.469 \times 10^{-5}$, 
with Eq.~(\ref{eq:A.15-02}) we obtain $l_A = 299$. 
By plugging this value and $\phi_1 = 0.265$~\cite{Doran:2001yw} 
into Eq.~(\ref{eq:IIC-add-03}), we find 
the first acoustic peak $l_1 = 220$, which is consistent with 
the CMB anisotropies observation. 

The data from Seven-Year Wilkinson Microwave Anisotropy Probe (WMAP)
observations~\cite{Komatsu:2010fb} on CMB can be used. 
The $\chi^{2}$ of the CMB data is expressed as 
$
\chi_{\mathrm{CMB}}^{2}=\left(x_{i,\mathrm{CMB}}^{\mathrm{th}}-x_{i,\mathrm{CMB}}^{\mathrm{obs}}\right)
\left({\mathcal{M}}_{\mathrm{CMB}}^{-1}\right)_{ij}
\left(x_{j,\mathrm{CMB}}^{\mathrm{th}}-x_{j,\mathrm{CMB}}^{\mathrm{obs}}
\right)
$
with $x_{i,\mathrm{CMB}}\equiv\left(l_{A}(z_{*}), 
\mathcal{R}(z_{*}), z_{*}\right)$
and ${\mathcal{M}}_{\mathrm{CMB}}^{-1}$ being the inverse covariance matrix. 
It follows from the WMAP7 data analysis~\cite{Komatsu:2010fb} 
that $l_{A}(z_{*})=302.09$, 
$\mathcal{R}(z_{*})=1.725$, 
$z_{*}=1091.3$, 
and 
\begin{equation}
{\mathcal{M}}_{\mathrm{CMB}}^{-1}=\left(\begin{array}{ccc}
2.305 & 29.698 & -1.333\\
29.698 & 6825.27 & -113.180\\
-1.333 & -113.180 & 3.414\end{array}\right)\,.
\label{eq:A.20} 
\end{equation} 

As a result, we find the $\chi_{\mathrm{total}}^{2}$ of all the observational data 
$
\chi_{\mathrm{total}}^{2}=\tilde{\chi}_{\mathrm{SN}}^{2}+\chi_{\mathrm{BAO}}^{2}+\chi_{\mathrm{CMB}}^{2}
$. 
It is known that 
there exists a fitting procedure called 
the Markov chain Monte Carlo (MCMC) approach, e.g., 
CosmoMC~\cite{Lewis:2002ah}. 

Finally, we mention an issue of the origin of the cosmological constant 
$\Lambda$, which is one of the most possible candidates of dark energy 
because its presence is consistent with a number of observations. 
In fact, however, the origin is not well understood 
yet~\cite{Weinberg:1988cp}. 
The vacuum energy density $\rho_{\mathrm{v}}$ is given by 
%
\begin{eqnarray}
\rho_{\mathrm{v}} 
\Eqn{=} \int_0^{k_{\mathrm{c}}} \frac{d^3 k}{
\left( 2\pi \right)^3} \frac{\sqrt{k^2 + m^2}}{2} 
\label{eq:IIC-add-04} \\ 
\Eqn{\approx} 
\frac{4\pi k^2 dk}{\left( 2\pi \right)^3} \frac{k}{2} 
= \frac{k_{\mathrm{c}}^4}{16\pi^2}\,. 
\label{eq:IIC-add-05}
\end{eqnarray}
%
In deriving Eq.~(\ref{eq:IIC-add-04}), 
we have used the zero-point energy of a field with its mass $m$, 
momentum $k$ and frequency $\omega$, $E = \omega/2 = \sqrt{k^2 + m^2}/2$, 
and $k_{\mathrm{c}} (\ll m)$ is a cut-off scale. 
The vacuum energy density is 
$\rho_{\mathrm{v}} \simeq 10^{74}\, \mathrm{GeV}^4$, provided that $k_{\mathrm{c}} = M_{\mathrm{Pl}}$. 
On the other hand, the current observed value of the energy density of 
dark energy is $\rho_{\mathrm{DE}} \simeq 10^{-47}\, \mathrm{GeV}^4$. 
Thus, the discrepancy between the theoretical estimation and observed value 
of the current energy density of dark energy is as large as $10^{121}$. 
This is one of the most difficult problems in terms of the cosmological 
constant. There also exists another problem why the ``present'' value of 
the energy density of vacuum in our universe is extremely small compared with 
its theoretical prediction. 

\section{Cosmology of dark fluid universe}

In this section, we present a description of dark fluid 
universe~\cite{Nojiri:2005sx, Nojiri:2005sr, Stefancic:2004kb}. 
We concentrate on the case in which 
there is only single fluid which corresponds to dark energy 
in general relativity.

\subsection{Basic equations} 

We represent 
the equation of state (EoS) of dark energy as 
\begin{equation} 
w_{\mathrm{DE}} \equiv \frac{P}{\rho} = 
-1 - \frac{f(\rho)}{\rho}\,,
\label{eq:2.20}
\end{equation}
with 
\begin{equation}  
P=-\rho -f(\rho)\,,
\label{eq:2.21}
\end{equation}
where $f(\rho)$ can be an arbitrary function of $\rho$. 
In what follows, for the time being, 
we will investigate the evolution of the universe at the dark energy dominated 
stage, so that 
we can regard $\rho = \rho_{\mathrm{DE}}$ and 
$P = P_{\mathrm{DE}}$. For simplicity, the ``DE'' subscription will be omitted 
as long as there is no need to mention. 
We also remark that $f(\rho)$ characterizes a deviation from 
the $\Lambda$CDM model. 

We assume the flat 
FLRW metric 
$ds^2 = - dt^2 + a^2(t) \sum_{i=1,2,3}\left(dx^i\right)^2$, 
which is equivalent to Eq.~(\ref{eq:2.2}) with $K=0$. 
%
%
In this 
background, 
the Einstein equations are given by
\begin{eqnarray}
H^2 \Eqn{=} \frac{\kappa^2}{3} \rho\,,
\label{eq:Add-2-01} \\ 
\dot{H} \Eqn{=} -\frac{\kappa^2}{2} \left( 
\rho + P \right)\,. 
\label{eq:Add-2-02}
\end{eqnarray}
%
Equations~(\ref{eq:Add-2-01}) and (\ref{eq:Add-2-02}) correspond to 
Eqs.~(\ref{eq:2.3}) and (\ref{eq:2.4}) 
with $\rho_{\phi} = \rho$, $P_{\phi} = P$, $\rho_{\mathrm{M}} = 0$, and 
$P_{\mathrm{M}} = 0$ (i.e., the case in which the scalar field $\phi$ is 
responsible for dark energy and there is no any other matter) 
shown in Sec.~II A, respectively. 
Furthermore, 
the continuity equation of the fluid is given by 
\begin{equation} 
\dot{\rho}+ 3H \left( \rho + P \right) = 0\,. 
\label{eq:Add-2-03} 
\end{equation}
%
{}From Eq.~(\ref{eq:Add-2-03}), we find that the scale factor is described as 
\begin{equation}  
a = a_{\mathrm{c}} \exp \left( \frac{1}{3} \int \frac{d \rho}{f(\rho)} 
\right)\,, 
\label{eq:Add-2-04} 
\end{equation}
where $a_{\mathrm{c}}$ is a constant. 
In addition, by combining Eqs.~(\ref{eq:Add-2-01}) and (\ref{eq:Add-2-03}) 
with Eq.~(\ref{eq:2.2}), we obtain 
\begin{equation} 
t = \int \frac{d \rho}{\kappa \sqrt{3\rho} f(\rho)}\,. 
\label{eq:Add-2-05} 
\end{equation}
Eqs.~(\ref{eq:Add-2-01}) and (\ref{eq:Add-2-02}) lead to 
\begin{equation} 
\frac{\ddot{a}}{a} = -\frac{\kappa^2}{6} \left( \rho + 3P \right) 
= \frac{\kappa^2}{6} \left( 2\rho + 3f(\rho) \right)\,. 
\label{eq:Add-2-06} 
\end{equation}
%

\subsection{EoS of dark energy in various cosmological models}

Provided that there exists only single fluid of dark energy, 
(i) for the $\Lambda$CDM model, $w_{\mathrm{DE}} = -1$ ($f(\rho) =0$), 
(ii) for a quintessence model, $-1 < w_{\mathrm{DE}} < -1/3$ 
($-2/3 < f(\rho)/\rho <0$), 
(iii) for a phantom model, $w_{\mathrm{DE}} < -1$ 
($f(\rho)/\rho >0$). 
If our universe lies beyond $w_{\mathrm{DE}} = -1$ region, 
then its future can be really dark. 
In other words, in finite-time phantom/quintessence universe 
may enter a future singularity.

\subsubsection{Finite-time future singularities and energy conditions}

In the FLRW background~(\ref{eq:2.2}), 
the effective EoS 
for the universe is given by~\cite{Review-Nojiri-Odintsov} 
\begin{equation} 
w_{\mathrm{eff}} \equiv \frac{P_{\mathrm{eff}}}{\rho_{\mathrm{eff}}} = 
-1 - \frac{2\dot{H}}{3H^2}\,, 
\label{eq:2.17}
\end{equation}
%
where 
$\rho_{\mathrm{eff}} \equiv 3H^2/\kappa^2$ 
and $P_{\mathrm{eff}} \equiv -\left(2\dot{H}+3H^2\right)/\kappa^2 $ correspond 
to the total energy density and pressure of the universe, respectively. 
When the energy density of dark energy becomes perfectly dominant over 
that of matter, 
one obtains $w_{\mathrm{DE}} \approx w_{\mathrm{eff}}$. 
%
%
%
For $\dot{H} < 0\ (>0)$, $w_\mathrm{eff} >-1\ (<-1)$, representing the non-phantom [i.e., quintessence] (phantom) phase, 
whereas 
$w_\mathrm{eff} =-1$ for $\dot{H} = 0$, corresponding to the 
cosmological constant. 
It is not clear from the very beginning how our universe evolves, 
and if it ends up in a singularity. 
This should always be checked. 

The finite-time future singularities can be classified 
into the following four types~\cite{Nojiri:2005sx}: 

%
{\large $\ast$}
Type I (``Big Rip''~\cite{Elizalde:2005ju, Big-Rip}) singularity:\ 
In the limit of $t\to t_{\mathrm{s}}$, 
all the scale factor, the effective energy density and pressure of the 
universe diverge as 
$a \to \infty$,
$\rho_{\mathrm{eff}} \to \infty$ and
$\abs{ P_{\mathrm{eff}} } \to \infty$. 
This also includes 
the case that 
$\rho_\mathrm{{eff}}$ and $P_{\mathrm{eff}}$ asymptotically approach finite 
values at $t = t_{\mathrm{s}}$. 

%
{\large $\ast$}
Type II (``sudden''~\cite{Barrow:2004xh, sudden}) singularity:\ 
In the limit of $t\to t_{\mathrm{s}}$, 
only the effective pressure of the universe becomes infinity as 
$a \to a_{\mathrm{s}}$, 
$\rho_{\mathrm{eff}} \to \rho_{\mathrm{s}}$ and 
$\abs{ P_{\mathrm{eff}} } \to \infty$. 

%
{\large $\ast$}
Type III singularity:\ 
In the limit of $t\to t_{\mathrm{s}}$, 
the effective energy density as well as the pressure of the 
universe diverge as 
$a \to a_{\mathrm{s}}$, 
$\rho_{\mathrm{eff}} \to \infty$ and
$\abs{ P_{\mathrm{eff}} } \to \infty$. 

%
{\large $\ast$} 
Type IV:\ 
In the limit of $t\to t_{\mathrm{s}}$, 
all the scale factor, the effective energy density and pressure of the 
universe do not diverge as 
$a \to a_{\mathrm{s}}$, 
$\rho_{\mathrm{eff}} \to 0$ 
and $\abs{ P_{\mathrm{eff}} } \to 0$. 
However, higher derivatives of $H$ become infinity.  
This also includes 
the case that $\rho_{\mathrm{eff}}$ and/or $\abs{ P_{\mathrm{eff}} }$ 
become finite values at $t = t_{\mathrm{s}}$~\cite{Shtanov:2002ek}. 
%

Here, $t_{\mathrm{s}}$, $a_{\mathrm{s}} (\neq 0)$ and $\rho_{\mathrm{s}}$ 
are constants. 
In Ref.~\cite{FS-F(R)-gravity}, 
the finite-time future singularities 
in $F(R)$ gravity have first been observed. 
Furthermore, 
the finite-time future singularities in various modified gravity 
theories have also been studied in Refs.~\cite{Nojiri:2008fk, Future-singularity-MG}. In particular, it has recently been demonstrated that 
the finite-time future singularities can occur in the framework of non-local gravity~\cite{Non-local-gravity-DW-NO} in Ref.~\cite{Bamba:2012ky} as well as in $f(T)$ gravity in Ref.~\cite{Bamba:2012vg}. 
Also, various studies on the finite-time future singularities have 
recently been executed, e.g., in Ref.~\cite{Recent-Studies-FTFS}. 

Moreover, 
there exist four energy conditions. 
\begin{list}{}{}
\item[(a)] 
The null energy condition (NEC): 
%
\begin{equation} 
\rho + P \geq 0\,. 
\label{eq:Add-2-B-1}
\end{equation}
\item[(b)] 
The dominant energy condition (DEC): 
%
\begin{equation} 
\rho \geq 0\,, 
\quad 
\rho \pm P \geq 0\,. 
\label{eq:Add-2-B-2-001}
\end{equation}
\item[(c)] 
The strong energy condition (SEC): 
%
\begin{equation} 
\rho + 3P \geq 0\,, 
\quad 
\rho + P \geq 0\,. 
\label{eq:Add-2-B-2}
\end{equation}
\item[(d)] 
The weak energy condition (WEC): 
%
\begin{equation} 
\rho \geq 0\,, 
\quad 
\rho + P \geq 0\,. 
\label{eq:Add-2-B-2-002}
\end{equation}
\end{list}
%
If $w_{\mathrm{DE}} = P/\rho < -1$, the Type I (Big Rip) singularity 
appears within a finite time. In such a case, 
all energy conditions are violated. 
On the other hand, 
even when 
the strong energy condition in Eq.~(\ref{eq:Add-2-B-2}) is 
satisfied, 
the Type II (sudden) singularity can appear~\cite{Barrow:2004xh}. 
We remark that 
if the Type I (Big Rip) singularity exists, 
all the four conditions 
are broken, 
whereas 
if there exist the Type II, III or IV singularity, 
only a part of four conditions is broken. 
Hence, it is expected that the origin of 
the finite-time future singularities 
is somehow related with the violation of all or a part of 
energy conditions. 

We mention that 
``$w$'' singularity has been studied in Refs.~\cite{Kiefer:2010zzb, Dabrowski:2009zzb, Dabrowski:2009pc} and 
parallel-propagated (p.p.) curvature 
singularities~\cite{L-Fernandez-Jambrina-PRD} have earlier been investigated. 
For the ``$w$'' singularity, 
when $t\to t_{\mathrm{s}}$, 
$a \to a_{\mathrm{s}}$, 
$\rho_{\mathrm{eff}} \to 0$, 
$\abs{ P_{\mathrm{eff}} } \to 0$, 
and the EoS for the universe becomes infinity.

\subsubsection{Example of a fluid with behavior very similar 
to the $\Lambda$CDM model}

To begin with, 
we investigate a fluid which behaves very similar to the $\Lambda$CDM model 
\begin{equation} 
f(\rho) = \rho^q -1\,, 
\label{eq:IIIB-add-01}
\end{equation}
where $q$ is a constant ($|q| \ll 1$). 
In this case, from Eq.~(\ref{eq:2.20}) we see that $\left|w_{\mathrm{DE}} -1 
\right| 
\ll 0$ and therefore this fluid model can be regarded as almost the 
$\Lambda$CDM model. 
{}For this fluid, by substituting Eq.~(\ref{eq:IIIB-add-01}) into 
Eqs.~(\ref{eq:Add-2-04}) and (\ref{eq:Add-2-05}), we obtain 
%
\begin{eqnarray}
a \Eqn{\approx} a_{\mathrm{c}} 
\left| \frac{\rho - 1}{\rho + 1} \right|^{1/\left( 6q \right)}\,, 
\label{eq:IIIB-add-02} \\ 
t \Eqn{\approx} \frac{1}{2\sqrt{3} \kappa p} 
\ln \left( \left| \rho - 1 \right| \left| \rho + 1 \right|^{\sqrt{-1}} 
\right)\,, 
\label{eq:IIIB-add-03}
\end{eqnarray}
%
where in deriving the approximate equalities we have used $\abs{ q } \ll 1$. 


\subsubsection{Generalized Chaplygin gas (GCG) model} 

Next, 
we examine the generalized Chaplygin gas (GCG) model proposed in 
Refs.~\cite{Kamenshchik:2001cp, Bento:2002ps}. 
This model is a proposal to explain the origin of dark energy as well as 
dark matter through a single fluid. 

\begin{equation} 
P = -\frac{\mathcal{A}}{\rho^u}\,, 
\label{eq:3-C-1} 
\end{equation}
where $\mathcal{A} (>0)$ is a positive constant and 
$u$ is a constant. 
For $u=1$, 
Eq.~(\ref{eq:3-C-1}) is for 
the original Chaplygin gas model~\cite{Kamenshchik:2001cp}. 


By combining Eq.~(\ref{eq:3-C-1}) and 
the continuity equation~(\ref{eq:Add-2-03}), we find 
\begin{equation} 
\rho = \left[\mathcal{A} + \frac{\mathcal{B}}{a^{3\left(1+u\right)}} 
\right]^{1/\left(1+u\right)}\,, 
\label{eq:IIIB-add-04} \\ 
\end{equation}
where $\mathcal{B}$ is an integration constant. 
It follows from Eq.~(\ref{eq:IIIB-add-04}) that the asymptotic behaviors 
in the early universe $a \ll 1$ as well as the late universe $a \gg 1$ 
are described as 
\begin{eqnarray}
\rho \Eqn{\sim} 
\mathcal{B} a^{-3}\,, 
\quad \mathrm{for} \,\, a \ll 1\,, 
\label{eq:IIIB-add-05} \\ 
\rho \Eqn{\sim} 
\mathcal{A}^{1/\left(1+u\right)}\,, 
\quad \mathrm{for} \,\, a \gg 1\,. 
\label{eq:IIIB-add-06} 
\end{eqnarray}
%
In the early universe, the energy density evolves $\rho \propto a^{-3}$ and 
hence this fluid behaves as non-relativistic matter, i.e., dark matter, 
whereas in the late universe, $\rho$ approaches a constant value of 
$\mathcal{A}^{1/\left(1+u\right)}$ and therefore it can correspond to 
dark energy. 
As a result, 
the generalized Chaplygin gas model 
can account for the origin of both dark matter and dark energy. 

It is known that 
the generalized Chaplygin gas model is very close to 
the $\Lambda$CDM model by analyzing the gauge-invariant matter perturbations 
$\delta_{\mathrm{M}}$ in Refs.~\cite{Carturan:2002si, Sandvik:2002jz}, 
which satisfy~\cite{Sandvik:2002jz}
\begin{equation} 
\ddot{\delta_{\mathrm{M}}} + \left(2+3c_{\mathrm{s}}^2 -6 w_{\mathrm{DE}} 
\right) H \dot{\delta_{\mathrm{M}}} - 
\left[
\frac{3}{2} \left(1-3c_{\mathrm{s}}^2 -3 w_{\mathrm{DE}}^2 +8 w_{\mathrm{DE}} 
\right) H^2 -\left(\frac{c_{\mathrm{s}} k}{a}\right)^2 
\right] \delta_{\mathrm{M}} = 0\,, 
\label{eq:IIIB-add-07} 
\end{equation}
where $k$ is a comoving wavenumber and $w_{\mathrm{DE}} = P/\rho$ with 
Eq.~(\ref{eq:3-C-1}) is the EoS of the generalized Chaplygin gas. 
In addition, $c_{\mathrm{s}}$ is the sound speed, defined by 
\begin{equation} 
c_{\mathrm{s}}^2 \equiv \frac{d P}{d \rho} = - u w_{\mathrm{DE}}\,.  
\label{eq:IIIB-add-08} 
\end{equation}
When $z \gg 1$, both $w_{\mathrm{DE}}$ and $c_{\mathrm{s}}$ approach to zero. 
Thus, at the matter-dominated stage $c_{\mathrm{s}} \ll 1$, and 
after it $c_{\mathrm{s}}$ grows. 
If $u >0 (< 0)$, $c_{\mathrm{s}}^2 >0 (< 0)$ because $w_{\mathrm{DE}} < 0$. 

It follows from Eq.~(\ref{eq:IIIB-add-07}) that $\delta_{\mathrm{M}}$ grows 
due to the gravitational instability, provided that 
\begin{equation} 
\left| c_{\mathrm{s}}^2 \right| < 
\frac{3}{2} \left(\frac{aH}{k}\right)^2\,. 
\label{eq:IIIB-add-09} 
\end{equation}
On the other hand, if $\left| c_{\mathrm{s}}^2 \right| > \left(3/2\right) 
\left(aH/k\right)^2$, the rapid growth or damped oscillation of 
$\delta_{\mathrm{M}}$ occurs, depending on the sign of $c_{\mathrm{s}}^2$. 
Around the present time when $\left| w_{\mathrm{DE}} \right| = \mathcal{O} 
(1)$ and thus $\left| c_{\mathrm{s}}^2 \right| \sim |u|$. 
The relation~(\ref{eq:IIIB-add-09}) imposes 
the following constraint on $u$~\cite{Sandvik:2002jz}: 
\begin{equation} 
|u| \lesssim 10^{-5}\,.  
\label{eq:IIIB-add-10} 
\end{equation}
Since the case of $u = 0$ is equivalent to the $\Lambda$CDM model, 
the viable generalized Chaplygin gas model would be very close to 
the $\Lambda$CDM model.

\subsubsection{Coupled dark energy with dark matter}

Related to the generalized Chaplygin gas model, in which 
dark energy and dark matter are interpolated each other, 
we explain a coupled dark energy with dark matter 
and its cosmological consequences (for recent discussions, see~\cite{Nojiri:2009pf, BB-CDE}). 

The equations corresponding to the conservation law are written as 
\begin{eqnarray} 
&&
\dot{\rho}_{\mathrm{DE}} + 3H \left( 1+w_{\mathrm{DE}} \right) 
\rho_{\mathrm{DE}} = -Q \rho_{\mathrm{DE}}\,, 
\label{eq:IIIB3-add-001} \\
&&
\dot{\rho}_{\mathrm{m}} + 3H \left( \rho_{\mathrm{m}} + P_{\mathrm{m}} \right)  = +Q \rho_{\mathrm{DE}}\,, 
\label{eq:IIIB3-add-002}
\end{eqnarray}
where $w_{\mathrm{DE}} \equiv P_{\mathrm{DE}}/\rho_{\mathrm{DE}}$. 
Here, 
$Q$ describes a coupling between dark energy and dark matter, 
which is assumed to be a constant. 
We note that the subscription ``m'' represents quantities of cold dark matter 
(CDM), and therefore $P_{\mathrm{m}} = 0$. 
It is important to emphasize that the continuity equation for the 
total energy density and pressure of dark energy and dark matter, which is 
the summation of Eqs.~(\ref{eq:IIIB3-add-001}) and (\ref{eq:IIIB3-add-002}), 
is satisfied. 

The solutions for Eqs.~(\ref{eq:IIIB3-add-001}) and (\ref{eq:IIIB3-add-002}) 
are given by 
\begin{eqnarray} 
\rho_{\mathrm{DE}} \Eqn{=} 
\rho_{\mathrm{DE} (0)} a^{-3\left( 1+w_{\mathrm{DE}} \right)} 
\e^{-Qt}\,, 
\label{eq:IIIB3-add-003} \\
\rho_{\mathrm{m}} \Eqn{=} 
Q a^{-3} 
\int^{t} d \tilde{t} \rho_{\mathrm{DE} (0)} 
a^{-3w_{\mathrm{DE}}} (\tilde{t}) \e^{-Q \tilde{t}}\,, 
\label{eq:IIIB3-add-004}
\end{eqnarray}
where $\rho_{\mathrm{DE} (0)}$ is an integration constant. 

In the flat FLRW background~(\ref{eq:2.2}), from 
Eqs.~(\ref{eq:Add-2-01}) and (\ref{eq:Add-2-01}) 
the Einstein equations are expressed as 
\begin{eqnarray}
\frac{3}{\kappa^2} H^2 \Eqn{=} \rho_{\mathrm{DE}} + \rho_{\mathrm{m}}\,,
\label{eq:IIIB3-add-005} \\ 
-\frac{1}{\kappa^2} \left( 2\dot{H} + 3H^2 \right) 
\Eqn{=} 
P_{\mathrm{DE}}
\label{eq:IIIB3-add-006} \\
\Eqn{=} 
w_{\mathrm{DE}} \rho_{\mathrm{DE}} 
= w_{\mathrm{DE}} \rho_{\mathrm{DE} (0)} 
a^{-3\left( 1+w_{\mathrm{DE}} \right)} \e^{-Qt}\,, 
\label{eq:IIIB3-add-007}
\end{eqnarray}
where in deriving the second equality in Eq.~(\ref{eq:IIIB3-add-007}) 
we have used Eq.~(\ref{eq:IIIB3-add-003}). 
We can obtain an exact de Sitter solution of Eq.~(\ref{eq:IIIB3-add-007}) as 
\begin{equation} 
a = a_0 \e^{H t}\,, 
\label{eq:IIIB3-add-008} 
\end{equation}
with 
\begin{equation}  
H = -\frac{Q}{3\left( 1+w_{\mathrm{DE}} \right)}\,. 
\label{eq:IIIB3-add-009} 
\end{equation}
Here, $a_0$ is a constant, and obeys 
\begin{equation} 
-\frac{3}{\kappa^2} 
\left[ \frac{Q}{3\left( 1+w_{\mathrm{DE}} \right)} \right]^2 
= w_{\mathrm{DE}} \rho_{\mathrm{DE} (0)} 
a_0^{-3\left( 1+w_{\mathrm{DE}} \right)}\,. 
\label{eq:IIIB3-add-010} 
\end{equation}
If the universe is in the phantom phase, i.e., 
$w_{\mathrm{DE}} < -1$, 
$H$ in Eq.~(\ref{eq:IIIB3-add-009}) is positive. 
Hence, Eq.~(\ref{eq:IIIB3-add-010}) has a real solution, so that 
$a_0$ can be a real number. 

On the other hand, 
when only dark energy exists and therefore there is no direct coupling of dark 
energy with dark matter, i.e., $\rho_{\mathrm{m}} = 0$ and $Q = 0$, 
from Eqs.~(\ref{eq:IIIB3-add-001}) and (\ref{eq:IIIB3-add-005}) we find 
$\dot{\rho}_{\mathrm{DE}} + 3H \left( 1+w_{\mathrm{DE}} \right) 
\rho_{\mathrm{DE}} = 0$
and 
$\left(3/\kappa^2\right) H^2 = \rho_{\mathrm{DE}}$, 
respectively. 
These equations can have an expression of $H$ as 
\begin{equation}  
H = -\frac{2/\left[ 3\left( 1+w_{\mathrm{DE}} \right) 
\right]}{t_{\mathrm{s}} -t}\,, 
\label{eq:IIIB3-add-011} 
\end{equation}
which leads to a Big Rip singularity at $t = t_{\mathrm{s}}$. 
Thus, one cosmological consequence of a coupling between dark energy and 
dark matter is that 
a de Sitter solution could be realized, and not a Big Rip singularity. 

Another cosmological consequence of a coupling between dark energy and 
dark matter is to present a solution for the so-called coincidence problem, 
i.e., the reason why the current energy density of dark matter is almost the 
same order of that of dark energy. 
$H$ in Eq.~(\ref{eq:IIIB3-add-009}) may be taken as the present 
value of the Hubble parameter, which is given by  
$H_{\mathrm{p}} = 2.1 h \times 10^{-42} \mathrm{GeV}$~\cite{Kolb and Turner}
with $h = 0.7$~\cite{Komatsu:2010fb, Freedman}. 
In this case, from Eqs.~(\ref{eq:IIIB3-add-003}) and (\ref{eq:IIIB3-add-008}) 
we see that the energy density of dark energy becomes constant 
\begin{equation} 
\rho_{\mathrm{DE}} = 
\rho_{\mathrm{DE} (0)} a_0^{-3\left( 1+w_{\mathrm{DE}} \right)}\,. 
\label{eq:IIIB3-add-012} 
\end{equation}
By substituting 
Eq.~(\ref{eq:IIIB3-add-012}) into Eq.~(\ref{eq:IIIB3-add-004}), we acquire 
\begin{equation}  
\rho_{\mathrm{m}} = \rho_{\mathrm{m}(0)} a^{-3} 
-\left( 1+w_{\mathrm{DE}} \right) 
\rho_{\mathrm{DE} (0)} a_0^{-3\left( 1+w_{\mathrm{DE}} \right)}\,, 
\label{eq:IIIB3-add-013} 
\end{equation}
where $\rho_{\mathrm{m}(0)}$ is an integration constant. 
By plugging Eq.~(\ref{eq:IIIB3-add-013}) into Eq.~(\ref{eq:IIIB3-add-005}) and 
using Eq.~(\ref{eq:IIIB3-add-010}), we find that $\rho_{\mathrm{m}(0)} = 0$. 
Hence, the energy density of dark matter is also constant as 
\begin{equation}  
\rho_{\mathrm{m}} = 
-\left( 1+w_{\mathrm{DE}} \right) 
\rho_{\mathrm{DE} (0)} a_0^{-3\left( 1+w_{\mathrm{DE}} \right)} 
= -\left( 1+w_{\mathrm{DE}} \right) \rho_{\mathrm{DE}}\,, 
\label{eq:IIIB3-add-014} 
\end{equation}
where the second equality follows from Eq.~(\ref{eq:IIIB3-add-012}). 
Thus, provided that the de Sitter solution in Eq.~(\ref{eq:IIIB3-add-008}) 
is an attractor one, by taking 
\begin{equation} 
\frac{\rho_{\mathrm{m}}}{\rho_{\mathrm{DE}}} 
= -\left( 1+w_{\mathrm{DE}} \right) 
\sim \frac{1}{3}\,, 
\label{eq:IIIB3-add-015} 
\end{equation}
from which we have 
\begin{equation} 
w_{\mathrm{DE}} \sim -\frac{4}{3}\,, 
\label{eq:IIIB3-add-016} 
\end{equation}
the coincidence problem could be resolved. 
Of course, this is rather qualitative presentation.

\subsection{Phantom scenarios} 

According to the present observations, there is the possibility that 
the EoS of dark energy $w_{\mathrm{DE}}$ would be less than $-1$. 
This is called the ``phantom phase''.  
It is known that 
if the phantom phase is described by 
a scalar field theory with the negative kinetic 
energy, which is the phantom model, the phantom field rolls up the 
potential due to the negative kinetic energy. 
For a potential unbounded from above, the energy density becomes infinity 
and eventually a Big Rip singularity appears.

\subsubsection{Phantom phase} 

In order to illustrate the phantom phase, 
we present a model in which the universe evolves from the non-phantom 
(quintessence) phase ($w_{\mathrm{DE}} > -1$) to 
the phantom phase ($w_{\mathrm{DE}} < -1$), 
namely, crossing of the phantom divide line of $w_{\mathrm{DE}} = -1$ 
occurs~\cite{Alam:2003fg}. 

The scale factor is expressed as 
\begin{equation}  
a = a_{\mathrm{c}} \left( \frac{t}{t_{\mathrm{s}} -t} \right)^n\,, 
\label{eq:III-P-p-1} 
\end{equation}
where $a_{\mathrm{c}}$ is a constant, $n (>0)$ is a positive constant, 
$t_{\mathrm{s}}$ is the time when a finite-time future singularity 
(a Big Rip singularity) appears, and we consider the period 
$0 < t < t_{\mathrm{s}}$. 
In this model, 
for $t \ll t_{\mathrm{s}}$, $a(t)$ behaves as $a \approx t^n$ and hence 
$w_{\mathrm{DE}} = -1 + 2/\left(3n\right) > -1$, 
whereas 
for $t \sim t_{\mathrm{s}}$, 
$w_{\mathrm{DE}} = -1 - 2/\left(3n\right) < -1$. 

It follows from Eq.~(\ref{eq:III-P-p-1}) that the Hubble parameter $H$ is 
given by 
\begin{equation}  
H = n \left( \frac{1}{t} + \frac{1}{t_{\mathrm{s}} -t} \right)\,. 
\label{eq:III-P-p-2} 
\end{equation}
By combining Eq.~(\ref{eq:Add-2-01}) with Eq.~(\ref{eq:III-P-p-2}), we find 
\begin{equation}  
\rho = \frac{3n^2}{\kappa^2} \left( \frac{1}{t} + \frac{1}{t_{\mathrm{s}} -t} 
\right)^2\,.
\label{eq:III-P-p-3} 
\end{equation}
$H$ as well as $\rho$ becomes minimum at $t = t_{\mathrm{s}}/2$ and 
those values are 
\begin{eqnarray}  
H_{\mathrm{min}} 
\Eqn{=} 
\frac{4n}{t_{\mathrm{s}}}\,, 
\label{eq:III-P-p-4} \\
\rho_{\mathrm{min}} 
\Eqn{=} 
\frac{48n^2}{\kappa^2 t_{\mathrm{s}}^2}\,.
\label{eq:III-P-p-5} 
\end{eqnarray}

Moreover, $\dot{\rho}$ is written as 
\begin{equation}  
\dot{\rho} = \pm 2 \rho \sqrt{ \frac{\kappa^2 \rho}{3n^2} - 
\frac{4}{n t_{\mathrm{s}}} \sqrt{\frac{\kappa^2 \rho}{3}} 
}\,, 
\label{eq:III-P-p-6} 
\end{equation}
where we have removed $t$ by using the relation between 
$\rho$ and $t$ in Eq.~(\ref{eq:III-P-p-3}). 
Here, the plus (minus) sign denotes the expression of $\dot{\rho}$ 
for $t > t_{\mathrm{s}}/2$ ($0 < t < t_{\mathrm{s}}/2$), i.e., 
the phantom (non-phantom) phase. This implies that 
the energy density of dark energy increases in the phantom phase. 
Substituting Eqs.~(\ref{eq:2.21}) and (\ref{eq:III-P-p-6}) into 
the continuity equation of dark energy (\ref{eq:Add-2-03}), 
we obtain 
\begin{equation}  
f(\rho) = \pm \frac{2 \rho}{3n} \sqrt{ 1 - 
\frac{4n}{t_{\mathrm{s}}} \sqrt{\frac{3}{\kappa^2 \rho}} 
}\,. 
\label{eq:III-P-p-7} 
\end{equation}

As a consequence, it is necessary for the EoS to be doubled valued, 
so that the transition from the non-phantom phase to the phantom one 
can occur. 
Furthermore, we see that $f(\rho_{\mathrm{min}}) = 0$. This means that 
at the phantom crossing point $w_{\mathrm{DE}} = -1$, 
both $H$ and $\rho$ have those minima, 
which can also be understood from the definition of $w_{\mathrm{DE}}$ in 
Eq.~(\ref{eq:2.20}). 
In addition, 
when a Big Rip singularity appears at $t = t_{\mathrm{s}}$, 
{}from Eq.~(\ref{eq:III-P-p-7}) we find that 
$f(\rho)$ evolves as $f(\rho) = 2 \rho/\left(3n\right)$, and 
by using this relation and Eqs.~(\ref{eq:2.21}) as well as (\ref{eq:2.20}) 
we see that 
$P = -\rho -2 \rho/\left(3n\right)$ and 
$w_{\mathrm{DE}} = -1 - 2/\left(3n\right)$, which is a constant. 
On the other hand, in the opposite limit of $t \to 0$, we also have 
a constant EoS as $w_{\mathrm{DE}} = -1 + 2/\left(3n\right)$. 

When the crossing of the phantom divide occurs, $f(\rho) = 0$. 
In order to realize this situation, 
the integration part of $\int d \rho/f(\rho)$ on the right-hand side (r.h.s.) 
of Eq.~(\ref{eq:Add-2-04}) should be finite. 
Thus, $f(\rho)$ need to behave as 
\begin{equation}  
f(\rho) \sim f_{\mathrm{c}} \left( \rho - \rho_{\mathrm{c}} 
\right)^{s}\,, 
\quad 
0 < s < 1\,, 
\label{eq:III-P-p-8} 
\end{equation}
where we have used the condition $f(\rho_{\mathrm{c}}) =0$. 
In general, it is necessary for $f(\rho)$ to be multi-valued around 
$\rho = \rho_{\mathrm{c}}$ because $0 < s < 1$.

\subsubsection{Coupled phantom scenario} 

Next, we explore another phantom scenario in which dark energy is coupled with 
dark matter. 
We suppose that dark energy can be regarded as a fluid satisfying 
Eq.~(\ref{eq:IIIB3-add-001}). 
Since the coupling $Q$ can be described by 
a function of $a$, $H$, $\rho_{\mathrm{m}}$, $\dot{\rho}_{\mathrm{m}}$, 
$\rho_{\mathrm{DE}}$ and $\dot{\rho}_{\mathrm{DE}}$, 
we consider the case in which the ratio of $\rho_{\mathrm{m}}$ to 
$\rho_{\mathrm{DE}}$ defined as 
$r \equiv \rho_{\mathrm{m}}/\rho_{\mathrm{DE}}$ is a constant, 
that is, $Q$ is represented by scaling solutions. 
In this case, $\rho_{\mathrm{m}}$ does not close to zero asymptotically. 
By using Eqs.~(\ref{eq:IIIB3-add-001}) and (\ref{eq:IIIB3-add-002}), 
we obtain 
\begin{equation} 
\dot{r} = r \left[ Q \left( \frac{1}{\rho_{\mathrm{m}}} + 
\frac{1}{\rho_{\mathrm{DE}}} \right) -3H \left( w_{\mathrm{m}} 
- w_{\mathrm{DE}} \right)
\right]\,, 
\label{eq:III-C-P-1} 
\end{equation}
where $w_{\mathrm{m}} \equiv P_{\mathrm{m}} / \rho_{\mathrm{m}}$. 
By taking $\dot{r} = 0$ in Eq.~(\ref{eq:III-C-P-1}), we acquire 
$Q$ in the presence of scaling solutions as 
\begin{equation} 
Q = 3H \left( w_{\mathrm{m}} - w_{\mathrm{DE}} \right) 
\frac{\rho_{\mathrm{DE}} \rho_{\mathrm{m}}}{\rho_{\mathrm{DE}} + 
\rho_{\mathrm{m}}}\,. 
\label{eq:III-C-P-2} 
\end{equation}

As a model which can be treated analytically, we examine $Q$ given by 
\begin{equation} 
Q = \delta H^2\,, 
\label{eq:III-C-P-3} 
\end{equation}
with $\delta$ being a constant. 
Here, cold dark matter is regarded as a dust, and hence $P_{\mathrm{m}} = 0$. 
Moreover, we provide that $f(\rho_{\mathrm{DE}}) = \rho_{\mathrm{DE}}$ in 
Eq.~(\ref{eq:2.21}), i.e., 
\begin{equation} 
P_{\mathrm{DE}} = -2 \rho_{\mathrm{DE}}\,. 
\label{eq:III-C-P-4} 
\end{equation}

By combining Eq.~(\ref{eq:IIIB3-add-005}) with Eqs.~(\ref{eq:IIIB3-add-001}) 
and (\ref{eq:IIIB3-add-002}), a solution is given by 
\begin{eqnarray}
H \Eqn{=} \frac{2}{3} \left( \frac{1}{t} + \frac{1}{t_{\mathrm{s}}- t} 
\right)\,,
\label{eq:III-C-P-5} \\ 
\rho_{\mathrm{m}} \Eqn{=} 
\frac{4}{3\kappa^2} \left( \frac{1}{t} + \frac{1}{t_{\mathrm{s}}- t} 
\right) \frac{1}{t}\,, 
\label{eq:III-C-P-6} \\ 
\rho_{\mathrm{DE}} \Eqn{=} 
\frac{4}{3\kappa^2} \left( \frac{1}{t} + \frac{1}{t_{\mathrm{s}}- t} 
\right) \frac{1}{t_{\mathrm{s}}- t}\,, 
\label{eq:III-C-P-7} 
\end{eqnarray}
with 
\begin{equation} 
t_{\mathrm{s}} \equiv \frac{9}{\delta \kappa^2}\,. 
\label{eq:III-C-P-8} 
\end{equation}

We note that 
the same solution can be derived if $Q$ is described as 
\begin{equation}  
Q = \frac{9H \rho_{\mathrm{DE}} \rho_{\mathrm{m}}}{2 \left( 
\rho_{\mathrm{DE}} + \rho_{\mathrm{m}} \right)}\,. 
\label{eq:III-C-P-8} 
\end{equation}
The expression of $H$ in 
Eq.~(\ref{eq:III-C-P-5}) is the same as that in Eq.~(\ref{eq:III-P-p-2}) with 
$n=2/3$, and thus a Big Rip singularity appears at $t = t_{\mathrm{s}}$. 
Incidentally, in this case $t_{\mathrm{s}}$ is not determined. 
We also mention that 
Eqs.~(\ref{eq:III-C-P-6}) and (\ref{eq:III-C-P-7}) lead to 
$r = \left( t_{\mathrm{s}} - t \right)/t$, which implies that 
this ratio is not a constant but changes dynamically and eventually becomes 
zero, and 
hence the solutions in Eqs.~(\ref{eq:III-C-P-5})--(\ref{eq:III-C-P-7}) 
do not correspond to scaling solutions. 
This is reasonable because around a Big Rip singularity, the energy density of 
dark energy becomes dominant over that of dark matter completely.

\subsubsection{Little Rip scenario}

We study a Little Rip scenario~\cite{Frampton:2011sp, Brevik:2011mm, 
Frampton:2011rh, Nojiri:2011kd, Astashenok:2012tv, GL-IT, Ito:2011ae, BBEO-XZL-MKO}, 
which corresponds to a mild phantom scenario. 
The Little Rip scenario has been proposed to avoid 
the finite-time future singularities, in particular, a Big Rip singularity within fluid dark energy. 
In this scenario, the energy density of dark energy increases in time with 
$w_{\mathrm{DE}}$ being less than $-1$ and then $w_{\mathrm{DE}}$ 
asymptotically approaches $w_{\mathrm{DE}} = -1$. 
However, its evolution 
eventually leads to a dissolution of 
bound structures at some time in the future. This process is called 
the ``Little Rip''.  

A sufficient condition in order to avoid a Big Rip singularity is that 
$a(t)$ should be a nonsingular function for all $t$. 
We suppose $a(t)$ is expressed as 
\begin{equation} 
a(t) = \e^{\tilde{f}(t)}\,, 
\label{eq:III-L-R-add-01} 
\end{equation}
where $\tilde{f}(t)$ is a nonsingular function. 
It follows from Eq.~(\ref{eq:Add-2-01}) that $\rho = 3 
\left(\dot{a}/a\right)^2 = 3 \dot{\tilde{f}}^2$. 
The condition for $\rho$ to be an increasing function of $a$ is that 
$d \rho/d a = \left(6/\dot{a}\right) \dot{\tilde{f}} \ddot{\tilde{f}} >0$. 
This condition can be met provided 
\begin{equation} 
\ddot{\tilde{f}} >0\,. 
\label{eq:III-L-R-add-02} 
\end{equation}
Thus, 
in all Little Rip scenarios, $a(t)$ is described by 
Eq.~(\ref{eq:Add-2-01}) with $\tilde{f}$ satisfying 
Eq.~(\ref{eq:III-L-R-add-02}). 

We derive the expression for $\rho$ as an increasing function of the scale 
factor $a$ and examine the upper and lower bounds on the growth rate of 
$\rho (a)$ which can be used to judge whether a Big Rig singularity appears. 
By defining $N \equiv \ln a$, Eq.~(\ref{eq:Add-2-01}) is rewritten 
to~\cite{Frampton:2011sp} 
\begin{equation} 
t = \int \sqrt{\frac{3}{\rho (N)}} d N\,. 
\label{eq:III-L-R-1} 
\end{equation}
The condition for the appearance of a finite-time future (Big Rig) singularity to be avoided is that it takes a Big Rig singularity infinite time to appear. 
This means that 
\begin{equation} 
\int_{N_{\mathrm{c}}}^{\infty} \frac{1}{\sqrt{\rho (N)}} d N \to \infty\,, 
\label{eq:III-L-R-2} 
\end{equation}
where $N_{\mathrm{c}} = \ln a_{\mathrm{c}}$. 
For the case $P=-\rho -A \rho^{1/2}$ with $A$ being a constant, 
we have 
\begin{equation}
\frac{\rho}{\rho_1} = \left[ \frac{3A}{2\sqrt{\rho_1}} 
\ln \left( \frac{a}{a_{\mathrm{c}}} \right) +1 \right]^2\,, 
\label{eq:III-L-R-3} 
\end{equation}
where $\rho_1$ is a constant and $\rho = \rho_1$ and 
$a = a_{\mathrm{c}}$ at a fixed time $t_1$. 
In this case, $A > 0$
is required, so that there can be exist 
the phantom phase as $w_{\mathrm{DE}} < -1$. 
Moreover, 
the expression of $\rho$ as a function of $t$ is given by 
\begin{equation}
\frac{\rho}{\rho_1} = \e^{\sqrt{3}A \left(t-t_1\right)}\,. 
\label{eq:III-L-R-4} 
\end{equation}
Furthermore, by using 
$w_{\mathrm{DE}} = -1 -A \rho^{-1/2}$ and Eq.~(\ref{eq:III-L-R-3}) 
we acquire 
\begin{equation} 
w_{\mathrm{DE}} = -1 -\left[ \frac{3}{2} 
\ln \left( \frac{a}{a_{\mathrm{c}}} \right) 
+ \frac{\rho_1}{A} \right]^{-1}\,, 
\label{eq:III-L-R-5} 
\end{equation}
which can also be derived from the relation 
$\left(a/\rho\right) \left(d \rho/d a\right) 
= -3 \left( 1+w_{\mathrm{DE}} \right)$. 
In addition, by eliminating $\rho$ from Eqs.~(\ref{eq:III-L-R-3}) 
and (\ref{eq:III-L-R-4}) we find 
\begin{equation}
\frac{a}{a_{\mathrm{c}}} = 
\exp \left\{ 
\frac{2\sqrt{\rho_1}}{3A} \left[ 
\e^{\sqrt{3}A/2 \left(t-t_1\right) -1} 
\right]
\right\}\,. 
\label{eq:III-L-R-6} 
\end{equation}
In this expression, by comparing Eq.~(\ref{eq:III-L-R-6}) with 
Eq.~(\ref{eq:III-L-R-add-01}), we obtain 
\begin{equation}
\tilde{f} 
= 2\sqrt{\rho_1}{3A} \left[ 
\e^{\sqrt{3}A/2 \left(t-t_1\right) -1} 
\right] + \ln a_{\mathrm{c}}\,. 
\label{eq:III-L-R-7} 
\end{equation}
In this case, 
$\ddot{\tilde{f}} = \left(\sqrt{\rho_1}A/2\right) 
\e^{\left(\sqrt{3}A/2\right) \left(t-t_1\right)} >0$ because of $A>0$, 
and hence the condition in Eq.~(\ref{eq:III-L-R-add-02}) is satisfied. 
We mention that the single model parameter $A$ satisfies the observational bounds estimated by 
using the Supernova Cosmology Project~\cite{Amanullah:2010vv}. 
The best fit value is given by $A = 3.46 \times 10^{-3}\, 
\mathrm{Gyr}^{-1}$ and the range for 95 \% Confidence Level fit is 
$-2.74 \times 10^{-3}\, \mathrm{Gyr}^{-1} \leq A \leq 9.67  \times 10^{-3}\, 
\mathrm{Gyr}^{-1}$~\cite{Frampton:2011sp}. 

{}From Eqs.~(\ref{eq:III-L-R-5}) and (\ref{eq:III-L-R-6}), we see that 
when $t \sim t_1$, $a \sim a_{\mathrm{c}}$ and $w_{\mathrm{DE}} < -1$, i.e., 
the universe is in the phantom phase. 
As the universe evolves, $a$ increases, and 
when $t \ll t_1$, $a$ becomes very large 
and thus $w_{\mathrm{DE}}$ asymptotically becomes close to $-1$. 
However, 
it takes $a$ as well as $\rho$ infinite time to diverge 
due to Eq.~(\ref{eq:III-L-R-2}), a Big Rip singularity cannot appear 
at a finite time in the future. This means that a Big Rip singularity can be 
avoided. 

As further recent observations on Little Rip cosmology, 
in Ref.~\cite{Brevik:2011mm} 
it has been demonstrated that a Little Rip scenario 
can be realized by viscous fluid. 
On the other hand, in Ref.~\cite{Frampton:2011rh} 
a new interpretation of a Little Rip scenario 
by means of an inertial force has been presented. 
It has been shown that a coupling of dark energy with dark matter can 
eliminate a little rip singularity and an asymptotic de Sitter space-time can appear. Moreover, a scalar field theory with realizing a little rip scenario has been reconstructed. 

We investigate the inertial force $F_{\mathrm{inert}}$ on 
a particle with mass $m$ in the context of the expanding universe. 
When the distance between two points is $l$, by using the scale factor 
$a$, the relative acceleration between those is described as 
$l \ddot{a}/a$. 
If a particle with mass $m$ exists at each of the points, 
an inertial force on the other mass would be measured by 
an observer at one of the masses. 
Thus, $F_{\mathrm{inert}}$ is expressed 
as~\cite{Frampton:2011sp, Frampton:2011rh}  
\begin{equation}
F_{\mathrm{inert}} = ml \frac{\ddot{a}}{a} 
= ml \left( \dot{H} + H^2 \right)\,, 
\label{eq:FT4-6-IVC4-add-006} 
\end{equation}
where in deriving the second equality in Eq.~(\ref{eq:FT4-6-IVC4-add-006}) 
we have used Eqs.~(\ref{eq:Add-2-01}) and (\ref{eq:Add-2-02}). 
If $F_{\mathrm{inert}} > 0$ and $F_{\mathrm{inert}} > F_{\mathrm{b}}$, 
where $F_{\mathrm{b}}$ is a constant force and bounds the two particles, 
the two particles becomes free, i.e., unbound, so that 
the bound structure can be dissociated. 

If the Hubble parameter is expressed as 
\begin{equation} 
H \sim \frac{h_{\mathrm{s}}}{ \left( t_{\mathrm{s}} - t 
\right)^{\tilde{q}}}\,, 
\label{eq:ED1-9-Add-IIIC3-01}
\end{equation}
where $h_{\mathrm{s}} (> 0)$ is a positive constant and 
$\tilde{q} (\geq 1)$ is larger than or equal to unity. 
In this case, a Big Rip singularity appears in the limit of 
$t \to t_{\mathrm{s}}$. 
By substituting Eq.~(\ref{eq:ED1-9-Add-IIIC3-01}) into 
Eq.~(\ref{eq:FT4-6-IVC4-add-006}), we obtain~\cite{Bamba:2012vg}
\begin{equation}
F_{\mathrm{inert}} 
= ml h_{\mathrm{s}} \left[ \frac{\tilde{q}}{\left( t_{\mathrm{s}} - t \right)^{\tilde{q}+1}} 
+ \frac{h_{\mathrm{s}}}{\left( t_{\mathrm{s}} - t \right)^{-2\tilde{q}}} 
\right] 
\longrightarrow \infty\,, 
\quad 
\mathrm{when} \,\,\,
t \to t_{\mathrm{s}}\,. 
\label{eq:FT4-6-IVC4-add-008} 
\end{equation}
Since in the limit of $t \to t_{\mathrm{s}}$, $H$ and $\dot{H}$ diverge, 
$F_{\mathrm{inert}}$ becomes infinity. 
The important point is that for a Big Rip singularity, $F_{\mathrm{inert}}$ 
diverges in the ``finite'' future time. 

On the other hand, 
a representation of the Hubble parameter realizing Little Rip scenario 
is described by~\cite{Frampton:2011rh} 
\begin{equation}
H = H_{\mathrm{LR}} \exp \left( \xi t \right)\,. 
\label{eq:FT4-6-IVC3-01}
\end{equation}
Here, $H_{\mathrm{LR}} (>0)$ and $\xi (>0)$ are positive constants. 
By plugging Eq.~(\ref{eq:FT4-6-IVC3-01}) into 
Eq.~(\ref{eq:FT4-6-IVC4-add-006}), we find~\cite{Bamba:2012vg}
\begin{equation}
F_{\mathrm{inert}} 
= ml H_{\mathrm{LR}} \left[ \xi + H_{\mathrm{LR}} \exp \left( \xi t \right) 
\right] \exp \left( \xi t \right)
\longrightarrow \infty\,, 
\quad 
\mathrm{when} \,\,\,
t \to \infty\,. 
\label{eq:FT4-6-IVC4-add-009} 
\end{equation}
In the limit of $t \to \infty$, 
the divergence of $H$ and $\dot{H}$ leads to the consequence of 
$F_{\mathrm{inert}} \to \infty$. 
Thus, a ``Rip'' phenomenon happens in both cases of a Big Rip singularity 
and Little Rip scenario. 
It is remarkable to note that for Little Rip cosmology, 
it takes $F_{\mathrm{inert}}$ 
the ``infinite'' future time to become infinity. 
This feature is different from that in case of 
a Big Rip singularity, which is the finite time future singularity. 

In Little Rip cosmology, as a demonstration 
we estimate the value of $F_{\mathrm{inert}}$ at the 
present time $t_0 \approx H_0^{-1}$, 
where 
$H_{0} = 2.1 h \times 10^{-42} \, \mathrm{GeV}$~\cite{Kolb and Turner}
with $h = 0.7$~\cite{Komatsu:2010fb, Freedman} 
is the present value of the Hubble parameter. 
The bound force $F_{\mathrm{b}}^{\mathrm{ES}}$ for the Earth-Sun system 
is given by $F_{\mathrm{b}}^{\mathrm{ES}} = G M_{\oplus} M_{\odot} / 
r_{\oplus-\odot}^2 = 4.37 \times 10^{16} \, 
\mathrm{GeV}^2$, 
where 
$M_{\oplus} = 3.357 \times 10^{51} \, 
\mathrm{GeV}$~\cite{Kolb and Turner} is the Earth mass and $M_{\odot} = 1.116 \times 10^{57} \, \mathrm{GeV}$~\cite{Kolb and Turner} is the mass of Sun, 
and $r_{\oplus-\odot} = 1 \mathrm{AU} = 7.5812 \times 10^{26} 
\, \mathrm{GeV}^{-1}$~\cite{Kolb and Turner} is the Astronomical unit 
corresponding to the distance between Earth and Sun. 
We take $\xi = H_0$, 
$m = M_{\oplus}$ and 
$l = r_{\oplus-\odot}$. 
In this case, from Eq.~(\ref{eq:FT4-6-IVC4-add-009}) we acquire 
$F_{\mathrm{inert}} = 2.545 \times 10^{78} e H_0^2 
\left[ 
\left( H_{\mathrm{LR}}/H_0 \right) + e \left( H_{\mathrm{LR}}/H_0 \right)^2 
\right]$, where $e = 2.71828$. 
In order for the current value of $F_{\mathrm{inert}}$ to be larger than or 
equal to $F_{\mathrm{b}}^{\mathrm{ES}}$, $H_{\mathrm{LR}}$ should be 
$H_{\mathrm{LR}} \geq 4.82 \times 10^{-30} \, \mathrm{GeV}$.

\subsubsection{Pseudo-Rip cosmology}

In addition, 
as an intermediate cosmology between 
the $\Lambda$CDM model, namely, the cosmological constant, 
and the Little Rip scenario, 
more recently the Pseudo-Rip model has been proposed in Ref.~\cite{Frampton:2011aa}. 
In this case, in the limit of $t \to \infty$
the Hubble parameter tends to a constant asymptotically. 
In other words, the Pseudo-Rip cosmology is a phantom scenario and 
has a feature of asymptotically de Sitter universe. 

A description realizing Pseudo-Rip cosmology is given by~\cite{Bamba:2012vg}
\begin{equation}
H(t) = H_{\mathrm{PR}} \tanh \left( \frac{t}{t_0} \right)\,. 
\label{eq:FT4-6-IVC4-add-005}
\end{equation}
Here, $H_{\mathrm{PR}} (>0)$ is a positive constant. 
It follows from Eq.~(\ref{eq:FT4-6-IVC4-add-005}) that we find 
\begin{equation}
a = a_{\mathrm{PR}} \cosh \left( \frac{t}{t_0} \right)\,, 
\label{eq:FT4-8-IVC4-2nd-adding-01}
\end{equation}
with $a_{\mathrm{PR}} (>0)$ being  a positive constant. 
In the limit of $t \to \infty$, 
$H(t) \to H_{\mathrm{PR}} < \infty$ and $H(t)$ increases monotonically 
in time. Hence, the universe evolves in the phantom phase and eventually 
goes to de Sitter space-time asymptotically. 
By combining Eqs.~(\ref{eq:FT4-6-IVC4-add-005}) and (\ref{eq:2.20}), 
we obtain 
\begin{equation}
w_{\mathrm{DE}} = 
-1 -\frac{2}{3 t_0 H_{\mathrm{PR}}} \frac{1}{\sinh^2 \left( t/t_0 \right)}\,.
\label{eq:FT4-8-IVC4-addition-01} 
\end{equation}
This implies that since $\dot{H}(t) = 
H_{\mathrm{PR}}/\left[ t_0 \cosh^2 \left( t/t_0 \right) \right] > 0$, 
the universe always evolves within the phantom phase 
as $w_{\mathrm{DE}} < -1$, 
and eventually, when $t \to \infty$, 
$w_{\mathrm{DE}} \to -1$, similarly to that in the Little Rip scenario. 
By comparing the observational present value of $w_{\mathrm{DE}}$ as 
$w_{\mathrm{DE}} = -1.10 \pm 0.14 \, 
(68 \% \, \mathrm{CL})$~\cite{Komatsu:2010fb} 
with $w_{\mathrm{DE}}$ in Eq.~(\ref{eq:FT4-8-IVC4-addition-01}) 
at $t = t_0 \approx H_0^{-1}$, 
we obtain the constraint 
$H_{\mathrm{PR}} \geq 
2.96 \times 10^{-42} \, \mathrm{GeV}$.  

Furthermore, by using Eq.~(\ref{eq:FT4-6-IVC4-add-006}) and (\ref{eq:FT4-6-IVC4-add-005}), we acquire~\cite{Bamba:2012vg} 
\begin{equation}
F_{\mathrm{inert}} 
= ml H_{\mathrm{PR}} \left[ 
\frac{1}{t_0 \cosh^2 \left( t/t_0 \right)} + H_{\mathrm{PR}} 
\tanh^2 \left( \frac{t}{t_0} \right)
\right] 
\longrightarrow 
F_{\mathrm{inert}\,,\infty}^{\mathrm{PR}}  
< \infty\,, 
\quad 
\mathrm{when} \,\,\, 
t \to \infty\,, 
\label{eq:FT4-6-IVC4-add-010} 
\end{equation}
with 
\begin{equation}
F_{\mathrm{inert}\,,\infty}^{\mathrm{PR}} \equiv ml H_{\mathrm{PR}}^2\,. 
\label{eq:FT4-6-IVC4-add-011} 
\end{equation}
This means that in Pseudo-Rip cosmology, 
$F_{\mathrm{inert}}$ becomes finite asymptotically 
in the limit of $t \to \infty$. 
This originates from the fact that when $t \to \infty$, 
$H \to H_{\mathrm{PR}}$ and $\dot{H} \to 0$. 
To realize the Pseudo-Rip scenario, the relation 
$F_{\mathrm{inert}\,,\infty}^{\mathrm{PR}} > F_{\mathrm{b}}^{\mathrm{ES}}$ 
should be satisfied, so that the ES system could be disintegrated 
much before the universe asymptotically goes to de Sitter space-time. 
For $m = M_{\oplus}$ and $l = r_{\oplus-\odot}$, 
we find $H_{\mathrm{PR}} > 
\sqrt{G M_{\odot} / r_{\oplus-\odot}^3} = 1.31 \times 10^{-31}\, 
\mathrm{GeV}$. This constraint is much stronger than that obtained from 
the present value of $w_{\mathrm{DE}}$ as 
$H_{\mathrm{PR}} \geq 2.96 \times 10^{-42} \, \mathrm{GeV}$ shown above. 
In Appendix A, it is examined how strong the inertial force can constrain 
the EoS of dark energy.

\subsection{Finite-time future singularities}

In the $\Lambda$CDM model, $f(\rho) = 0$ in Eq.~(\ref{eq:2.20}) 
and hence $w_{\mathrm{DE}} = -1$. 
In a quintessence model, 
the type II, III and IV singularities can occur. 
On the other hand, 
in a phantom model, 
the singularities of Type I and type II can appear. 
We explicitly demonstrate that the EoS in Eq.~(\ref{eq:2.20}) 
can lead to all the four types of the finite-time future singularities.

\subsubsection{Type I and III singularities} 

We examine the case in which $f(\rho)$ is given by
\begin{equation}  
f(\rho) = A \rho^{\alpha}\,,
\label{eq:Add-2-B-1-01}
\end{equation}
where $A$ and $\alpha$ are constants. 
By using Eq.~(\ref{eq:Add-2-04}), 
we obtain 
\begin{equation}  
a = a_{\mathrm{c}} \exp \left[ 
\frac{\rho^{1-\alpha}}{3\left(1-\alpha\right)A}  
\right]\,. 
\label{eq:Add-2-B-1-02} 
\end{equation}
For $\alpha > 1$, when $\rho \to \infty$, $a$ approaches a finite value, 
whereas, 
for $\alpha < 1$, when $\rho \to \infty$, if $A>0$ $(A<0)$, $a$ diverges 
(vanishes). 
We note that in this sub-subsection (i.e., Sec.~III D 1), we concentrate on 
the limit of $\rho \to \infty$\footnote{It has also been examined in 
Ref.~\cite{Stefancic:2004kb} that for $\alpha < 0$, when $\rho \to 0$, 
there can appear the Type II singularity.}. 
We also mention the case in the opposite 
limit of $\rho \to 0$ in Sec.~III E 2. 

Moreover, from Eq.~(\ref{eq:Add-2-05}) we have 
\begin{eqnarray}
t \Eqn{=} t_{\mathrm{s}} + \frac{2}{\sqrt{3}\kappa A} 
\frac{\rho^{-\alpha+1/2}}{1-2\alpha}\,, 
\quad 
\mathrm{for} \,\,\, \alpha \neq \frac{1}{2}\,, 
\label{eq:Add-2-B-1-03} \\
t \Eqn{=} t_{\mathrm{s}} + \frac{\ln \rho}{\sqrt{3}\kappa A}\,, 
\quad 
\mathrm{for} \,\,\, \alpha = \frac{1}{2}\,.  
\label{eq:Add-2-B-1-04}
\end{eqnarray} 
For $\alpha \geq 1/2$, $\rho$ can diverge within the finite future or past 
$t = t_{\mathrm{s}}$. 
Meanwhile, 
for $\alpha \leq 1/2$, it takes infinite time for $\rho$ to diverge 
and therefore $\rho \to \infty$ in the infinite future or past. 

It follows from Eq.~(\ref{eq:2.21}) with Eq.~(\ref{eq:Add-2-B-1-01}) that 
$P=-\rho -A \rho^{\alpha}$. When $\rho$ diverges, $P$ also becomes infinity. 
By substituting Eq.~(\ref{eq:Add-2-B-1-01}) into Eq.~(\ref{eq:2.20}), we find 
\begin{equation} 
w_{\mathrm{DE}} = \frac{P}{\rho} = 
-1 -A\rho^{\alpha - 1}\,. 
\label{eq:Add-2-B-1-05}
\end{equation}
For $\alpha > 1$, when $\rho \to \infty$, if $A>0$ $(A<0)$, 
$w_{\mathrm{DE}} \to +\infty$ ($w_{\mathrm{DE}} \to -\infty$). 
On the other hand, 
for $\alpha < 1$, when $\rho \to \infty$, if $A>0$ $(A<0)$, 
$w_{\mathrm{DE}} \to -1+0$ ($w_{\mathrm{DE}} \to -1-0$), 
i.e., $w_{\mathrm{DE}}$ approaches $-1$. 

The above considerations are summarized as follows. 
\begin{list}{}{}
\item[(a)] 
For $\alpha > 1$, 
the Type III singularity can exist. 
If $A>0$ $(A<0)$, 
$w_{\mathrm{DE}} \to +\infty$ ($w_{\mathrm{DE}} \to -\infty$). 
\item[(b)] 
For $1/2 < \alpha < 1$, if $A>0$, 
there can exist the Type I (Big Rip) singularity. 
If $A<0$, $\rho \to \infty$, $a \to 0$. 
Such a singularity in the past (future) may be called a Big Bang (Big Crunch) 
singularity. 
When $\rho \to \infty$, if $A>0$ $(A<0)$, 
$w_{\mathrm{DE}} \to -1+0$ ($w_{\mathrm{DE}} \to -1-0$). 
\item[(c)] 
For $0 < \alpha \leq 1/2$, 
there does not exist any ``finite-time'' future singularity. 
\end{list}
%

\subsubsection{Type II singularity} 

Next, we investigate the following form of $f(\rho)$: 
\begin{equation}  
f(\rho) = \frac{C}{\left(\rho_{\mathrm{c}} -\rho \right)^{\gamma}}\,,
\label{eq:Add-2-B-2-01}
\end{equation}
where $C$, $\rho_{\mathrm{c}}$ and $\gamma (>0)$ are constants. 
We note that for $\rho_{\mathrm{c}} =0$, $(-1)^{-\gamma} C = A$, 
$\gamma = -\alpha$, the expression in Eq.~(\ref{eq:Add-2-B-2-01}) 
is equivalent to that in Eq.~(\ref{eq:Add-2-B-1-01}). 
We concentrate on the case in which $\rho < \rho_{\mathrm{c}}$. 
Since $P=-\rho-C/\left(\rho_{\mathrm{c}} -\rho \right)^{\gamma}$, 
when $\rho \to \rho_{\mathrm{c}}$, $P$ diverges. Hence, 
$R = 2\kappa^2 \left(\rho -3P \right)$ also becomes infinite. 
{}From Eq.~(\ref{eq:Add-2-04}), we obtain 
\begin{equation}  
a = a_{\mathrm{c}} \exp \left[ 
-\frac{\left(\rho_{\mathrm{c}} -\rho \right)^{\gamma +1}}{3C\left(
\gamma +1\right)} \right]\,. 
\label{eq:Add-2-B-2-02} 
\end{equation}
Thus, when $\rho \to \rho_{\mathrm{c}}$, $a$ is finite as 
$a \to a_{\mathrm{c}}$ and hence $\dot{a}$ is also finite because
$H = \dot{a}/a \propto \sqrt{\rho}$ given by Eq.~(\ref{eq:Add-2-01}), 
whereas $\ddot{a}$ diverges. 
Furthermore, by using Eq.~(\ref{eq:Add-2-05}), we find
\begin{equation}
t \simeq t_{\mathrm{s}} -\frac{\left(\rho_{\mathrm{c}} -\rho \right)^{\gamma +1}}{\kappa C \sqrt{3\rho_{\mathrm{c}}} \left(\gamma +1\right)}\,, 
\label{eq:Add-2-B-2-03}
\end{equation}
where $t_{\mathrm{s}}$ is an integration constant. 
Therefore, $t \to t_{\mathrm{s}}$ when $\rho \to \rho_{\mathrm{c}}$. 
By combining Eq.~(\ref{eq:Add-2-B-2-01}) into Eq.~(\ref{eq:2.20}), we have 
\begin{equation} 
w_{\mathrm{DE}} = -1 -\frac{C}{\rho \left(\rho_{\mathrm{c}} -\rho 
\right)^{\gamma}}\,. 
\label{eq:Add-2-B-2-04}
\end{equation}
When $\rho \to \rho_{\mathrm{c}}$, $w_{\mathrm{DE}} \to -\infty$ 
($w_{\mathrm{DE}} \to \infty$) for $C>0$ ($C<0$). 
As a consequence, there can exist the Type II (sudden) singularity if 
$f(\rho)$ is given by Eq.~(\ref{eq:Add-2-B-2-01}). 
It is interesting to remark that 
for $C<0$, when $\rho$ becomes around $\rho_{\mathrm{c}}$, 
the strong energy condition in Eq.~(\ref{eq:Add-2-B-2}) is satisfied. 
This means that 
the Type II (sudden) singularity can appear in the quintessence era. 

Next, we demonstrate a quintessence model as well as a phantom one 
in which the Type II singularity occurs~\cite{Astashenok:2012kb}. 
We consider the case that $f(\rho)$ is given by 
\begin{equation}  
f(\rho) = \frac{\zeta^2}{1-\rho/\rho_{\mathrm{s}}}\,,
\label{eq:ED1-13-IIID2-Add-1}
\end{equation}
with $\zeta (>0)$ being a positive constant. 
In the limit of $t \to t_{\mathrm{s}}$, $\rho$ approaches $\rho_{\mathrm{s}} 
(>0)$. It follows from Eq.~(\ref{eq:2.20}) that if $f(\rho) >0$ with $\rho>0$, 
which is realized for $\rho_0 < \rho_{\mathrm{s}}$ with $\rho_0$ being 
the present energy density of the universe, 
$w_{\mathrm{DE}} < -1$, and hence the universe is in the phantom phase and 
$\rho$ increases until the pressure of the universe diverges. 
While, for $\rho >0$, if $-2/3 < f(\rho)/\rho < 0$, 
$-1 < w_{\mathrm{DE}} <-1/3$ and therefore the universe is in the non-phantom 
(quintessence) phase. 
For $\rho_0 > \rho_{\mathrm{s}}$, since $f(\rho) <0$, this case can correspond 
to a quintessence model. The energy density $\rho$ 
becomes small in time and finally a Big crunch happens 
at $\rho = \rho_{\mathrm{s}}$. 
We estimate how long it takes 
the Type II singularity to appear from the present time $t_0$. 
The remaining time is described by 
\begin{eqnarray}  
t_{\mathrm{s}} - t_0 \Eqn{=} \frac{2}{\sqrt{3}} 
\int_{\bar{x}_0}^{\bar{x}_{\mathrm{s}}} \frac{d \bar{x}}{\zeta^2} 
\left[ 1-\left( \frac{\bar{x}}{\bar{x}_{\mathrm{s}}} \right) \right]\,, 
\label{eq:ED1-13-IIID2-Add-2} \\
\bar{x} \Eqn{\equiv} \sqrt{\rho}\,. 
\label{eq:ED1-13-IIID2-Add-3}
\end{eqnarray}
By using Eq.~(\ref{eq:Add-2-04}), we acquire 
\begin{eqnarray} 
\frac{\rho}{\rho_{\mathrm{s}} } \Eqn{=} 
1 \pm \sqrt{ \left( 1-\Theta \right)^2 + 6\xi \ln \left( 1+z \right) }\,, 
\label{eq:ED1-13-IIID2-Add-4} \\ 
\Theta \Eqn{\equiv} \frac{\rho_0}{\rho_{\mathrm{s}}}\,,  
\label{eq:ED1-13-IIID2-Add-5} \\  
\xi \Eqn{\equiv} \frac{\zeta^2}{\bar{x}_{\mathrm{s}}^2}\,. 
\label{eq:ED1-13-IIID2-Add-6}
\end{eqnarray}
Here, in Eq.~(\ref{eq:ED1-13-IIID2-Add-4}) the ``$+$ $(-)$'' sign denotes 
the case that $\Theta > 1$ $(\Theta < 1)$ describing a quintessence 
(phantom) model. 
Moreover, the EoS for dark energy at the present time is expressed as 
\begin{equation}
w_{\mathrm{DE}(0)} = -1-\frac{\xi}{\Theta \left( 1-\Theta \right)}\,.
\label{eq:ED1-13-IIID2-Add-7}
\end{equation}
In addition, from Eqs.~(\ref{eq:A.2}), 
the luminosity distance $d_L$ is described by 
\begin{eqnarray} 
d_L \Eqn{=} \frac{1}{H_0} \left(1+z\right) \int_{0}^{z}\frac{dz'}{E(z')}\,,
\label{eq:ED1-13-IIID2-Add-8} \\ 
E(z) \Eqn{=}
\sqrt{\Omega_{\mathrm{m}}^{(0)}\left(1+z\right)^{3} 
+\Omega_{\mathrm{DE}}^{(0)} \varrho (z)}\,,
\label{eq:ED1-13-IIID2-Add-9} \\ 
\varrho (z) \Eqn{\equiv} 
\frac{1}{\Theta} 
\left[ 1 \pm \sqrt{ \left( 1-\Theta \right)^2 + 6\xi \ln \left( 1+z \right) } \right]\,. 
\label{eq:ED1-13-IIID2-Add-10}
\end{eqnarray}
Furthermore, the remaining time from the present time to the appearance of 
the finite-time future singularity is given by 
\begin{equation}  
t_{\mathrm{s}} - t_0 = \frac{1}{H_0} 
\int_{\bar{u}}^{0} 
\frac{d \bar{u}'}{\left(1+\bar{u}'\right) \sqrt{\Omega_{\mathrm{DE}}^{(0)}\left(1+z'\right)^{3} +\Omega_{\mathrm{DE}}^{(0)} \varrho (\bar{u}')}}\,, 
\label{eq:ED1-13-IIID2-Add-11}  
\end{equation}
where $\bar{u} \equiv a_0/a - 1$. This variable changes from $\bar{u} = 0$ 
at $t=t_0$ to $\bar{u} = \exp \left[- \left( 1-\Theta \right)^2/\left( 6\xi \right) \right] - 1$ at the time when $P$ diverges, i.e., 
in the limit of $t \to t_{\mathrm{s}}$. 
We note that $\bar{u}$ is equivalent to the redshift $z$, 
and hence for $\bar{u} = z$, 
$\varrho (\bar{u}')$ in Eq.~(\ref{eq:ED1-13-IIID2-Add-11}) becomes equal to 
$\varrho (z)$ in Eq.~(\ref{eq:ED1-13-IIID2-Add-10}). 
As numerical estimations, 
for the phantom phase with 
$w_{\mathrm{DE} (0)} =  -1.10$ and $\Theta = 0.95$, 
$t_{\mathrm{s}} - t_0 = 1.1$Gyr, 
whereas 
for the quintessence phase with 
$w_{\mathrm{DE} (0)} =  -0.98$ and $\Theta = 1.05$, 
$t_{\mathrm{s}} - t_0 = 5.79$Gyr and $t_{\mathrm{d}} - t_0 = 5.78$Gyr, 
where $t_{\mathrm{d}}$ is the time when the cosmic deceleration starts, 
i.e., $\ddot{a}$ becomes zero 
from its positive value~\cite{Astashenok:2012kb}. 
Here, $H_0^{-1} =13.6$Gyr has been used. 

The deceleration parameter $q_{\mathrm{dec}}$ and 
the jerk parameter $j$ are defined by~\cite{Sahni:2002fz} 
\begin{eqnarray}
q_{\mathrm{dec}} \Eqn{\equiv} -\frac{1}{aH^2} \frac{d^2 a}{dt^2} 
= -\frac{1}{H^2} \left[ \frac{1}{2} \frac{d \left( H^2 \right)
}{d N} \right]\,, 
\label{eq:ED1-9-Add-IIIE-13} \\ 
j \Eqn{\equiv} 
\frac{1}{aH^3} \frac{d^3 a}{dt^3}
= \frac{1}{2H^2} \left[ \frac{d^2 \left( H^2 \right)
}{d N^2} + 3\frac{d \left( H^2 \right)
}{d N} + 2H^2 \right]\,,
\label{eq:ED1-9-Add-IIIE-14} 
\end{eqnarray}
where $N \equiv -\ln \left( 1+z \right)$ is the number of $e$-folds and 
$N=0$ at the present time $t=t_0$. 
The values of $q_{\mathrm{dec}}$ and $j$ at the present time $t=t_0$ are 
written as 
\begin{eqnarray} 
q_{\mathrm{dec}} (t=t_0) \Eqn{\equiv} 
q_{\mathrm{dec}(0)} = 
\frac{9\xi}{2\Theta \left(1-\Theta\right)} \Omega_{\mathrm{DE}}^{(0)} 
+\frac{3}{2}\Omega_{\mathrm{DE}}^{(0)} - 1 
\label{eq:ED1-13-IIID2-Add-14} \\
\Eqn{=} 
-\frac{9}{2} \left( w_{\mathrm{DE}} + 1\right) \Omega_{\mathrm{DE}}^{(0)} 
+\frac{3}{2}\Omega_{\mathrm{DE}}^{(0)} - 1\,,
\label{eq:ED1-13-IIID2-Add-15} \\ 
j (t=t_0) \Eqn{\equiv} 
j_0 = 
-\frac{9\xi}{2\Theta \left(1-\Theta\right)} 
\left[ 1+\frac{\xi}{\left(1-\Theta\right)^2} \right] 
\Omega_{\mathrm{DE}}^{(0)} + 1 
\label{eq:ED1-13-IIID2-Add-16} \\ 
\Eqn{=} 
\frac{9}{2} \left( w_{\mathrm{DE}} + 1\right) \left[1 - \frac{\Theta}{1-\Theta} \left(w_{\mathrm{DE}} + 1\right) \right] \Omega_{\mathrm{DE}}^{(0)} + 1\,. 
\label{eq:ED1-13-IIID2-Add-17}
\end{eqnarray}
If $w_{\mathrm{DE} (0)} \approx -1$, except the limit of $\Theta \to 1$, 
$j_0$ is not so different from its value for the $\Lambda$CDM model, where 
$j_0$ is unity. 
Moreover, in the range $ -1.05 < w_{\mathrm{DE} (0)} < -0.95$, we find 
$q_{\mathrm{dec}(0)}^{(\Lambda)} -0.16 < q_{\mathrm{dec}(0)} < q_{\mathrm{dec}(0)}^{(\Lambda)} + 0.16$~\cite{Astashenok:2012kb}, 
where $q_{\mathrm{dec}(0)}^{(\Lambda)} = - 1$ is 
the value of $q_{\mathrm{dec}(0)}$ for the $\Lambda$CDM model. 
Thus, it can be considered that the model in Eq.~(\ref{eq:ED1-13-IIID2-Add-1}) 
describing both the quintessence and phantom phases, where the Big Crunch and 
the Type II singularity eventually happens, respectively, fits the latest supernova data from the Supernova Cosmology project~\cite{Amanullah:2010vv} well. Similar results have been obtained also in Refs.~\cite{Ghodsi:2011wu, Balcerzak:2012ae}. 

We remark that for the case describing the quintessence phase, 
the dissolution of the bound structure before the appearance of the Big Crunch 
cannot be realized. 
The inertial force is given by 
\begin{equation} 
F_{\mathrm{inert}} = -\frac{ml}{2} \left( w_{\mathrm{DE}} + \frac{1}{3}\right) 
\rho\,. 
\label{eq:ED1-13-IIID2-Add-18}  
\end{equation}
For the quintessence phase, the maximum value of $F_{\mathrm{inert}}$ in Eq.~(\ref{eq:ED1-13-IIID2-Add-18}) is 
$
F_{\mathrm{inert}}^{\mathrm{max}} =ml\rho/3    
$. The inertial force becomes small in time because the energy density of the universe in the quintessence phase decreases. 

\subsubsection{Type IV singularity} 

We further explore the case in which 
\begin{equation}  
f(\rho) = \frac{AB \rho^{\alpha + \beta}}{A \rho^{\alpha} + B \rho^{\beta}}\,,
\label{eq:Add-2-B-3-01}
\end{equation}
where $A$, $B$, $\alpha$ and $\beta$ are constants. 
For $\alpha > \beta$, we see that 
$f(\rho) \to A \rho^{\alpha}$ when $\rho \to 0$, while 
$f(\rho) \to B \rho^{\beta}$ when $\rho \to \infty$. 

For $\alpha \neq 1$ and $\beta \neq 1$, Eq.~(\ref{eq:Add-2-04}) yields 
\begin{equation}  
a = a_{\mathrm{c}} \exp \left\{ 
-\frac{1}{3} \left[ \frac{\rho^{-\alpha+1}}{\left(\alpha -1 \right)A} 
+ \frac{\rho^{-\beta+1}}{\left(\beta -1 \right)B} 
\right]
\right\}\,. 
\label{eq:Add-2-B-3-03} 
\end{equation}
In addition, 
for $\alpha = 2\beta - 1$, 
by using Eq.~(\ref{eq:Add-2-05}), we acquire 
\begin{equation}
\tau \equiv 
-\sqrt{3}\kappa A \left(t -t_{\mathrm{s}} \right) 
= \frac{2}{4\beta -3} \rho^{-\left(4\beta -3\right)/2} 
+ \frac{2A}{\left(2\beta -1\right)B} \rho^{-\left(2\beta -1\right)/2}\,, 
\label{eq:Add-2-B-3-04}
\end{equation}
%
where $t_{\mathrm{s}}$ is an integration constant. 
Equation 
(\ref{eq:Add-2-B-3-04}) is 
available for 
$\beta \neq 1$, $\beta \neq 3/4$, and $\beta \neq 1/2$. 
Furthermore, from Eq.~(\ref{eq:2.20}) with Eq.~(\ref{eq:Add-2-B-3-01}), 
we have
\begin{equation} 
w_{\mathrm{DE}} = -1 - 
\frac{AB \rho^{\alpha + \beta -1}}{A \rho^{\alpha} + B \rho^{\beta}}\,. 
\label{eq:Add-2-B-3-05}
\end{equation}
%

For $0< \beta < 1/2$, when $\rho \to 0$, 
\begin{equation}  
P \to -\rho -B \rho^{\beta}\,. 
\label{eq:Add-2-B-3-06}
\end{equation}
Equation (\ref{eq:Add-2-B-3-04}) implies that 
when $\rho \to 0$, $t \to t_{\mathrm{s}}$. 
Hence, from Eq.~(\ref{eq:Add-2-B-3-06}) we see that 
in the limit $t \to t_{\mathrm{s}}$, 
$\rho \to 0$ and $P \to 0$. 
For $\alpha = 2\beta - 1$, from Eq.~(\ref{eq:Add-2-B-3-01}) we find 
\begin{equation}  
\ln \left( \frac{a}{a_{\mathrm{c}}} \right) 
= -\frac{1}{3\left( \beta - 1 \right)} 
\left[ \frac{\rho^{1-\beta}}{2A} + \frac{1}{B} \right] \rho^{1-\beta} 
\sim -\frac{1}{3\left( \beta - 1 \right)B} 
\left[ \frac{\left( 2\beta - 1 \right)B}{2A} \right]^q \tau^q\,, 
\label{eq:Add-2-B-3-06}
\end{equation}
with 
\begin{equation} 
q \equiv 1 -\frac{1}{2\beta - 1}\,.
\label{eq:Add-2-B-3-07}
\end{equation}
Since $q >2$, $a \to a_{\mathrm{c}}$ in the limit $t \to t_{\mathrm{s}}$. 
Moreover, $H$ and $\dot{H}$ are also finite. 
However, unless $q$ is an integer, 
higher derivatives of $H$, 
$d^n H/dt^n$ where $n > -1/\left( 2\beta - 1 \right)$, diverges. 
As a result, there can exist the Type IV singularity. 
In this case, i.e., 
for $\alpha = 2\beta - 1$ with $0< \beta < 1/2$, when $\rho \to 0$, 
from Eq.~(\ref{eq:Add-2-B-3-05}) we see that 
$w_{\mathrm{DE}} \to \infty$ ($w_{\mathrm{DE}} \to -\infty$) for $B<0$ 
($B>0$). 

It has been examined in Ref.~\cite{Nojiri:2005sx} that 
if $f(\rho)$ is expressed as Eq.~(\ref{eq:Add-2-B-3-01}), 
there can also exist the Type I, II and III singularities. 
For $3/4 < \beta <1$ with $A >0$, the Type I singularity can appear. 
For $A/B <0$ with any value of $\beta$ or 
for $\beta <0$ irrespective the sign of $A/B$, 
the Type II singularity can exist. 
For $\beta >1$, the Type III singularity can occur. 

\subsection{Asymptotically de Sitter phantom universe} 

We study 
an example of a fluid realizing asymptotically de Sitter phantom universe.  
We present an important model constructed in Ref.~\cite{Astashenok:2012tv} 
in which the observational data consistent with the $\Lambda$CDM model are 
satisfied, but it develops the dissolution of the bound structure. 

For convenience, we introduce a new variable $x \equiv \sqrt{\rho}$. 
By using $x$, Eqs.~(\ref{eq:Add-2-04}) and (\ref{eq:Add-2-05}) are 
rewritten to 
\begin{eqnarray}
a \Eqn{=} a_{\mathrm{c}} \exp \left( \frac{2}{3} \int_{x_{\mathrm{c}}}^{x} 
\frac{x^{\prime} d x^{\prime}}{f(x^{\prime})} \right)\,,
\label{eq:ED1-9-Add-IIIE-1} \\ 
t \Eqn{=} \frac{2}{\sqrt{3} \kappa} \int_{x_{\mathrm{c}}}^{x} \frac{d x^{\prime}}{f(x^{\prime})}\,, 
\label{eq:ED1-9-Add-IIIE-2}
\end{eqnarray}
where $x_{\mathrm{c}}$ is the value of $x$ at $t=0$. 
If at $x = x_{\mathrm{f}} < \infty$, 
the integration in Eq.~(\ref{eq:ED1-9-Add-IIIE-2}) becomes infinity. 
Hence, it takes the energy density infinite time to 
arrive at $\rho_{\mathrm{f}} \equiv x_{\mathrm{f}}^2$. 
In other words, 
the cosmic expansion becomes exponential behavior asymptotically. 
In addition, the energy density approaches a constant value, i.e., 
the cosmological constant, whereas the EoS $w$ for the universe always 
evolves within the phantom phase $w < -1$. 
We examine a model in which $f(x)$ is given by 
\begin{equation} 
f(x) = \beta \sqrt{x} \left[1 - \left(\frac{x}{x_{\mathrm{f}}}\right)^{3/2} 
\right]\,. 
\label{eq:ED1-9-Add-IIIE-3} 
\end{equation}
We suppose that the energy density of dark energy varies from zero 
to $\rho_{\mathrm{f}} = x_{\mathrm{f}}^2$. 
It follows from Eq.~(\ref{eq:ED1-9-Add-IIIE-1}) that 
$\rho$ is expressed as a function of the redshift $z$ 
\begin{eqnarray}
\rho(z) \Eqn{=} \rho_{\mathrm{f}} 
\left[1 - \left( 1+z \right)^{\gamma} \left( 1-\Delta \right) 
\right]^{4/3}\,, 
\label{eq:ED1-9-Add-IIIE-4} \\ 
\gamma \Eqn{\equiv} \frac{3\beta}{2} \rho_{\mathrm{f}}^{-3/4}\,, 
\label{eq:ED1-9-Add-IIIE-5} \\ 
\Delta \Eqn{\equiv} \left( \frac{\rho_0}{\rho_{\mathrm{f}}} \right)^{3/4}\,,  
\label{eq:ED1-9-Add-IIIE-6}
\end{eqnarray}
where $\rho_0$ is the energy density of dark energy at the present time $t_0$. 
Furtheremore, the current EoS for dark energy is given by 
\begin{equation}  
w_{\mathrm{DE} (0)} = -1-\frac{2\gamma}{3} \frac{1-\Delta}{\Delta}\,. 
\label{eq:ED1-9-Add-IIIE-7} 
\end{equation}
By using a parmeter $v$ varing from zero at $t=t_0$ to unity in the limit of 
$t \to \infty$, we can obtain the parametric description of the scale factor 
as 
\begin{eqnarray}
\hspace{-5mm}
a(v) \Eqn{=} \frac{a_{\mathrm{c}}}{\left( 1-v^{3/2} \right)^{2/\left(3\gamma\right)}}\,, 
\label{eq:ED1-9-Add-IIIE-8} \\  
\hspace{-5mm}
t(v) \Eqn{=} t_0 + \frac{2\sqrt{3}}{\gamma x_{\mathrm{f}}} 
\left[ \frac{\ln \left( v^2+v+1 \right)}{6} + \frac{1}{\sqrt{3}} 
\arctan \frac{2v+1}{\sqrt{3}} -\frac{\pi}{6\sqrt{3}} -\frac{\ln \left( 1-v 
\right)}{3} \right]\,,  
\label{eq:ED1-9-Add-IIIE-9}
\end{eqnarray}
where in the limit of $v \to 1$ 
the last term in Eq.~(\ref{eq:ED1-9-Add-IIIE-9}) becomes dominant. 
With these parametric descriptions, the scale factor can accurately be 
expressed as 
\begin{equation}   
a(t) = a_{\mathrm{c}} 
\left( \frac{2}{3} \right)^{2/\left(3\gamma\right)}
\exp \left[ \frac{x_{\mathrm{f}} \left( t-t_0 \right)}{\sqrt{3}}\right]\,. 
\label{eq:ED1-9-Add-IIIE-10} 
\end{equation}
The energy density of non-relativistic matter and baryons behaves as 
$\rho_{\mathrm{m}} = \rho_{\mathrm{m}(0)} \left( 1+z \right)^3$, 
where $\rho_{\mathrm{m}(0)}$ is the current value of 
the energy density of non-relativistic matter and baryons. 
Thus, it follows from Eq.~(\ref{eq:A.2}) that the luminosity distance $d_L$ 
is written as 
\begin{eqnarray}
d_L \Eqn{=} \frac{1}{H_0} \left( 1+z \right) \int_{0}^{z} \left[
\Omega_{\mathrm{m}}^{(0)}\left(1+z'\right)^{3}
+\Omega_{\mathrm{DE}}^{(0)} \tilde{h}(z') \right]^{-1/2}dz'\,, 
\label{eq:ED1-9-Add-IIIE-11} \\  
\tilde{h}(z) \Eqn{\equiv} \Delta^{-4/3} 
\left[1 - \left( 1+z \right)^{\gamma} \left( 1-\Delta \right) 
\right]^{4/3}\,,  
\label{eq:ED1-9-Add-IIIE-12}
\end{eqnarray}
where we have neglected the contribution from radiation because 
it is much smaller than those from non-relativistic matter and baryons 
and dark energy. 
In case of the $\Lambda$CDM model, $\Omega_{\mathrm{DE}}^{(0)} = \Omega_{\Lambda}$ and $\tilde{h}(z) = 1$ in Eq.~(\ref{eq:ED1-9-Add-IIIE-11}), where 
$\Omega_{\Lambda} \equiv \rho_{\Lambda}/\rho_{\mathrm{crit}}^{(0)}$. 
Provided that $\gamma \approx 0$, where $\beta$ is very small, 
and $\Delta \approx 0$, we find $\tilde{h}(z) \approx 1$. 
Thus, in this case it is impossible to distinguish this model with 
the $\Lambda$CDM model. {}From Eq.~(\ref{eq:ED1-9-Add-IIIE-7}), we also 
have $w_{\mathrm{DE} (0)} \approx -1$.  
On the other hand, if we take $\Delta = 0.5$ and $\gamma = 0.075$, 
$w_{\mathrm{DE} (0)} = -1.05$. 
According to the numerical analysis in Ref.~\cite{Astashenok:2012tv}, 
the difference between the distance modulus $\mu$ in this model 
and that in the $\Lambda$CDM model is estimated as 
$\delta \mu \equiv 5\log \left(d_L/d_L^{(\Lambda \mathrm{CDM})} \right) 
< 0.016$, where $d_L^{(\Lambda \mathrm{CDM})}$ is the luminosity distance 
in the $\Lambda$CDM model, 
for $\Omega_{\mathrm{m}}^{(0)} = 0.28$ and 
$\Omega_{\mathrm{DE}}^{(0)} = \Omega_{\Lambda} = 0.72$. 
Since errors of the modulus of SNe are $\sim 0.15$, which are larger than 
the above estimation of $\delta$, 
this model can fit the observational data as well as the $\Lambda$CDM model. 

In this model, the Friedmann eqution (\ref{eq:Add-2-01}) is given by 
\begin{equation} 
H^2= \frac{\kappa^2}{3} \left( \rho + \rho_{\mathrm{m}} \right) 
= \frac{\kappa^2}{3} \left\{
\rho_{\mathrm{f}} 
\left[1 - \e^{-\gamma N} \left( 1-\Delta \right) \right]^{4/3} + 
\rho_{\mathrm{m}(0)} \e^{-3N} \right\}\,. 
\label{eq:ED1-9-Add-IIIE-15} 
\end{equation}
By substituting Eq.~(\ref{eq:ED1-9-Add-IIIE-15}) into 
Eqs.~(\ref{eq:ED1-9-Add-IIIE-13}) and (\ref{eq:ED1-9-Add-IIIE-14}), we 
acquire
\begin{eqnarray}
q_{\mathrm{dec} (0)} \Eqn{=} 
\frac{2}{3} \gamma \left( \Delta^{-1} -1 \right) \Omega_{\Lambda} 
+ \frac{3}{2} \Omega_{\mathrm{m}}^{(0)} -1\,, 
\label{eq:ED1-9-Add-IIIE-16} \\ 
j_0 \Eqn{=} 
\left[
\frac{2}{9} \gamma^2 \left( \Delta^{-1} -1 \right)^2 
+ \left( -\frac{2}{3} \gamma^2 +2\gamma \right)
\left( \Delta^{-1} -1 \right) 
\right]\Omega_{\Lambda} 
+1\,. 
\label{eq:ED1-9-Add-IIIE-17}  
\end{eqnarray}
%
For $\Delta = 0.5$, $\gamma = 0.075$, 
$\Omega_{\mathrm{m}}^{(0)} = 0.28$ and 
$\Omega_{\mathrm{DE}}^{(0)} = \Omega_{\Lambda} = 0.72$, 
we obtain $q_{\mathrm{dec} (0)} = -0.54$ and $j_0 = 1.11$. 
In the $\Lambda$CDM model, where $\Delta = 0$, 
we have $q_{\mathrm{dec} (0)} = -0.58$ and $j_0 = 1$. 
As a consequence, the values of $q_{\mathrm{dec} (0)}$ and $j_0$ 
in this model are near to those in the $\Lambda$CDM model.  

In addition, we explore whether the dissolution of the bound structure can 
occur. 
By using Eqs.~(\ref{eq:Add-2-01}), (\ref{eq:Add-2-02}) and 
(\ref{eq:FT4-6-IVC4-add-006}) 
$F_{\mathrm{inert}}$ is also represented as 
\begin{equation}
F_{\mathrm{inert}} 
= -ml \frac{\kappa^2}{6} \left( \rho_{\mathrm{DE}} (a) 
+ 3 P_{\mathrm{DE}} (a) \right) 
= ml \frac{\kappa^2}{6} \left( 2\rho_{\mathrm{DE}} (a) 
+ \frac{d \rho_{\mathrm{DE}} (a)}{d a} a \right)\,. 
\label{eq:ED1-9-Add-IIIE-18} 
\end{equation}
The dimensionless discription for $F_{\mathrm{inert}}$ is 
given by~\cite{Frampton:2011aa}
\begin{equation}
\bar{F}_{\mathrm{inert}} \equiv \frac{1}{\rho_{\mathrm{DE}(0)}} 
\left( 2\rho_{\mathrm{DE}} (a) 
+ \frac{d \rho_{\mathrm{DE}} (a)}{d a} a \right)\,,
\label{eq:ED1-9-Add-IIIE-19} 
\end{equation}
where $\rho_{\mathrm{DE}(0)}$ is the current energy density of 
dark energy. 

The parametric discription of the dimensionless inertial force 
is given by 
\begin{equation}  
\bar{F}_{\mathrm{inert}} (u) = 
2\Delta^{-4/3} 
\left[1 - \left( 1+u \right)^{\gamma} \left( 1-\Delta \right)
\right]^{1/3}
\left[ 1+\left( \frac{2\gamma}{3}  -1 \right) \left( 1+u \right)^{\gamma} 
\left( 1-\Delta \right)
\right]\,,  
\label{eq:ED1-9-Add-IIIE-20} 
\end{equation}
and $t-t_0$ is written by 
\begin{equation}  
t-t_0 = \frac{1}{H_0} \int_u^{0} \frac{du'}{\left( 1+u' \right) 
\sqrt{\Omega_{\mathrm{m}}^{(0)}\left(1+u'\right)^{3}
+\Omega_{\mathrm{DE}}^{(0)} \tilde{h}(u')}}
\,.
\label{eq:ED1-9-Add-IIIE-21} 
\end{equation}
Here, we have introduced 
the new variable $u \equiv a_0/a -1$, 
which varies from 0 at the present time $t=t_0$ to $-1$ in the limit of 
$t \to \infty$. 
It follows from Eq.~(\ref{eq:ED1-9-Add-IIIE-20}) that 
in the limit of $t \to \infty$, we find 
$\bar{F}_{\mathrm{inert}} (u) \to 2\Delta^{-4/3}$. 
For the Earth-Sun system, the necessary value of $\bar{F}_{\mathrm{inert}}$ 
for the disintegration of the bound structure is estimated as 
$\bar{F}_{\mathrm{inert}} \gtrsim 10^{23}$~\cite{Astashenok:2012tv}, 
 which leads to the condition for the dissolution 
$\Delta \leq \Delta_{\mathrm{min}} = 10^{-17}$. 
Moreover, from Eq.~(\ref{eq:ED1-9-Add-IIIE-7}) we see that 
for $\Delta = \Delta_{\mathrm{min}} = 10^{-17}$, 
$\gamma$ can change from 0 (in which $w_{\mathrm{DE} (0)} = -1$) 
to $3.6 \times 10^{-18}$ (where $w_{\mathrm{DE} (0)} = -1.24$, which is 
the observational lowest constraint~\cite{Komatsu:2010fb}). 
As a consequence, if $\Delta < \Delta_{\mathrm{min}} = 10^{-17}$ 
and $0 < \gamma < 10^{-18}$, this model can be compatible with 
the observational data of SNe. 
Accordingly, in this model the dissolution of the bound structure 
can be realized, although it satisfies the observational data which 
is consistent with the $\Lambda$CDM model.

\subsection{Inhomogeneous (imperfect) dark fluid universe} 

In this subsection, 
we explain inhomogeneous (imperfect) dark fluid universe by 
following Refs.~\cite{Nojiri:2005sr, Capozziello:2005pa}. 
We investigate so-called inhomogeneous EoS of dark energy, 
which the pressure has the dependence not only on the energy density 
but also on the Hubble parameter $H$. This idea comes from, e.g., 
a time dependent bulk viscosity in the ideal fluid~\cite{Viscous-DE, RMH-VC} or a modification of gravity. 
For a recent study of imperfect fluids, see~\cite{Pujolas:2011he}.

\subsubsection{Inhomogeneous EoS} 

An inhomogeneous expression of the pressure is described by 
\begin{equation}  
P=-\rho + f(\rho) + G(H)\,,
\label{eq:III-E-1}
\end{equation}
where $G(H)$ is a function of $H$. 
We note that generally speaking, $G$ is a function of the Hubble parameter $H$, its derivatives and the scale factor $a$. However, for simplicity we 
consider mainly the case that $G$ depends on only $H$. 
By substituting Eq.~(\ref{eq:III-E-1}) into 
Eq.~(\ref{eq:Add-2-03}), the continuity equation of the inhomogeneous fluid is 
represented by 
\begin{equation} 
0= \dot{\rho}+ 3H \left( f(\rho) + G(H) \right)\,. 
\label{eq:III-E-2} 
\end{equation}

By plugging $H = \kappa \sqrt{\rho/3}$, which follows from 
Eq.~(\ref{eq:Add-2-01}) for the expanding universe ($H \geq 0$), 
into Eq.~(\ref{eq:III-E-2}), we obtain 
\begin{equation}  
\dot{\rho} = 
\mathcal{F} (\rho) 
\equiv -3\kappa 
\sqrt{\frac{\rho}{3}} \left( f(\rho) + G(\kappa 
\sqrt{\frac{\rho}{3}}) \right)\,. 
\label{eq:III-E-3} 
\end{equation}
On the other hand, 
by combining Eq.~(\ref{eq:III-E-1}) with Eq.~(\ref{eq:Add-2-02}), 
we have a representation of $G(H)$ in terms of $f$ as 
\begin{equation}  
G(H) = -f(\frac{3H^2}{\kappa^2}) + \frac{2}{\kappa^2} \dot{H}\,. 
\label{eq:III-E-4} 
\end{equation}
%

\subsubsection{
Influences on the structure of the finite-time future singularities
} 

First, 
as a concrete example, we examine the case in which the relation between 
$P$ and $\rho$ in Eq.~(\ref{eq:2.21}) is changed as 
%
\begin{eqnarray}
P \Eqn{=} w_{\mathrm{h}} \rho + w_1 H^2\,,
\label{eq:III-E-5} \\ 
\Eqn{=} \left( w_{\mathrm{h}} + \frac{\kappa^2 w_1}{3} \right) \rho\,, 
\label{eq:III-E-6}
\end{eqnarray}
%
where $w_{\mathrm{h}}$ is the homogeneous EoS, $w_1$ is a constant, 
and in deriving Eq.~(\ref{eq:III-E-6}) we have used Eq.~(\ref{eq:Add-2-01}). 
Thus, for the inhomogeneous fluid, the inhomogeneous EoS $w_{\mathrm{ih}}$ 
is shifted from the homogeneous EoS $w_{\mathrm{h}} (= w)$ as 
\begin{equation}  
w \longrightarrow 
w_{\mathrm{ih}} \equiv w_{\mathrm{h}} + \frac{\kappa^2 w_1}{3}\,.
\label{eq:III-E-7}  
\end{equation}
{}From Eq.~(\ref{eq:III-E-7}), we see that 
if $w_{\mathrm{ih}} > -1$, 
there does not appear a Big Rip singularity, 
even for $w_{\mathrm{h}} < -1$. 
In other words, provided that if $w_1 \ll -1$, the universe 
can evolve to the phantom phase $w_{\mathrm{ih}} < -1$, even though 
in the beginning, the universe is in the non-phantom (quintessence) phase 
with $w_{\mathrm{h}} > -1$. 

Furthermore, 
in order to demonstrate 
how the inhomogeneous modification in EoS influences on 
the structure of the finite-time future singularities, 
as another example, we investigate the case that 
%
\begin{eqnarray} 
f_{\mathrm{ih}} (\rho) \equiv f(\rho) + G(H) 
\Eqn{=} -A \rho^{\alpha} -BH^{2\beta} 
\label{eq:III-E-8} \\
\Eqn{=} -A \rho^{\alpha} -\tilde{B} \rho^{\beta}\,,
\label{eq:III-E-9}
\end{eqnarray} 
%
where $B$ and $\beta$ are constants, and by using Eq.~(\ref{eq:Add-2-01}) 
$\tilde{B}$ is defined as 
\begin{equation}   
\tilde{B} \equiv B \left( \frac{\kappa^2}{3} \right)^{\beta}\,.
\label{eq:III-E-10}  
\end{equation}
For $\beta > \alpha$, when $\rho$ is large, the inhomogeneous part, i.e., 
the second term, of Eq.~(\ref{eq:III-E-9}) becomes dominant as 
\begin{equation} 
f_{\mathrm{ih}} \to -\tilde{B} \rho^{\beta}\,.
\label{eq:III-E-11}  
\end{equation}
On the other hand, for $\beta < \alpha$, when $\rho \to 0$, 
the inhomogeneous part of Eq.~(\ref{eq:III-E-9}) becomes also dominant 
as in Eq.~(\ref{eq:III-E-11}). 

We explicitly describe the cases in which 
the structure of the finite-time future singularities are changed 
due to the presence of the inhomogeneous term. 
First, we examine the limit that $\rho$ diverges. 
For $1/2 < \alpha < 1$ with $A>0$, if there only exists the homogeneous term 
without the in homogeneous term $G(H)$ in Eq.~(\ref{eq:III-E-1}), i.e., 
the case in Eq.~(\ref{eq:Add-2-B-1-01}) in Sec.~III D 1, 
there can appear the Type I singularity. However, in the presence of 
the inhomogeneous term as in Eq.~(\ref{eq:III-E-9}), 
for $\beta > 1 (> \alpha)$, in which the situation described by 
Eq.~(\ref{eq:III-E-11}) is realized because $\beta > \alpha$, 
in the limit of $\rho \to \infty$, 
$|P|$ also diverges because $P \to - \rho - \tilde{B} \rho^{\beta}$. 
Thus, the Type III singularity appears instead of the Type I singularity. 
For $\alpha = 1/2$, 
if $\beta > 1 (> \alpha)$,the Type III singularity appears, 
whereas if $(\alpha < ) 1/2 < \beta <1$, the Type I singularity occurs. 
For $0< \alpha < 1/2$, if $\beta > 1 (> \alpha)$ with $\tilde{B} (B>0)$,  
the Type III singularity appears, 
while if $(\alpha <) 1/2 < \beta < 1$ with $\tilde{B} (B>0)$, 
the Type I singularity occurs. 

Second, we consider the opposite limit that $\rho$ tends to zero. 
In case of the homogeneous EoS in Eq.~(\ref{eq:Add-2-B-1-01}), 
for $0< \alpha < 1/2$, in the limit of $t \to t_{\mathrm{s}}$, 
$a \to a_{\mathrm{s}}$, $\rho \to 0$ and $|P| \to 0$. 
The substitution of Eq.~(\ref{eq:Add-2-B-1-03}) into 
Eq.~(\ref{eq:Add-2-B-1-02}) leads to 
\begin{equation} 
a = a_{\mathrm{c}} 
\exp \left\{ 
\frac{1}{3\left( 1-\alpha \right) A} \left[ 
\frac{\sqrt{3} \left( 1-2\alpha \right)}{2} \kappa A 
\right]^{\left( \alpha - 1 \right)/\left( 
\alpha - 1/2 \right)} 
|t-t_{\mathrm{s}}|^{\left( \alpha - 1 \right)/\left( 
\alpha - 1/2 \right)} \right\}\,.
\label{eq:III-E2-add-01}  
\end{equation}
The exponent $\left( \alpha - 1 \right)/\left( \alpha - 1/2 \right)$ 
in terms of $|t-t_{\mathrm{s}}|$ in Eq.~(\ref{eq:III-E2-add-01}) 
is not always an integer. 
Hence, there is the possibility that the higher derivatives of $H$ diverge, 
even though $a$ becomes finite. 
As a result, the Type IV singularity can appear. 
For $\alpha < 0$, 
in the limit of $t \to t_{\mathrm{s}}$, $a \to a_{\mathrm{s}}$, 
$\rho \to 0$ and $|P| \to \infty$. 
Thus, the Type II singularity can occur. 

We explain the case in presence of 
the inhomogeneous term as in Eq.~(\ref{eq:III-E-9}). 
For $\alpha = 1/2$, 
if $0 < \beta < 1/2 (< \alpha)$, 
in which the inhomogeneous term becomes dominant over the homogeneous one as 
in Eq.~(\ref{eq:III-E-11}) due to the relation $\beta < \alpha$, 
the Type IV singularity appears, 
or if $\beta <0$, the Type IV singularity occurs. 
For $0< \alpha < 1/2$, if $\beta < 0 (< \alpha)$,  
the Type II singularity appears instead of the Type IV singularity. 

\subsubsection{Implicit inhomogeneous EoS} 

Next, in a general case, we suppose that 
in a proper limit, such as $\rho$ being large or small, 
$\mathcal{F} (\rho)$ in Eq.~(\ref{eq:III-E-3}) is described by 
\begin{equation}  
\mathcal{F} (\rho) \sim \bar{\mathcal{F}} \rho^{\gamma}\,, 
\label{eq:III-E-12} 
\end{equation}
where $\bar{\mathcal{F}}$ and $\gamma$ are constants. 
For $\gamma \neq 1$, we can integrate Eq.~(\ref{eq:III-E-3}) as 
\begin{equation}  
\bar{\mathcal{F}} \left( t - t_{\mathrm{c}} \right) 
\sim \frac{\rho^{1-\gamma}}{1-\gamma}\,, 
\label{eq:III-E-13} 
\end{equation}
which is rewritten to 
\begin{equation} 
\rho \sim \left[ \left( 1-\gamma \right) \bar{\mathcal{F}} \left( t - t_{\mathrm{c}} \right) \right]^{1/\left( 1-\gamma \right)}\,, 
\label{eq:III-E-14} 
\end{equation}
where $t_{\mathrm{c}}$ is a constant of the integration. 
It follows from Eq.~(\ref{eq:Add-2-01}) that the scale factor is given by 
\begin{equation} 
a = a_{\mathrm{c}} \exp \left\{ 
\pm \frac{2\kappa}{\left( 3-2\gamma \right)\sqrt{3} \bar{\mathcal{F}}}
\left[ \left( 1-\gamma \right) \bar{\mathcal{F}} \left( t - t_{\mathrm{c}} 
\right) \right]^{\left( 3-2\gamma \right)/\left[ 
2 \left(1-\gamma \right) \right]} 
\right\}\,. 
\label{eq:III-E-15} 
\end{equation}
If $\gamma = 1$, from Eq.~(\ref{eq:III-E-3}) we find 
\begin{equation}  
\rho = \bar{\rho} \e^{\bar{\mathcal{F}}t}\,, 
\label{eq:III-E-16} 
\end{equation}
where $\bar{\rho}$ is a constant. 
By substituting Eq.~(\ref{eq:III-E-16}) into Eq.~(\ref{eq:Add-2-01}), 
we obtain 
\begin{equation} 
a = a_{\mathrm{c}} \exp \left[ 
\pm \frac{2\kappa}{\bar{\mathcal{F}}} \sqrt{\frac{\bar{\rho}}{3}} 
\e^{\left( \bar{\mathcal{F}}/2 \right)t}
\right]\,. 
\label{eq:III-E-17} 
\end{equation}

We propose an implicit inhomogeneous EoS by generalizing 
the expression of $\mathcal{F} (\rho)$ in Eq.~(\ref{eq:III-E-3}) as 
\begin{equation}  
\mathcal{F} (P, \rho, H) = 0\,. 
\label{eq:III-E-18} 
\end{equation}
In order to understand the cosmological consequences 
of the implicit inhomogeneous EoS, we present the following example: 
\begin{equation}  
\left( P + \rho \right)^2 - C_{\mathrm{c}} \rho^2 
\left( 1 - \frac{H_{\mathrm{c}}}{H} \right) = 0\,, 
\label{eq:III-E-19} 
\end{equation}
with $C_{\mathrm{c}} (> 0)$ and $H_{\mathrm{c}} (> 0)$ are positive constants. 

Plugging Eq.~(\ref{eq:III-E-19}) into Eq.~(\ref{eq:Add-2-02}) and using 
Eq.~(\ref{eq:Add-2-01}), we acquire 
\begin{equation} 
\dot{H}^2 = \frac{9}{4} C_{\mathrm{c}} H^4 \left( 1 - \frac{H_{\mathrm{c}}}{H} 
\right)\,. 
\label{eq:III-E-20} 
\end{equation}
We can integrate Eq.~(\ref{eq:III-E-20}) as 
\begin{equation} 
H = \frac{16}{9 C_{\mathrm{c}}^2 H_{\mathrm{c}} \left( t - t_- \right) 
\left( t_+ - t \right)}\,, 
\label{eq:III-E-20} 
\end{equation}
with 
\begin{equation} 
t_\pm \equiv t_{\mathrm{c}} \pm \frac{4}{3 C_{\mathrm{c}} H_{\mathrm{c}}}\,, 
\label{eq:III-E-21} 
\end{equation}
where $t_{\mathrm{c}}$ is a constant of integration. 
By combining Eq.~(\ref{eq:III-E-19}) with Eq.~(\ref{eq:III-E-20}) and 
substituting Eq.~(\ref{eq:III-E-20}) into Eq.~(\ref{eq:Add-2-01}), we find 
\begin{eqnarray}
P \Eqn{=} - \rho \left[ 1 + \frac{3 C_{\mathrm{c}}^2}{4 H_{\mathrm{c}}} 
\left( t - t_{\mathrm{c}} \right) \right]\,,  
\label{eq:III-E-22} \\ 
\rho \Eqn{=} \frac{256}{27 C_{\mathrm{c}}^4 H_{\mathrm{c}}^2 \kappa^2 
\left( t - t_- \right)^2 \left( t_+ - t \right)^2}\,. 
\label{eq:III-E-23} 
\end{eqnarray}
Thus, by using Eq.~(\ref{eq:III-E-22}), we have 
\begin{equation}  
w = \frac{P}{\rho} = -1 - \frac{3 C_{\mathrm{c}}^2}{4 H_{\mathrm{c}}} 
\left( t - t_{\mathrm{c}} \right)\,.
\label{eq:III-E-24} 
\end{equation}
{}From Eq.~(\ref{eq:III-E-20}), we see that 
if $t_- < t < t_+$, $H > 0$ because $t_- < t_{\mathrm{c}} < t_+$. 
At $t = t_{\mathrm{c}} = \left( t_- + t_+ \right)/2$, 
$H$ becomes the minimum of $H =H_{\mathrm{c}}$. 
On the other hand, in the limit of $t \to t_\pm$, $H \to \infty$. 
This may be interpreted that at $t = t_-$, there exists 
a Big Bang singularity, whereas at $t = t_-$, 
a Big Rip singularity appears. 
It is clearly seen from Eq.~(\ref{eq:III-E-20}) that 
when $t_- < t < t_{\mathrm{c}}$, 
$w > -1$ (the non-phantom (quintessence) phase, 
and when $t_{\mathrm{c}} < t < t_+$, 
$w < -1$ (the phantom phase), 
namely, at $t = t_{\mathrm{c}}$, there can occur the 
crossing of phantom divide from the non-phantom phase to the phantom one. 
This is realized by an inhomogeneous term in the EoS. 

We present another example in which 
de Sitter universe is asymptotically realized. 
\begin{equation}  
\left( P + \rho \right)^2 
+ \frac{16}{\kappa^4 t_{\mathrm{c}}^2} \left( h_{\mathrm{c}} - H \right) 
\ln \left( \frac{h_{\mathrm{c}} - H}{h_1} \right) = 0\,, 
\label{eq:III-E-25} 
\end{equation}
where $t_{\mathrm{c}}$, $h_{\mathrm{c}}$ and $h_1$ are constants 
and $h_{\mathrm{c}} > h_1 >0$. 

\begin{eqnarray} 
H \Eqn{=} h_{\mathrm{c}} - h_1 \e^{-t^2/t_{\mathrm{c}}^2}
\label{eq:III-E-26} \\ 
\rho \Eqn{=} 
\frac{3}{\kappa^2} \left( h_{\mathrm{c}} - h_1 \e^{-t^2/t_{\mathrm{c}}^2} 
\right)^2 \,,
\label{eq:III-E-27} \\
P \Eqn{=} -\frac{3}{\kappa^2} \left( h_{\mathrm{c}} - h_1 \e^{-t^2/t_{\mathrm{c}}^2} \right)^2 - \frac{4 h_1 t}{\kappa^2 t_{\mathrm{c}}^2} 
\e^{-t^2/t_{\mathrm{c}}^2}\,. 
\label{eq:III-E-28} 
\end{eqnarray} 
It follows from Eq.~(\ref{eq:III-E-26}) that 
\begin{equation}   
\dot{H} = \frac{2 h_1 t}{t_{\mathrm{c}}^2} 
\e^{-t^2/t_{\mathrm{c}}^2}\,. 
\label{eq:III-E-29} 
\end{equation}
By using Eqs.~(\ref{eq:Add-2-01}), (\ref{eq:III-E-1}) and (\ref{eq:III-E-2}), 
we obtain Eq.~(\ref{eq:Add-2-02}). 
For $t < 0$, $\dot{H} <0$, whereas for $t > 0$, $\dot{H} >0$. 
For $t < 0$, $w_{\mathrm{eff}} > -1$, while for $t > 0$, 
$w_{\mathrm{eff}} < -1$. 
In the limit of $t \to \pm \infty$, the universe approaches to 
de Sitter one asymptotically, and hence there appears 
a Big Rip singularity nor a Big Bang singularity. 

We note that in principle, the implicit inhomogeneous EoS can be 
expressed in more general form by including higher derivatives of 
$H$ as 
\begin{equation}  
\mathcal{F} (P, \rho, H, \dot{H}, \ddot{H}, \ldots) = 0\,. 
\label{eq:III-E-30} 
\end{equation}
There is a simple example 
\begin{equation}   
P = w_{\mathrm{eff}} \rho 
- \frac{2}{\kappa^2} \dot{H} 
- \frac{3 \left( 1 + w_{\mathrm{eff}} \right)}{\kappa^2} H^2 \,. 
\label{eq:III-E-31} 
\end{equation}
{}From Eqs.~(\ref{eq:Add-2-01}) and (\ref{eq:Add-2-02}), we find 
\begin{equation}   
\rho = \frac{3}{\kappa^2} H^2 \,, 
\quad 
P = w_{\mathrm{eff}} \rho 
- \frac{2}{\kappa^2} \dot{H} 
- \frac{3}{\kappa^2} H^2 \,. 
\label{eq:III-E-31-(2)} 
\end{equation}
It follows from Eqs.~(\ref{eq:III-E-31}) and (\ref{eq:III-E-31-(2)}) that 
Eq.~(\ref{eq:III-E-31}) is an identity. 
This implies that there exists a solution for any cosmology, provided that 
Eq.~(\ref{eq:III-E-31}) is satisfied. 

We can also find another example 
\begin{equation}   
P = w_{\mathrm{eff}} \rho - G_1 
- \frac{2}{\kappa^2} \dot{H}  
+ G_2 \dot{H}^2\,, 
\label{eq:III-E-32} 
\end{equation}
where $G_1$ and $G_2$ are constants, and 
$G_1 \left( 1 + w_{\mathrm{eff}} \right) > 0$ is supposed to be 
satisfied. 
For $G_2 \left( 1 + w_{\mathrm{eff}} \right) > 0$, 
we have a solution expressing an oscillating universe 
\begin{eqnarray} 
H \Eqn{=} h_{\mathrm{c}} \cos \bar{\omega} t 
\label{eq:III-E-33} \\ 
a \Eqn{=}  
a_{\mathrm{c}} \exp \left( \frac{h_{\mathrm{c}}}{\bar{\omega}} 
\sin \omega t \right)\,,
\label{eq:III-E-34} 
\end{eqnarray} 
with 
a coefficient $h_{\mathrm{c}}$ and a frequency $\bar{\omega}$, given by 
\begin{eqnarray} 
h_{\mathrm{c}} \Eqn{\equiv} 
\kappa \sqrt{\frac{G_1}{3 \left( 1 + w_{\mathrm{eff}} \right)}}\,,
\label{eq:III-E-35} \\ 
\bar{\omega} \Eqn{=} 
\sqrt{\frac{3 \left( 1 + w_{\mathrm{eff}} \right)}{G_1 \kappa^2}}\,. 
\label{eq:III-E-36} 
\end{eqnarray} 
On the other hand, for $G_2 \left( 1 + w_{\mathrm{eff}} \right) < 0$, 
\begin{eqnarray} 
H \Eqn{=} 
h_{\mathrm{c}} \cosh \tilde{\omega} t\,, 
\label{eq:III-E-37} \\ 
a \Eqn{=}  
a_{\mathrm{c}} \exp \left( \frac{h_{\mathrm{c}}}{\tilde{\omega}} 
\sinh \tilde{\omega} t \right)\,,
\label{eq:III-E-38} 
\end{eqnarray} 
where $h_{\mathrm{c}}$ is given by Eqs.~(\ref{eq:III-E-37}) and 
a frequency $\tilde{\omega}$ is described as 
\begin{equation}
\tilde{\omega} = 
\sqrt{-\frac{3 \left( 1 + w_{\mathrm{eff}} \right)}{G_1 \kappa^2}}\,.
\label{eq:III-E-39} 
\end{equation}
Hence, the above investigations show that a number of models describing cosmology with an inhomogeneous EoS can be constructed. 

Finally, we mention that 
in Ref.~\cite{Balcerzak:2012ae}, 
cosmological density perturbations around the finite-time future singularities 
with the scale factor being finite have been examined. 
At the present stage, it seems that 
a number of models with a finite-time future singularity are not considered to 
be distinguishable with the $\Lambda$CDM model by using the observational 
test. As a recent investigation, 
the cosmological density perturbations in k-essence scenario has been investigated in Ref.~\cite{Bamba:2011ih}. 

\section{Scalar field theory as dark energy of the universe}

In this section, we explore scalar field theories in general relativity. 

\subsection{Scalar field theories}

The action of scalar field theories in general relativity is given by
\begin{equation} 
S = 
\int d^4 x \sqrt{-g} \left( \frac{R}{2\kappa^2} - 
\frac{1}{2} \omega (\phi) 
g^{\mu\nu} {\partial}_{\mu} \phi {\partial}_{\nu} \phi 
- V(\phi) \right) + 
\int d^4 x 
{\mathcal{L}}_{\mathrm{M}} 
\left( g_{\mu\nu}, {\Psi}_{\mathrm{M}} \right)\,,
\label{eq:2.1}
\end{equation}
where  
$\omega (\phi)$ is a function of the scalar field $\phi$ and 
$V(\phi)$ is the potential of $\phi$.  
(i) For $\omega (\phi) = 0$ and $V(\phi) = \Lambda/\kappa^2$ with $\Lambda$ 
being a cosmological constant, the action in Eq.~(\ref{eq:2.1}) describes 
the $\Lambda$CDM model in Eq.~(\ref{eq:Add-2-0-1}). 
(ii) For $\omega (\phi) = +1$, this action corresponds to the one for a 
quintessence model with a canonical kinetic term. 
(iii) For $\omega (\phi) = -1$, this action expresses a phantom model. 
We consider the case in which the scalar field $\phi$ is 
a spatially homogeneous one, i.e., it depends only on time $t$. 

In the FLRW background~(\ref{eq:2.2}), the Einstein equations are given by
\begin{eqnarray}
H^2 \Eqn{=} \frac{\kappa^2}{3} \left( \rho_{\phi} + \rho_{\mathrm{M}} 
\right)\,,
\label{eq:2.3} \\ 
\dot{H} \Eqn{=} -\frac{\kappa^2}{2} \left( 
\rho_{\phi} + P_{\phi} + \rho_{\mathrm{M}} + P_{\mathrm{M}} \right)\,,
\label{eq:2.4}
\end{eqnarray}
where 
$\rho_{\phi}$ and $P_{\phi}$ are the energy density and 
pressure of the scalar field $\phi$, respectively, given by  
\begin{eqnarray}
\rho_{\phi} \Eqn{=} \frac{1}{2} \omega (\phi) \dot{\phi}^2
+V(\phi)\,,
\label{eq:2.5} \\
P_{\phi} \Eqn{=} \frac{1}{2} \omega (\phi) \dot{\phi}^2
-V(\phi)\,. 
\label{eq:2.6}
\end{eqnarray}
In addition, $\rho_{\mathrm{M}}$ and $P_{\mathrm{M}}$ are 
the energy density and pressure of matter, respectively. 
Equations (\ref{eq:2.5}) and (\ref{eq:2.6}) 
give the way to express scalar dark energy as fluid dark energy, 
where scalar field equation becomes just the conservation law for such fluid 
description. 

Here, for clear understanding, 
by using Eqs.~(\ref{eq:2.5}) and (\ref{eq:2.6}), 
we describe the explicit expression of $w_{\mathrm{DE}}$ in 
the case that there exists only single scalar field of 
dark energy, i.e., dark energy sufficiently dominates over matter. 
(i) For the $\Lambda$CDM model 
with $\omega (\phi) = 0$ and $V(\phi) = \Lambda/\kappa^2$, 
$w_{\mathrm{DE}} = -1$.  
(ii) For a quintessence model with $\omega (\phi) = +1$, 
\begin{equation}  
w_{\mathrm{DE}} = \frac{\dot{\phi}^2 - 2V(\phi)}{\dot{\phi}^2 + 2V(\phi)}\,, 
\label{eq:IV-A-addition-01}
\end{equation}
it follows from which that $-1 < w_{\mathrm{DE}} < -1/3$. 
(iii) For a phantom model with $\omega (\phi) = -1$
\begin{equation}  
w_{\mathrm{DE}} = \frac{\dot{\phi}^2 + 2V(\phi)}{\dot{\phi}^2 - 2V(\phi)}\,, 
\label{eq:IV-A-addition-02}
\end{equation}
which leads to $w_{\mathrm{DE}} < -1$. 

By using Eqs.~(\ref{eq:2.3})--(\ref{eq:2.6}), we obtain 
\begin{eqnarray}
\omega (\phi) \dot{\phi}^2 \Eqn{=} -\frac{2}{\kappa^2}
\dot{H} - \left( \rho_{\mathrm{M}} + P_{\mathrm{M}} \right)\,,
\label{eq:2.7} \\
V(\phi) \Eqn{=} \frac{1}{\kappa^2}
\left( 3H^2 + \dot{H} \right) -\frac{1}{2} 
\left( \rho_{\mathrm{M}} - P_{\mathrm{M}} \right)\,.
\label{eq:2.8}
\end{eqnarray}

If there is no coupling between matter and the scalar field $\phi$, 
the continuity equations for matter and $\phi$ are given by 
\begin{eqnarray}
&&
\dot{\rho}_{\mathrm{M}}+3H \left( 
\rho_{\mathrm{M}} + P_{\mathrm{M}}
\right)  
= 0\,. 
\label{eq:2.9}\\ 
&&
\dot{\rho}_{\phi}+3H \left( 
\rho_{\phi} + P_{\phi}
\right)  
= 0\,. 
\label{eq:2.10}
\end{eqnarray} 

For a matter with a constant EoS 
$w_{\mathrm{M}} \equiv P_{\mathrm{M}}/\rho_{\mathrm{M}}$, 
from Eq.~(\ref{eq:2.9}) we find
\begin{equation} 
\rho_{\mathrm{M}} = \rho_{\mathrm{M} \mathrm{c}} 
a^{-3\left(1+w_{\mathrm{M}}\right)}\,, 
\label{eq:2.11}
\end{equation}
where $\rho_{\mathrm{M} \mathrm{c}}$ is a constant. 

It is the interesting case that 
$\omega (\phi)$ and $V(\phi)$ are defined 
in terms of a single function $I(\phi)$ as~\cite{Capozziello:2005mj}
\begin{eqnarray}
\omega (\phi) \Eqn{=} -\frac{2}{\kappa^2}
\frac{dI(\phi)}{d \phi} 
-\left(1+w_{\mathrm{M}}\right) 
\mathcal{I}_{\mathrm{c}} \e^{-3\left(1+w_{\mathrm{M}}\right) \mathcal{I}(\phi)}\,,
\label{eq:2.12} \\
V(\phi) \Eqn{=} \frac{1}{\kappa^2}
\left( 3I^2 (\phi) + \frac{dI(\phi)}{d \phi} \right)
-\frac{1}{2} \left(1-w_{\mathrm{M}}\right) 
\mathcal{I}_{\mathrm{c}} \e^{-3\left(1+w_{\mathrm{M}}\right) \mathcal{I}(\phi)}\,. 
\label{eq:2.13}
\end{eqnarray} 
Here, $I(\phi)$ is defined as 
\begin{equation} 
I(\phi) \equiv \frac{d \mathcal{I}(\phi)}{d \phi}\,, 
\label{eq:2.14}
\end{equation}
with $\mathcal{I}(\phi)$ and $\mathcal{I}_{\mathrm{c}}$ 
being an arbitrary twice differentiable 
function of $\phi$ and an integration constant, 
respectively. 
Thus, we can acquire the following solutions: 
\begin{equation} 
\phi = t\,,
\quad
H = I(t)\,.
\label{eq:2.15}
\end{equation}
For this solution, the equation of motion for $\phi$ is derived 
from the variation of the action in Eq.~(\ref{eq:2.1}) 
over $\phi$ as 
\begin{equation} 
\omega (\phi) \ddot{\phi} + \frac{1}{2} \frac{\partial \omega (\phi)}{\partial \phi} \dot{\phi}^2 + 3H\omega (\phi) \dot{\phi} + \frac{\partial V (\phi)}{\partial \phi} = 0\,. 
\label{eq:ED1-10-Add-IVA-1}
\end{equation}
{}From these solutions in Eq.~(\ref{eq:2.15}), we have 
\begin{equation} 
a(t) = a_{\mathrm{c}} \e^{\mathcal{I}(t)}\,,
\quad
a_{\mathrm{c}} = \left(\frac{\rho_{\mathrm{M} \mathrm{c}} }{\mathcal{I}_{\mathrm{c}}}
\right)^{1/\left[3\left(1+w_{\mathrm{M}}\right)\right]}\,.
\label{eq:2.16}
\end{equation}
In what follows, we consider the case in which these solutions are satisfied.

For this case, 
in the FLRW background~(\ref{eq:2.2}), 
the effective EoS 
for the universe is given by Eq.~(\ref{eq:2.17}) 
with 
\begin{eqnarray}
\rho_{\mathrm{eff}} \Eqn{=} \rho_{\phi} + \rho_{\mathrm{M}}\,,
\label{eq:2.18} \\ 
P_{\mathrm{eff}} \Eqn{=} P_{\phi} + P_{\mathrm{M}}\,.
\label{eq:2.19} 
\end{eqnarray}
%

If we define a new scalar field $\Phi$ as
\begin{equation}
\Phi \equiv \int\limits^{\phi} d\phi \sqrt{\abs{ \omega (\phi)}}\,,
\label{eq:2.22}
\end{equation}
the action in Eq.~(\ref{eq:2.1}) can be rewritten to the form 
\begin{equation} 
S_{\chi} =
\int d^{4}x \sqrt{-g}
\left[
\frac{R}{2\kappa^2} \mp \frac{1}{2} 
g^{\mu\nu} {\partial}_{\mu} \Phi {\partial}_{\nu} \Phi
-\tilde{V}(\Phi) \right] 
+ \int d^4 x 
{\mathcal{L}}_{\mathrm{M}} 
\left( g_{\mu\nu}, {\Psi}_{\mathrm{M}} \right)\,. 
\label{eq:2.23}
\end{equation}
where the sign in front of the kinetic term depends on that of
$\omega (\phi)$. If the sign of $\omega (\phi)$
is positive (negative), that of the kinetic term is $-$ ($+$).
In the non-phantom phase, the sign of the kinetic term is always $-$, and
in the phantom one it is always $+$.
In principle, it follows from Eq.~(\ref{eq:2.22}) that $\phi$ can be solved
with respect to $\chi$ as $\phi = \phi (\Phi)$. Hence, the potential
$\tilde{V}(\Phi)$ is given by
$\tilde{V}(\Phi) = V \left( \phi (\Phi) \right)$.

\subsection{Equivalence between fluid descriptions and scalar field theories}

In this subsection, we show the equivalence between fluid descriptions and 
scalar field theories. 
We first take a fluid and then construct a scalar field theory with the same 
EoS as that in a fluid description. This process leads to constraints on a 
coefficient function of the kinetic term $\omega (\phi)$ and the potential 
$V (\phi)$ of the scalar field $\phi$ in the action in Eq.~(\ref{eq:2.1}). 
Through this procedure, 
we propose a way of expressing a fluid model as an explicit scalar field 
theory. In other words, we can obtain the explicit expressions of 
$\omega (\phi)$ and $V (\phi)$ in the corresponding scalar field theory 
for a fluid model. 

For simplicity, we suppose the dark energy dominated stage, so that we 
can neglect matter and therefore $w_{\mathrm{DE}} \approx w_{\mathrm{eff}}$, 
namely, $\rho_{\mathrm{eff}} \approx \rho = \rho_{\phi}$ in 
Eq.~(\ref{eq:2.18}) 
and $\rho_{\mathrm{eff}} \approx \rho = \rho_{\phi}$ in Eq.~(\ref{eq:2.19}). 
In a fluid description, from Eq.~(\ref{eq:2.20}) we find 
\begin{equation} 
w_{\mathrm{eff}} = \frac{P}{\rho} = 
-1 - \frac{f(\rho)}{\rho}\,. 
\label{eq:IVC-1}
\end{equation}
While, in a scalar field theory with the solutions in (\ref{eq:2.15}), 
we obtain 
\begin{eqnarray}
\omega \Eqn{=} \rho + P = - f(\rho)\,,
\label{eq:IVC-2} \\ 
V \Eqn{=} \frac{1}{2} \left( \rho - P \right) = \rho + \frac{f(\rho)}{2}\,, 
\label{eq:IVC-3} 
\end{eqnarray}
where the second equalities follow from Eq.~(\ref{eq:2.21}). 
Since Eq.~(\ref{eq:Add-2-01}) presents $\rho = 3H^2/\kappa^2$, 
if $H (= I(t))$ is given, we acquire the expression of $\rho$ as a function of 
$t (= \phi)$ as $\rho = \rho (t) = \rho (\phi)$. 
Hence, by substituting this relation into 
Eqs.~(\ref{eq:IVC-2}) and (\ref{eq:IVC-3}), 
in principle we have $\omega = \omega (\phi)$ and $V = V (\phi)$. 
Thus, this procedure yields 
an explicit scalar field theory, which corresponds to an original fluid model. 
On the other hand, as the opposite direction, 
provided that we have a scalar field theory described by 
$\omega (\phi)$ and $V (\phi)$ in the action in Eq.~(\ref{eq:2.1}). 
By using Eqs.~(\ref{eq:2.5}) and (\ref{eq:2.6}) and 
the solution $\phi = t$ and $H = I(t)$ in Eq.~(\ref{eq:2.15}), 
we get the explicit expression of $w_{\mathrm{eff}}$ in 
Eq.~(\ref{eq:IVC-1}). Hence, by combining this expression 
and $\rho$ in Eq.~(\ref{eq:2.5}) and comparing 
the resultant expression with the representation of $w_{\mathrm{eff}}$ in 
Eq.~(\ref{eq:2.20}), we acquire $f(\rho)$ in the fluid description. 
Consequently, 
it can be interpreted that 
the considerations on both these directions 
imply the equivalence between the representation of 
a scalar field theory and the description of a fluid model. 

As concrete examples for a fluid description, 
we first consider the case of 
Eq.~(\ref{eq:IIIB-add-01}). 
By substituting Eq.~(\ref{eq:IIIB-add-01}) into 
Eqs.~(\ref{eq:IVC-1})--(\ref{eq:IVC-3}), we have 
\begin{eqnarray}
w_{\mathrm{eff}} \Eqn{=} 
-1 - \frac{1}{\rho} \left( \rho^q - 1 \right)\,. 
\label{eq:IVC-4} \\
\omega \Eqn{=} 
- \rho^q + 1\,,
\label{eq:IVC-5} \\ 
V \Eqn{=} 
\rho + \frac{\rho^q - 1}{2}\,.
\label{eq:IVC-6} 
\end{eqnarray}
Second, for the case of Eq.~(\ref{eq:3-C-1}), 
by combining Eq.~(\ref{eq:3-C-1}) and Eqs.~(\ref{eq:IVC-1})--(\ref{eq:IVC-3}) 
we find 
\begin{eqnarray}
w_{\mathrm{eff}} \Eqn{=} 
-\frac{\mathcal{A}}{\rho^{u+1}}\,. 
\label{eq:IVC-7} \\
\omega \Eqn{=} 
\rho \left( 1 - \frac{\mathcal{A}}{\rho^{u+1}} \right)\,,
\label{eq:IVC-8} \\ 
V \Eqn{=} 
\frac{\rho}{2} \left( 1 + \frac{\mathcal{A}}{\rho^{u+1}} \right)\,.
\label{eq:IVC-9} 
\end{eqnarray}
As the third case described in Eq.~(\ref{eq:III-P-p-7}), 
by plugging Eq.~(\ref{eq:III-P-p-7}) into Eqs.~(\ref{eq:IVC-1})--(\ref{eq:IVC-3}) we acquire 
\begin{eqnarray}
w_{\mathrm{eff}} \Eqn{=} 
-1 \mp \frac{2}{3n} \sqrt{ 1 - 
\frac{4n}{t_{\mathrm{s}}} \sqrt{\frac{3}{\kappa^2 \rho}} 
}\,. 
\label{eq:ED1-9-Add-IVB-1} \\
\omega \Eqn{=}  
\mp \frac{2 \rho}{3n} \sqrt{ 1 - 
\frac{4n}{t_{\mathrm{s}}} \sqrt{\frac{3}{\kappa^2 \rho}} 
}\,,
\label{eq:ED1-9-Add-IVB-2} \\ 
V \Eqn{=} 
\rho \left( 1 + \pm \frac{2 \rho}{3n} \sqrt{ 1 - 
\frac{4n}{t_{\mathrm{s}}} \sqrt{\frac{3}{\kappa^2 \rho}} 
} \right)\,.
\label{eq:ED1-9-Add-IVB-3} 
\end{eqnarray}
As the last example, we consider a model of Little Rip scenario given in 
Eq.~(\ref{eq:FT4-6-IVC3-01}). By using Eqs.~(\ref{eq:Add-2-01}) and (\ref{eq:Add-2-02}), we obtain 
\begin{eqnarray}
w_{\mathrm{eff}} \Eqn{=} 
-1 -\frac{2 \xi}{3 H_{\mathrm{LR}}} \exp \left( -\xi t \right)\,,
\label{eq:ED1-9-Add-IVB-4} \\
\omega \Eqn{=} 
-\frac{2 \xi}{\kappa^2} H_{\mathrm{LR}} \exp \left( \xi t \right)\,, 
\label{eq:ED1-9-Add-IVB-5} \\ 
V \Eqn{=}  
\frac{H_{\mathrm{LR}} \exp \left( \xi t \right)}{\kappa^2} 
\left[ 3H_{\mathrm{LR}} \exp \left( \xi t \right) + \xi \right]\,.
\label{eq:ED1-9-Add-IVB-6} 
\end{eqnarray}
It is meaningful to remark that 
if the Hubble parameter $H$ can be represented by $t$, 
by using the relations (\ref{eq:2.12}) and (\ref{eq:2.13}) and 
the solutions (\ref{eq:2.15}) 
we can find the explicit expressions $\omega = \omega (\phi)$ and 
$V = V (\phi)$. 
This is clearly demonstrated in the following Sec.~IV C.

\subsection{Cosmological models}

In this subsection, 
we reconstruct scalar field theories corresponding to 
(i) the $\Lambda$CDM model, 
(ii) quintessence model, 
(iii) phantom model 
and 
(iv) unified scenario of inflation and late-time cosmic 
acceleration. 
In addition, 
(v) scalar field models with realizing the crossing the phantom divide 
is also considered including its stability issue.

\subsubsection{The $\Lambda$CDM model}

For the $\Lambda$CDM model, we have 
\begin{equation} 
I(\phi) = H_{\mathrm{c}}\,, 
\label{eq:3-A-1.1}
\end{equation}
where $H_{\mathrm{c}}$ is a constant. 
{}From Eq.~(\ref{eq:2.15}), we find 
\begin{eqnarray}
H \Eqn{=} H_{\mathrm{c}}\,,
\label{eq:3-A-1.2} \\
a(t) \Eqn{=} a_{\mathrm{c}} \e^{H_{\mathrm{c}} t}\,. 
\label{eq:3-A-1.3} 
\end{eqnarray} 
Moreover, by using Eqs.~(\ref{eq:2.12}) and (\ref{eq:2.13}), we obtain 
\begin{eqnarray}
\omega (\phi) \Eqn{=}  
-\left(1+w_{\mathrm{M}}\right) 
\mathcal{I}_{\mathrm{c}} \e^{-3\left(1+w_{\mathrm{M}}\right) \mathcal{I}(\phi)}\,,
\label{eq:3-A-1.4} \\
V(\phi) \Eqn{=} \frac{3H_{\mathrm{c}}^2}{\kappa^2} 
-\frac{1}{2} \left(1-w_{\mathrm{M}}\right) 
\mathcal{I}_{\mathrm{c}} \e^{-3\left(1+w_{\mathrm{M}}\right) \mathcal{I}(\phi)}\,, 
\label{eq:3-A-1.5}
\end{eqnarray} 
with 
\begin{equation} 
\mathcal{I}(\phi) = H_{\mathrm{c}} \phi\,.
\label{eq:3-A-1.6}
\end{equation}
In the $\Lambda$CDM model, we acquire 
\begin{equation} 
w_{\mathrm{eff}} = -1\,.
\label{eq:3-A-1.7}
\end{equation}
%

\subsubsection{Quintessence model}

As an example of a quintessence model, we investigate the following model: 
\begin{equation} 
I(\phi) = H_{\mathrm{c}} + \frac{H_1}{\phi^n}\,, 
\label{eq:3-A-2.1}
\end{equation}
where $H_{\mathrm{c}}$ and $H_1 (>0)$ are constants and $n (>1)$ is a positive 
(constant) integer larger than unity. 
By using Eq.~(\ref{eq:2.15}), we find 
\begin{eqnarray}
H \Eqn{=} H_{\mathrm{c}} + \frac{H_1}{t^n}\,,
\label{eq:3-A-2.2} \\
a(t) \Eqn{=} a_{\mathrm{c}} \exp \left[ H_{\mathrm{c}} t - \frac{H_1}{\left(n-1\right) t^{n-1}} 
\right]\,. 
\label{eq:3-A-2.3} 
\end{eqnarray} 
In addition, from Eqs.~(\ref{eq:2.12}) and (\ref{eq:2.13}) we obtain 
\begin{eqnarray}
\omega (\phi) \Eqn{=} \frac{2}{\kappa^2} \frac{n H_1}{\phi^{n+1}} 
-\left(1+w_{\mathrm{M}}\right) 
\mathcal{I}_{\mathrm{c}} 
\e^{-3\left(1+w_{\mathrm{M}}\right) \mathcal{I}(\phi)}\,, 
\label{eq:3-A-2.4} \\
V(\phi) \Eqn{=} \frac{1}{\kappa^2} \frac{3}{\phi^{n+1}} 
\left[ 
\frac{\left( H_{\mathrm{c}} \phi^{n} + H_1 \right)^2}{\phi^{n-1}} 
-\frac{n H_1}{3} 
\right] 
-\frac{1}{2} \left(1-w_{\mathrm{M}}\right) 
\mathcal{I}_{\mathrm{c}} 
\e^{-3\left(1+w_{\mathrm{M}}\right) \mathcal{I}(\phi)}\,, 
\label{eq:3-A-2.5}
\end{eqnarray} 
with 
\begin{equation} 
\mathcal{I}(\phi) = 
H_{\mathrm{c}} \phi - \frac{H_1}{\left(n-1\right) \phi^{n-1}}\,.
\label{eq:3-A-2.6}
\end{equation}
Using Eq.~(\ref{eq:2.17}), the effective EoS is written as 
\begin{equation} 
w_{\mathrm{eff}} = -1 + 
\frac{2n H_1 t^{n-1}}{3\left( H_{\mathrm{c}} t^{n} + H_1 \right)^2}\,.
\label{eq:3-A-2.7}
\end{equation}
{}From Eq.~(\ref{eq:3-A-2.7}), we see that $w_{\mathrm{eff}} > -1$ because 
$H_1 >0$ and $n >1$. 
Thus, this model corresponds to a quintessence model.

\subsubsection{Phantom model}

As an example of a phantom model, we explore the following model: 
\begin{equation} 
I(\phi) = \frac{H_2}{t_{\mathrm{s}} - \phi} + \frac{H_3}{\phi^2}\,, 
\label{eq:3-A-3.1}
\end{equation}
where $H_2$ and $H_3$ are constants and $t_{\mathrm{s}}$ is the time 
when a Big Rip singularity appears. 
We examine the range $0 < t <t_{\mathrm{s}}$. 
{}From Eq.~(\ref{eq:2.15}), we have 
\begin{eqnarray}
H \Eqn{=} \frac{H_2}{t_{\mathrm{s}} - t} + \frac{H_3}{t^2}\,,
\label{eq:3-A-3.2} \\
a(t) \Eqn{=} a_{\mathrm{c}} \left(t_{\mathrm{s}} - t \right)^{-H_2} 
\e^{-\left(H_3/t\right)}\,. 
\label{eq:3-A-3.3} 
\end{eqnarray} 
Furthermore, it follows from Eqs.~(\ref{eq:2.12}) and (\ref{eq:2.13}) 
that 
\begin{eqnarray}
\omega (\phi) \Eqn{=} 
-\frac{2}{\kappa^2} \left[ \frac{H_2}{\left(t_{\mathrm{s}} - \phi \right)^2} 
-\frac{2H_3}{\phi^3} \right] 
-\left(1+w_{\mathrm{M}}\right) 
\mathcal{I}_{\mathrm{c}} 
\left(t_{\mathrm{s}} - \phi \right)^{3\left(1+w_{\mathrm{M}}\right)H_2} 
\e^{\left[3\left(1+w_{\mathrm{M}}\right)H_3/\phi \right]}\,,
\label{eq:3-A-3.4} \\
V(\phi) \Eqn{=} 
\frac{1}{\kappa^2} \left[ 
\frac{H_2 \left(3H_2 + 1\right)}{\left(t_{\mathrm{s}} - \phi \right)^2} 
+\frac{H_3}{\phi^3} \left( \frac{3H_3}{\phi} - 2 \right) 
+\frac{6 H_2 H_3}{\left(t_{\mathrm{s}} - \phi \right) \phi^2} 
\right] 
\nonumber \\ 
&&
{}-\frac{1}{2} \left(1-w_{\mathrm{M}}\right) 
\mathcal{I}_{\mathrm{c}} 
\left(t_{\mathrm{s}} - \phi \right)^{3\left(1+w_{\mathrm{M}}\right)H_2} 
\e^{\left[3\left(1+w_{\mathrm{M}}\right)H_3/\phi \right]}\,. 
\label{eq:3-A-3.5}
\end{eqnarray} 
Using Eq.~(\ref{eq:2.17}), the effective EoS is expressed as 
\begin{equation} 
w_{\mathrm{eff}} = -1 
-\frac{2 \left[H_2 t^3 -2H_3 \left(t_{\mathrm{s}} - t \right)^2 
\right] t}{3\left[ H_2 t^2 + H_3 \left(t_{\mathrm{s}} - t \right) 
\right]^2}\,.
\label{eq:3-A-3.6}
\end{equation}
It follows from Eq.~(\ref{eq:3-A-3.6}) that 
when the time $t$ approaches $t_{\mathrm{s}}$, the second term 
on the right-hand side (r.h.s.) of Eq.~(\ref{eq:3-A-3.6}) becomes 
negative because $H_2 t^3 -2H_3 \left(t_{\mathrm{s}} - t \right)^2 >0$ 
and therefore $w_{\mathrm{eff}} < -1$. 
Hence, there exists the phantom phase in this model and 
this consequence originates from the realization of $\dot{H} > 0$, 
i.e., superacceleration. 
We note that when $0< t \ll t_{\mathrm{s}}$, 
for $0 < t \lesssim \left(2H_3 t_{\mathrm{s}}^2/H_2 \right)^{1/3}$, 
$w_{\mathrm{eff}} > -1$ and therefore the non-phantom phase exists 
before the phantom phase appears.

\subsubsection{Unified scenario of inflation and late-time cosmic 
acceleration}

As an example of a unified scenario of inflation and the late-time 
acceleration of the universe, we study the following model: 
\begin{equation} 
I(\phi) = h_{\mathrm{c}}^2 \left( \frac{1}{t_{\mathrm{s}}^2 - \phi^2} \right) + 
\frac{1}{t_1^2 + \phi^2}\,, 
\label{eq:3-A-4.1}
\end{equation}
where $h_{\mathrm{c}}$ is a constant, $t_{\mathrm{s}}$ corresponds to 
the time when a Big Rip singularity appears, 
and $t_1$ is a time. 
{}From Eq.~(\ref{eq:2.15}), we find 
\begin{eqnarray}
H \Eqn{=} h_{\mathrm{c}}^2 \left( \frac{1}{t_{\mathrm{s}}^2 - t^2} 
+ \frac{1}{t_1^2 + t^2} \right)\,,
\label{eq:3-A-4.2} \\
a(t) \Eqn{=} a_{\mathrm{c}} 
\left( \frac{t_{\mathrm{s}} + t}{t_{\mathrm{s}} - t} 
\right)^{h_{\mathrm{c}}^2/\left(2t_{\mathrm{s}}\right)} 
\e^{\left(h_{\mathrm{c}}^2/t_1\right) \arctan \left(t/t_1\right)}\,. 
\label{eq:3-A-4.3} 
\end{eqnarray} 
Moreover, by using Eqs.~(\ref{eq:2.12}) and (\ref{eq:2.13}), we obtain 
\begin{eqnarray}
\omega (\phi) \Eqn{=} 
-\frac{8}{\kappa^2} \frac{h_{\mathrm{c}}^2 \left( t_1^2 + t_{\mathrm{s}}^2 \right) 
\left[ \phi^2 + \left( t_1^2 - t_{\mathrm{s}}^2 \right)/2 \right]\phi}{\left( t_1^2 
+ \phi^2 \right)^2 \left( t_{\mathrm{s}}^2 - \phi^2 \right)^2} 
-\left(1+w_{\mathrm{M}}\right) 
\mathcal{I}_{\mathrm{c}} \e^{-3\left(1+w_{\mathrm{M}}\right) \mathcal{I}(\phi)}\,,
\label{eq:3-A-4.4} \\
V(\phi) \Eqn{=} 
\frac{1}{\kappa^2} \frac{h_{\mathrm{c}}^2 \left( t_1^2 + t_{\mathrm{s}}^2 \right)}{\left( t_1^2 
+ \phi^2 \right)^2 \left( t_{\mathrm{s}}^2 - \phi^2 \right)^2} 
\left[ 3h_{\mathrm{c}}^2 \left( t_1^2 + t_{\mathrm{s}}^2 \right) + 4\phi 
\left( \phi^2 + \frac{t_1^2 - t_{\mathrm{s}}^2}{2} \right) \right] 
\nonumber \\ 
&&
{}-\frac{1}{2} \left(1-w_{\mathrm{M}}\right) 
\mathcal{I}_{\mathrm{c}} \e^{-3\left(1+w_{\mathrm{M}}\right) \mathcal{I}(\phi)}\,, 
\label{eq:3-A-4.5}
\end{eqnarray} 
with 
\begin{equation} 
\mathcal{I}(\phi) = \frac{h_{\mathrm{c}}^2}{2t_{\mathrm{s}}} \ln \left( 
\frac{t_{\mathrm{s}} + \phi}{t_{\mathrm{s}} - \phi} \right) 
+ \frac{h_{\mathrm{c}}^2}{t_1} \arctan \left( \frac{\phi}{t_1} \right)\,.
\label{eq:3-A-4.6}
\end{equation}
Using Eq.~(\ref{eq:2.17}), 
the effective EoS is expressed as 
\begin{equation} 
w_{\mathrm{eff}} = -1 
-\frac{8}{3h_{\mathrm{c}}^2} \frac{t \left( t - t_{+} \right) 
\left( t - t_{-} \right)}{t_1^2 + t_{\mathrm{s}}^2}\,, 
\label{eq:3-A-4.7}
\end{equation}
where $t_{\pm} \equiv \pm \sqrt{\left( t_{\mathrm{s}}^2 - t_1^2 \right)/2}$. 
There exist two phantom phases: 
$t_{-} < t < 0$ and $t > t_{+}$, in which $w_{\mathrm{eff}} < -1$. 
On the other hand, there are also two non-phantom phases: 
$-t_{\mathrm{s}} < t < t_{-}$ and $0 < t < t_{+}$, in which $w_{\mathrm{eff}} > -1$. 

The history of the universe in this model can be interpreted as follows. 
The universe is created at $t = -t_{\mathrm{s}}$ because the value of 
the scale factor $a(t)$ in Eq.~(\ref{eq:3-A-4.3}) becomes zero 
$a(-t_{\mathrm{s}}) = 0$. During $-t_{\mathrm{s}} < t < t_{-}$, there is 
the first non-phantom phase. 
The first phantom phase in $t_{-} < t < 0$ corresponds to 
the inflationary stage. 
After inflation, the second non-phantom phase in $0 < t < t_{+}$ 
becomes the radiation/matter-dominated stages. 
Then, the second phantom phase in $t > t_{+}$ plays a role of 
the dark energy dominated stage, i.e., the late-time accelerated 
expansion of the universe. Finally, a Big Rip singularity occurs 
at $t = t_{\mathrm{s}}$. 
As a result, 
this model can present a unified scenario of 
inflation in the early universe and the late-time cosmic acceleration. 
Incidentally, phantom inflation has been studied in Ref.~\cite{PZZ-PIP}.

\subsubsection{Scalar field models with the crossing the phantom divide}

The instability of a single scalar field theory with 
the crossing the phantom divide was examined in Ref.~\cite{Vikman:2004dc}. 
In addition, 
the stability issue in a single scalar field theory as well as a 
two scalar field theory in which the crossing the phantom divide 
can be realized has recently been discussed in Ref.~\cite{Saitou:2012xw}. 
In this subsection, we first examine the stability of a single scalar field 
theory when the crossing the phantom divide occurs. 
The considerations for a two scalar field theory with 
the crossing the phantom divide are presented in Sec.~VI A 2. 

We define the new variables $Z \equiv \dot{\phi}$ and $Y \equiv I(\phi)/H$. 
With these variables, in the flat FLRW background 
the Friedmann equation (\ref{eq:2.3}) with 
Eq.~(\ref{eq:2.5})
and the equation of motion for $\phi$ (\ref{eq:ED1-10-Add-IVA-1}) are 
rewritten to 
\begin{eqnarray}
\frac{dZ}{dN} \Eqn{=} -\frac{I_{, \phi \phi} (\phi) \left(Z^2-Y\right)}{
2I_{, \phi} (\phi) H} -3\left(Z-Y\right)\,,
\label{eq:ED1-10-IVC5-Add-1} \\ 
\frac{dY}{dN} \Eqn{=} \frac{I_{, \phi} (\phi) \left(1-ZY\right)Z}{H^2}  
\,,
\label{eq:ED1-10-IVC5-Add-2}
\end{eqnarray}
where $I_{, \phi} (\phi) \equiv \partial I (\phi) /\partial \phi$ and 
$I_{, \phi \phi} (\phi) \equiv \partial^2 I (\phi) /\partial \phi^2$, 
$N$ is the number of $e$-folds and the scale factor is 
expressed as $a = \e^{N-N_0}$ with $N_0$ being the current value of $N$. 
For the solutions (\ref{eq:2.15}), we find $Z=1$ and $Y=1$. 
Hence, we examine the small perturbations $\delta Z$ and $\delta Y$ 
around this solution as follows 
\begin{eqnarray} 
Z \Eqn{=} 1+\delta Z\,,
\label{eq:ED1-10-IVC5-Add-3} \\   
Y \Eqn{=} 1+\delta Y\,,
\label{eq:ED1-10-IVC5-Add-4}
\end{eqnarray}
By using the solutions (\ref{eq:2.15}), we see that 
these perturbations obey the equation 
\begin{equation}
\frac{d}{dN}
\left(
\begin{array}{c} 
\delta X_{\phi} \\ 
\delta X_{\chi}  \\
\end{array} 
\right)
= {\mathcal M}_1
\left(
\begin{array}{c} 
\delta X_{\phi} \\ 
\delta X_{\chi}  \\
\end{array} 
\right)\,, 
\label{eq:ED1-10-IVC5-Add-5}
\end{equation}
where ${\mathcal M}_1$ is a matrix, given by 
\begin{equation} 
{\mathcal M}_1 \equiv 
\left( 
\begin{array}{cc} 
-\frac{\ddot{H}}{\dot{H} H}-3 & -3  \\ 
-\frac{\dot{H}}{H^2} & -\frac{\dot{H}}{H^2} \\ 
\end{array} 
\right)\,.
\label{eq:ED1-10-IVC5-Add-6} 
\end{equation}
Thus, we can obtain the eigenvalues of the matrix ${\mathcal M}_1$ as 
\begin{equation}  
m_{\pm} = \frac{1}{2} \left[ 
-\left( \frac{\ddot{H}}{\dot{H}H} + \frac{\dot{H}}{H^2} +3 \right) 
\pm \sqrt{
\left( \frac{\ddot{H}}{\dot{H}H} + \frac{\dot{H}}{H^2} +3 \right)^2 
-4\frac{\ddot{H}}{H^3}}\right]\,.
\label{eq:ED1-10-IVC5-Add-6} 
\end{equation}
The stability condition for the solutions (\ref{eq:2.15}) is that 
both of the eigenvalues $m_{+}$ and $m_{-}$ are negative. 
When the crossing of the phantom divide occurs around $\dot{H} \sim 0$, 
for $\ddot{H} >0$, the transition from the non-phantom (quintessence) 
phase ($\dot{H} <0$) to the phantom one ($\dot{H} >0$) 
occurs, whereas for $\ddot{H} <0$, the opposite direction transition from 
the phantom phase to the non-phantom one happens. 
In case of the expanding universe ($H>0$), the term 
$\ddot{H}/\left(\dot{H}H\right)$ is negative. 
Thus, around the crossing of the phantom divide, {}from 
Eq.~(\ref{eq:ED1-10-IVC5-Add-6})
we find $m_{+} \sim - \ddot{H}/\left(\dot{H}H\right) > 0$ and 
$m_{-} \sim 0$. As a result, at the crossing time when $\dot{H} =0$ 
$m_{+}$ becomes $+ \infty$ and therefore the solution in Eq.~(\ref{eq:2.15}) 
is unstable when the crossing of the phantom divide occurs. 
In other words, in a single scalar field theory 
the crossing of the phantom divide cannot be realized.

\section{Tachyon scalar field theory}

In this section, we examine a tachyon scalar by following 
Ref.~\cite{Book-Amendola-Tsujikawa}. 
The effective 4-dimensional action for the tachyon field which is an unstable 
mode of D-branes [non-Bobomol'nyi-Prasad-Sommerfield (non-BPS) branes] 
is given by 
\begin{equation}  
S=-\int d^4x V(\phi) \sqrt{-\det \left( g_{\mu\nu} + 
{\partial}_{\mu} \phi {\partial}_{\nu} \phi \right)}\,, 
\label{eq:5.1}
\end{equation}
where $\phi$ is a tachyon scalar field and $V(\phi)$ is a 
potential of $\phi$. 
{}From the action in Eq.~(\ref{eq:5.1}), 
the energy-momentum tensor of $\phi$ is derived as 
\begin{equation} 
T_{\mu\nu}^{(\phi)} 
= \frac{V(\phi) {\partial}_{\mu} \phi {\partial}_{\nu} \phi}{
\sqrt{1+g^{\alpha \beta} {\partial}_{\alpha} \phi {\partial}_{\beta}}} 
-g_{\mu\nu} V(\phi) \sqrt{1+g^{\alpha \beta} {\partial}_{\alpha} \phi 
{\partial}_{\beta}}\,. 
\label{eq:5.2}
\end{equation}
In the flat FLRW background~(\ref{eq:2.2}) (with $K=0$), 
$\rho_{\phi} = -T_0^{0\,(\phi)}$ and $P_{\phi} = T_i^{i\,(\phi)}$ are given by 
\begin{eqnarray}
\rho_{\phi} \Eqn{=} 
\frac{V(\phi)}{\sqrt{1-\dot{\phi}^2}}\,,
\label{eq:5.3} \\  
P_{\phi} \Eqn{=} 
-V(\phi) \sqrt{1-\dot{\phi}^2}\,.
\label{eq:5.4}
\end{eqnarray}
%
We remark that through the use of Eqs.~(\ref{eq:5.3}) and (\ref{eq:5.4}), 
this model may be written as a fluid. 
By using $\ddot{a}/{a} = H^2 + \dot{H}$ and 
plugging Eqs.~(\ref{eq:5.3}) and (\ref{eq:5.4}) into 
Eqs.~(\ref{eq:Add-2-01}) and (\ref{eq:Add-2-02}), we find 
\begin{equation}  
\frac{\ddot{a}}{a} = \frac{\kappa^2}{3} \frac{V(\phi)}{\sqrt{1-\dot{\phi}^2}} 
\left( 1- \frac{3}{2} \dot{\phi}^2 \right)\,. 
\label{eq:5.5}
\end{equation}
It follows from Eq.~(\ref{eq:5.5}) that the condition for the accelerated 
expansion $\ddot{a} > 0$ is given by $\dot{\phi}^2 <2/3$. 
Furthermore, the EoS for $\phi$ is represented by 
\begin{equation} 
w_{\phi} \equiv \frac{P_{\phi}}{\rho_{\phi}} = \dot{\phi}^2 - 1\,. 
\label{eq:5.6}
\end{equation}
Hence, from the above condition $\dot{\phi}^2 <2/3$ and Eq.~(\ref{eq:5.6}) 
we see that the possible range of the value of $w_{\phi}$ with realizing 
the cosmic acceleration is $-1 < w_{\phi} < -1/3$, which corresponds to 
the non-phantom (quintessence) phase. 

We derive expressions of $V(\phi)$ and $\phi$ in terms of $H$ and $\dot{H}$. 
By using $H^2 = \left(\kappa^2/3\right) V(\phi)/\sqrt{1-\dot{\phi}^2}$, which 
follows from Eq.~(\ref{eq:Add-2-01}) with Eq.~(\ref{eq:5.3}), and 
Eq.~(\ref{eq:5.5}), 
we have $\dot{H}/H^2 = - \left(3/2\right) \dot{\phi}^2$. 
The combination of these equations leads to 
\begin{eqnarray} 
V(\phi) \Eqn{=} \frac{3 H^2}{\kappa^2} 
\sqrt{1 + \frac{2 \dot{H}}{3 H^2}}\,,
\label{eq:5.7} \\  
\phi \Eqn{=} 
\int dt \sqrt{- \frac{2 \dot{H}}{3 H^2}}\,. 
\label{eq:5.8}
\end{eqnarray}
We suppose that the scale factor is expressed by a power-law expansion as 
$a(t) \propto t^p$ with $p > 1$ being a constant larger than unity. 
By combining this relation with Eqs.~(\ref{eq:5.7}) and (\ref{eq:5.8}), 
we obtain 
\begin{eqnarray} 
V(\phi) \Eqn{=} \frac{2p}{\kappa^2} \sqrt{1-\frac{2}{3p}} \frac{1}{\phi^2}\,. 
\label{eq:5.9} \\
\phi \Eqn{=} \sqrt{\frac{2}{3p}} t\,, 
\label{eq:5.10}
\end{eqnarray}
where we have taken the integration constant as zero. 
As a result, we acquire an inverse square power-law tachyon potential 
$V(\phi) \propto \phi^{-2}$. 

In the open string theory, the form of a tachyon potential $V(\phi)$ 
is given by~\cite{Kutasov:2003er}
\begin{equation} 
V(\phi) = \frac{V_0}{\cosh \left( \phi/\phi_0 \right)}\,, 
\label{eq:5.11}
\end{equation}
with $\phi_0 = \sqrt{2}$ for the non-BPS D-branes in the superstring 
and $\phi_0 = 2$ for the bosonic string. This form has a ground state 
in the limit $\phi \to \infty$. Here, $V_0$ and $\phi_0$ are constants and 
$V(\phi = \phi_0) = V_0$. 
Moreover, when a tachyon potential appears as 
the excitation of massive scalar fields on the anti D-branes, 
$V(\phi)$ is given by~\cite{T-P-2}
\begin{equation} 
V(\phi) = V_0 \e^{m^2 \phi^2 / 2}\,, 
\label{eq:5.12}
\end{equation}
where $m$ is the mass of $\phi$ and there exists a minimum of $V(\phi)$ at 
$\phi = 0$. 

We mention that a tachyon scalar field can be generalized to 
so-called k-essence~\cite{k-essence-DE}, which is a scalar field with non-canonical kinetic terms (for its application to inflation, 
so-called k-inflation, see~\cite{ArmendarizPicon:1999rj, Garriga:1999vw}), 
and for a unified scenario between inflation and late-time cosmic acceleration 
in the framework of k-essence model, see, e.g.,~\cite{Saitou:2011hv}). 
The action of k-essence is described by 
\begin{equation} 
S = 
\int d^4 x \sqrt{-g} \left( \frac{R}{2\kappa^2} + P(\phi, X) \right) + 
\int d^4 x 
{\mathcal{L}}_{\mathrm{M}} 
\left( g_{\mu\nu}, {\Psi}_{\mathrm{M}} \right)\,,
\label{eq:5.13}
\end{equation}
where $P(\phi, X)$ is a function of a scalar field $\phi$ and its kinetic term 
$X \equiv -\left(1/2\right) g^{\mu\nu} {\partial}_{\mu} \phi {\partial}_{\nu} 
\phi$. 
It is known that the accelerated expansion can be realized by the kinetic 
energy even without the scalar field potential. 
An effective 4-dimensional Lagrangian describing a tachyon field is 
given by~\cite{k-essence-Tachyon} 
\begin{equation} 
P = -V(\phi) \sqrt{1-2X}\,. 
\label{eq:5.14}
\end{equation}

Finally, we discuss a stability issue. 
We now consider the action in Eq.~(\ref{eq:5.13}). 
In this case, the energy density and pressure of $\phi$ are given by 
$\rho_{\phi} = 2X P_{,X}$ and $P_{\phi} = P$, respectively, 
where $P_{,X} \equiv \partial P/\partial X$. The EoS for $\phi$ is written as 
\begin{equation}  
w_{\phi} = \frac{P}{2X P_{,X} -P}\,. 
\label{eq:5.15}
\end{equation}
If the relation $|2X P_{,X}| \ll P$ is realized, $w_{\phi}$ can be 
close to $-1$. 

We investigate the stability conditions for k-essence in the 
ultra-violet (UV) regime. 
In the Minkowski background, 
we write $\phi$ as $\phi (t, \Vec{x}) = \phi_{\mathrm{b}} (t) + 
\delta \phi (t, \Vec{x})$, where $\phi_{\mathrm{b}} (t)$ is 
the background part and $\delta \phi (t, \Vec{x})$ is 
the perturbed one. 
By deriving the Lagrangian and Hamiltonian for $\delta \phi (t, \Vec{x})$, 
the second-order Hamiltonian is obtained as~\cite{Piazza:2004df}
\begin{equation}  
\delta \mathcal{H} = 
\xi_1 \frac{\left(\delta \dot{\phi}\right)^2}{2} 
+ \xi_2 \frac{\left(\nabla \delta \phi\right)^2}{2} 
+ \xi_3 \frac{\left(\delta \phi\right)^2}{2}\,,
\label{eq:5.16}
\end{equation}
with 
\begin{eqnarray}
\xi_1 \Eqn{\equiv} 
\left(P_{,X} + 2X P_{,XX} \right) \geq 0\,, 
\label{eq:5.17} \\ 
\xi_2 \Eqn{\equiv} P_{,X} \geq 0\,, 
\label{eq:5.18} \\
\xi_3 \Eqn{\equiv} 
- P_{,\phi \phi} \geq 0\,, 
\label{eq:5.19}
\end{eqnarray} 
where $P_{,XX} \equiv \partial^2P/\partial X^2$ and 
$P_{,\phi \phi} \equiv \partial^2P/\partial \phi^2$. 
The stability conditions are represented by the positivity of 
each three terms on the r.h.s. of Eq.~(\ref{eq:5.16}). 
The propagation speed of $\phi$ is defined as 
\begin{equation}
c_{\mathrm{s}}^{2} \equiv \frac{P_{\phi,X}}{\rho_{\phi,X}} 
=  \frac{\xi_2}{\xi_1}\,. 
\label{eq:5.20}
\end{equation}
As long as the stability conditions $\xi_1 \geq 0$ in Eq.~(\ref{eq:5.17}) and 
$\xi_2 \geq 0$ in Eq.~(\ref{eq:5.18}) are satisfied, 
$c_{\mathrm{s}}^2 \geq 0$. 
In addition, the condition that the propagation speed should be sub-luminal 
is given by~\cite{Garriga:1999vw}
\begin{equation} 
P_{,XX} >0\,. 
\label{eq:5.21}
\end{equation}
We note that 
the finite-time future singularities in tachyon cosmology 
have also been examined in Ref.~~\cite{GKMP-KGGMK-TS}.

\section{Multiple scalar field theories}

In this section, we describe multiple scalar field theories.  

\subsection{Two scalar field theories}

To begin with, 
in this subsection we investigate two scalar field theories. 
First, we explain the standard type of two scalar field theories. 
Next, we discuss a new type of two scalar field theories 
with realizing the crossing of the phantom divide, 
which has recently been constructed in Ref.~\cite{Saitou:2012xw}.

\subsubsection{Standard two scalar field theories}

First, we explore the standard type two scalar field theories~\cite{Nojiri:2005pu, Capozziello:2005tf, Elizalde:2008yf}. 
The action of two scalar field theories in general relativity is given by 
\begin{equation} 
S = 
\int d^4 x \sqrt{-g} \left( \frac{R}{2\kappa^2} 
- \frac{1}{2} \omega (\phi) 
g^{\mu\nu} {\partial}_{\mu} \phi {\partial}_{\nu} \phi 
- \frac{1}{2} \sigma (\chi) 
g^{\mu\nu} {\partial}_{\mu} \chi {\partial}_{\nu} \chi 
- V(\phi, \chi) \right)\,,
\label{eq:6.1}
\end{equation}
where $\sigma (\chi)$ is a functions of the scalar field $\chi$ and 
$V(\phi, \chi)$ is the potential term of $\phi$ and $\chi$. 
Here, we concentrate on the scalar-field part of the action and 
do not take into account the matter part of it. 
If there does not exist the second scalar field $\chi$, the action in 
Eq.~(\ref{eq:6.1}) is the same as the action in Eq.~(\ref{eq:2.1}) without 
the matter part of it. 

In the FLRW background~(\ref{eq:2.2}), the Einstein equations are given by
\begin{eqnarray}
H^2 \Eqn{=} \frac{\kappa^2}{3} \rho_{\mathrm{t}}\,,
\label{eq:6.2} \\ 
\dot{H} \Eqn{=} -\frac{\kappa^2}{2} \left( 
\rho_{\mathrm{t}} + P_{\mathrm{t}} \right)\,,
\label{eq:6.3}
\end{eqnarray}
where 
$\rho_{\mathrm{t}}$ and $P_{\mathrm{t}}$ are the total energy density and 
pressure of the two scalar fields $\phi$ and $\chi$, respectively, given by 
\begin{eqnarray}
\rho_{\mathrm{t}} \Eqn{=} \frac{1}{2} \omega (\phi) \dot{\phi}^2 
+\frac{1}{2} \sigma (\chi) \dot{\chi}^2 
+V(\phi, \chi)\,,
\label{eq:6.4} \\
P_{\mathrm{t}} \Eqn{=} \frac{1}{2} \omega (\phi) \dot{\phi}^2 
+\frac{1}{2} \sigma (\chi) \dot{\chi}^2 
-V(\phi, \chi)\,. 
\label{eq:6.5}
\end{eqnarray}
Moreover, 
equations of motion for the scalar fields are given by 
\begin{eqnarray} 
\omega (\phi) \ddot{\phi} + \frac{1}{2} \frac{d \omega (\phi)}{d \phi} 
\dot{\phi}^2 + 3H \omega (\phi) \dot{\phi} 
+ \frac{\partial V(\phi, \chi)}{\partial \phi} \Eqn{=} 0\,,
\label{eq:6.6} \\ 
\sigma (\phi) \ddot{\chi} + \frac{1}{2} \frac{d \sigma (\chi)}{d \chi} 
\dot{\chi}^2 + 3H \sigma (\chi) \dot{\chi} 
+ \frac{\partial V(\phi, \chi)}{\partial \chi} \Eqn{=} 0\,. 
\label{eq:6.7}
\end{eqnarray}

Since it is possible to redefine the scalar fields through 
a convenient transformation, 
we take $\phi = \chi = t$. 
If a solution $H(t) = I(t)$ is given, where $I(t)$ is a 
function of $t$, by plugging 
Eq.~(\ref{eq:6.2}) into Eqs.~(\ref{eq:6.6}) and (\ref{eq:6.7}), 
we obtain 
\begin{equation}
\omega (\phi) = -\frac{2}{\kappa^2} 
\frac{\partial I(\phi, \chi)}{\partial \phi}\,, 
\quad 
\sigma (\chi) = -\frac{2}{\kappa^2} 
\frac{\partial I(\phi, \chi)}{\partial \chi}\,.
\label{eq:6.8} 
\end{equation}
Here, by taking $\phi = \chi = t$ into consideration, we can interpret 
$I(\phi, \chi)$ as $I(t, t) \equiv I(t)$. 
Thus, we define $I(\phi, \chi)$ as 
\begin{equation}
I(\phi, \chi) = -\frac{\kappa^2}{2} 
\left( \int \omega (\phi) d \phi + \int \sigma (\chi) d \chi \right)\,.
\label{eq:6.9} 
\end{equation}
We can represent the potential term $V(\phi, \chi)$ by 
\begin{equation}
V(\phi, \chi) = \frac{1}{\kappa^2} 
\left( 3 I^2 (\phi, \chi) + \frac{\partial I(\phi, \chi)}{\partial \phi} 
+ \frac{\partial I(\phi, \chi)}{\partial \chi} \right)\,.
\label{eq:6.10} 
\end{equation}
Furthermore, Eq.~(\ref{eq:6.3}) is rewritten to 
\begin{equation}
-\frac{2}{\kappa^2} \frac{d I(t)}{d t} 
= \omega (t) + \sigma (t)\,.
\label{eq:6.11} 
\end{equation}
In addition, we can describe 
a part of coefficient of the kinetic terms, which is a function of 
a scalar field, 
$\omega (\phi)$ and 
$\sigma (\chi)$, as 
%
\begin{eqnarray}
\omega (\phi) \Eqn{=} 
-\frac{2}{\kappa^2} 
\left( \frac{d I(\phi)}{d \phi} + \tilde{g} (\phi) \right)\,, 
\label{eq:6.12} \\
\sigma (\chi) \Eqn{=} 
\frac{2}{\kappa^2} \tilde{g} (\chi)\,, 
\label{eq:6.13}
\end{eqnarray} 
%
where $\tilde{g} (\phi)$ is an arbitrary function of $\phi$. 
Thus, by substituting Eq.~(\ref{eq:6.9}) with Eqs.~(\ref{eq:6.12}) and 
(\ref{eq:6.13}) into Eq.~(\ref{eq:6.10}), we acquire 
(for details, see~\cite{Elizalde:2008yf})
\begin{equation} 
V(\phi, \chi) = \frac{1}{\kappa^2} 
\left( 3 I^2 (\phi, \chi) + \frac{d I(\phi)}{d \phi} 
+ \tilde{g} (\phi) - \tilde{g} (\chi) \right)\,.
\label{eq:6.14} 
\end{equation}

In order to demonstrate an example of two scalar field theories, we examine 
the following model in Eq.~(\ref{eq:3-A-3.1}): 
\begin{equation} 
I(t) = \frac{H_2}{t_{\mathrm{s}} - t} + \frac{H_3}{t^2}\,. 
\label{eq:6.15}
\end{equation}
In this case, by using Eqs.~(\ref{eq:6.12}) and (\ref{eq:6.15}) 
we find that $\omega (\phi)$ is expressed as 
%
\begin{equation} 
\omega (\phi) = 
-\frac{2}{\kappa^2} 
\left[ \frac{H_2}{\left(t_{\mathrm{s}} - t\right)^2} 
- \frac{2 H_3}{\phi^3} + \tilde{g} (\phi) \right]\,, 
\label{eq:6.16} 
\end{equation}
%
and $\sigma (\chi)$ is described as in Eqs.~(\ref{eq:6.13}). 
In addition, it follows from Eqs.~(\ref{eq:6.9}), (\ref{eq:6.10}) and 
(\ref{eq:6.15}) that $I(\phi, \chi)$ and $V(\phi, \chi)$ are 
represented as 
%
\begin{eqnarray}
I(\phi, \chi) \Eqn{=} \frac{H_2}{t_{\mathrm{s}} - \phi} + \frac{H_3}{\phi^2} 
+ \int d \phi \omega (\phi)  - \int d \chi \tilde{g} (\chi)\,.
\label{eq:6.17} \\ 
V(\phi, \chi) \Eqn{=} \frac{1}{\kappa^2} 
\left[ 3 I^2 (\phi, \chi) + \frac{H_2}{\left(t_{\mathrm{s}} - t\right)^2} 
- \frac{2 H_3}{\phi^3} 
+ \tilde{g} (\phi) - \tilde{g} (\chi) \right]\,. 
\label{eq:6.18}
\end{eqnarray}
%

We investigate the stability of the system described by the solution in 
Eq.~(\ref{eq:6.15}). 
We define the following variables: 
\begin{equation} 
X_{\phi} \equiv \dot{\phi}\,, 
\quad 
X_{\chi} \equiv \dot{\chi}\,, 
\quad 
Y \equiv \frac{I(\phi, \chi)}{H}\,. 
\label{eq:6.19}
\end{equation}
By using the variables in Eq.~(\ref{eq:6.19}), 
Eqs.~(\ref{eq:6.2}), (\ref{eq:6.6}) and 
(\ref{eq:6.7}) are rewritten to 
\begin{eqnarray}  
\frac{d X_{\phi}}{d N}
\Eqn{=} -\frac{1}{2H} \frac{d \omega (\phi) /d \phi}{\omega (\phi)} 
\left(X_{\phi}^2 - 1 \right) -3\left(X_{\phi} - Y \right)\,,
\label{eq:6.20} \\  
\frac{d X_{\chi}}{d N}
\Eqn{=} -\frac{1}{2H} \frac{d \sigma (\chi) /d \chi}{\sigma (\chi)} 
\left(X_{\chi}^2 - 1 \right) -3\left(X_{\chi} - Y \right)\,, 
\label{eq:6.21} \\
\frac{d Y}{d N}
\Eqn{=} \frac{\kappa^2}{2H^2} \left[ \omega (\phi) X_{\phi} 
\left(Y X_{\phi} - 1 \right) + \sigma (\chi) X_{\chi} 
\left(Y X_{\chi} - 1 \right)
\right]\,,
\label{eq:6.22}
\end{eqnarray}
where we have used $d/dN = \left( 1/H \right) d/dt$. 
We analyze the perturbations $|\delta X_{\phi}| \ll 1$, 
$|\delta X_{\chi}| \ll 1$ and $|\delta Y| \ll 1$ 
around $(X_{\phi}, X_{\chi}, Y) = (1, 1, 1)$ as 
\begin{equation} 
X_{\phi} = 1 + \delta X_{\phi}\,, 
\quad 
X_{\chi} = 1 + \delta X_{\chi}\,, 
\quad 
Y = 1 + \delta Y\,. 
\label{eq:6.23}
\end{equation}

The perturbations satisfy the equation 
\begin{equation}
\frac{d}{dN}
\left(
\begin{array}{c} 
\delta X_{\phi} \\ 
\delta X_{\chi}  \\
\delta Y \\
\end{array} 
\right)
= M 
\left(
\begin{array}{c} 
\delta X_{\phi} \\ 
\delta X_{\chi}  \\
\delta Y \\
\end{array} 
\right)\,, 
\label{eq:6.24}
\end{equation}
where $M$ is the matrix, defined by 
\begin{equation} 
M \equiv 
\left( 
\begin{array}{ccc} 
-\frac{d \omega (\phi) /d \phi}{H \omega (\phi)}-3 & 0 & 3 \\ 
0 & -\frac{d \sigma (\chi) /d \chi}{H \sigma (\chi)}-3 & 3 \\
\kappa^2 \frac{\omega (\phi)}{2H^2} & \kappa^2 \frac{\sigma (\chi)}{2H^2} & 
\kappa^2 \frac{\omega (\phi) + \sigma (\chi)}{2H^2} \\
\end{array} 
\right)\,.
\label{eq:6.25} 
\end{equation}
The characteristic equation for $M$ is given by 
\begin{eqnarray} 
&&
\det |M - \lambda E| = 
\nonumber \\
&& 
\left( \frac{d \omega (\phi) /d \phi}{H \omega (\phi)}+3 + \lambda \right) 
\left( \frac{d \sigma (\chi) /d \chi}{H \sigma (\chi)}+3 + \lambda \right) 
\left( \kappa^2 \frac{\omega (\phi) + \sigma (\chi)}{2H^2} - \lambda \right) 
\nonumber \\ 
&& 
{}+ \frac{3 \kappa^2 \omega (\phi)}{2H^2} 
\left( \frac{d \sigma (\chi) /d \chi}{H \sigma (\chi)}+3 + \lambda \right) 
+ \frac{3 \kappa^2 \sigma (\phi)}{2H^2} 
\left( \frac{d \omega (\phi) /d \phi}{H \omega (\phi)}+3 + \lambda \right) 
=0\,,
\label{eq:6.26}
\end{eqnarray}
where $\lambda$ denotes an eigenvalue of $M$ and $E$ is a unit matrix. 
We impose the following conditions 
\begin{equation} 
\omega (\phi) \neq 0\,, 
\quad 
\sigma (\chi) \neq 0\,,  
\label{eq:6.27}
\end{equation}
so that the eigenvalues can be finite without diverging. 
As a result, if the conditions Eq.~(\ref{eq:6.27}) are satisfied, 
the solution in Eq.~(\ref{eq:6.15}) does not have infinite instability 
at the crossing of the phantom divide from the non-phantom phase to 
the phantom one. 
For an illustration, e.g., 
we take, $\tilde{g} (t) = \tilde{g}_{\mathrm{c}}/t^3$ with 
$\tilde{g}_{\mathrm{c}}$ being a constant and satisfying 
$\tilde{g}_{\mathrm{c}} > 2H_3$. {}From Eq.~(\ref{eq:6.17}), we find 
\begin{equation} 
I(\phi, \chi) = \frac{H_2}{t_{\mathrm{s}} - \phi} 
- \frac{\tilde{g}_{\mathrm{c}} -2 H_3}{2\phi^2} 
+ \frac{\tilde{g}_{\mathrm{c}}}{2\chi^2}\,.
\label{eq:6.28}
\end{equation}
Moreover, it follows from Eqs.~(\ref{eq:6.12}) and (\ref{eq:6.13}) that 
\begin{eqnarray}
\omega (\phi) \Eqn{=} 
-\frac{2}{\kappa^2} 
\left[ \frac{H_2}{\left( t_{\mathrm{s}} - \phi \right)^2} 
- \frac{2 H_3}{\phi^3} 
+ \frac{\tilde{g}_{\mathrm{c}}}{\phi^3} 
\right]\,, 
\label{eq:6.29} \\
\sigma (\chi) \Eqn{=} 
\frac{2}{\kappa^2} \frac{\tilde{g}_{\mathrm{c}}}{\chi^3} \,. 
\label{eq:6.30}
\end{eqnarray} 
Furthermore, by using Eq.~(\ref{eq:6.18}) we obtain 
\begin{equation} 
V(\phi, \chi) = \frac{1}{\kappa^2} 
\left[ 3 I^2 (\phi, \chi) 
+ \frac{H_2}{\left( t_{\mathrm{s}} - \phi \right)^2} 
+ \frac{\tilde{g}_{\mathrm{c}} -2 H_3}{\phi^3} 
- \frac{\tilde{g}_{\mathrm{c}}}{\phi^3} 
\right]\,.
\label{eq:6.31}
\end{equation}
%
Examples of two scalar field theories are an oscillating quintom 
model~\cite{FLPZ-ENO-Q} or a quintom with two scalar 
fields~\cite{Zhang:2005eg} in the framework of general relativity 
(see also~\cite{M-Li, Cai:2009zp}). 

\subsubsection{New type of two scalar field theories} 

A new type of two scalar field theories is given by~\cite{Saitou:2012xw} 
\begin{equation} 
S = 
\int d^4 x \sqrt{-g} \left( \frac{R}{2\kappa^2} 
+X +\omega (\phi) X^2  -U +\eta(\chi) U^2 - V(\phi, \chi)
\right)
+ \int d^4 x 
{\mathcal{L}}_{\mathrm{M}} 
\left( g_{\mu\nu}, {\Psi}_{\mathrm{M}} \right)\,,
\label{eq:ED1-10-Add-VIA2-1}
\end{equation}
where $X \equiv -\left(1/2\right) g^{\mu\nu} {\partial}_{\mu} \phi {\partial}_{\nu} \phi$, $U \equiv -\left(1/2\right) g^{\mu\nu} {\partial}_{\mu} \chi {\partial}_{\nu} \chi$. 
We suppose that there is no direct interaction between the scalar fields and 
matters, so that the continuity equation of matter can be satisfied 
$\rho^{\prime}_{\mathrm{M}} (N) + 3 \left( \rho_{\mathrm{M}} (N) + P_{\mathrm{M}} (N) \right) = 0$, where a prime denotes the partial derivative with respect to $N$ of $\partial/\partial N$. 
In the flat FLRW background, 
for the action in Eq.~(\ref{eq:ED1-10-Add-VIA2-1}) 
the gravitational field equations are given by 
\begin{eqnarray} 
X +\omega (\phi) X^2  -U +\eta(\chi) U^2 - V(\phi, \chi) 
\Eqn{=} 
-\frac{1}{\kappa^2} \left( 2HH^{\prime} + 3H^2 \right) 
+ \rho_{\mathrm{M}} + \frac{\rho^{\prime}_{\mathrm{M}}}{3}\,,
\label{eq:ED1-10-Add-VIA2-2} \\ 
X +2\omega (\phi) X^2  -U +2\eta(\chi) U^2 
\Eqn{=} 
-\frac{1}{\kappa^2} HH^{\prime}
+ \rho_{\mathrm{M}} + \frac{\rho^{\prime}_{\mathrm{M}}}{6}\,. 
\label{eq:ED1-10-Add-VIA2-3}
\end{eqnarray}
On the other hand, the Friedmann equation is also represented as 
$H^2 (N) = \left( \kappa^2/3 \right) \left( \rho_{\mathrm{s}} (N) + \rho_{\mathrm{M}} (N) \right)$, where $\rho_{\mathrm{s}}$ is the energy density of the 
scalar fields, namely, we have expressed the right-hand side by dividing the 
energy density into the contributions from the scalar fields and matter. 
By using this expression, the gravitational field equations 
(\ref{eq:ED1-10-Add-VIA2-2}) and (\ref{eq:ED1-10-Add-VIA2-3}) are 
rewritten to 
\begin{eqnarray}
V(\phi, \chi) - \frac{1}{2} \left( X-U \right) 
\Eqn{=} \rho_{\mathrm{s}} (N) + \frac{\rho^{\prime}_{\mathrm{s}} (N)}{4}\,,
\label{eq:ED1-10-Add-VIA2-4} \\ 
\omega (\phi) X^2  +\eta(\chi) U^2 + \frac{1}{2} \left( X-U \right) 
\Eqn{=} 
-\frac{\rho^{\prime}_{\mathrm{s}} (N)}{12}\,.
\label{eq:ED1-10-Add-VIA2-5}
\end{eqnarray}
Furthermore, the equation of motion for the scalar fields are given by 
\begin{eqnarray}
\hspace{-10mm}
V_{, \phi}(\phi, \chi) +3\omega_{, \phi}(\phi)X^2 
\Eqn{=} 
-H^2 \left( 1 + 6 \omega X \right) \phi^{\prime \prime} 
-\left[ HH^{\prime} \left( 1 + 6 \omega X \right) + 3H^2 + 6H^2 \omega X
\right] \phi^{\prime}\,, 
\label{eq:ED1-10-Add-VIA2-6} \\ 
\hspace{-10mm}
V_{, \chi}(\phi, \chi) +3\eta_{, \chi}(\chi)U^2 
\Eqn{=} 
H^2 \left( 1 - 6 \eta U \right) \chi^{\prime \prime} 
+\left[ HH^{\prime} \left( 1 - 6 \eta U \right) + 3H^2 - 6H^2 \eta U 
\right] \chi^{\prime}\,, 
\label{eq:ED1-10-Add-VIA2-7}
\end{eqnarray}
where the subscription ``${}_{, \phi}$'' denotes a partial derivative 
with respect to $\phi$, e.g., 
$V_{, \phi}(\phi, \chi) \equiv \partial V(\phi, \chi)/\partial \phi$  
and
$V_{, \chi}(\phi, \chi) \equiv \partial V(\phi, \chi)/\partial \chi$. 
Thus, we acquire the following solutions 
\begin{eqnarray} 
\phi \Eqn{=} \chi = \bar{m} N\,, 
\label{eq:ED1-10-Add-VIA2-8} \\ 
H^2 (N)
\Eqn{=} 
\frac{\kappa^2}{3} \left( J(N) + \rho_{\mathrm{M}} (N) \right)\,, 
\label{eq:ED1-10-Add-VIA2-9}
\end{eqnarray}
provided that $V(\phi, \chi)$, $\omega (\phi)$ and $\eta (\chi)$ 
satisfy the equations 
\begin{eqnarray} 
\hspace{-10mm}
V(\bar{m} N, \bar{m} N) \Eqn{=} J(N) + \frac{J^{\prime} (N)}{4}\,, 
\label{eq:ED1-10-Add-VIA2-10} \\ 
\omega (\bar{m} N) + \eta (\bar{m} N) 
\Eqn{=} 
-\frac{J^{\prime} (N)}{3\bar{m}^4 H^4 (N)}\,, 
\label{eq:ED1-10-Add-VIA2-11} \\ 
\hspace{-10mm}
V_{, \phi}(\bar{m} N, \bar{m} N) + \frac{3}{4}\bar{m}^4 H^4 (N) 
\omega_{, \phi} (\bar{m} N) 
\Eqn{=} 
-\bar{m} \left( 3H^2 (N) + HH^{\prime} (N) \right) 
\nonumber \\ 
\hspace{-10mm}
&&
{}- 3\bar{m}^2 H^2 (N) \omega (\bar{m} N) 
\left( H^2 (N) + HH^{\prime} (N) \right)\,, 
\label{eq:ED1-10-Add-VIA2-12} \\ 
\hspace{-10mm}
V_{, \chi}(\bar{m} N, \bar{m} N) + \frac{3}{4}\bar{m}^4 H^4 (N) 
\eta_{, \chi} (\bar{m} N) 
\Eqn{=} 
\bar{m} \left( 3H^2 (N) + HH^{\prime} (N) \right) 
\nonumber \\ 
\hspace{-10mm}
&&
{}- 3\bar{m}^2 H^2 (N) \eta (\bar{m} N) 
\left( H^2 (N) + HH^{\prime} (N) \right)\,, 
\label{eq:ED1-10-Add-VIA2-13}
\end{eqnarray}
where $J(N)$ is an arbitrary function and 
$\bar{m}$ is a constant. 
We derive the explicit forms of 
$\omega (\phi)$, $\eta (\chi)$ and $V(\phi, \chi)$ in the action in 
Eq.~(\ref{eq:ED1-10-Add-VIA2-1}) so that these should obey 
Eqs.~(\ref{eq:ED1-10-Add-VIA2-10})--(\ref{eq:ED1-10-Add-VIA2-13}). 
As an example, by using an arbitrary function $\bar{\alpha}$ and 
$J$ we can express $\omega (\phi)$ and $\eta (\chi)$ as 
\begin{eqnarray} 
\omega (\phi) \Eqn{=} 
-\frac{1}{3\bar{m}^4 H^4 (\phi/\bar{m})} 
\frac{\partial J(\phi/\bar{m})}{\partial \phi} + \bar{\alpha} (\phi)\,, 
\label{eq:ED1-10-Add-VIA2-14} \\ 
\eta(\chi) \Eqn{=} -\bar{\alpha} (\chi)\,. 
\label{eq:ED1-10-Add-VIA2-15} 
\end{eqnarray}
In addition, we define another function $\tilde{J} (\phi, \chi)$ as 
\begin{eqnarray} 
\tilde{J} (\phi, \chi) \Eqn{\equiv}  
-\bar{m} \left[ 
\int d\phi^{\prime} \left( 3\bar{m}^2 \omega (\phi^{\prime}) H^4 (\phi^{\prime}/\bar{m}) +2H^2 (\phi^{\prime}/\bar{m}) \right) 
\right. \nonumber \\ 
&& \left. 
{}+\int d\chi^{\prime} \left( 3\bar{m}^2 \eta (\chi^{\prime}) H^4 (\chi^{\prime}/\bar{m}) -2H^2 (\chi^{\prime}/\bar{m}) \right) \right]\,. 
\label{eq:ED1-10-Add-VIA2-16} 
\end{eqnarray}
It follows from Eq.~(\ref{eq:ED1-10-Add-VIA2-16}) that 
$
\tilde{J} (\bar{m} N, \bar{m} N) = J (N)
$. 
With Eq.~(\ref{eq:ED1-10-Add-VIA2-16}), we determine the form of 
$V(\phi, \chi)$ as 
\begin{equation}  
V(\phi, \chi) = -\bar{m} \left( 
\int d\phi^{\prime} H^2 (\phi^{\prime}/\bar{m}) - 
\int d\chi^{\prime} H^2 (\chi^{\prime}/\bar{m})
\right) + \tilde{J} (\phi, \chi) + \frac{\bar{m}}{4} 
\left( \frac{\partial \tilde{J} (\phi)}{\partial \phi} 
+ \frac{\partial \tilde{J} (\chi)}{\partial \chi} 
\right)\,. 
\label{eq:ED1-10-Add-VIA2-17}
\end{equation}
Thus, through this procedure, it is possible to reconstruct 
any expanding history of the universe by using two arbitrary functions 
$J$ of $\phi^{\prime}/\bar{m}$ or $\chi^{\prime}/\bar{m}$
and $\bar{\alpha}$ of $\phi$ or $\chi$. 

Next, we explore the stability of the solutions 
(\ref{eq:ED1-10-Add-VIA2-8}) and 
(\ref{eq:ED1-10-Add-VIA2-9}). 
There are two approaches to study the stability. 
One is the perturbative analysis. Another is to examine the 
sound speed of the scalar fields. 
First, we investigate the first order perturbations 
from the solutions, given by 
\begin{equation} 
\phi = \phi_0 + \delta \phi (N)\,, 
\quad 
\chi = \chi_0 + \delta \chi (N)\,,
\quad
\dot{\phi} = \dot{\phi}_0 + \delta x (N)\,,
\quad  
\dot{\chi} = \dot{\chi}_0 + \delta y (N)\,,
\label{eq:ED1-10-Add-VIA2-18}
\end{equation}
where we have defined $\delta x (N) \equiv \delta \dot{\phi} (N)$ and 
$\delta y (N) \equiv \delta \dot{\chi} (N)$. 
{}From the Friedmann equation, there exists the following constraint 
between the perturbations 
\begin{eqnarray}
\hspace{-13mm}
&&
\delta H = 
\frac{\kappa^2 \delta \rho_{\mathrm{s}}}{6H} 
\nonumber \\
\hspace{-13mm} 
&&
= \dot{\phi} \left( 1+3\omega \dot{\phi}^2 \right) \delta x 
+\left( \frac{3\omega_{, \phi} \dot{\phi}^4}{4}+V_{, \phi} \right) 
\delta \phi 
+\dot{\chi} \left( -1+3\eta \dot{\chi}^2 \right) \delta y 
+\left( \frac{3\eta_{, \chi} \dot{\chi}^4}{4}+V_{, \chi} \right) 
\delta \chi\,.
\label{eq:ED1-10-Add-VIA2-19}
\end{eqnarray}
Hence, in this system there are four degrees of freedom $(\phi, \chi, 
\dot{\phi}, \dot{\chi})$. 
By combining Eqs.~(\ref{eq:ED1-10-Add-VIA2-4})--(\ref{eq:ED1-10-Add-VIA2-7}) and(\ref{eq:ED1-10-Add-VIA2-18}) 
and using Eq.~(\ref{eq:ED1-10-Add-VIA2-19}), 
we obtain 
\begin{equation}
\frac{d}{dN}
\left(
\begin{array}{ccc} 
\delta \phi \\ 
\delta \chi \\
\delta x \\ 
\delta y \\
\end{array} 
\right)
= {\mathcal M}_2
\left(
\begin{array}{ccc} 
\delta \phi \\ 
\delta \chi \\
\delta x \\ 
\delta y \\
\end{array} 
\right)\,, 
\label{eq:ED1-10-Add-VIA2-20}
\end{equation}
where ${\mathcal M}_1$ is a matrix, given by 
\begin{equation} 
{\mathcal M}_2 \equiv 
\left( 
\begin{array}{cccc} 
0 & 0 & H^{-1} & 0 \\ 
0 & 0 & 0 & H^{-1} \\ 
{\mathcal M}_2^{(31)} & {\mathcal M}_2^{(32)} & {\mathcal M}_2^{(33)} & {\mathcal M}_2^{(34)} \\ 
{\mathcal M}_2^{(41)} & {\mathcal M}_2^{(42)} & {\mathcal M}_2^{(43)} & {\mathcal M}_2^{(44)} \\ 
\end{array} 
\right)\,,
\label{eq:ED1-10-Add-VIA2-21} 
\end{equation}
with 
\begin{eqnarray}
{\mathcal M}_2^{(31)} \Eqn{=} 
\frac{2HH^{\prime} -2H^{\prime 2} +3\bar{m}^3 \omega_{, \phi} H^3 H^{\prime} 
+ \bar{m}^2 \kappa^2 H^2 \left(1+\bar{m}^2 \omega H^2 \right)}{H 
\left(1+3\bar{m}^2 \omega H^2 \right)} 
\nonumber \\ 
&&
{}+\frac{1}{H} 
\left[ 4HH^{\prime} +3H^{\prime 2} +HH^{\prime \prime} + 
\frac{\bar{m}^2 \kappa^2}{2} \left(1+\bar{m}^2 \omega H^2 \right) 
\left(H^2+HH^{\prime} \right)
\right]\,,
\label{eq:ED1-10-Add-VIA2-22} \\
{\mathcal M}_2^{(32)} \Eqn{=} 
\frac{\bar{m}^2 \kappa^2 \left(1+\bar{m}^2 \omega H^2 \right)}{H
\left(1+3\bar{m}^2 \omega H^2 \right)}
\left[ 
\frac{1}{2}\left(H^2+HH^{\prime} \right)
\left(-1+3\bar{m}^2 \eta H^2 \right) -H^2
\right]
\,,
\label{eq:ED1-10-Add-VIA2-23} \\
{\mathcal M}_2^{(33)} \Eqn{=} 
-3-\frac{\bar{m}^2 \kappa^2}{2} \left(1+\bar{m}^2 \omega H^2 \right) 
- \frac{3\bar{m}^2 H \left(2\omega H^{\prime} +\bar{m} \omega_{, \phi} H 
\right)}{1+3\bar{m}^2 \omega H^2}
\,,
\label{eq:ED1-10-Add-VIA2-24} \\
{\mathcal M}_2^{(34)} \Eqn{=} 
-\frac{\bar{m}^2 \kappa^2}{2} 
\frac{1+\bar{m}^2 \omega H^2}{1+3\bar{m}^2 \omega H^2} 
\left(-1+3\bar{m}^2 \eta H^2 \right)\,,
\label{eq:ED1-10-Add-VIA2-25} \\
{\mathcal M}_2^{(41)} \Eqn{=} 
\frac{\bar{m}^2 \kappa^2 \left(-1+\bar{m}^2 \eta H^2 \right)}{H
\left(-1+3\bar{m}^2 \eta H^2 \right)} 
\left[ \frac{1}{2}\left(H^2+HH^{\prime} \right)
\left(1+3\bar{m}^2 \omega H^2 \right) +H^2 \right]\,,
\label{eq:ED1-10-Add-VIA2-26} \\
{\mathcal M}_2^{(42)} \Eqn{=} 
-\frac{2HH^{\prime} -2H^{\prime 2} -3\bar{m}^3 \eta_{, \chi} H^3 H^{\prime} 
+ \bar{m}^2 \kappa^2 H^2 \left(-1+\bar{m}^2 \eta H^2 \right)}{H 
\left(-1+3\bar{m}^2 \eta H^2 \right)} 
\nonumber \\ 
&&
{}+\frac{1}{H} 
\left[ 4HH^{\prime} +3H^{\prime 2} +HH^{\prime \prime} + 
\frac{\bar{m}^2 \kappa^2}{2} \left(-1+\bar{m}^2 \eta H^2 \right) 
\left(H^2+HH^{\prime} \right)
\right]\,,
\label{eq:ED1-10-Add-VIA2-27} \\
{\mathcal M}_2^{(43)} \Eqn{=} 
-\frac{\bar{m}^2 \kappa^2}{2} 
\frac{-1+\bar{m}^2 \eta H^2}{-1+3\bar{m}^2 \eta H^2} 
\left(1+3\bar{m}^2 \omega H^2 \right)
\,,
\label{eq:ED1-10-Add-VIA2-28} \\
{\mathcal M}_2^{(44)} \Eqn{=} 
-3-\frac{\bar{m}^2 \kappa^2}{2} \left(-1+\bar{m}^2 \eta H^2 \right) 
- \frac{3\bar{m}^2H \left(2\eta H^{\prime} +\bar{m} \eta_{, \chi} H 
\right)}{-1+3\bar{m}^2 \eta H^2}
\,.
\label{eq:ED1-10-Add-VIA2-29} 
\end{eqnarray}
The condition for the solutions (\ref{eq:ED1-10-Add-VIA2-8}) and 
(\ref{eq:ED1-10-Add-VIA2-9}) to be stable is 
that the real part of all the eigenvalues of the matrix ${\mathcal M}_2$ 
(\ref{eq:ED1-10-Add-VIA2-21}) should be negative. 
The characteristic equation for ${\mathcal M}_2$ is given by 
$\det |{\mathcal M}_2 - \lambda E| = 
\lambda^4 + A_1 \lambda^3 + A_2 \lambda^2 + A_3 \lambda + A_4 = 0$, 
where 
$
A_1 = -{\mathcal M}_2^{(33)} -{\mathcal M}_2^{(44)}
$, 
$
A_2 = {\mathcal M}_2^{(33)}{\mathcal M}_2^{(44)} - {\mathcal M}_2^{(34)}{\mathcal M}_2^{(43)} -\left( {\mathcal M}_2^{(31)} +{\mathcal M}_2^{(42)} 
\right)H^{-1}
$, 
$
A_3 = \left( {\mathcal M}_2^{(31)}{\mathcal M}_2^{(44)} + 
{\mathcal M}_2^{(33)}{\mathcal M}_2^{(42)} 
-{\mathcal M}_2^{(34)}{\mathcal M}_2^{(41)}
-{\mathcal M}_2^{(32)}{\mathcal M}_2^{(43)}
\right)H^{-1}
$, 
and 
$
A_4 = \left( {\mathcal M}_2^{(31)}{\mathcal M}_2^{(42)} 
-{\mathcal M}_2^{(32)}{\mathcal M}_2^{(41)} 
\right)H^{-2}
$. 
The solutions of this equation are given by 
$
\lambda = \left( \pm \sqrt{2\Xi -B_1} \pm \sqrt{-B_1 -2\Xi \pm 4\sqrt{\Xi-B_3}} 
\right)/2 - A_1/4
$, 
where $\Xi$ satisfies the cubic equation 
$
\Xi^3 - \left(B_1/2 \right)\Xi^2 -B_3 \Xi +B_1B_3/2 - B_2^2/8 = 0
$. 
Here, 
$
B_1 = -3A_1^2/8 + A_2
$, 
$
B_2 = A_1^3/8 - A_1 A_2/2 + A_3
$, 
and 
$
B_3 = -3A_1^4/256 + A_1^2 A_2/16 -A_1 A_3/4 + A_4
$.
The solution of this cubic equation is given by 
$
\Xi = \left(-{\mathcal P} - \sqrt{{\mathcal P}^2 + {\mathcal Q}^3} 
\right)^{1/3} + \left(-{\mathcal P} + \sqrt{{\mathcal P}^2 + {\mathcal Q}^3 } 
\right)^{1/3} + B_1/6
$, 
with 
$
{\mathcal P} \equiv -B_1^3/216 + B_1 B_3/6 - B_2^2/16 
$
and 
$
{\mathcal Q} \equiv -B_1^2/36 - B_2/3 
$. 
We explore the real solution of the cubic equation so that 
the maximum of the real parts of $\lambda$, which is described by 
${\mathrm {Re}} \lambda_{\mathrm{max}}$, can be negative. 
As a result, $\Xi$ is expressed as follows. 
(i) For ${\mathcal Q} < 0$ and ${\mathcal P}^2 < |{\mathcal Q}^3|$, 
$
\Xi = 2\sqrt{|{\mathcal Q}|} \cos \left[ 
\left(1/3\right) \arccos\left(-{\mathcal P} |{\mathcal Q}|^{-3/2}\right) 
\right] + B_1/6
$. 
(ii) For ${\mathcal Q} < 0$, ${\mathcal P}^2 > |{\mathcal Q}^3|$ and 
${\mathcal P} \geq 0$, 
$
\Xi = -\left({\mathcal P} +\sqrt{{\mathcal P}^2 + {\mathcal Q}^3} 
\right)^{1/3} -\left({\mathcal P} - \sqrt{{\mathcal P}^2 + {\mathcal Q}^3} 
\right)^{1/3} + B_1/6
$. 
(iii) For ${\mathcal Q} < 0$, ${\mathcal P}^2 > |{\mathcal Q}^3|$ and 
${\mathcal P} < 0$, 
$
\Xi = \left(-{\mathcal P} +\sqrt{{\mathcal P}^2 + {\mathcal Q}^3} 
\right)^{1/3} +\left(-{\mathcal P} - \sqrt{{\mathcal P}^2 + {\mathcal Q}^3} 
\right)^{1/3} + B_1/6
$. 
(iv) For ${\mathcal Q} > 0$, 
$
\Xi = -\left({\mathcal P} +\sqrt{{\mathcal P}^2 + {\mathcal Q}^3} 
\right)^{1/3} +\left(-{\mathcal P} +\sqrt{{\mathcal P}^2 + {\mathcal Q}^3} 
\right)^{1/3} + B_1/6
$. 

Moreover, ${\mathrm {Re}} \lambda_{\mathrm{max}}$ is 
represented as follows. 
If $\Xi^2 \geq B_3$, 
\begin{eqnarray}
{\mathrm {Re}} \lambda_{\mathrm{max}}
\Eqn{=} 
\frac{1}{2} \left( \sqrt{2\Xi-B_1} + \sqrt{-2\Xi-B_1 +4\sqrt{\Xi-B_3}} 
\right) - \frac{A_1}{4} 
\nonumber \\ 
\hspace{10mm}
&&
\mathrm{for}\, \,
\Xi \geq \frac{B_1}{2}\, \, 
\mathrm{and}\, \, 
-2\Xi-B_1+4l \geq 0 
\nonumber \\ 
\Eqn{=} 
\frac{1}{2} \sqrt{2\Xi-B_1} - \frac{A_1}{4} \leq 0 
\quad 
\quad 
\mathrm{for}\, \, 
\Xi \geq \frac{B_1}{2}\, \, 
\mathrm{and}\, \, 
-2\Xi-B_1+4l < 0 
\nonumber \\ 
\Eqn{=} 
\frac{1}{2} \sqrt{-2\Xi-B_1 +4\sqrt{\Xi-B_3}} - \frac{A_1}{4} 
\quad 
\quad 
\mathrm{for}\, \, 
\Xi < \frac{B_1}{2}\, \, 
\mathrm{and}\, \, 
-2\Xi-B_1+4l \geq 0
\nonumber \\ 
\Eqn{=} 
-\frac{A_1}{4}
\quad 
\quad 
\mathrm{for}\, \, 
\Xi < \frac{B_1}{2}\, \,  
\mathrm{and}\, \,  
-2\Xi-B_1+4l < 0
\nonumber \\ 
\Eqn{\leq} 0\,.
\label{eq:ED1-10-Add-VIA2-30} 
\end{eqnarray}
On the other hand, if $\Xi^2 < B_3$,  
\begin{eqnarray}
{\mathrm {Re}} \lambda_{\mathrm{max}}
\Eqn{=} 
\frac{1}{2} \left\{ \sqrt{2\Xi-B_1} + 
\sqrt{
\frac{1}{2} \left[ 
-2\Xi-B_1 +\sqrt{ \left(2\Xi+B_1\right)^2 + 16 \left( \Xi-B_3 \right)}
\right]
} \right\} -\frac{A_1}{4} 
\nonumber \\ 
\hspace{10mm}
&&
\mathrm{for}\, \,
\Xi \geq \frac{B_1}{2}
\nonumber \\ 
\Eqn{=} 
\frac{1}{2} \sqrt{
\frac{1}{2} \left[ 
-2\Xi-B_1 +\sqrt{ \left(2\Xi+B_1\right)^2 + 16 \left( \Xi-B_3 \right)}
\right]} -\frac{A_1}{4} 
\nonumber \\ 
\hspace{10mm}
&&
\mathrm{for}\, \,
\Xi < \frac{B_1}{2}
\nonumber \\ 
\Eqn{\leq} 0\,.
\label{eq:ED1-10-Add-VIA2-31} 
\end{eqnarray} 
Thus, 
if the conditions described by Eqs.~(\ref{eq:ED1-10-Add-VIA2-30}) 
and (\ref{eq:ED1-10-Add-VIA2-31}) are satisfied, 
the solutions (\ref{eq:ED1-10-Add-VIA2-8}) and 
(\ref{eq:ED1-10-Add-VIA2-9}) can be stable. 
Since these solutions (\ref{eq:ED1-10-Add-VIA2-8}) and 
(\ref{eq:ED1-10-Add-VIA2-9}) correspond to those 
(\ref{eq:2.15}) for a single scalar field theory discussed in 
Sec.~IV, 
through the investigations in Sec.~IV C 5, 
the stability of these solutions implies 
that the crossing of the phantom divide can be realized 
in the new type of two scalar field theories described by 
the action in Eq.~(\ref{eq:ED1-10-Add-VIA2-1}). 

In principle, the stability condition from the perturbative analysis can 
yield constraints on the forms of, e.g., $\bar{\alpha}^{\prime} (\bar{m} N)$, 
$\bar{\alpha} (\bar{m} N)$, $J(N)$ and $\omega_{, \phi} (\bar{m} N )$. 
However, it is difficult to derive the explicit analytical expressions 
of such a constraint on $\bar{\alpha}^{\prime} (\bar{m} N)$ or 
$J(N)$. 
On the other hand, the stability condition obtained by 
the sound speed of the scalar fields can present 
the analytical representations of constraints on 
$\bar{\alpha} (\bar{m} N)$. 
Therefore, we explore the sound speed of the scalar fields. 
Its square has to be positive for the stability of the universe. 
The sound speed $c_{\mathrm{s} j}^2$, where $j =\phi\,, \chi$, 
of the scalar fields are defined as  
%
$
c_{\mathrm{s} \phi}^2 \equiv 
P_{\phi, X}/\rho_{\phi, X} = 
\left(1+2\omega X\right)/\left(1+6\omega X\right)
$ 
and 
$
c_{\mathrm{s} \chi}^2 \equiv 
P_{\chi, X}/\rho_{\chi, X} = 
\left(-1+2\eta X\right)/\left(-1+6\eta X\right)
$. 
Hence, the stability condition is expressed as 
$0 \leq c_{\mathrm{s} j}^2 (\leq 1)$. 
For the solutions (\ref{eq:ED1-10-Add-VIA2-8}) and 
(\ref{eq:ED1-10-Add-VIA2-9}), 
this condition can lead to constraints on the function $\bar{\alpha}$ 
as 
$
\bar{\alpha} (\phi = \bar{m} N) \geq J^{\prime}/\left( 3\bar{m}^4H^4 \right) 
$ 
or 
$
\bar{\alpha} (\phi = \bar{m} N) \leq J^{\prime}/\left( 3\bar{m}^4H^4 \right) 
-1/\left( \bar{m}^2H^2 \right) 
$, 
and 
$
\bar{\alpha} (\phi = \bar{m} N) \geq 0
$ 
or 
$
\bar{\alpha} (\phi = \bar{m} N) \leq 
-1/\left( \bar{m}^2H^2 \right) 
$. 

It is interesting to mention that 
in Sec.~IV C 5, we have shown that 
the crossing of the phantom divide cannot occur 
in a single scalar field theory represented by 
the action in Eq.~(\ref{eq:2.1}) 
because of the instability of the solutions 
(\ref{eq:2.15}), 
whereas in the new type of two scalar field theories whose 
action is given by Eq.~(\ref{eq:ED1-10-Add-VIA2-1}), 
the crossing of the phantom divide can happen 
due to the stability of the solutions (\ref{eq:ED1-10-Add-VIA2-8}) and 
(\ref{eq:ED1-10-Add-VIA2-9}). 
This result can be a proposal of a clue for the searches on the 
non-equivalence of dark energy models on a theoretical level.

\subsection{$n (\geq 2)$ scalar field theories} 

We generalize the investigations for two scalar field theories in Sec.~VI B. 
The action of $n$ scalar field theories in general relativity is given by 
\begin{equation} 
S = 
\int d^4 x \sqrt{-g} \left( \frac{R}{2\kappa^2} 
- \frac{1}{2} \sum_{i=1}^n \omega_i (\phi_i) 
g^{\mu\nu} {\partial}_{\mu} \phi_i {\partial}_{\nu} \phi_i 
- V(\phi_1, \phi_2, \dotsb, \phi_n) \right)\,,
\label{eq:6.32}
\end{equation}
where $\omega_i (\phi_i)$ is a function of a scalar field $\phi_i$ and 
$V(\phi_1, \phi_2, \dotsb, \phi_n)$ is the potential term of $n$ scalar 
fields $\phi_1, \phi_2, \dotsb, \phi_n$. 
For $n=2$, i.e., two scalar field theories, this action in Eq.~(\ref{eq:6.32}) 
with $\phi_1 = \phi$, $\phi_2 = \chi$, $\omega_1 (\phi_1) = \omega (\phi)$ 
and $\omega_2 (\phi_2) = \sigma (\chi)$ 
is equivalent to that in Eq.~(\ref{eq:6.1}). 
In the FLRW background~(\ref{eq:2.2}), the Einstein equations are given by 
Eqs.~(\ref{eq:6.2}) and (\ref{eq:6.3}) with $\rho_{\mathrm{t}}$ and 
$P_{\mathrm{t}}$ being 
\begin{eqnarray}
\rho_{\mathrm{t}} \Eqn{=} 
\frac{1}{2} \sum_{i=1}^n \omega_i (\phi_i) \dot{\phi}_i^2 + 
V(\phi_1, \phi_2, \dotsb, \phi_n)\,,
\label{eq:6.33} \\
P_{\mathrm{t}} \Eqn{=}  
\sum_{i=1}^n \omega_i (\phi_i) \dot{\phi}_i^2\,. 
\label{eq:6.34}
\end{eqnarray}

We can apply the same procedure executed in the case of 
two scalar field theories for the $n$ scalar field ones. 
{}From Eq.~(\ref{eq:6.11}), we analogously find 
\begin{equation} 
\sum_{i=1}^n \omega_i (t) = 
-\frac{2}{\kappa^2} \frac{d I(t)}{d t}\,. 
\label{eq:6.35}
\end{equation}
Thus, from Eqs.~(\ref{eq:6.12}) and (\ref{eq:6.14}), we obtain 
\begin{eqnarray}
\omega_i (\phi_i) 
\Eqn{=} 
-\frac{2}{\kappa^2} 
\frac{\partial I(\phi_1, \dotsb, \phi_n)}{\partial \phi_i}\,,
\label{eq:6.36} \\ 
V(\phi_1, \dotsb, \phi_n)
\Eqn{=} 
\frac{1}{\kappa^2} \left( 3 I^2 (\phi_1, \dotsb, \phi_n) 
+ \sum_{i=1}^n \frac{\partial I(\phi_1, \dotsb, \phi_n)}{\partial \phi_i} 
\right)\,, 
\label{eq:6.37}
\end{eqnarray}
where $I(\phi_1, \dotsb, \phi_n)$ can be regarded as 
$I(t, \dotsb, t) \equiv I(t)$. 
Furthermore, we acquire the following solutions: 
\begin{equation} 
\phi_i = t\,, 
\quad 
H(t) = I (t)\,. 
\label{eq:6.38}
\end{equation}

In addition, from Eq.~(\ref{eq:6.35}) we take 
\begin{eqnarray}
\omega_1 (\phi_1) 
\Eqn{=} 
-\frac{2}{\kappa^2} 
\left(
\frac{d I(\phi_1)}{d \phi_1} + \tilde{g}_2 (\phi_1) + \dotsb 
+ \tilde{g}_n (\phi_1) \right)\,, 
\label{eq:6.39} \\  
\omega_j (\phi_j) 
\Eqn{=} 
\frac{2}{\kappa^2} \tilde{g}_j (\phi_j)\,, 
\quad 
j = 2, 3, \dots, n\,.
\label{eq:6.40}
\end{eqnarray}
Hence, 
there exist $n-1$ arbitrary functions $\tilde{g}_2, \dotsb, \tilde{g}_n$ 
which can reproduce the solution (\ref{eq:6.38}), 
and therefore the reconstruction can be executed through the procedure 
explained above. 

Here, we explicitly demonstrate the equivalence of multiple scalar field 
theories to fluid descriptions. 
By combining the consequences in Eqs.~(\ref{eq:6.36})--(\ref{eq:6.38}) 
and the investigations in Eqs.~(\ref{eq:IVC-1})--(\ref{eq:IVC-3}) 
in terms of the equivalence between fluid descriptions and 
scalar field theories, we obtain 
\begin{eqnarray}
f(\rho) \Eqn{=} -\sum_{i=1}^n \omega_i (t)\,, 
\label{eq:ED1-9-Add-VIB-1} \\
P_{\mathrm{t}} \Eqn{=} -\rho_{\mathrm{t}} -f(\rho) 
= -V(t) + \frac{1}{2} \sum_{i=1}^n \omega_i (t)\,,
\label{eq:ED1-9-Add-VIB-2} \\ 
\rho_{\mathrm{t}} \Eqn{=} 
V - \frac{f(\rho)}{2} 
= V(t) + \frac{1}{2} \sum_{i=1}^n \omega_i (t)\,,
\label{eq:ED1-9-Add-VIB-3} \\
w_{\mathrm{eff}} \Eqn{=} \frac{P_{\mathrm{t}}}{\rho_{\mathrm{t}}} 
= \frac{-2V(t) + \sum_{i=1}^n \omega_i (t)}{2V(t) + \sum_{i=1}^n \omega_i (t)}
\,,
\label{eq:ED1-9-Add-VIB-4}
\end{eqnarray}
where in deriving the second equalities in Eqs.~(\ref{eq:ED1-9-Add-VIB-2}) and 
(\ref{eq:ED1-9-Add-VIB-3}) we have used Eq.~(\ref{eq:ED1-9-Add-VIB-1}), 
and Eq.~(\ref{eq:ED1-9-Add-VIB-4}) follows from 
Eqs.~(\ref{eq:ED1-9-Add-VIB-2}) and 
(\ref{eq:ED1-9-Add-VIB-3}). 
Accordingly, these results imply that 
multiple scalar field theories can also be represented by 
a fluid description, similarly to that for a scalar field theory 
as shown in Sec.~IV B. 

\section{Holographic dark energy}

In this section, we study holographic dark energy scenario and 
its generalization by following Ref.~\cite{Nojiri:2005pu} 
through the analogy with anti de Sitter (AdS)/ conformal field theory (CFT) 
correspondence. 
In particular, 
we investigate the case in which the infrared (IR) cut-off scale 
is represented by a combination of 
the particle and future horizons, the time when a Big Rip singularity 
appears (namely, the life time of the universe), 
the Hubble parameter and the length scale coming from the cosmological constant
.

\subsection{Model of holographic dark energy}

First, we make an overview for a model of holographic dark energy~\cite{Li:2004rb}. 
The energy density of holographic dark energy is proposed as 
\begin{equation} 
\rho_{\mathrm{h}} \equiv 
\frac{3 C_{\mathrm{h}}^2}{\kappa^2 L_{\mathrm{h}}^2}\,, 
\label{eq:VII-1}
\end{equation}
with $C_{\mathrm{h}}$ being a numerical constant. 
Here, $L_{\mathrm{h}}$ is the IR 
cut-off scale with a dimension of 
length. 
At the dark energy dominated stage, we suppose that the contribution of 
matter is negligible. 
{}From Eq.~(\ref{eq:Add-2-01}), we have the following Friedmann equation: 
$
3H^2/\kappa^2 = \rho_{\mathrm{h}}
$. By combining this equation and Eq.~(\ref{eq:VII-1}), we find 
\begin{equation} 
H = \frac{C_{\mathrm{h}}}{L_{\mathrm{h}}}\,, 
\label{eq:VII-2}
\end{equation}
where $C_{\mathrm{h}} (> 0)$ is assumed to be positive in order to 
describe the expanding universe. 

It is necessary for us to discuss how to take the IR cut-off $L_{\mathrm{h}}$ 
because if the IR cut-off is identified with the Hubble parameter, 
the cosmic acceleration cannot be realized. 
We define the particle $L_{\mathrm{ph}}$ and future $L_{\mathrm{fh}}$ 
horizons as 
\begin{eqnarray}
L_{\mathrm{ph}} \Eqn{\equiv} 
a(t) \int_{0}^{t} \frac{d t^{\prime}}{a (t^{\prime})}\,, 
\label{eq:VII-3} \\ 
L_{\mathrm{fh}} \Eqn{\equiv} a(t) \int_{t}^{\infty}  
\frac{d t^{\prime}}{a (t^{\prime})}\,. 
\label{eq:VII-4}
\end{eqnarray}
In the flat FLRW space-time in Eq.~(\ref{eq:2.2}) with $K=0$, 
provided that the IR cut-off scale $L_{\mathrm{h}}$ is 
identified with the particle horizon $L_{\mathrm{ph}}$ or the future 
horizon $L_{\mathrm{fh}}$, we obtain 
\begin{equation}  
\frac{d }{d t} \left( \frac{C_{\mathrm{h}}}{a H} \right) = \pm 
\frac{1}{a}\,, 
\label{eq:VII-5} 
\end{equation}
where the plus $+$ (minus $-$) sign corresponds to $L_{\mathrm{ph}}$ 
($L_{\mathrm{fh}}$). We can solve Eq.~(\ref{eq:VII-5}) as 
\begin{eqnarray}
a \Eqn{=} a_{\mathrm{c}} t^{h_{\mathrm{c}}}\,, 
\label{eq:VII-6} \\ 
h_{\mathrm{c}} \Eqn{\equiv} \frac{1}{1 \pm 1/C_{\mathrm{h}}}\,. 
\label{eq:VII-7}
\end{eqnarray}
Thus, 
for $L_{\mathrm{h}} = L_{\mathrm{fh}}$, an accelerated expansion of 
the universe can be realized, because the power index $h_{\mathrm{c}}$ in terms of $t$ in Eq.~(\ref{eq:VII-7}) is larger than unity and hence power-law inflation can occur. 
On the other hand, for $L_{\mathrm{h}} = L_{\mathrm{ph}}$, 
the universe will shrink due to $h_{\mathrm{c}} < 0$. 

We suppose that the theory is invariant under the change of the time direction 
as $t \to -t$. In addition, by shifting the origin of time appropriately, 
we have the following expression for $a$ instead of Eq.~(\ref{eq:VII-6}): 
\begin{equation}  
a = a_{\mathrm{c}} \left( t_{\mathrm{s}} - t \right)^{h_{\mathrm{c}}}\,. 
\label{eq:VII-8} 
\end{equation}
Thus, for $h_{\mathrm{c}} < 0$, a Big Rip singularity will appear at 
$t = t_{\mathrm{s}}$ because $a$ diverges at that time. 

If we change the direction of time, the particle horizon becomes like 
a future one as 
\begin{equation} 
L_{\mathrm{ph}} \rightarrow 
\tilde{L}_{\mathrm{fh}} \equiv 
a(t) \int_{t}^{t_{\mathrm{s}}}  
\frac{d t^{\prime}}{a (t^{\prime})}
= a(t) \int_{0}^{\infty}  
\frac{d a}{H a^2}\,. 
\label{eq:VII-9} 
\end{equation}
{}From Eq.~(\ref{eq:2.17}) with Eq.~(\ref{eq:VII-6}) or Eq.~(\ref{eq:VII-8}), 
we obtain 
\begin{equation}  
w_{\mathrm{eff}} = -1 + \frac{2}{3 h_{\mathrm{c}}}\,. 
\label{eq:VII-10} 
\end{equation}

We remark that when we take $L_{\mathrm{h}}$ as the future horizon 
$L_{\mathrm{fh}}$ in Eq.~(\ref{eq:VII-4}), we can acquire a de Sitter solution 
\begin{eqnarray}
a \Eqn{=} a_{\mathrm{c}} \e^{H_{\mathrm{dS}} t}\,, 
\label{eq:VII-11} \\ 
H_{\mathrm{dS}} \Eqn{\equiv} \frac{1}{l}\,, 
\label{eq:VII-12}  
\end{eqnarray}
where $l$ is a constant denoting a length scale and 
hence $H_{\mathrm{dS}}$ is a constant 
Hubble parameter describing de Sitter space. 
It follows from Eq.~(\ref{eq:VII-1}) and $L_{\mathrm{fh}} = l$ 
that $\rho_{\mathrm{h}} = 3 C_{\mathrm{h}}^2/\left( \kappa^2 l^2 \right)$. 
For $C_{\mathrm{h}} = 1$, by using Eq.~(\ref{eq:VII-12}) we identically find 
the Friedmann equation (\ref{eq:Add-2-01}) as 
$3H^2/\kappa^2 = \rho_{\mathrm{h}}$, 
whereas for $C_{\mathrm{h}} \neq 1$, the de Sitter solution (\ref{eq:VII-12}) 
cannot be satisfied. 
When we choose $L_{\mathrm{h}}$ as the particle horizon, 
there does not exist the de Sitter solution 
because the particle horizon in Eq.~(\ref{eq:VII-3}) varies in time as 
$L_{\mathrm{ph}} = \left( 1 - \e^{t/l} \right)/l$ 
and not a constant.

\subsection{Generalized holographic dark energy}

Next, we explain generalized holographic dark energy~\cite{Nojiri:2005pu, Elizalde:2005ju}. 
It has been pointed out in Ref.~\cite{H-ES} that 
if $L_{\mathrm{h}}$ is taken as $L_{\mathrm{ph}}$, 
the EoS vanishes because $L_{\mathrm{ph}}$ behaves as being 
proportional to $a^{3/2}(t)$, 
although the value of the energy density is compatible with 
the observations. 
Therefore, we examine the generalization of holographic dark energy. 
In more general, $L_{\mathrm{h}}$ could be represented 
as a function of 
both $L_{\mathrm{ph}}$ and $L_{\mathrm{fh}}$. 
Provided that the life time of the universe is finite, 
then $t_{\mathrm{s}}$ can correspond to an IR cut-off, 
and $L_{\mathrm{fh}}$ in Eq.~(\ref{eq:VII-4}) is not well-defined 
because of the finiteness of the cosmic time $t$. 
Thus, the future horizon may be re-defined by 
\begin{equation} 
L_{\mathrm{fh}} \rightarrow 
\tilde{L}_{\mathrm{fh}} \equiv 
a(t) \int_{t}^{t_{\mathrm{s}}}  
\frac{d t^{\prime}}{a (t^{\prime})}
= a(t) \int_{0}^{\infty}  
\frac{d a}{H a^2}\,, 
\label{eq:VII-13} 
\end{equation}
as in (\ref{eq:VII-9}). 
By analogy with AdS/CFT correspondence, 
we suppose that $L_{\mathrm{h}}$ may be described by 
\begin{equation} 
L_{\mathrm{h}}  
= L_{\mathrm{h}} (L_{\mathrm{ph}}, \tilde{L}_{\mathrm{fh}}, t_{\mathrm{s}})\,, 
\label{eq:VII-14} 
\end{equation}
as long as $L_{\mathrm{ph}}$, $\tilde{L}_{\mathrm{fh}}$ and $t_{\mathrm{s}}$ are finite, because there exist a lot of possible choices for 
the IR cut-off~\cite{G-O-IRCO}. 
We note that holographic dark energy from the Ricci scalar curvature, the 
so-called Ricci dark energy has been explored 
in Ref.~\cite{Ricci-dark-energy}. 
Moreover, there exist a number of studies on holographic dark energy 
in theoretical aspects as well as observational constraints~\cite{H-D-E}. 
A concrete example is given by~\cite{Elizalde:2005ju}
\begin{eqnarray}
\frac{L_{\mathrm{h}}}{C_{\mathrm{h}}} 
\Eqn{=} \frac{\mathcal{X}}{h_{\mathrm{c}} \left( 1+\mathcal{X} \right)^2}\,, 
\label{eq:VII-15} \\ 
\mathcal{X} \Eqn{\equiv} 
\left( \frac{t_{\mathrm{s}}}{L_{\mathrm{ph}} + \tilde{L}_{\mathrm{fh}}} 
B(\bar{p}, \bar{q}) \right)^{1/h_{\mathrm{c}}}\,, 
\label{eq:VII-16}
\end{eqnarray}
where $h_{\mathrm{c}} > 0$, and $B(\bar{p}, \bar{q})$ is a beta-function 
defined by 
\begin{eqnarray} 
&& 
B(\bar{p}, \bar{q}) \equiv
\int_0^{\infty} dt \frac{t^{\bar{p}-1}}{\left( 1+t \right)^{\bar{p} + \bar{q}}}
\,, 
\label{eq:VII-17} \\ 
&& 
\bar{p} \equiv
1+h_{\mathrm{c}}\,, 
\quad 
\bar{q} \equiv 1-h_{\mathrm{c}}\,. 
\label{eq:VII-18}
\end{eqnarray}
It follows from Eq.~(\ref{eq:VII-15}) that the solution is given by 
\begin{eqnarray}
H \Eqn{=} h_{\mathrm{c}} \left( \frac{1}{t} + \frac{1}{t_{\mathrm{s}} - t} 
\right)\,, 
\label{eq:VII-19} \\ 
a \Eqn{=} 
a_{\mathrm{c}} \left( \frac{t}{t_{\mathrm{s}} - t} \right)^{h_{\mathrm{c}}}\,.
\label{eq:VII-20}
\end{eqnarray}
Furthermore, since we have 
\begin{equation} 
L_{\mathrm{fh}} + \tilde{L}_{\mathrm{fh}} 
= a(t) \int_0^{t_{\mathrm{s}}} \frac{dt}{a}
= t_{\mathrm{s}} \left( \frac{t}{t_{\mathrm{s}} - t} \right)^{h_{\mathrm{c}}} 
B(\bar{p}, \bar{q})\,, 
\label{eq:VII-21} 
\end{equation}
by substituting Eq.~(\ref{eq:VII-21}) into Eq.~(\ref{eq:VII-15}), we acquire 
\begin{equation} 
\frac{C_{\mathrm{h}}}{L_{\mathrm{h}}} 
= h_{\mathrm{c}} \left( \frac{1}{t} + \frac{1}{t_{\mathrm{s}} - t} \right) 
= H\,, 
\label{eq:VII-22} 
\end{equation}
where the second equality follows from Eq.~(\ref{eq:VII-20}). 
Clearly, Eq.~(\ref{eq:VII-22}) is equivalent to Eq.~(\ref{eq:VII-2}). 

We mention that 
for the solution (\ref{eq:VII-20}), from Eq.~(\ref{eq:2.17}) we find 
$
w_{\mathrm{eff}} = -1+ 2\left( t_{\mathrm{s}} - 2t \right)/ 
\left( 3 h_{\mathrm{c}} t_{\mathrm{s}} \right)
$. 
When $t \to 0$, $w_{\mathrm{eff}} \to -1 + 2/\left( 3 h_{\mathrm{c}} \right) 
> -1$, i.e., the universe is in the non-phantom phase. 
At $t = t_{\mathrm{s}}/2$, $w_{\mathrm{eff}} = -1$. 
After that, in the limit of $t \to t_{\mathrm{s}}$, 
$w_{\mathrm{eff}} \to -1 - 2/\left( 3 h_{\mathrm{c}} \right) 
< -1$, i.e., the universe is in the phantom phase. 
Consequently, it can occur the crossing of the phantom divide. 

We also investigate the case that there exists matter with its EoS being 
$w_{\mathrm{m}} \equiv P_{\mathrm{m}}/ \rho_{\mathrm{m}}$. 
In what follows, with $w_{\mathrm{m}}$ we define $h_{\mathrm{c}}$ as 
$h_{\mathrm{c}} \equiv \left( 2/3 \right) / 
\left( 1 + w_{\mathrm{m}} \right)$. 
We assume the existence of an interaction between holographic 
matters~\cite{interaction between holographic matters}. 
The equation for matter corresponding to the continuity equation is given by 
\begin{equation} 
\dot{\rho}_{\mathrm{m}} + 3H \left( \rho_{\mathrm{m}} + 
P_{\mathrm{m}} \right) = 3H \frac{4 \rho_{\mathrm{c}}}{3 h_{\mathrm{c}}} 
\frac{\left( 1 + \mathcal{X} \right)^3}{\mathcal{X}^2}\,, 
\label{eq:VII-23} 
\end{equation}
where $\rho_{\mathrm{c}}$ is a constant. 
We also suppose 
\begin{equation}
\frac{L_{\mathrm{h}}}{C_{\mathrm{h}}} 
= \left( 1 - \frac{\kappa^2 \rho_{\mathrm{c}}}{3 h_{\mathrm{c}}^2} 
\right)^{-1/2} 
\frac{\mathcal{X}}{h_{\mathrm{c}} \left( 1+\mathcal{X} \right)^2}\,. 
\label{eq:VII-24} 
\end{equation}
In this case, 
the Friedmann equation (\ref{eq:Add-2-01}) becomes 
$3H^2/\kappa^2 = \rho_{\mathrm{h}} + \rho_{\mathrm{m}}$. 
Thus, we obtain the solution (\ref{eq:VII-20}) as well as 
$\rho_{\mathrm{m}}$, given by
\begin{equation} 
\rho_{\mathrm{m}} = \rho_{\mathrm{c}} \left( \frac{1}{t} + 
\frac{1}{t_{\mathrm{s}} - t} \right)^2\,. 
\label{eq:VII-25} 
\end{equation}
By using the Friedmann equation 
$3H^2/\kappa^2 = \rho_{\mathrm{h}} + \rho_{\mathrm{m}}$, 
Eqs.~(\ref{eq:VII-20}) and (\ref{eq:VII-25}), we find that 
the ratio of the energy density of holographic dark energy in 
Eq.~(\ref{eq:VII-1}) to that of matter becomes constant and 
it is given by 
\begin{equation} 
\frac{\rho_{\mathrm{h}}}{\rho_{\mathrm{m}}} = 
\frac{3 h_{\mathrm{c}}}{\kappa^2 \rho_{\mathrm{c}}} 
\left( 1 - \frac{\kappa^2 \rho_{\mathrm{c}}}{3 h_{\mathrm{c}}} \right)\,. 
\label{eq:VII-26} 
\end{equation}
This consequence may be a resolution of coincidence problem between 
the current energy density of dark energy and that of dark matter. 

In Ref.~\cite{Pavon:2005yx}, a naive model of such an interaction scenario 
between dark energy and dark matter in (\ref{eq:VII-23})
realizing a similar result with a constant $w_{\mathrm{eff}}$ 
has been proposed, although in the present case of $H$ in 
Eq.~(\ref{eq:VII-19}) we have a dynamical $w_{\mathrm{eff}}$. 
In addition, it has been examined in Ref.~\cite{Wang:2005jx} that 
in more general case, the ratio of the energy density of dark energy to 
that of matter is not constant. 

In addition, by extending the relation (\ref{eq:VII-14}), 
we examine more general cases that 
$L_{\mathrm{h}}$ depends on the Hubble parameter $H$ and 
the length scale which originates from the cosmological constant as 
$\Lambda = 12/l^2$, i.e., $L_{\mathrm{h}}$ is expressed as 
\begin{equation} 
L_{\mathrm{h}} = 
L_{\mathrm{h}} (L_{\mathrm{ph}}, \tilde{L}_{\mathrm{fh}}, t_{\mathrm{s}}, 
H, l)\,, 
\label{eq:VII-27} 
\end{equation}
or 
\begin{equation} 
L_{\mathrm{h}} =
L_{\mathrm{h}} (L_{\mathrm{ph}}, L_{\mathrm{fh}}, t_{\mathrm{s}}, 
H, l)\,.
\label{eq:VII-28}
\end{equation}
%
The proposal of generalized holographic dark energy in the form in Eqs.~(\ref{eq:VII-27}) or (\ref{eq:VII-28}) has been made in Ref.~\cite{Nojiri:2005pu} where instead of $H$, the scale factor $a$ was used, supposing that such a cut-off may depend on the scale factor and its derivatives (i.e., also from $H$). Hence, this is the most general proposal for the IR cut-off which eventually covers all known proposals. 

An example in such an extended class of generalized holographic dark energy 
is given by 
\begin{equation} 
\frac{C_{\mathrm{h}}}{L_{\mathrm{h}}} 
= \frac{1}{\alpha_{\mathrm{h}} L_{\mathrm{fh}}}
\left[ \alpha_{\mathrm{h}} + 1 + 2 \left( \alpha_{\mathrm{h}}^2 - \alpha_{\mathrm{h}} -1 \right) \left( \frac{L_{\mathrm{h}}}{\alpha_{\mathrm{h}} l} \right) + 
2 \left( \alpha_{\mathrm{h}}^3 - 2\alpha_{\mathrm{h}}^2 + \alpha_{\mathrm{h}} + 1 \right) \left( \frac{L_{\mathrm{h}}}{\alpha_{\mathrm{h}} l} \right)^2 
\right]\,, 
\label{eq:VII-29} 
\end{equation}
where $\alpha_{\mathrm{h}} (> 0)$ is a positive dimensionless parameter. 
Here, we find
$L_{\mathrm{h}} > 0$ due to $C_{\mathrm{h}} > 0$ and 
$\alpha_{\mathrm{h}} (> 0)$. This can be seen from the following relations: 
$
\left( \alpha_{\mathrm{h}}^2 - \alpha_{\mathrm{h}} -1 \right)^2 
-2 \left( \alpha_{\mathrm{h}}^3 - 2\alpha_{\mathrm{h}}^2 + \alpha_{\mathrm{h}} + 1 \right) = - \left( \alpha_{\mathrm{h}}^2 - 1/2 \right)^2 - 
2\alpha_{\mathrm{h}} -3/4 <0
$
and 
$
\alpha_{\mathrm{h}}^3 - 2\alpha_{\mathrm{h}}^2 + \alpha_{\mathrm{h}} + 1 
= \alpha_{\mathrm{h}} \left( \alpha_{\mathrm{h}} - 1 \right)^2 + 1 >0
$. 
\begin{eqnarray}
a \Eqn{=} \frac{t^{\alpha_{\mathrm{h}} + 1} \e^{t/l}}{L_{\mathrm{c}} 
\alpha_{\mathrm{h}} \left[ 1 + t/\left( \alpha_{\mathrm{h}} l \right) \right]}
\,, 
\label{eq:VII-30} \\ 
L_{\mathrm{fh}} \Eqn{=} 
\frac{t}{\alpha_{\mathrm{h}} \left[ 1 + t/\left( \alpha_{\mathrm{h}} l \right) \right]}\,, 
\label{eq:VII-31} 
\end{eqnarray}
with $L_{\mathrm{c}}$ being an integration constant. 
The solution in (\ref{eq:VII-30}) with (\ref{eq:VII-31}) yields 
\begin{equation} 
H = \frac{1 + \alpha_{\mathrm{h}} \left[ 1 + t/\left( \alpha_{\mathrm{h}} l 
\right) \right]^2}{t \left[ 1 + t/\left( \alpha_{\mathrm{h}} l \right) 
\right]}\,.
\label{eq:VII-32} 
\end{equation}

We mention the case that $\alpha_{\mathrm{h}}$ is negative. 
{}From the denominator of Eq.~(\ref{eq:VII-30}), we see that 
at $t = - \alpha_{\mathrm{h}} l$ 
a Big Rip singularity appears because $a$ diverges. 
Furthermore, $L_{\mathrm{h}}$ can be negative, so that 
when matter is not included, 
$H = C_{\mathrm{h}}/L_{\mathrm{h}}$ can also be negative and 
therefore the universe will be shrink. 

When $t$ is small, i.e., in the limit of $t \to 0$, 
from Eq.~(\ref{eq:VII-32}) we find 
\begin{equation} 
H \rightarrow \frac{\alpha_{\mathrm{h}} + 1}{t}\,. 
\label{eq:VII-33} 
\end{equation}
This means that in this limit, the energy of the universe is 
dominated by that of a fluid whose Eos is given by 
$w_{\mathrm{m}} = -\left( \alpha_{\mathrm{h}} - 1 \right)/ 
\left[ 3 \left( \alpha_{\mathrm{h}} + 1 \right) \right]$. 
While, when $t$ is large, i.e., in the opposite limit of $t \to \infty$, 
from Eq.~(\ref{eq:VII-32}) we see that $H$ becomes a constant as 
\begin{equation} 
H \rightarrow \frac{1}{l}\,. 
\label{eq:VII-34} 
\end{equation}
This implies that the universe asymptotically approaches to de Sitter space. 
We explore the following model: 
\begin{equation} 
\frac{C_{\mathrm{h}}}{L_{\mathrm{h}}} 
= \frac{H}{\alpha_{\mathrm{h}} + 1} 
+ \frac{1}{L_{\mathrm{fh}}} 
\left\{ 1 + \frac{2 \left( \alpha_{\mathrm{h}}^2 - \alpha_{\mathrm{h}} -1 
\right)}{\alpha_{\mathrm{h}} + 1} 
\left( \frac{L_{\mathrm{h}}}{\alpha_{\mathrm{h}} l} \right) + 
\left[ 1 - \frac{\alpha_{\mathrm{h}}^2 \left( \alpha_{\mathrm{h}} + 2 
\right)}{\alpha_{\mathrm{h}} + 1} \right] 
\left( \frac{L_{\mathrm{h}}}{\alpha_{\mathrm{h}} l} \right)^2 
\right\}\,.
\label{eq:VII-35} 
\end{equation}
Also in this case, the solution (\ref{eq:VII-30}) with (\ref{eq:VII-31}) is 
again satisfied.

\subsection{The Hubble entropy in the holographic principle}

It follows from Eqs.~(\ref{eq:Add-2-01}) and (\ref{eq:VII-1}) that 
the Friedmann equation is described by 
$
3H^2/\kappa^2 = 3 C_{\mathrm{h}}^2/\left( \kappa^2 L_{\mathrm{h}}^2 \right) 
+ \rho_{\mathrm{m}}
$. 
We define the energy of matter $E_{\mathrm{m}}$, the Casimir energy 
$E_{\mathrm{C}}$ and the Hubble entropy $S_{\mathrm{H}}$
as 
\begin{eqnarray}
E_{\mathrm{m}} \Eqn{\equiv} 
\rho_{\mathrm{m}} L_{\mathrm{h}}^3\,,  
\label{eq:VII-36} \\ 
E_{\mathrm{C}} \Eqn{\equiv} \frac{3 C_{\mathrm{h}}^2 L_{\mathrm{h}}}{\kappa^2}
\,. 
\label{eq:VII-37} \\
S_{\mathrm{H}} \Eqn{\equiv} \frac{18 \pi C_{\mathrm{h}} L_{\mathrm{h}}^3}{
\kappa^2} H\,. 
\label{eq:VII-38} 
\end{eqnarray}
By substituting Eqs.~(\ref{eq:VII-36})--(\ref{eq:VII-38}) into 
the Friedmann equation shown above, we have~\cite{Verlinde:2000wg} 
the Cardy-Verlinde holographic formula of  
the Friedmann equation~\cite{Verlinde:2000wg}  
\begin{equation}
S_{\mathrm{H}}^2 = \left( \kappa^2 L_{\mathrm{h}}^2 \right)^2 
E_{\mathrm{C}} \left( E_{\mathrm{C}} + E_{\mathrm{m}} \right)\,.
\label{eq:VII-39} 
\end{equation}

Suppose that $\rho_{\mathrm{m}}$ can be neglected, Eq.~(\ref{eq:VII-38}) is 
rewritten to 
\begin{equation}
S_{\mathrm{H}} = \frac{18 \pi C_{\mathrm{h}}^2 L_{\mathrm{h}}^2}{
\kappa^2} = \frac{18 \pi C_{\mathrm{h}}^4}{\kappa^2 H^2}\,, 
\label{eq:VII-40} 
\end{equation}
where in deriving the second and third equalities, we have eliminated 
$H$ and $L_{\mathrm{h}}$ by using Eq.~(\ref{eq:VII-3}), respectively. 
It is seen from Eq.~(\ref{eq:VII-40}) that 
$S_{\mathrm{H}}$ depends on time, not a constant as in 
case of de Sitter universe in Ref.~\cite{Cases of de Sitter universe}. 
In the model in Eq.~(\ref{eq:VII-15}) with the solution (\ref{eq:VII-19}), 
from Eq.~(\ref{eq:VII-40}) we acquire 
\begin{equation}
S_{\mathrm{H}} = \frac{18 \pi C_{\mathrm{h}}^4 t^2 \left( t_{\mathrm{s}} - t 
\right)^2}{\kappa^2 h_{\mathrm{c}}^2 t_{\mathrm{s}}^2}\,. 
\label{eq:VII-41} 
\end{equation}
Therefore, in this model $S_{\mathrm{H}} = 0$ at $t=0$ and $t=t_{\mathrm{s}}$, 
and $S_{\mathrm{H}}$ becomes maximum at $t=t_{\mathrm{s}}/2$. 

On the other hand, for a generalized model in Eq.~(\ref{eq:VII-24}) with 
the interaction represented in Eq.~(\ref{eq:VII-23}), we have 
\begin{equation}
S_{\mathrm{H}} = 
\left( 1 - \frac{\kappa^2 \rho_{\mathrm{c}}}{3 h_{\mathrm{c}}} \right)^{-3} 
\frac{18 \pi C_{\mathrm{h}}^4 t^2 \left( t_{\mathrm{s}} - t 
\right)^2}{\kappa^2 h_{\mathrm{c}}^2 t_{\mathrm{s}}^2}\,. 
\label{eq:VII-42} 
\end{equation}
In comparison with Eq.~(\ref{eq:VII-41}), the form in Eq.~(\ref{eq:VII-42}) 
is multiplied by the first constant factor on the right-hand side. 
In case of Eq.~(\ref{eq:VII-29}) or the Friedmann equation 
$
3H^2/\kappa^2 = 3 C_{\mathrm{h}}^2/\left( \kappa^2 L_{\mathrm{h}}^2 \right) 
+ \rho_{\mathrm{m}}
$, we have 
\begin{equation}
S_{\mathrm{H}} =  
\frac{18 \pi C_{\mathrm{h}}^4 t^2 \left[ 1 + t/\left( \alpha_{\mathrm{c}} l 
\right) \right]^2}{\kappa^2 
\left\{1 + \alpha_{\mathrm{h}} \left[ 1 + t/\left( \alpha_{\mathrm{h}} l 
\right) \right]^2\right\}^2}\,. 
\label{eq:VII-43} 
\end{equation}
Therefore, in this model $S_{\mathrm{H}} = 0$ at $t=0$, 
whereas, in the limit of $t \to \infty$, $S_{\mathrm{H}}$ approaches to 
a constant as follows. 
\begin{equation}
S_{\mathrm{H}} \rightarrow 
\frac{18 \pi C_{\mathrm{h}}^4 l^2}{\kappa^2}\,.
\label{eq:VII-44} 
\end{equation}

We remark that $S_{\mathrm{H}}$ in Eq.~(\ref{eq:VII-43}) 
is positive, even though $\alpha_{\mathrm{c}} < 0$. 
Moreover, $S_{\mathrm{H}}$ in Eq.~(\ref{eq:VII-40}) is always positive. 
However, for the case that $S_{\mathrm{H}}$ is given by Eq.~(\ref{eq:VII-42}), 
if $\kappa^2 \rho_{\mathrm{c}}/\left( 3 h_{\mathrm{c}} \right) >1$, 
$S_{\mathrm{H}}$ can be negative. 
This implies that entropy of the universe should be negative, 
provided that $S_{\mathrm{H}}$ corresponds to the upper bound on the 
entropy of the universe. 
In Ref.~\cite{Brevik:2004sd}, negative entropy in the phantom phase has been 
observed. While, if the phantom phase is transient in the late time, 
the entropy of the universe may remain positive~\cite{Nojiri:2005sr}.

To connect the holographic dark energy scenario with 
the reconstruction of the corresponding scalar field theory 
in Sec.~IV, we explore another model. 
\begin{equation} 
\frac{C_{\mathrm{h}}}{L_{\mathrm{h}}} 
= \frac{\bar{\alpha}}{3} \Upsilon^3 - \bar{\beta} \Upsilon + \bar{\gamma}\,, 
\label{eq:VII-45} 
\end{equation}
where $\bar{\alpha} (>0)$, $\bar{\beta} (>0)$ and $\bar{\gamma} (>0)$ are 
positive constants and satisfy 
\begin{equation} 
\bar{\gamma} > \frac{2 \bar{\beta}}{3} \sqrt{\frac{\bar{\beta}}{\bar{\alpha}}}
\,. 
\label{eq:VII-46} 
\end{equation}
Here, $\Upsilon$ is defined by 
\begin{eqnarray}
\Upsilon \Eqn{\equiv} \frac{H - \bar{\gamma} + 4 \bar{\alpha}/\left(3 \bar{\beta}^2 \right) \left( H + 3 \bar{\gamma} \right) \ln \left( \Xi/\Theta \right)}{\bar{\alpha}/\left(3 \bar{\beta}^2 \right) \left( H + 3 \bar{\gamma} \right)^2 - \bar{\beta} - 4 \bar{\alpha}/\left(3 \bar{\beta}^2 \right) \ln \left( \Xi/\Theta \right)}\,, 
\label{eq:VII-47} \\ 
\Xi \Eqn{\equiv} \frac{a(t)}{a(0)} \Theta\,, 
\label{eq:VII-48} \\
\Theta \Eqn{\equiv} \int_{-\infty}^{+\infty} dt 
\exp \left( -\frac{\bar{\alpha}}{12} t^4 + \frac{\bar{\beta}}{2} t^2 - 
\bar{\gamma} t \right)\,. 
\label{eq:VII-49}
\end{eqnarray}
Equation (\ref{eq:VII-47}) informs us that 
$H$ becomes zero only once as a function of $\Upsilon$. 
We also have 
\begin{equation} 
\bar{\gamma} = -\frac{\bar{\alpha}}{3} \Upsilon_{\mathrm{c}}^3 + 
\bar{\beta} \Upsilon_{\mathrm{c}}\,. 
\label{eq:VII-50} 
\end{equation}
Here, $\Upsilon_{\mathrm{c}} (<0)$ is a negative constant and it is 
determined that 
when $\Upsilon = \Upsilon_{\mathrm{c}}$, 
$H(\Upsilon = \Upsilon_{\mathrm{c}}) = 0$. 

We reconstruct a corresponding scalar field theory 
in the holographic dark energy scenario. 
We take a concrete form of $I(\phi)$ in Eq.~(\ref{eq:2.15}) as 
%
\begin{equation} 
I(\phi) = 
\frac{\bar{\alpha}}{3} \left(\Upsilon_{\mathrm{c}} + \phi\right)^3 
- \bar{\beta} \left(\Upsilon_{\mathrm{c}} + \phi\right) + \bar{\gamma}\,, 
\label{eq:VII-add-01} 
\end{equation}
%
where $\bar{\gamma}$ is given by Eq.~(\ref{eq:VII-50}). 
In this model, 
$\omega(\phi)$ and $V(\phi)$ in the action in Eq.~(\ref{eq:2.1}) 
is expressed as 
\begin{eqnarray}
\omega(\phi) \Eqn{=} 
-\frac{2}{\kappa^2} \left[ 
\bar{\alpha} \left(\Upsilon_{\mathrm{c}} + \phi\right)^2 
- \bar{\beta} \right] 
\label{eq:VII-add-02} \\ 
V(\phi) \Eqn{=} 
\frac{3}{\kappa^2} \left[ 
\frac{\bar{\alpha}^2}{3} \left(\Upsilon_{\mathrm{c}} + \phi\right)^6 
-2 \bar{\alpha} \bar{\beta} \left(\Upsilon_{\mathrm{c}} + \phi\right)^4 
+ \bar{\alpha} \bar{\gamma} \left(\Upsilon_{\mathrm{c}} + \phi\right)^3 
\right. \nonumber \\
&& \left. 
+ \left( \bar{\alpha} + 3 \bar{\beta}^2 \right) 
\left(\Upsilon_{\mathrm{c}} + \phi\right)^2 
-2 \bar{\beta} \bar{\gamma} \left(\Upsilon_{\mathrm{c}} + \phi\right) 
+ 3 \bar{\gamma}^2 \right]\,. 
\label{eq:VII-add-03}
\end{eqnarray}

It follows from Eq.~(\ref{eq:2.15}) that we have the solution 
$\phi = t$, $H = I(t)$, and hence $a$ is given by
\begin{equation}  
a = a_{\mathrm{c}} \exp \left[ 
\frac{\bar{\alpha}}{12} \left(\Upsilon_{\mathrm{c}} + t\right)^4 
- \frac{\bar{\beta}}{2} \left(\Upsilon_{\mathrm{c}} + t\right)^2 
+ \bar{\gamma} \left(\Upsilon_{\mathrm{c}} + t\right)
\right]\,.
\label{eq:VII-add-04} 
\end{equation}
The scale factor $a$ has a minimum because $H = 0$ at $t=0$. 
Therefore, for $t<0$ the universe will shrink, whereas 
for $t>0$ it will expand. 
{}From $\dot{H} = \bar{\alpha} \left(\Upsilon_{\mathrm{c}} + t\right)^2 
- \bar{\beta}$, we see that $\dot{H} =0$ at 
\begin{equation}
t = t_{\pm} \equiv -\Upsilon_{\mathrm{c}} \pm 
\sqrt{\frac{\bar{\beta}}{\bar{\alpha}}} >0\,.
\label{eq:VII-add-05} 
\end{equation}
By using Eqs.~(\ref{eq:2.17}) and $H=I(t)$ with Eq.~(\ref{eq:2.15}) and 
$\phi = t$, $w_{\mathrm{eff}}$ is written as 
\begin{equation}
w_{\mathrm{eff}} = -1 - 
\frac{2\left[ \bar{\alpha} \left(\Upsilon_{\mathrm{c}} + t\right)^2 
- \bar{\beta} \right]}{\bar{\alpha} \left(\Upsilon_{\mathrm{c}} + \phi\right)^3 - 3\bar{\beta} \left(\Upsilon_{\mathrm{c}} + \phi\right) + 
3\bar{\gamma}}\,.
\label{eq:VII-add-06} 
\end{equation}
Accordingly, 
it is seen from Eq.~(\ref{eq:2.17}) that 
for $t_-<t<t_+$, $w_{\mathrm{eff}} > -1$, i.e., 
the universe is in the non-phantom phase, 
whereas for $0<t<t_-$ or $t_+<t$, $w_{\mathrm{eff}} < -1$, i.e., 
the universe is in the phantom phase, 
and that at $t = t_\pm$, $w_{\mathrm{eff}} = -1$. 
In summary, at $t = t_+ (t_-)$, 
the crossing of the phantom divide can 
occur from the non-phantom (phantom) phase to 
the phantom (non-phantom) one. 

In the model in Eq.~(\ref{eq:VII-45}), the Hubble entropy in 
Eq.~(\ref{eq:VII-38}) is given by 
\begin{equation} 
S_{\mathrm{H}} = \frac{18 \pi C_{\mathrm{h}}^4}{
\kappa^2 \left[ 
\left(\bar{\alpha}/3\right) \left(t+\Upsilon_{\mathrm{c}}\right)^3 
- \bar{\beta} \left(t+\Upsilon_{\mathrm{c}}\right) + \bar{\gamma}
\right]^2}\,.
\label{eq:VII-51} 
\end{equation}
Thus, $S_{\mathrm{H}}$ is always positive. 
It follows from Eqs.~(\ref{eq:VII-50}) and (\ref{eq:VII-51}) that at $t=0$, 
$S_{\mathrm{H}}$ diverges. 
{}From Eq.~(\ref{eq:VII-40}), we see that $S_{\mathrm{H}} \propto H^{-2}$. 
Thus, 
for $0<t<t_-$ (the phantom phase with $\dot{H} > 0$), 
$S_{\mathrm{H}}$ decreases. 
For $t_-<t<t_+$ (the non-phantom phase with $\dot{H} < 0$), 
$S_{\mathrm{H}}$ increases. 
For $t_+<t$ (the phantom phase with $\dot{H} > 0$), 
$S_{\mathrm{H}}$ again becomes small. 
Eventually, 
in the limit of $t \to \infty$, $S_{\mathrm{H}} \to 0$. 
As a result, 
in the non-phantom phase $S_{\mathrm{H}}$ grows, 
whereas in the phantom phase, $S_{\mathrm{H}}$ decreases. 

\section{Accelerating cosmology in $F(R)$ gravity}

In this section, we study an accelerating cosmology in $F(R)$ gravity. 
First, 
we consider relations between 
a scalar field theory in the Einstein frame 
and an $F(R)$ theory in the Jordan frame. 
Furthermore, 
we show how to obtain the $\Lambda$CDM, phantom-like or quintessence-like 
cosmologies in $F(R)$ gravity by following 
Refs.~\cite{Nojiri:2006gh, Nojiri:2006be, Bamba:2008hq, 
BGNO-Phantom-Crossing}. 
We mention that the reconstrucrion of $F(R, T_{\mathrm{st}})$ gravity 
has also been investigated, where $T_{\mathrm{st}}$ is the trace of the 
stress-energy tensor, e.g., in Ref.~\cite{Momeni:2011am}.

\subsection{$F(R)$ gravity and a corresponding scalar field theory}

The action describing $F(R)$ gravity with matter is given by
\begin{equation} 
S = \int d^4 x \sqrt{-g} 
\frac{F(R)}{2\kappa^2} + 
\int d^4 x 
{\mathcal{L}}_{\mathrm{M}} \left( g_{\mu\nu}, {\Psi}_{\mathrm{M}} \right)\,,
\label{eq:VIII-A-2.1}
\end{equation}
where ${\mathcal{L}}_{\mathrm{M}}$ is the Lagrangian of matter. 
By making a conformal transformation, we move to the Einstein frame: 
\begin{eqnarray} 
\tilde{g}_{\mu\nu} \Eqn{=} \Omega^2 g_{\mu\nu}\,,
\label{eq:VIII-A-2.2} \\
\Omega^2 \Eqn{\equiv} F_{,R}\,, 
\label{eq:VIII-A-2.3} \\ 
F_{,R} \Eqn{\equiv} \frac{d F(R)}{d R}\,,
\label{eq:VIII-A-2.4} 
\end{eqnarray}
where a tilde  
denotes quantities in the Einstein frame. 
We define a new scalar field $\phi$ as 
\begin{equation} 
\phi \equiv \sqrt{\frac{3}{2}} \frac{1}{\kappa} \ln F_{,R}\,.
\label{eq:VIII-A-2.5}
\end{equation}
Moreover, $R$ is represented by using $\tilde{R}$ as 
\begin{equation} 
R = e^{1/\sqrt{3} \kappa \phi} 
\left[ \tilde{R} + \sqrt{3} \tilde{\Box} \left( \kappa \phi \right) 
- \frac{1}{2} \tilde{g}^{\mu\nu}
{\partial}_{\mu} \left( \kappa \phi \right) 
{\partial}_{\nu} \left( \kappa \phi \right)
\right]\,, 
\label{eq:VIII-A-2.6}
\end{equation}
with 
\begin{equation} 
\tilde{\Box} \left( \kappa \phi \right) 
= \frac{1}{\sqrt{-\tilde{g}}} {\partial}_{\mu}
\left[ \sqrt{-\tilde{g}} \tilde{g}^{\mu\nu} {\partial}_{\nu} 
\left( \kappa \phi \right) \right]\,.
\label{eq:VIII-A-2.7}
\end{equation}
As a result, we acquire the action in the Einstein frame~\cite{F-M} 
\begin{equation} 
S_{\mathrm{E}} = 
\int d^4 x \sqrt{-\tilde{g}} \left( \frac{\tilde{R}}{2\kappa^2} - 
\frac{1}{2} \tilde{g}^{\mu\nu} {\partial}_{\mu} \phi {\partial}_{\nu} \phi 
- V(\phi) \right) + 
\int d^4 x 
{\mathcal{L}}_{\mathrm{M}} 
\left( \left( F_{,R} \right)^{-1}(\phi) \tilde{g}_{\mu\nu}, 
{\Psi}_{\mathrm{M}} \right)\,,
\label{eq:VIII-A-2.8}
\end{equation}
where the potential $V(\phi)$ is represented by 
\begin{equation} 
V(\phi) = \frac{F_{,R}\tilde{R}-F}{2\kappa^2 \left(F_{,R}\right)^2}\,. 
\label{eq:VIII-A-2.9}
\end{equation}
We note that an important cosmological application of the 
relations between $F(R)$ gravity in the Jordan frame and 
its corresponding scalar field theory in the Einstein frame 
to the time variation of the fine structure constant in non-minimal Maxwell-$F(R)$ gravity~\cite{NM-MYM-BNO} has recently been executed in Ref.~\cite{Bamba:2011nm} by using a novel consequence of a static domain wall solution~\cite{Toyozato:2012zh} in $F(R)$ gravity. 
Such non-minimal Maxwell theories with its coupling to a scalar field or 
the scalar curvature break the conformal invariance of the electromagnetic fields, so that the large-scale magnetic fields from inflation can be 
generated~\cite{B-Y-S-MF}. 
This would be considered to be a significant cosmological 
implication of the investigations of the present section.

\subsection{Reconstruction method of $F(R)$ gravity}

To begin with, 
we explain the reconstruction method of $F(R)$ gravity~\cite{Nojiri:2006gh, Nojiri:2006be}. 
We introduce proper functions $P(\phi)$ and $Q(\phi)$ of a scalar field $\phi$ 
and rewrite the action in Eq.~(\ref{eq:VIII-A-2.1}) to 
the following form 
\begin{equation}
S=\int d^4 x \sqrt{-g} 
\frac{1}{2\kappa^2} \left( P(\phi) R + Q(\phi)
\right) 
+ \int d^4 x 
{\mathcal{L}}_{\mathrm{M}} \left( g_{\mu\nu}, {\Psi}_{\mathrm{M}} \right)
\,.
\label{eq:VIII-B-2}
\end{equation}
Since the scalar field $\phi$ does not have the kinetic term, 
it may be regarded as an auxiliary scalar field.  
It follows from Eq.~(\ref{eq:VIII-B-2}) that 
the equation of motion of $\phi$ reads 
\begin{equation}
0=\frac{d P(\phi)}{d \phi} R + \frac{d Q(\phi)}{d \phi}\,.
\label{eq:VIII-B-3}
\end{equation}
Hence, by solving Eq.~(\ref{eq:VIII-B-3}) in terms of $R$ 
in principle we find the expression $\phi=\phi(R)$. 
By combining this expression and the action in Eq.~(\ref{eq:VIII-B-2}), 
we obtain the representation of $F(R)$ as 
\begin{equation}
F(R) = P(\phi(R)) R + Q(\phi(R))\,.
\label{eq:VIII-B-4}
\end{equation}
Furthermore, the variation of the action in Eq.~(\ref{eq:VIII-B-2}) leads to 
the gravitational field equation
\begin{equation}
\frac{1}{2}g_{\mu \nu} \left( P(\phi) R + Q(\phi) \right) 
-R_{\mu \nu} P(\phi) -g_{\mu \nu} \Box P(\phi) +
{\nabla}_{\mu} {\nabla}_{\nu}P(\phi) + \kappa^2
T^{(\mathrm{M})}_{\mu \nu} = 0\,. 
\label{eq:VIII-B-5}
\end{equation}
Here, ${\nabla}_{\mu}$ is the covariant derivative operator associated with 
$g_{\mu \nu}$ and $\Box \equiv g^{\mu \nu} {\nabla}_{\mu} {\nabla}_{\nu}$ 
is the covariant d'Alembertian for a scalar field. 
Moreover, 
$T^{(\mathrm{M})}_{\mu \nu}$ is the energy-momentum tensor of matter. 
We take the flat FLRW space-time 
$ds^2 = - dt^2 + a^2(t) \sum_{i=1,2,3}\left(dx^i\right)^2$. 
In this background, 
the $(\mu,\nu)=(0,0)$ and $(\mu,\nu)=(i,j)$ $(i,j=1,\cdots,3)$ 
components of Eq.~(\ref{eq:VIII-B-5}) become 
\begin{eqnarray}
\hspace{-5mm}
&&
-6H^2P(\phi(t)) -Q(\phi(t)) -6H \frac{dP(\phi(t))}{dt} 
+ 2\kappa^2\rho_{\mathrm{M}} = 0\,,
\label{eq:VIII-B-7} \\
\hspace{-5mm}
&&
2\frac{d^2P(\phi(t))}{dt^2}+4H\frac{dP(\phi(t))}{dt}+
\left(4\dot{H}+6H^2 \right)P(\phi(t)) +Q(\phi(t)) 
+ 2\kappa^2 P_{\mathrm{M}} = 0\,. 
\label{eq:VIII-B-8}
\end{eqnarray} 
Here, we have expressed the sum of the energy density and
pressure of matters with a constant EoS 
$w_{\mathrm{M} i}$ as $\rho_{\mathrm{M}}$ and $P_{\mathrm{M}}$, 
respectively, where the subscription ``$i$'' denotes a component of matters. 
By using Eqs.~(\ref{eq:VIII-B-7}) and (\ref{eq:VIII-B-8}), we 
eliminate $Q(\phi)$, and eventually we have 
\begin{equation}
\frac{d^2P(\phi(t))}{dt^2} -H\frac{dP(\phi(t))}{dt} +2\dot{H}P(\phi(t)) +
\kappa^2 \left( \rho_{\mathrm{M}} + P_{\mathrm{M}} \right) = 0\,.
\label{eq:VIII-B-9}
\end{equation}
If we appropriately redefine 
the scalar field $\phi$, it can be taken as $\phi = t$. 
In addition, we describe the form of $a(t)$ by 
\begin{equation}
a(t) = \bar{a} \exp \left( \tilde{g}(t) \right)\,, 
\label{eq:VIII-B-10}
\end{equation}
with $\bar{a}$ being a constant and $\tilde{g}(t)$ 
being a proper function of $t$. 
In this case, the Hubble parameter is given by 
$H= d \tilde{g}(\phi)/\left(d \phi \right)$. 
By using this expression, 
Eq.~(\ref{eq:VIII-B-9}) can be rewritten to 
\begin{eqnarray}
&&
\frac{d^2P(\phi)}{d\phi^2} -\frac{d \tilde{g}(\phi)}{d\phi}
\frac{dP(\phi)}{d\phi} +2 \frac{d^2 \tilde{g}(\phi)}{d \phi^2}
P(\phi) \nonumber \\
&& \hspace{10mm}
{}+
\kappa^2 \sum_i \left( 1+w_{\mathrm{M} i} \right) \bar{\rho}_{\mathrm{M} i} 
\bar{a}^{-3\left( 1+w_{\mathrm{M} i} \right)} \exp
\left[ -3\left( 1+w_{\mathrm{M} i} \right) \tilde{g}(\phi) \right] = 0\,. 
\label{eq:VIII-B-11}
\end{eqnarray}
Here, $\bar{\rho}_{\mathrm{M} i}$ is a constant. 
In addition, by solving Eq.~(\ref{eq:VIII-B-7}) in terms of $Q(\phi)$, 
we acquire 
\begin{eqnarray}
Q(\phi) \Eqn{=} -6 \left[ \frac{d \tilde{g}(\phi)}{d\phi} \right]^2 P(\phi)
-6\frac{d \tilde{g}(\phi)}{d\phi} \frac{dP(\phi)}{d\phi} \nonumber \\
&& \hspace{10mm}
{}+
2\kappa^2 \sum_i \bar{\rho}_i \bar{a}^{-3\left( 1+w_{\mathrm{M} i} \right)}
\exp
\left[ -3\left( 1+w_{\mathrm{M} i} \right) \tilde{g}(\phi) \right]\,.
\label{eq:VIII-B-12}
\end{eqnarray}
%

It is significant to emphasize that 
by redefining the auxiliary scalar field $\phi$ as 
$\phi=\Phi(\varphi)$ with a proper function $\Phi$ and 
defining $\tilde P(\varphi) \equiv P(\Phi(\varphi))$ and
$\tilde Q(\varphi) \equiv Q(\Phi(\varphi))$, we obtain the new form of 
the action 
\begin{equation}
S = \int d^4 x \sqrt{-g}
\frac{\tilde F(R)}{2\kappa^2} 
+ \int d^4 x 
{\mathcal{L}}_{\mathrm{M}} \left( g_{\mu\nu}, {\Psi}_{\mathrm{M}} \right)
\,, 
\label{eq:VIII-B-A1} 
\end{equation}
where 
\begin{equation}
\tilde F(R) \equiv \tilde P(\varphi) R + \tilde Q(\varphi)\,. 
\label{eq:VIII-B-A2}
\end{equation}
Since $\tilde F(R) = F(R)$, 
this action in Eq.~(\ref{eq:VIII-B-A2}) 
is equivalent to that in Eq.~(\ref{eq:VIII-B-2}). 
Furthermore, 
$\varphi$ is the inverse function of $\Phi$, and therefore 
by using $\phi = \phi(R)$ 
$\varphi$ can be solved with respect to $R$ as 
$\varphi=\varphi(R) = \Phi^{-1}(\phi(R))$. 
Accordingly, there exist the choices in $\phi$ as a gauge symmetry, 
and hence $\phi$ can be identified with time $t$ as $\phi=t$. 
This can be considered as a gauge condition which corresponds to 
the reparameterization of $\phi=\phi(\varphi)$~\cite{Bamba:2008hq}. 
As a result, 
if we obtain the solution $t = t(R)$, 
by solving Eq.~(\ref{eq:VIII-B-11}) and 
(\ref{eq:VIII-B-12}) and substituting these solutions into 
Eq.~(\ref{eq:VIII-B-4}), 
the explicit expression of $F(R)$ can be acquired. 
It should be noted that in a naive model of $F(R)$ gravity 
the crossing of the phantom divide cannot be realized 
because $F(R)$ has to be a double-valued function in order that 
the crossing of the phantom divide can occur. 
In fact, however, if the action of $F(R)$ gravity is extended to 
the form of $P(\phi) R + Q(\phi)$, the crossing of the phantom divide 
can happen. We show an explicit example to realize the 
crossing of the phantom divide in Sec.~VIII C 3.

\subsection{Reconstructed $F(R)$ forms and its cosmologies}

\subsubsection{The $\Lambda$CDM cosmology} 

We demonstrate the reconstruction process of $F(R)$ gravity in which 
the $\Lambda$CDM cosmology is realized~\cite{Nojiri:2006gh} (for other 
method of the reconstruction of $F(R)$ gravity reproducing 
the $\Lambda$CDM model, see~\cite{delaCruzDombriz:2006fj}). 
In the flat FLRW background, if there exist a matter with 
its EoS $w_{\mathrm{M}}$ and the cosmological constant $\Lambda$, 
the Friedmann equation (\ref{eq:Add-2-01}) is written as 
\begin{equation} 
H^2 = \frac{\kappa^2}{3} \rho_{\mathrm{M} \mathrm{c}} a^{-3\left( 
1 + w_{\mathrm{M}} \right)} + \frac{1}{l_{\Lambda}^2}\,, 
\label{eq:ED1-10-VIIIC1-1} 
\end{equation}
where $\rho_{\mathrm{M} \mathrm{c}}$ is a constant and 
$l_{\Lambda}$ is a length scale related to 
the cosmological constant $\Lambda$. 
{}From Eq.~(\ref{eq:ED1-10-VIIIC1-1}), we find that the scale factor can be expressed by Eq.~(\ref{eq:VIII-B-10}) with $\tilde{g}(t)$ being 
%
\begin{equation} 
\tilde{g}(t) = 
\frac{2}{3\left( 1 + w_{\mathrm{M}} \right)} 
\ln \left( 
\sqrt{
\frac{\kappa^2 l_{\Lambda}^2}{3} \rho_{\mathrm{M} \mathrm{c}} 
\bar{a}^{-3\left( 
1 + w_{\mathrm{M}} \right)}
} 
\sinh \left( 
\frac{3\left( 1 + w_{\mathrm{M}} \right)}{2l_{\Lambda}} 
\left( t-t_{\Lambda} \right)
\right) 
\right)\,, 
\label{eq:ED1-10-VIIIC1-2} 
\end{equation}
%
where $t_{\Lambda}$ corresponds to an integration constant. 
We suppose that geometrical dark energy originating from $F(R)$ gravity 
is dominant over matter and therefore matter contribution can be neglected. 
In this case, by using Eq.~(\ref{eq:ED1-10-VIIIC1-2}), 
Eq.~(\ref{eq:VIII-B-11}) is expressed as 
\begin{eqnarray}
&&
\frac{d^2P(\phi)}{d\phi^2} -\frac{1}{l_{\Lambda}} 
\coth \left( 
\frac{3\left( 1 + w_{\mathrm{M}} \right)}{2l_{\Lambda}} 
\left( t-t_{\Lambda} \right)
\right) \frac{dP(\phi)}{d\phi} 
\nonumber \\
&& \hspace{20mm}
{}-\frac{3\left( 1 + w_{\mathrm{M}} \right)}{l_{\Lambda}^2} 
\sinh^{-2} \left( 
\frac{3\left( 1 + w_{\mathrm{M}} \right)}{2l_{\Lambda}} 
\left( t-t_{\Lambda} \right)
\right) P(\phi) = 0\,. 
\label{eq:ED1-10-VIIIC1-3}
\end{eqnarray}
This expression can further been rewritten to 
a Gauss's hypergeometric differential equation 
by replacing the variable $\phi$ with $z$ as 
\begin{eqnarray}
&&
z\left(1-z\right)\frac{d^2P(z)}{dz^2} + 
\left[ \tilde{\gamma} - \left(\tilde{\alpha}+\tilde{\beta}+1\right)z \right]
\frac{dP(z)}{dz} -\tilde{\alpha} \tilde{\beta} P(z) = 0\,, 
\label{eq:ED1-10-VIIIC1-4} \\
&&
z \equiv -\sinh^{-2} \left( 
\frac{3\left( 1 + w_{\mathrm{M}} \right)}{2l_{\Lambda}} 
\left( t-t_{\Lambda} \right)
\right)\,, 
\label{eq:ED1-10-VIIIC1-5}
\end{eqnarray} 
with 
\begin{eqnarray}
\tilde{\gamma} \Eqn{\equiv} 4+\frac{1}{3\left( 1 + w_{\mathrm{M}} \right)}\,, 
\label{eq:ED1-10-VIIIC1-6} \\
\tilde{\alpha}+\tilde{\beta}+1 \Eqn{\equiv} 
6+\frac{1}{3\left( 1 + w_{\mathrm{M}} \right)}\,,
\label{eq:ED1-10-VIIIC1-7} \\
\tilde{\alpha} \tilde{\beta} \Eqn{\equiv} 
-\frac{1}{3\left( 1 + w_{\mathrm{M}} \right)}\,.
\label{eq:ED1-10-VIIIC1-8}
\end{eqnarray}
By using the Gauss's hypergeometric function, 
a solution of Eq.~(\ref{eq:ED1-10-VIIIC1-4}) is described as 
\begin{equation} 
P(z) = P_{\mathrm{c}1}
F_{\mathrm{G}} (\tilde{\alpha}, \tilde{\beta}, \tilde{\gamma}; z)
+ P_{\mathrm{c}2} 
\left(1-z\right)^{\tilde{\gamma}-\tilde{\alpha}-\tilde{\beta}} 
F_{\mathrm{G}} (\tilde{\gamma}-\tilde{\alpha}, \tilde{\gamma}-\tilde{\beta}, 
\tilde{\gamma}; z)\,, 
\label{eq:ED1-10-VIIIC1-9}
\end{equation}
where $P_{\mathrm{c}1}$ and $P_{\mathrm{c}2}$ are constants and 
$F_{\mathrm{G}}$ is the Gauss's hypergeometric function, defined by 
\begin{equation}  
F_{\mathrm{G}} (\tilde{\alpha}, \tilde{\beta}, \tilde{\gamma}; z) \equiv 
\frac{\Gamma(\tilde{\gamma})}{\Gamma(\tilde{\alpha}) \Gamma(\tilde{\beta})} 
\sum_{n=0}^{\infty} \frac{\Gamma(\tilde{\alpha}+n) \Gamma(\tilde{\beta}+n)}{\Gamma(\tilde{\gamma}+n)} \frac{z^n}{n!}\,, 
\label{eq:ED1-10-VIIIC1-10}
\end{equation}
with $\Gamma$ being the $\Gamma$ function. 
For simplicity in order to obtain the form of $F(R)$, 
we set $P_{\mathrm{c}2} = 0$, namely, we take only the first linearly independent solution in Eq.~(\ref{eq:ED1-10-VIIIC1-9}). 
It follows from Eq.~(\ref{eq:VIII-B-12}) that we obtain the form of $Q$ as 
\begin{equation} 
Q(z) = 
-\frac{3\left(1-z\right)P_{\mathrm{c}1}}{l_{\Lambda}^2} 
\left(2F_{\mathrm{G}} (\tilde{\alpha}, \tilde{\beta}, \tilde{\gamma}; z) 
+\frac{1+w_{\mathrm{M}}}{13+12w_{\mathrm{M}}} z 
F_{\mathrm{G}} (\tilde{\alpha}+1, \tilde{\beta}+1, \tilde{\gamma}+1; z) 
\right)\,. 
\label{eq:ED1-10-VIIIC1-11}
\end{equation}
In the limit of $t = \phi \to \infty$, from Eq.~(\ref{eq:ED1-10-VIIIC1-5}) 
we see $z \to 0$. Thus, in this limit, by substituting 
Eqs.~(\ref{eq:ED1-10-VIIIC1-9}) and (\ref{eq:ED1-10-VIIIC1-11}) into 
Eq.~(\ref{eq:VIII-B-4}) we find 
\begin{equation}
F(R) = P(\phi(R)) R + Q(\phi(R)) 
\to P_{\mathrm{c}1} \left( R - \frac{6}{l_{\Lambda}^2} \right)\,.
\label{eq:ED1-10-VIIIC1-12}
\end{equation}
As a consequence, by comparing Eq.~(\ref{eq:ED1-10-VIIIC1-12}) with 
Eq.~(\ref{eq:Add-2-0-1}), we see that if we take
\begin{eqnarray}
P_{\mathrm{c}1} \Eqn{\equiv} \frac{1}{2\kappa^2}\,, 
\label{eq:ED1-10-VIIIC1-13} \\ 
\frac{6}{l_{\Lambda}^2} \Eqn{\equiv} 2\Lambda\,, 
\label{eq:ED1-10-VIIIC1-14} 
\end{eqnarray}
the general relativity with the cosmological constant is realized. 
Moreover, by plugging Eq.~(\ref{eq:ED1-10-VIIIC1-12}) with 
Eqs.~(\ref{eq:ED1-10-VIIIC1-13}) and (\ref{eq:ED1-10-VIIIC1-14}) 
into the action in Eq.~(\ref{eq:VIII-A-2.8}), we acquire the 
action describing the $\Lambda$CDM cosmology which corresponds to 
the one in Eq.~(\ref{eq:Add-2-0-1}).

\subsubsection{Quintessence cosmology} 

Next, we reconstruct $F(R)$ gravity in which quintessence-like cosmology 
is produced~\cite{Nojiri:2006gh}. 
We investigate the case that $\tilde{g}(t)$ in Eq.~(\ref{eq:VIII-B-10}) is 
given by 
\begin{equation} 
\tilde{g}(\phi) = \bar{h}(\phi) \ln \left( \frac{\phi}{\phi_{\mathrm{c}}} 
\right)\,, 
\label{eq:ED1-10-VIIIC2-1}  
\end{equation}
where $\phi_{\mathrm{c}}$ is a constant and 
$\bar{h}(\phi)$ is a function varying slowly in $\phi$. 
Therefore, an adiabatic approximation is applied to $\bar{h}(\phi)$, 
so that the derivative of $\bar{h}(\phi)$ can be neglected, 
i.e., $d\bar{h}(\phi)/d\phi \sim d^2\bar{h}(\phi)/d\phi^2 \sim 0$. 
Through the same procedure as in Sec.~VIII C 1, we derive the solutions 
of $P(\phi)$ and $Q(\phi)$. 
The equation of $P(\phi)$ in (\ref{eq:VIII-B-11}) is given by 
\begin{eqnarray}
&&
\frac{d^2P(\phi)}{d\phi^2} -\frac{\bar{h}(\phi)}{\phi}
\frac{dP(\phi)}{d\phi} -\frac{2\bar{h}(\phi)}{\phi^2}
P(\phi) 
\nonumber \\
&& \hspace{10mm}
{}+
\kappa^2 \sum_i \left( 1+w_{\mathrm{M} i} \right) \bar{\rho}_{\mathrm{M} i} 
\bar{a}^{-3\left( 1+w_{\mathrm{M} i} \right)} 
\left( \frac{\phi}{\phi_{\mathrm{c}}} \right)^{-3\left( 1+w_{\mathrm{M} i} \right) \bar{h}(\phi)} 
= 0\,. 
\label{eq:ED1-10-VIIIC2-2}
\end{eqnarray}
We acquire a solution of Eq.~(\ref{eq:ED1-10-VIIIC2-2}) as 
\begin{eqnarray}
P(\phi) \Eqn{=} P_{+} \phi^{n_+} + P_{-} \phi^{n_-} + 
\sum_{i} P_i (\phi) \phi^{-3\left( 1+w_{\mathrm{M} i} \right) \bar{h}(\phi)+2}\,,
\label{eq:ED1-10-VIIIC2-3} \\ 
n_{\pm} (\phi) \Eqn{=} \frac{1}{2} \left( \bar{h}(\phi)-1 \pm 
\sqrt{\bar{h}^2(\phi) +6\bar{h}(\phi) +1} \right)\,, 
\label{eq:ED1-10-VIIIC2-4} \\ 
P_i (\phi) \Eqn{=} 
-\frac{ 
\left( 1+w_{\mathrm{M} i} \right) \bar{\rho}_{\mathrm{M} i} 
\bar{a}^{-3\left( 1+w_{\mathrm{M} i} \right)} 
\phi_{\mathrm{c}}^{3\left( 1+w_{\mathrm{M} i} \right) \bar{h}(\phi)}
}{6\left( 1+w_{\mathrm{M} i} \right)\left( 4+3w_{\mathrm{M} i} \right)\bar{h}^2(\phi)-2\left( 13+9w_{\mathrm{M} i} \right)\bar{h}(\phi)+4}\,, 
\label{eq:ED1-10-VIIIC2-5}
\end{eqnarray}
where $P_{\pm}$ are arbitrary constants, and 
$w_{\mathrm{M} \mathrm{r}} = 1/3$ for radiation and 
$w_{\mathrm{M} \mathrm{m}} = 0$ for non-relativistic matter. 
Furthermore, by using Eq.~(\ref{eq:VIII-B-12}), $Q$ is written by 
\begin{eqnarray} 
Q(\phi) \Eqn{=} -6\bar{h}(\phi) \left( 
P_+ (\bar{h}(\phi) + n_+ (\phi)) \phi^{n_+ (\phi) -2} 
+ P_- (\bar{h}(\phi) + n_- (\phi)) \phi^{n_- (\phi) -2} 
\right) 
\nonumber \\
&& \hspace{0mm}
{}+ \sum_{i} \left\{ 
-6\bar{h}(\phi) \left[ -\left( 2+3w_{\mathrm{M} i} \right)\bar{h}(\phi) 
+2\right]P_i (\phi) +\bar{\rho}_{\mathrm{M} i} 
\bar{a}^{-3\left( 1+w_{\mathrm{M} i} \right)} 
\phi_{\mathrm{c}}^{3\left( 1+w_{\mathrm{M} i} \right) \bar{h}(\phi)}
\right\} 
\nonumber \\
&& \hspace{10mm}
{}\times 
\phi^{-3\left( 1+w_{\mathrm{M} i} \right) \bar{h}(\phi)}
\,. 
\label{eq:ED1-10-VIIIC2-6}
\end{eqnarray}
On the other hand, from Eq.~(\ref{eq:ED1-10-VIIIC2-1}) we find 
$H \sim \bar{h}(t)/t$ and hence 
$R=6\left(\dot{H} + 2H^2 \right) \sim 6\left(-\bar{h}(t) + 2\bar{h}^2(t) 
\right)/t^2$. 
Here, we provided that in the limit of $\phi \to 0$, $\bar{h}(\phi) \to 
\bar{h}_{\mathrm{i}}$, and in the opposite limit of $\phi \to \infty$, 
$\bar{h}(\phi) \to \bar{h}_{\mathrm{f}}$. 
As a form of $\bar{h}(\phi)$, we take 
\begin{equation} 
\bar{h}(\phi) = \frac{\bar{h}_{\mathrm{i}} + \bar{h}_{\mathrm{f}} \vartheta 
\phi^2}{1+\vartheta \phi^2}\,,
\label{eq:ED1-10-VIIIC2-7}
\end{equation}
where $\vartheta$ is a small constant enough for $\bar{h}(\phi)$ to 
be a function varying slowly in $\phi$. 
The substitution of 
Eqs.~(\ref{eq:ED1-10-VIIIC2-5}) and (\ref{eq:ED1-10-VIIIC2-6}) into 
Eq.~(\ref{eq:VIII-B-4}) yields
\begin{equation}
F(R) = P(\Phi_{\mathrm{c}} (R)) R + Q(\Phi_{\mathrm{c}} (R))\,,
\label{eq:ED1-10-VIIIC2-8}
\end{equation}
where 
\begin{eqnarray}  
\hspace{-15mm}
\Phi_{\mathrm{c}} (R) \Eqn{\equiv} \sqrt{\Phi_+^{1/3} + \Phi_-^{1/3}}\,,
\label{eq:ED1-10-VIIIC2-9} \\
\Phi_ {\pm} \Eqn{\equiv} \frac{1}{2} \left( -\beta_1 \pm 
\sqrt{\beta_1^2 - \frac{4\beta_2^3}{27}} \right)\,,
\label{eq:ED1-10-VIIIC2-10} \\
\hspace{-15mm}
\beta_1 \Eqn{\equiv} \frac{16 \left[R+3\bar{h}_{\mathrm{f}} 
\left( 1-2\bar{h}_{\mathrm{f}} \right) \vartheta 
\right]^3}{27\vartheta^3 R^3}
\nonumber \\
\hspace{-15mm}
&&
{}-\frac{2\left[R+3\bar{h}_{\mathrm{f}} 
\left( 1-2\bar{h}_{\mathrm{f}} \right) \vartheta 
\right] 
\left[R+2\left( 3\bar{h}_{\mathrm{i}} + 3\bar{h}_{\mathrm{f}} -2\bar{h}_{\mathrm{i}} \bar{h}_{\mathrm{f}} \right) \vartheta 
\right]}{3\vartheta R} 
+ 6\bar{h}_{\mathrm{i}} 
\left( 1-2\bar{h}_{\mathrm{f}} \right)\,,
\label{eq:ED1-10-VIIIC2-11} \\
\hspace{-15mm}
\beta_2 \Eqn{\equiv} 
-\frac{4 \left[R+3\bar{h}_{\mathrm{f}} 
\left( 1-2\bar{h}_{\mathrm{f}} \right) \vartheta 
\right]^2}{3\vartheta^2 R^2}
-\frac{R+2\left( 3\bar{h}_{\mathrm{i}} + 3\bar{h}_{\mathrm{f}} -2\bar{h}_{\mathrm{i}} \bar{h}_{\mathrm{f}} \right) \vartheta}{\vartheta^2 R}\,.
\label{eq:ED1-10-VIIIC2-12}
\end{eqnarray}

At the dark energy dominated stage, $w_{\mathrm{DE}} \approx w_{\mathrm{eff}}$ 
because the energy density of matter can be negligible compared with that of 
dark energy. In the present model in Eq.~(\ref{eq:ED1-10-VIIIC2-7}), from 
Eq.~(\ref{eq:2.17}) we have 
$w_{\mathrm{DE}} = -1+2/\left( 3\bar{h}_{\mathrm{f}} \right)$. 
Thus, for $0<\bar{h}_{\mathrm{f}}<1$, $-1< w_{\mathrm{DE}}< -1/3$. 
As a result, this means that in the reconstructed $F(R)$ gravity model, 
quintessence-like cosmology can be realized. 
In addition, 
for the reconstructed $F(R)$ gravity model in Eq.~(\ref{eq:ED1-10-VIIIC2-8}), 
in the late limit of $\phi \to \infty$, which can be regarded as 
the limit of the present time, 
the asymptotic behavior is given by a power-law description as 
$F(R) \sim R^{\bar{s}}$ with $\bar{s} \equiv -\left( \bar{h}_{\mathrm{f}} -5 
+\sqrt{\bar{h}_{\mathrm{f}}^2 +6\bar{h}_{\mathrm{f}} +1} \right)/4$. 
By using this expression, we find that in 
the action of a scalar field theory 
in Eq.~(\ref{eq:VIII-A-2.8}), the potential $V(\phi)$ in Eq.~(\ref{eq:VIII-A-2.9}) 
is written as 
\begin{equation} 
V(\phi) \sim \frac{1}{2\kappa^2} \frac{\bar{s}-1}{\bar{s}^2} R^{-\left( \bar{s}-2 \right)} 
= \frac{1}{2\kappa^2} \frac{\bar{s}-1}{\bar{s}^2} 
\left[\frac{1}{\bar{s}}\e^{ \left(2/3\right) \kappa \phi}\right]^{-\left(\bar{s}-2\right)/\left(\bar{s}-1\right)}\,. 
\label{eq:ED1-10-VIIIC2-13}
\end{equation}
Here, in deriving the second equality, we have used 
$R= \left[ \left(1/\bar{s}\right) \e^{ \left(2/3\right) \kappa \phi} \right]^{1/\left(\bar{s}-1\right)}$, which follows from Eq.~(\ref{eq:VIII-A-2.5}). 
This can be interpreted as quintessence potential.

\subsubsection{Phantom cosmology} 

Furthermore, we reconstruct $F(R)$ gravity in which 
the crossing of the phantom divide is realized~\cite{BGNO-Phantom-Crossing} 
and eventually phantom-like cosmology is produced. 

For matter to be neglected because of the dark energy domination, 
as an example, we examine the case that $\tilde{g}(t)$ in 
Eq.~(\ref{eq:VIII-B-10}) is described by 
\begin{equation}
\tilde{g}(\phi) = - 10 \ln \left[ \left(\frac{\phi}{t_0}\right)^{-\bar{\gamma}}
 - C_{\mathrm{p}} \left(\frac{\phi}{t_0}\right)^{\bar{\gamma}+1} \right]\,, 
\label{eq:ED1-10-VIIIC3-1}
\end{equation}
where $\bar{\gamma} (> 0)$ and $C_{\mathrm{p}} (> 0)$ are positive constants 
and $t_0$ is the present time. 
We note that since there occurs a Big Rip singularity at $\phi = t_{\mathrm{s}} \equiv t_0 C_{\mathrm{p}}^{-1/\left(2\bar{\gamma}+1\right)}$, we investigate 
the period $0<t<t_{\mathrm{s}}$ in order for $\tilde{g}(\phi)$ to be a 
real number. 
{}From Eq.~(\ref{eq:ED1-10-VIIIC3-1}), the expression of the Hubble parameter 
$H(t)= d \tilde{g}(\phi)/d\phi$ is given by 
\begin{eqnarray} 
H(t) \Eqn{=} \left(\frac{10}{t_0}\right) \left[ \frac{ \bar{\gamma} \left(\phi/t_0\right)^{-\left(\bar{\gamma}+1\right)}
 + \left(\bar{\gamma}+1\right) C_{\mathrm{p}} \left(\phi/t_0\right)^{\bar{\gamma}} }{\left(\phi/t_0\right)^{-\bar{\gamma}}
 - C_{\mathrm{p}} \left(\phi/t_0\right)^{\bar{\gamma}+1}}\right] 
\label{eq:ED1-10-VIIIC3-2} \\
\Eqn{=}
\left[\frac{10}{t_{\mathrm{s}} C_{\mathrm{p}}^{1/\left(2\bar{\gamma}+1\right)}}\right] \left[ \frac{\bar{\gamma} + \left(\bar{\gamma}+1\right) \left(t/t_{\mathrm{s}}\right)^{2\bar{\gamma}+1}}{1-\left(t/t_{\mathrm{s}}\right)^{2\bar{\gamma}+1}}\right]\,,
\label{eq:ED1-10-VIIIC3-3}
\end{eqnarray}
where in deriving Eq.~(\ref{eq:ED1-10-VIIIC3-3}) we have used the relation 
$t_{\mathrm{s}} = t_0 C_{\mathrm{p}}^{-1/\left(2\bar{\gamma}+1\right)}$ 
and $\phi=t$. 
A solution of equation (\ref{eq:VIII-B-11}) for $P(\phi)$ is 
derived as 
\begin{eqnarray} 
P(\phi) \Eqn{=} \exp \left( \frac{\tilde{g}(\phi)}{2} \right) 
\sum_{j=\pm} \bar{p}_j \phi^{\bar{\beta}_j}\,, 
\label{eq:ED1-10-VIIIC3-4} \\ 
\bar{\beta}_{\pm} \Eqn{\equiv} 
\frac{1 \pm \sqrt{1 + 100 \bar{\gamma} (\bar{\gamma} + 1)}}{2}\,, 
\label{eq:ED1-10-VIIIC3-5}
\end{eqnarray}
with $\bar{p}_{\pm}$ being arbitrary constants. 

It follows from Eq.~(\ref{eq:2.17}) that 
$w_{\mathrm{DE}} \approx w_{\mathrm{eff}}$ at the dark energy dominated stage 
is expressed as 
\begin{equation} 
w_{\mathrm{DE}} = -1 -\frac{-\bar{\gamma} + 4\bar{\gamma} \left( \bar{\gamma}+1 \right) \left( t/t_{\mathrm{s}} \right)^{2\bar{\gamma}+1} +
\left( \bar{\gamma}+1 \right) \left( t/t_{\mathrm{s}}
\right)^{2\left( 2\bar{\gamma}+1 \right)} }
{15 \left[ \bar{\gamma} + \left( \bar{\gamma}+1 \right)
\left( t/t_{\mathrm{s}} \right)^{2\bar{\gamma}+1} \right]^2}\,. 
\label{eq:ED1-10-VIIIC3-6}
\end{equation}
In the limit of $t \to 0$, namely, $t/t_{\mathrm{s}} \ll 1$, 
from Eqs.~(\ref{eq:ED1-10-VIIIC3-2}) and (\ref{eq:ED1-10-VIIIC3-6}) 
we find $H(t) \sim 10\bar{\gamma}/t$ 
and $w_\mathrm{DE} \sim -1 + 1/\left(15\bar{\gamma}\right) (> -1)$, 
respectively. This is the non-phantom (quintessence) phase. 
While, in the opposite limit of $t \to t_{\mathrm{s}}$ 
it follows from 
Eqs.~(\ref{eq:ED1-10-VIIIC3-2}) and (\ref{eq:ED1-10-VIIIC3-6}) 
that $H(t) \sim 10/\left( t_{\mathrm{s}} - t \right)$, 
which leads to $a(t) \sim \bar{a} \left( t_{\mathrm{s}} - t \right)^{-10}$, 
and $w_\mathrm{DE} \sim -16/15 (< -1)$, respectively. 
Moreover, 
$d \left( w_\mathrm{DE} +1 \right)/dt$ monotonously decreases in time. 
Hence, first the universe is in the non-phantom phase. 
As the time passes, when $t$ closes to $t_{\mathrm{s}}$, the universe enters the phantom phase. Thus, the crossing of the phantom divide line of  
$w_\mathrm{DE} = -1$ can occur at 
$t = t_{\mathrm{c}} \equiv t_\mathrm{s} \left[-2\bar{\gamma} +
\sqrt{4\bar{\gamma}^2 + \bar{\gamma}/\left(\bar{\gamma}+1\right)}
\right]^{1/\left( 2\bar{\gamma} + 1 \right)}$. 

In addition, by using 
Eq.~(\ref{eq:ED1-10-VIIIC3-4}) and substituting it into 
Eq.~(\ref{eq:VIII-B-12}), we acquire the forms of $P(t)$ and $Q(t)$ as 
\begin{eqnarray}
P(t) \Eqn{=} 
\left[ \frac{\left( t/t_0  \right)^{\bar{\gamma}}}
{1-\left( t/t_{\mathrm{s}} \right)^{2\bar{\gamma}+1}} \right]^5
\sum_{j=\pm} \bar{p}_j t^{\bar{\beta}_j}\,,
\label{eq:ED1-10-VIIIC3-7} \\ 
Q(t) \Eqn{=} -6H
\left[ \frac{\left( t/t_0 \right)^{\bar{\gamma}}}
{1-\left( t/t_{\mathrm{s}} \right)^{2\bar{\gamma}+1}} \right]^5
\sum_{j=\pm} \left( \frac{3}{2}H + \frac{\bar{\beta}_j}{t} \right)
\bar{p}_j t^{\bar{\beta}_j}\,. 
\label{eq:ED1-10-VIIIC3-8}
\end{eqnarray}
On the other hand, from $R=6\left(\dot{H} + 2H^2 \right)$, 
the relation between $R$ and $t$ is given by 
\begin{equation}  
R = 
\frac{60
\left[ 
\bar{\gamma} \left( 20\bar{\gamma} -1 \right) + 44\bar{\gamma} \left( \bar{\gamma}+1 \right)
\left( t/t_{\mathrm{s}} \right)^{2\bar{\gamma}+1} 
+ \left( \bar{\gamma}+1 \right) \left( 20\bar{\gamma}+21 \right)
\left( t/t_{\mathrm{s}} \right)^{2\left( 2\bar{\gamma}+1 \right)}
\right]
}
{t^2 \left[ 1- \left( t/t_{\mathrm{s}} \right)^{2\bar{\gamma}+1} \right]^2}
\,. 
\label{eq:ED1-10-VIIIC3-9}
\end{equation}
Accordingly, in principle, if we obtain the relation $t=t(R)$ by 
solving Eq.~(\ref{eq:ED1-10-VIIIC3-9}) reversely, 
the substitution of it into Eqs.~(\ref{eq:ED1-10-VIIIC3-7}) and 
(\ref{eq:ED1-10-VIIIC3-8}) and the combination of those with 
Eq.~(\ref{eq:VIII-B-4}) yield the explicit form of $F(R)$. 
We show analytically solvable cases below. 
In the limit of $t \to 0$ $(t/t_{\mathrm{s}} \ll 1)$, from 
Eq.~(\ref{eq:ED1-10-VIIIC3-9}) we obtain 
$t \sim \sqrt{60\bar{\gamma} \left( 20\bar{\gamma} -1 \right)/R}$. 
By using this asymptotic relation, we acquire~\cite{Bamba:2009ay} 
\begin{eqnarray}
F(R) \Eqn{\approx} 
\left[ \frac{1}{t_0} \sqrt{60\bar{\gamma} \left( 20\bar{\gamma} -1 \right)} 
\right]^{5\bar{\gamma}} R^{-5\bar{\gamma}/2 +1} 
\nonumber \\
&& \hspace{10mm}
{}\times
\sum_{j=\pm}
\left\{ \left( \frac{5\bar{\gamma} -\bar{\beta}_j -1}{20\bar{\gamma} -1} \right) \bar{p}_j
\left[60\bar{\gamma} \left( 20\bar{\gamma} -1 \right) 
\right]^{\bar{\beta}_j /2} R^{-\bar{\beta}_j /2}
\right\}\,.
\label{eq:ED1-10-VIIIC3-10}
\end{eqnarray}
In the opposite limit of $t \to t_{\mathrm{s}}$, it follows 
from Eq.~(\ref{eq:ED1-10-VIIIC3-9}) that 
$t \sim t_{\mathrm{s}} - 3\sqrt{140/R}$. 
With this asymptotic relation, for the large curvature regime as 
$t_{\mathrm{s}}^2 R \gg 1$ we find 
\begin{eqnarray}  
F(R) \Eqn{\approx} \bar{F} R^{7/2}\,, 
\label{eq:ED1-10-VIIIC3-11} \\ 
\bar{F} \Eqn{\equiv} 
\frac{2}{7}
\left[
\frac{1}{3\sqrt{140} \left( 2\bar{\gamma} +1 \right)}
\left( \frac{t_{\mathrm{s}}}{t_0} \right)^{\bar{\gamma}} \, \right]^5
\left(
\sum_{j=\pm}
\bar{p}_j t_{\mathrm{s}}^{\bar{\beta}_j} \right)
t_{\mathrm{s}}^5\,.
\label{eq:ED1-10-VIIIC3-12}
\end{eqnarray}
For the power-law form of $F(R)$ in Eq.~(\ref{eq:ED1-10-VIIIC3-12}), 
the potential $V(\phi)$ in Eq.~(\ref{eq:VIII-A-2.9}) of the 
action for a scalar field theory in Eq.~(\ref{eq:VIII-A-2.8}) 
is described as 
\begin{equation} 
V(\phi) \sim \frac{5}{49\bar{F} \kappa^2} 
\left[\frac{2}{7}\e^{\left(2/3\right) \kappa \phi}\right]^{-3/5}\,. 
\label{eq:ED1-10-VIIIC3-13}
\end{equation}
This can be interpreted as phantom potential. 

Using the reconstruction program with auxiliary scalar fields as 
discussed in this section, or without the use of auxiliary scalar fields, 
following Ref.~\cite{Nojiri:2009kx}, 
one can eventually reconstruct any dark energy cosmology studied in this review. For instance, Little rip cosmology for modified gravity has 
been presented in Refs.~\cite{Brevik:2011mm, Nojiri:2011kd}.

\subsection{Dark energy cosmology in $F(R)$ Ho\v{r}ava-Lifshitz gravity} 

As a candidate for a renormalizable gravitational theory in four dimensions, 
the Ho\v{r}ava-Lifshitz gravity has been proposed 
in Ref.~\cite{Horava:2009uw} (for a review on the Ho\v{r}ava-Lifshitz cosmology, see, e.g.,~\cite{Mukohyama:2010xz}), although it cannot maintain the Lorentz invariance. 
In addition, its extension to an $F(R)$ formalism has been executed in Ref.~\cite{Chaichian:2010yi}. 
In this subsection, we study cosmology for dark energy in 
$F(R)$ Ho\v{r}ava-Lifshitz gravity~\cite{Elizalde:2010ep}.

\subsubsection{$F(R)$ Ho\v{r}ava-Lifshitz gravity} 

The model action describing $F(R)$ Ho\v{r}ava-Lifshitz gravity 
with the matter action is given by~\cite{Chaichian:2010yi} 
\begin{eqnarray} 
S \Eqn{=} 
\int dt d^3x \sqrt{g^{(3)}} 
\frac{1}{2\bar{\kappa}^2} N_{\mathrm{l}} F(\bar{R}) 
+
\int d^4 x 
{\mathcal{L}}_{\mathrm{M}}\,,
\label{eq:ED1-15-VIIID1-Add-1} \\
\bar{R} \Eqn{\equiv} 
K_{ij}K^{ij} -\lambda_{\mathrm{HL}} \bar{K}^2 +R^{(3)} + 
2\mu_{\mathrm{HL}} \nabla_{\mu} \left( n^{\mu} \nabla_{\nu} n^{\nu} 
- n^{\nu} \nabla_{\nu} n^{\mu} \right) -L^{(3)}(g_{ij}^{(3)})\,, 
\label{eq:ED1-15-VIIID1-Add-2} \\
\bar{K} \Eqn{=} g^{ij} K_{ij} = g^{ij} \frac{1}{N_{\mathrm{l}}} 
\left( \dot{g}_{ij}^{(3)} -\nabla_{i}^{(3)} N_j -\nabla_{j}^{(3)} N_i 
\right)\,, 
\label{eq:ED1-15-VIIID1-Add-3}
\end{eqnarray}
with 
\begin{eqnarray} 
L^{(3)}(g_{ij}^{(3)}) \Eqn{\equiv} E^{ij} G_{ijkl} E^{kl}\,, 
\label{eq:ED1-15-VIIID1-Add-4} \\
G_{ijkl} \Eqn{=} \frac{1}{2} \left( g_{ik}^{(3)} g_{jl}^{(3)} + g_{il}^{(3)} g_{jk}^{(3)} \right) - \bar{\lambda}_{\mathrm{HL}} g_{ij}^{(3)} g_{kl}^{(3)}\,, 
\label{eq:ED1-15-VIIID1-Add-5} \\ 
\bar{\lambda}_{\mathrm{HL}} \Eqn{\equiv} 
\frac{\lambda_{\mathrm{HL}}}{3\lambda_{\mathrm{HL}} -1}\,, 
\label{eq:ED1-15-VIIID1-Add-6} \\ 
E^{ij} \Eqn{\equiv} 
\frac{1}{\sqrt{g^{(3)}}} 
\frac{\delta W [g_{kl}^{(3)}]}{\delta g_{ij}^{(3)}}\,, 
\label{eq:ED1-15-VIIID1-Add-7}
\end{eqnarray}
where, $i, j, k, l$ run over $1, 2, 3$, $N_{\mathrm{l}}$ is the lapse variable in the Arnowitt-Deser-Misner (ADM) decomposition in $(3+1)$ space-time, $N^{i}$ is the shift $3$-vector~\cite{M-T-W, ADM-D}, 
${\mathcal{L}}_{\mathrm{M}}$ is 
the Lagrangian of matter which is a perfect fluid, 
$K_{ij}$ is the extrinsic curvature, $R^{(3)}$ is the spatial scalar 
curvature, $n^{\mu}$ is a unit vector perpendicular to a 
constant time hypersurface, 
$\bar{\kappa}$ is the dimensionless gravitational coupling, and 
$\lambda_{\mathrm{HL}}$ and $\mu_{\mathrm{HL}}$ are constants and 
these break the full diffeomorphism invariance. 
Furthermore, $G_{ijkl}$ is the inverse of the generalized De Witt metric 
and it exists only for $\bar{\lambda}_{\mathrm{HL}} \neq 1/3$ because 
it follows from Eq.~(\ref{eq:ED1-15-VIIID1-Add-6}) that 
if $\bar{\lambda}_{\mathrm{HL}} = 1/3$, $G_{ijkl}$ becomes singular. 
Moreover, 
$E^{ij}$ is constructed so that the detailed balance principle restricting 
the number of free model parameters can be met~\cite{Horava:2009uw}. 
In the Ho\v{r}ava-Lifshitz gravity, there exists the difference of the scaling properties between the space and time coordinates as 
$x^{i} \to b x^{i}$ and $t \to b^{\bar{z}} t$, with 
$b$ being a constant and $\bar{z}$ being a dynamical critical exponent. 
Here, if $z = 3$ in $(3+1)$ space-time dimensions, the theory is 
renormalizable, whereas for $\bar{z}=1$, it is the general relativity. 
Such scaling properties give the theory only the foliation preserving 
diffeomorphisms, given by $\delta x^i = \bar{\zeta} (x^i, t)$ and 
$\delta t = \bar{\xi} (t)$, where $\bar{\zeta}$ and $\bar{\xi}$ are 
functions of $x^i$ as well as $t$ and $t$, respectively. 
We also mention that 
in the Ho\v{r}ava-Lifshitz gravity~\cite{Horava:2009uw}, 
$N_{\mathrm{l}}$ is supposed to depend only on time in order for 
the projectability condition to be satisfied and it 
is set to be unity ($N_{\mathrm{l}} = 1$) by applying 
the foliation preserving diffeomorphisms. 
For $\bar{z} = 2$ and $\bar{z} = 3$, the expression of $W [g_{kl}^{(3)}]$ in 
Eq.~(\ref{eq:ED1-15-VIIID1-Add-7}) are presented in Ref.~\cite{Horava:2008ih}. 
In the spatially flat FLRW space-time 
$ds^2= -N_{\mathrm{l}}^2 dt^2 +\sum_{i=1,2,3}\left(dx^i\right)^2$, we have 
\begin{equation} 
\bar{R} = 
\frac{3\left(1-3\lambda_{\mathrm{HL}} +6\mu_{\mathrm{HL}} \right)H^ 2}{N_{\mathrm{l}}^2} + \frac{6\mu_{\mathrm{HL}}}{N_{\mathrm{l}}} \frac{d}{dt} 
\left( \frac{H}{N_{\mathrm{l}}} \right)\,.
\label{eq:ED1-15-VIIID1-Add-8}
\end{equation}
By varying the action in Eq.~(\ref{eq:ED1-15-VIIID1-Add-1}) 
with respect to $N_{\mathrm{l}}$ and $g_{ij}^{(3)}$, we acquire 
\begin{eqnarray} 
\hspace{-10mm}
&&
\int d^3x \left[ 
F(\bar{R}) -6\left(1-3\lambda_{\mathrm{HL}} +3\mu_{\mathrm{HL}} \right)H^2 
-6\mu_{\mathrm{HL}} \left( \dot{H} -H \dot{\bar{R}} F^{''}(\bar{R}) \right) 
-\kappa^2 \rho_\mathrm{M} \right] = 0\,, 
\label{eq:ED1-15-VIIID1-Add-9} \\
\hspace{-10mm}
&&
F(\bar{R}) -2\left(1-3\lambda_{\mathrm{HL}} +3\mu_{\mathrm{HL}} \right) 
\left(\dot{H} + 3H^2\right) F^{'}(\bar{R}) 
-2\left(1-3\lambda_{\mathrm{HL}}\right) \dot{\bar{R}} F^{''}(\bar{R}) 
\nonumber \\ 
\hspace{-10mm}
&&
{}+2\mu_{\mathrm{HL}} \left( \dot{\bar{R}}^2 F^{'''}(\bar{R}) + \ddot{\bar{R}} 
F^{''}(\bar{R}) \right) + \kappa^2 P_\mathrm{M} = 0\,, 
\label{eq:ED1-15-VIIID1-Add-10}
\end{eqnarray}
where the prime denotes the derivative with respect to $\bar{R}$, and 
$\rho_\mathrm{M}$ and $P_\mathrm{M}$ are the energy density and pressure of a perfect fluid, respectively. 
In deriving Eq.~(\ref{eq:ED1-15-VIIID1-Add-9}), we have used the projectability condition for $N_{\mathrm{l}}$ to have only the time dependence, 
and in obtaining Eq.~(\ref{eq:ED1-15-VIIID1-Add-10}), we have 
taken $N_{\mathrm{l}} = 1$. 
For $\lambda_{\mathrm{HL}} = \mu_{\mathrm{HL}} = 1$, 
these resultant equations are reduced to those in ordinary $F(R)$ gravity. 

By using the continuity equation $\dot{\rho}_\mathrm{M} + 3H\left( \rho_\mathrm{M} + P_\mathrm{M} \right) = 0$ in terms of a perfect fluid and 
executing the integration of Eq.~(\ref{eq:ED1-15-VIIID1-Add-9}), we find 
\begin{equation} 
F(\bar{R}) -6\left[
\left(1-3\lambda_{\mathrm{HL}} +3\mu_{\mathrm{HL}} \right)H^2 
+ \mu_{\mathrm{HL}} \dot{H} \right]F^{'}(\bar{R}) 
+6\mu_{\mathrm{HL}} H \dot{\bar{R}} F^{''}(\bar{R}) -\kappa^2 \rho_\mathrm{M} 
-\frac{C_{\mathrm{HL}}}{a^3} = 0\,,
\label{eq:ED1-15-VIIID1-Add-11}
\end{equation}
with $C_{\mathrm{HL}}$ being a constant of integration, which has to be 
chosen to $0$ so that the constraint equation (\ref{eq:ED1-15-VIIID1-Add-9}) 
can be satisfied. 

We note that for $F(\bar{R}) = \bar{R}$, 
in the flat FLRW background the gravitational field equations are written as 
$H^2 = \left\{ \kappa^2/\left[3\left(3\lambda_{\mathrm{HL}}-1\right)\right]\right\} \rho_\mathrm{M}$
and 
$\dot{H} = -\left\{ \kappa^2/\left[2\left(3\lambda_{\mathrm{HL}}-1\right)\right]\right\}\left( \rho_\mathrm{M} + P_\mathrm{M} \right)$ 
with $\lambda_{\mathrm{HL}} > 1/3$ due to the consistency, 
and for $\lambda_{\mathrm{HL}} \to 1$ these equations become 
the ordinary Einstein equations in general relativity.

\subsubsection{Reconstruction of $F(R)$ form} 

We further analyze Eq.~(\ref{eq:ED1-15-VIIID1-Add-11}). 
It follows from the continuity equation of a perfect fluid with 
its constant EoS $w_\mathrm{M} \equiv P_\mathrm{M}/\rho_\mathrm{M}$ 
that $\rho_\mathrm{M} = \rho_{\mathrm{M} \, \mathrm{c}} 
a^{-3\left( 1+w_\mathrm{M} \right)} \e^{-3\left( 1+w_\mathrm{M} \right) N}$ 
with $\rho_{\mathrm{M} \, \mathrm{c}}$ being a constant, 
where $N \equiv \ln \left(a/a_\mathrm{c}\right)$ with $a_\mathrm{c}$ being 
a constant is the number of $e$-folds. 
We replace the cosmic time $t$ as a variable with $N$, 
so that Eq.~(\ref{eq:ED1-15-VIIID1-Add-8}) can be rewritten to 
$\bar{R} 
= 3\left(1-3\lambda_{\mathrm{HL}} +6\mu_{\mathrm{HL}} \right) \bar{G}(N) 
+3\mu_{\mathrm{HL}} \left(d\bar{G}(N)/dN\right)
$, where we have defined $\bar{G}(N) \equiv H^2$ and used it 
in order to analyze Eq.~(\ref{eq:ED1-15-VIIID1-Add-8}) easier. 
Since this equation can be solved as $N =N (\bar{R})$ and 
$H$ can be represented as $H=H(N)$, Eq.~(\ref{eq:ED1-15-VIIID1-Add-8}) 
can be described as an equation of $F(\bar{R})$ in terms of $\bar{R}$. 
Accordingly, 
Eq.~(\ref{eq:ED1-15-VIIID1-Add-8}) can be rewritten to 
\begin{eqnarray} 
&&
F(\bar{R}) -6\left[
\left(1-3\lambda_{\mathrm{HL}} +3\mu_{\mathrm{HL}} \right) \bar{G} 
+ \frac{\mu_{\mathrm{HL}}}{2} \frac{d\bar{G}(N)}{dN} \right] 
\frac{dF(\bar{R})}{d\bar{R}} 
\nonumber \\ 
&&
{}+18\mu_{\mathrm{HL}} 
\left[
\left(1-3\lambda_{\mathrm{HL}} +6\mu_{\mathrm{HL}} \right) \bar{G} 
\frac{d\bar{G}}{dN} 
+ \mu_{\mathrm{HL}} \mathcal{G} \frac{d^2\bar{G}(N)}{dN^2} \right] 
\frac{d^2F(\bar{R})}{d\bar{R}^2} 
\nonumber \\ 
&&
{}
-\kappa^2 \rho_{\mathrm{M} \, \mathrm{c}} 
a_{\mathrm{c}}^{-3\left( 1+w_\mathrm{M} \right)} \e^{-3\left( 1+w_\mathrm{M} \right) N} 
= 0\,,
\label{eq:ED1-15-VIIID1-Add-12}
\end{eqnarray}

First, we reconstruct the form of $F(R)$ with realizing 
the $\Lambda$CDM cosmology, in which 
the Friedmann equation (\ref{eq:Add-2-01}) can be described by 
$\bar{G}(N) = H^2 = H_\mathrm{c}^2  + \left(\kappa^2/3\right) 
\rho_{\mathrm{M} \, \mathrm{c}} 
a_{\mathrm{c}}^{-3} \e^{-3N}$ with $H_\mathrm{c}$ being a constant, 
where we have used $w_\mathrm{M} = 0$ because a perfect fluid is 
considered to a non-relativistic matter. 
For general relativity, $H_\mathrm{c}^2 = \Lambda/3$ as seen 
in Eq.~(\ref{eq:II.02}). 
Moreover, we have 
$
\e^{-3N} = 
\left[
\bar{R}-3\left(1-3\lambda_{\mathrm{HL}} +6\mu_{\mathrm{HL}} \right)
H_\mathrm{c}^2 \right]/\left\{ \kappa^2 \rho_{\mathrm{M} \, \mathrm{c}} 
a_{\mathrm{c}}^{-3} \left[1+3\left(\mu_{\mathrm{HL}}-\lambda_{\mathrm{HL}}\right) \right] \right\}
$. 
With these equations and $w_\mathrm{M} = 0$, 
Eq.~(\ref{eq:ED1-15-VIIID1-Add-12}) is represented as 
\begin{eqnarray} 
&&
\left(1-3\lambda_{\mathrm{HL}} +3\mu_{\mathrm{HL}} \right)F(\bar{R}) 
-2\left(1-3\lambda_{\mathrm{HL}}+\frac{3}{2}\mu_{\mathrm{HL}} \right)\bar{R} 
+9\mu_{\mathrm{HL}}\left(1-3\lambda_{\mathrm{HL}}\right) H_\mathrm{c}^2 
\frac{dF(\bar{R})}{d\bar{R}} 
\nonumber \\ 
&&
{}-6\mu_{\mathrm{HL}} \left(\bar{R}-9\mu_{\mathrm{HL}}H_\mathrm{c}^2\right) 
\left[ \bar{R}-3H_\mathrm{c}^2 
\left(1-3\lambda_{\mathrm{HL}} +6\mu_{\mathrm{HL}} \right) 
\right] \frac{d^2F(\bar{R})}{d\bar{R}^2} 
\nonumber \\ 
&&
{}-\bar{R}-3\left(1-3\lambda_{\mathrm{HL}} +6\mu_{\mathrm{HL}} 
\right)H_\mathrm{c}^2 = 0\,,
\label{eq:ED1-15-VIIID1-Add-13}
\end{eqnarray}
The homogeneous part of Eq.~(\ref{eq:ED1-15-VIIID1-Add-13}) is 
rewritten to 
\begin{eqnarray}
&&
\tau\left(1-\tau\right)\frac{d^2F(\bar{R})}{d\tau^2} + 
\left[ \bar{\gamma} - \left(\bar{\alpha}+\bar{\beta}+1\right) \tau\right]
\frac{dF(\bar{R})}{d\tau} -\bar{\alpha} \bar{\beta} F(\bar{R}) = 0\,, 
\label{eq:ED1-15-VIIID1-Add-14} \\
&&
\tau \equiv \frac{\bar{R}-9\mu_{\mathrm{HL}}H_\mathrm{c}^2}{3H_\mathrm{c}^2
\left[ 1+3\left(\mu_{\mathrm{HL}} -\lambda_{\mathrm{HL}}\right)\right]}\,, 
\label{eq:ED1-15-VIIID1-Add-15}
\end{eqnarray} 
where 
\begin{eqnarray}
\bar{\gamma} \Eqn{\equiv} -\frac{1}{2}\,, 
\label{eq:ED1-15-VIIID1-Add-16} \\
\bar{\alpha}+\bar{\beta}+1 \Eqn{\equiv} \frac{1-3\left(\lambda_{\mathrm{HL}}+\mu_{\mathrm{HL}}/2\right)}{3\mu_{\mathrm{HL}}}\,,
\label{eq:ED1-15-VIIID1-Add-17} \\
\bar{\alpha} \bar{\beta} \Eqn{\equiv} 
-\frac{1+3\left(\mu_{\mathrm{HL}}-\lambda_{\mathrm{HL}}\right)}{6\mu_{\mathrm{HL}}}\,.
\label{eq:ED1-15-VIIID1-Add-18}
\end{eqnarray}
We describe the complete solution of Eq.~(\ref{eq:ED1-15-VIIID1-Add-15}) 
with the Gauss's hypergeometric function $F_{\mathrm{G}}$ as 
\begin{eqnarray} 
F(\bar{R}) \Eqn{=} F_{\mathrm{c}1}
F_{\mathrm{G}} (\bar{\alpha}, \bar{\beta}, \bar{\gamma}; \tau)
+ F_{\mathrm{c}2} \tau^{1-\bar{\gamma}} 
F_{\mathrm{G}} (\bar{\alpha}-\bar{\gamma}+1, \bar{\beta}-\bar{\gamma}+1, 
-\bar{\gamma}+2; \tau) 
\nonumber \\
&&
{}+\frac{1}{3\lambda_{\mathrm{HL}}-1}\bar{R} -2\Lambda \,, 
\label{eq:ED1-15-VIIID1-Add-19} \\
\Lambda \Eqn{=} 
-\frac{3\left[1-3\left(\lambda_{\mathrm{HL}}-3\mu_{\mathrm{HL}}\right)\right]H_\mathrm{c}^2}{2
\left[ 1-3\left(\lambda_{\mathrm{HL}}-\mu_{\mathrm{HL}}\right) \right]}\,, 
\label{eq:ED1-15-VIIID1-Add-20}
\end{eqnarray}
with $F_{\mathrm{c}1}$ and $F_{\mathrm{c}2}$ being constants. 
Thus, a class of the reconstructed $F(R)$ theories in Eq.~(\ref{eq:ED1-15-VIIID1-Add-20}) can represent the $\Lambda$CDM cosmology. 
We remark that for $\mu_{\mathrm{HL}} = \lambda_{\mathrm{HL}} -1/3$, 
$\bar{R}$ becomes a constant, so that the solution in Eq.~(\ref{eq:ED1-15-VIIID1-Add-20}) can be expressed as 
$F(\bar{R}) = F(\bar{R})/\left(3\lambda_{\mathrm{HL}} -1\right) - 2\Lambda$, 
where $\Lambda = \left(3/2\right)\left(3\lambda_{\mathrm{HL}} -1\right)H_\mathrm{c}^2$. 

Next, we reconstruct an $F(R)$ form describing the phantom cosmology. 
Since we examine the dark energy dominated stage, for simplicity, non-relativistic matter contributions are neglected.
At the dark energy dominated stage, by using the continuity equation 
in terms of dark energy, the Hubble parameter can be represented as 
$H=H_{\mathrm{ph}}/\left(t_{\mathrm{s}}-t\right)$ with 
$H_{\mathrm{ph}} \equiv -1/3\left(1+w_{\mathrm{DE}}\right)$, 
where at $t=t_{\mathrm{s}}$, a Big Rip singularity appears. 
In what follows, 
{}From this expression, we have 
$\bar{G}(N) = H^2 (N) = H_{\mathrm{ph}} \e^{2N/H_{\mathrm{ph}}}$. 
In this case, the relation of $N$ to $\bar{R}$ is described by 
$
\e^{2N/H_{\mathrm{ph}}} = \bar{R}/
\left[ H_{\mathrm{ph}} \left( A_{\mathrm{ph}} H_{\mathrm{ph}} +6\mu_{\mathrm{HL}} \right) 
\right]
$ with $A_{\mathrm{ph}}$ being a constant. 
By combining these relations and Eq.~(\ref{eq:ED1-15-VIIID1-Add-12}), 
we acquire the Euler equation 
\begin{eqnarray} 
&&
\bar{R}^2 \frac{d^2F(\bar{R})}{d\bar{R}^2} 
+\mathcal{W}_{\mathrm{ph} 1} \bar{R} \frac{dF(\bar{R})}{d\bar{R}} 
+\mathcal{W}_{\mathrm{ph} 2} F(\bar{R}) =0\,,
\label{eq:ED1-15-VIIID1-Add-21} \\
&&
\mathcal{W}_{\mathrm{ph} 1} \equiv 
-\frac{\left( A_{\mathrm{ph}} H_{\mathrm{ph}} +6\mu_{\mathrm{HL}} \right) \left( A_{\mathrm{ph}} H_{\mathrm{ph}} +3\mu_{\mathrm{HL}} \right) }{6\mu_{\mathrm{HL}}\left( A_{\mathrm{ph}} H_{\mathrm{ph}} +12\mu_{\mathrm{HL}} \right)}
\,,
\label{eq:ED1-15-VIIID1-Add-22} \\
&&
\mathcal{W}_{\mathrm{ph} 2} \equiv
\frac{\left( A_{\mathrm{ph}} H_{\mathrm{ph}} +6\mu_{\mathrm{HL}} \right)^2}{12\mu_{\mathrm{HL}}\left( A_{\mathrm{ph}} H_{\mathrm{ph}} +12\mu_{\mathrm{HL}} \right)}\,.
\label{eq:ED1-15-VIIID1-Add-23} 
\end{eqnarray}
A solution of Eq.~(\ref{eq:ED1-15-VIIID1-Add-21}) is given by 
\begin{eqnarray} 
F(\bar{R}) \Eqn{=} F_{\mathrm{ph} +} \bar{R}^{\bar{y}_{\mathrm{ph} +}} 
+ F_{\mathrm{ph} -} \bar{R}^{\bar{y}_{\mathrm{ph} -}} 
\,,
\label{eq:ED1-15-VIIID1-Add-24} \\
\bar{y}_{\mathrm{ph} \pm} \Eqn{\equiv} 
\frac{1-\mathcal{W}_{\mathrm{ph} 1} \pm \sqrt{\left(\mathcal{W}_{\mathrm{ph} 1}-1\right)^2 -4 \mathcal{W}_{\mathrm{ph} 2}}}{2}\,, 
\label{eq:ED1-15-VIIID1-Add-25}
\end{eqnarray}
where $F_{\mathrm{ph} +}$ and $F_{\mathrm{ph} -}$ are constants. 
Hence, a reconstructed form of $F(R)$ in Eq.~(\ref{eq:ED1-15-VIIID1-Add-24}) can describe the phantom cosmology. 
As a result, it is considered that by using the procedure explained above, 
in principle, 
an $F(R)$ form with representing any cosmology could be reconstructed.

\subsection{$F(R, \mathcal{T})$ gravity}

In Ref.~\cite{Harko:2011kv}, the formulations of a novel modified gravitational theory, the so-called $F(R, \mathcal{T})$ gravity with 
$\mathcal{T}$ being the trace of the stress-energy tensor, 
which can explain the late-time cosmic acceleration, 
have been investigated. 
In this subsection, we review this latest theory.

\subsubsection{Formulations}

The action of $F(R, \mathcal{T})$ gravity is given by~\cite{Harko:2011kv} 
\begin{equation} 
S = 
\int d^4 x \sqrt{-g} 
\frac{F(R, \mathcal{T})}{16\pi} + 
\int d^4 x \sqrt{-g} 
L_{\mathrm{M}}\,,
\label{eq:VIIIE-001}
\end{equation}
where $\mathcal{T} = g^{\mu\nu} \mathcal{T}_{\mu\nu}$ is the trace of the stress-energy tensor of matter, defined as~\cite{Landau-Lifshitz1998} 
$\mathcal{T}_{\mu\nu} \equiv - \left(2/\sqrt{-g}\right) 
\delta \left( \sqrt{-g} L_{\mathrm{M}} \right)/\delta g\Theta_{\mu\nu}$, 
$L_{\mathrm{M}}$ is the Lagrangian density of matter, 
and $F(R, \mathcal{T})$ is an arbitrary function of $R$ and 
$T$. 
Here and in this subsection, we use the unit of $G=c=1$. 
As past related studies, a theory whose 
Lagrangian density is described by an arbitrary function of $R$ and 
the Lagrangian density of matter as $F(R, L_{\mathrm{M}})$ 
has been explored in Ref.~\cite{Harko:2010mv}. 
Moreover, in Ref.~\cite{Poplawski:2006ey} a theory in which the cosmological 
constant is written by a function of the trace of the stress-energy tensor as 
$\Lambda (\mathcal{T})$ has been investigated. 

{}From the action in Eq.~(\ref{eq:VIIIE-001}), the gravitational field 
equation is given by
\begin{equation} 
F_R(R, \mathcal{T}) -\frac{1}{2} F(R, \mathcal{T}) g_{\mu\nu} 
+ \left(g_{\mu\nu} \Box - \nabla_\mu \nabla_\nu \right) F_R(R, \mathcal{T}) 
= 8\pi\left(-F_\mathcal{T}(R, \mathcal{T})\right) \mathcal{T}_{\mu\nu} 
- F_\mathcal{T}(R, \mathcal{T}) \Theta_{\mu\nu}\,,
\label{eq:VIIIE-002}
\end{equation}
with 
%
$
\Theta_{\mu\nu} \equiv g^{\alpha \beta} 
\left( \delta \mathcal{T}_{\alpha \beta}/\delta g^{\mu\nu} \right)
$, which follows from the relation 
$\delta \left( g^{\alpha \beta} \mathcal{T}_{\alpha \beta}/\delta g^{\mu\nu} \right) = \mathcal{T}_{\mu\nu} + \Theta_{\mu\nu}$, 
%
and $F_R(R, \mathcal{T}) \equiv \partial F(R, \mathcal{T})/ \partial R$, 
$F_\mathcal{T} (R, \mathcal{T}) \equiv \partial F(R, \mathcal{T})/ \partial \mathcal{T}$. 
The contraction of Eq.~(\ref{eq:VIIIE-002}) yields 
$
F_R(R, \mathcal{T}) R + 3\Box F_R(R, \mathcal{T}) -2F(R, \mathcal{T}) 
= \left(8\pi - F_\mathcal{T}(R, \mathcal{T})\right) \mathcal{T} - F_\mathcal{T}(R, \mathcal{T}) \Theta$ 
with $\Theta \equiv g^{\mu\nu} \Theta_{\mu\nu}$. 
Combining Eq.~(\ref{eq:VIIIE-002}) and the contracted equation and 
eliminating the $\Box F_R(R, \mathcal{T})$ term from these equations, 
we find 
\begin{eqnarray}
\hspace{-18mm}
&&
F_R(R, \mathcal{T}) 
\left( R_{\mu\nu} -\frac{1}{3}R g_{\mu\nu} \right) 
+\frac{1}{6}F(R, \mathcal{T}) g_{\mu\nu} 
\nonumber \\ 
\hspace{-18mm}
&&
{}
= \left(8\pi - F_\mathcal{T}(R, \mathcal{T}) \right) 
\left( \mathcal{T}_{\mu\nu} -\frac{1}{3} \mathcal{T} g_{\mu\nu} \right) 
-F_\mathcal{T}(R, \mathcal{T})
\left( \Theta_{\mu\nu} -\frac{1}{3} \Theta g_{\mu\nu} \right) 
+ \nabla_\mu \nabla_\nu F_R(R, \mathcal{T})\,.
\label{eq:VIIIE-003}
\end{eqnarray}

On the other hand, the covariant divergence of Eq.~(\ref{eq:VIIIE-001}) 
as well as the energy-momentum conservation law 
$
\nabla^{\mu} \left[ 
F_R(R, \mathcal{T}) -\left(1/2\right) F(R, \mathcal{T}) g_{\mu\nu} 
+ \left(g_{\mu\nu} \Box - \nabla_\mu \nabla_\nu \right) F_R(R, \mathcal{T}) 
\right] \equiv 0$, 
which corresponds to the divergence of the left-hand side of Eq.~(\ref{eq:VIIIE-001}), we acquire the divergence of $\mathcal{T}_{\mu\nu}$ as 
\begin{equation}  
\nabla^{\mu} \mathcal{T}_{\mu\nu} = 
\frac{F_\mathcal{T}(R, \mathcal{T})}{8\pi - F_\mathcal{T}(R, \mathcal{T})} 
\left[ \left( \mathcal{T}_{\mu\nu} + \Theta_{\mu\nu} \right) \nabla^{\mu} 
\ln F_\mathcal{T}(R, \mathcal{T}) + \nabla^{\mu} \Theta_{\mu\nu} 
\right]\,.
\label{eq:VIIIE-004}
\end{equation}

In addition, from $\mathcal{T}_{\mu\nu} = g_{\mu\nu} L_{\mathrm{M}} -2\left( 
\partial L_{\mathrm{M}}/ \partial g^{\mu\nu} \right)$ 
we have 
\begin{equation} 
\frac{\delta \mathcal{T}_{\alpha \beta}}{\delta g^{\mu\nu}} 
= \left( \frac{\delta g_{\alpha \beta}}{\delta g^{\mu\nu}} +\frac{1}{2} g_{\alpha \beta} g_{\mu\nu} \right) L_{\mathrm{M}} -\frac{1}{2} g_{\alpha \beta} 
\mathcal{T}_{\mu\nu} -2 \frac{\partial^2 L_{\mathrm{M}}}{\partial g^{\mu\nu} \partial g^{\alpha \beta}}\,.
\label{eq:VIIIE-005}
\end{equation}
Using the relation $\delta g_{\alpha \beta}/\delta g^{\mu\nu} = -g_{\alpha \rho}g_{\beta \sigma} \delta_{\mu\nu}^{\rho \sigma}$ 
with $\delta_{\mu\nu}^{\rho \sigma} = 
\delta g^{\rho \sigma}/\delta g^{\mu\nu}$, 
which follows from 
$g_{\alpha \rho} g^{\rho \beta} = \delta_{\alpha}^{\beta}$, 
we obtain 
\begin{equation}  
\Theta_{\mu\nu} = -2 \mathcal{T}_{\mu\nu} + g_{\mu\nu} L_{\mathrm{M}} -2g^{\alpha \beta} \frac{\partial^2 L_{\mathrm{M}}}{\partial g^{\mu\nu} \partial g^{\alpha \beta}}\,.
\label{eq:VIIIE-006}
\end{equation}
Provided that matter is regarded as a perfect fluid, $\mathcal{T}_{\mu\nu}$ is 
expressed as $\mathcal{T}_{\mu\nu} = \left( \rho_{\mathrm{M}} + P_{\mathrm{M}} \right) u_\mu u_\nu -P_{\mathrm{M}} g_{\mu\nu}$, where $u_\mu$ being the four velocity satisfying $g^{\mu\nu} u_\mu u_\nu =-1$, and $\rho_{\mathrm{M}}$ and $P_{\mathrm{M}}$ are the energy density and pressure of the perfect fluid, respectively, we acquire $\Theta_{\mu\nu} = -2 \mathcal{T}_{\mu\nu} 
- P_{\mathrm{M}} g_{\mu\nu}$.

\subsubsection{Example}

As an example, we examine the case that $F(R, \mathcal{T}) = R + 2F_1 (\mathcal{T})$ with $F_1 (\mathcal{T})$ being an arbitrary function of $\mathcal{T}$ 
and matter is a perfect fluid. In this case, Eq.~(\ref{eq:VIIIE-002}) 
becomes
\begin{equation} 
G_{\mu\nu} 
= \left( 8\pi+2 \frac{d F_1 (\mathcal{T})}{d \mathcal{T}} \right) \mathcal{T}_{\mu\nu} + \left( 2P_{\mathrm{M}} \frac{d F_1 (\mathcal{T})}{d \mathcal{T}} + F_1 (\mathcal{T}) \right) g_{\mu\nu}\,, 
\label{eq:VIIIE-007}
\end{equation}
where $G_{\mu\nu} = R_{\mu\nu} - \left(1/2\right)Rg_{\mu\nu}$ is the 
Einstein tensor. 
In the flat FLRW background, 
for the matter to be a dust, i.e., $P_{\mathrm{M}} = 0$, 
and $F_1 (\mathcal{T}) = \lambda \mathcal{T}$ with $\lambda$ being a constant, 
the gravitational field equations are given by 
\begin{eqnarray}
\left(\frac{\dot{a}}{a}\right)^2 \Eqn{=} \left( \frac{8\pi}{3} + \lambda \right) \rho_{\mathrm{M}}\,,
\label{eq:VIIIE-008} \\  
2\frac{\ddot{a}}{a} + \left(\frac{\dot{a}}{a}\right)^2 \Eqn{=} \lambda \rho_{\mathrm{M}}\,. 
\label{eq:VIIIE-009}
\end{eqnarray}
{}From these equations, we have 
\begin{equation} 
\dot{H} + \frac{3\left(8\pi + 2\lambda \right)}{2\left(8\pi + 3\lambda\right)} 
H^2 =0\,.
\label{eq:VIIIE-010}
\end{equation}
The solution of this equation is given by 
$H= \tilde{p}/t$ with $\tilde{p} \equiv \left[2\left(8\pi + 3\lambda\right)\right]/\left[3\left(8\pi + 2\lambda\right)\right]$. 
Thus, we obtain $a = t^{\tilde{p}}$ with $\tilde{p} > 1$, 
and consequently the accelerated expansion of the universe can be realized.

\section{$f(T)$ gravity}

In this section, we explore $f(T)$ gravity\footnote{
For clarity, we use the notation ``$F(R)$'' gravity and 
``$f(T)$'' gravity throughout this review.}. 
It is known that 
as a candidate of an alternative 
gravitational theory to general relativity, 
there exists ``teleparallelism" in which 
the Weitzenb\"{o}ck connection is used~\cite{Teleparallelism}. 
In this theory, 
there is only torsion $T$ and 
the curvature $R$ defined by the Levi-Civita connection 
does not exist. 
Recently, 
to account for the late time accelerated expansion of the universe as well as 
inflation in the early universe~\cite{Inflation-F-F}, 
by extending 
the teleparallel Lagrangian density described by the torsion scalar $T$ 
to a function of $T$ as $f(T)$~\cite{Bengochea:2008gz, Linder:2010py}, 
various studies in $f(T)$ gravity have been executed. 
This concept has the same origin as the idea of $f(R)$ gravity. 
In order to examine 
whether $f(T)$ gravity can be worthy of being 
an alternative theory of gravitation to general relativity, 
recently a number of aspects of $f(T)$ gravity have widely been investigated 
in the literature~\cite{hossein, f(T)-Refs, BGL-Comment, 
Local-Lorentz-invariance, C-T-f(T), Thermodynamics-f(T), Bamba:2011pz, Bamba:2012vg, hossein, Setare:2012vs}. 
For example, 
the local Lorentz invariance~\cite{Local-Lorentz-invariance}, 
non-trivial conformal frames~\cite{C-T-f(T)}, 
thermodynamics~\cite{Thermodynamics-f(T), Bamba:2011pz}, 
and finite-time future singularities~\cite{Bamba:2012vg, Setare:2012vs}. 
In this review, we concentrate on the issues on the finite-time future singularities in $f(T)$ gravity and review the results in Ref.~\cite{Bamba:2012vg}.

\subsection{Basic formalism and fundamental equations}

We use 
orthonormal tetrad components $e_A (x^{\mu})$ in the teleparallelism. 
At each point $x^{\mu}$ of the manifold, 
an index $A$ runs over $0, 1, 2, 3$ for the tangent space. 
The relation to the metric $g^{\mu\nu}$ is given by 
%
$
g_{\mu\nu}=\eta_{A B} e^A_\mu e^B_\nu 
$,  
%
where $\mu$ and $\nu$ are coordinate indices on the manifold 
and run over $0, 1, 2, 3$. 
Moreover, $e_A^\mu$ forms the tangent vector of the manifold. 
We define 
the torsion $T^\rho_{\verb| |\mu\nu}$ and contorsion 
$K^{\mu\nu}_{\verb|  |\rho}$ tensors as 
\begin{equation}
T^\rho_{\verb| |\mu\nu} \equiv e^\rho_A 
\left( \partial_\mu e^A_\nu - \partial_\nu e^A_\mu \right) 
\label{eq:IXA-2.2} 
\end{equation}
and 
\begin{equation}
K^{\mu\nu}_{\verb|  |\rho} 
\equiv 
-\frac{1}{2} 
\left(T^{\mu\nu}_{\verb|  |\rho} - T^{\nu \mu}_{\verb|  |\rho} - 
T_\rho^{\verb| |\mu\nu}\right)\,,
\label{eq:IXA-2.3}
\end{equation}
respectively. 
In general relativity, the Lagrangian 
density is described by the Ricci scalar $R$, 
whereas the teleparallel Lagrangian density is represented by 
the torsion scalar $T$. 
With the torsion tensor as well as the contorsion tensor, 
we first define the following quantity 
\begin{equation}
S_\rho^{\verb| |\mu\nu} \equiv \frac{1}{2}
\left(K^{\mu\nu}_{\verb|  |\rho}+\delta^\mu_\rho \ 
T^{\alpha \nu}_{\verb|  |\alpha}-\delta^\nu_\rho \ 
T^{\alpha \mu}_{\verb|  |\alpha}\right)\,. 
\label{eq:IXA-2.5}
\end{equation}
By using this quantity in Eq.~(\ref{eq:IXA-2.5}), 
the torsion scalar $T$ is described as 
%
\begin{equation}
T \equiv S_\rho^{\verb| |\mu\nu} T^\rho_{\verb| |\mu\nu}\,. 
\label{eq:IXA-2.4}
\end{equation}
The modified teleparallel action of $f(T)$ 
gravity with the matter Lagrangian ${\mathcal{L}}_{\mathrm{M}}$ 
is expressed as~\cite{Linder:2010py} 
\begin{equation}
I= 
\int d^4x \abs{e} \left[ \frac{f(T)}{2{\kappa}^2} 
+{\mathcal{L}}_{\mathrm{M}} \right]\,. 
\label{eq:IXA-2.6}
\end{equation}
Here, $\abs{e}= \det \left(e^A_\mu \right)=\sqrt{-g}$. 
By varying the action in Eq.~(\ref{eq:2.6}) with respect to 
the vierbein vector field $e_A^\mu$, 
we acquire the gravitational field equation~\cite{Bengochea:2008gz} 
\begin{equation}
\frac{1}{e} \partial_\mu \left( eS_A^{\verb| |\mu\nu} \right) f^{\prime} 
-e_A^\lambda T^\rho_{\verb| |\mu \lambda} S_\rho^{\verb| |\nu\mu} 
f^{\prime} +S_A^{\verb| |\mu\nu} \partial_\mu \left(T\right) f^{\prime\prime} 
+\frac{1}{4} e_A^\nu f = \frac{{\kappa}^2}{2} e_A^\rho 
{T^{(\mathrm{M})}}_\rho^{\verb| |\nu}\,, 
\label{eq:IXA-2.7}
\end{equation}
with ${T^{(\mathrm{M})}}_\rho^{\verb| |\nu}$ 
being the energy-momentum tensor of all 
perfect fluids of ordinary matter such as 
radiation and non-relativistic matter. 

We assume the flat FLRW 
space-time with the metric 
$ds^2 = dt^2 - a^2(t) \sum_{i=1,2,3}\left(dx^i\right)^2$\footnote{In this section, the metric signature of $(+, -, -, -)$ is adopted.}. 
Therefore, we have 
$g_{\mu \nu}= \mathrm{diag} (1, -a^2, -a^2, -a^2)$ and 
the tetrad components $e^A_\mu = (1,a,a,a)$. 
{}From these relations, we find $T=-6H^2$. 
In this flat FLRW universe, we can write 
the gravitational field equations in the same forms as 
those in general relativity 
\begin{eqnarray}
H^2 \Eqn{=} \frac{{\kappa}^2}{3} \left(\rho_{\mathrm{M}}+\rho_{\mathrm{DE}} 
\right)\,, 
\label{eq:IXA-4.1} \\ 
%
%
\dot{H}
\Eqn{=} -\frac{{\kappa}^2}{2} \left(\rho_{\mathrm{M}} + P_{\mathrm{M}} + 
\rho_{\mathrm{DE}} + P_{\mathrm{DE}} \right)\,, 
\label{eq:IXA-4.2} \\ 
\rho_{\mathrm{DE}} 
\Eqn{=} 
\frac{1}{2{\kappa}^2} J_1\,,
\label{eq:IXA-4.3} \\ 
P_{\mathrm{DE}} 
\Eqn{=} 
-\frac{1}{2{\kappa}^2} 
\left( 
4J_2 
+ J_1 
\right)\,,
\label{eq:IXA-4.4} \\ 
J_1 \Eqn{\equiv} -T -f +2TF\,, 
\label{eq:IXA-IIB-Add-01} \\ 
J_2 \Eqn{\equiv} \left( 1 -F -2TF^{\prime} \right) \dot{H}\,. 
\label{eq:IXA-IIB-Add-02}
\end{eqnarray}
Here, $F\equiv df/dT$ and $F^{\prime} \equiv dF/dT$. 
Moreover, we express the energy density and pressure of all perfect fluids of generic matter as $\rho_{\mathrm{M}}$ and $P_{\mathrm{M}}$, respectively. 
These perfect fluids satisfy the continuity equation 
%
$
\dot{\rho}_{\mathrm{M}}+3H\left( \rho_{\mathrm{M}} + P_{\mathrm{M}} \right)
=0 
$.
%
In addition, 
for the representations of $\rho_{\mathrm{DE}}$ in Eq.~(\ref{eq:IXA-4.3}) and $P_{\mathrm{DE}}$ in Eq.~(\ref{eq:IXA-4.4}), the standard continuity equation 
can be met as 
%
$
\dot{\rho}_{\mathrm{DE}}+3H \left( 
\rho_{\mathrm{DE}} + P_{\mathrm{DE}}
\right)  
= 0
$.

\subsection{Reconstruction of $f(T)$ gravity with realizing the 
finite-time future singularities}

\subsubsection{Finite-time future singularities in $f(T)$ gravity}

We provided that an expression of the Hubble parameter~\cite{Nojiri:2008fk} 
realizing the finite-time future singularities 
and the resultant scale factor obtained from the form of the Hubble parameter 
are given by
\begin{eqnarray} 
H \Eqn{\sim} \frac{h_{\mathrm{s}}}{ \left( t_{\mathrm{s}} - t 
\right)^{q}}\,, 
\quad 
\mathrm{for} \,\,\, q > 0\,,
\label{eq:ED1-10-IXB-2.13} \\ 
H \Eqn{\sim} H_{\mathrm{s}} + \frac{h_{\mathrm{s}}}{ 
\left( t_{\mathrm{s}} - t \right)^{q}}\,, 
\quad 
\mathrm{for} \,\,\, q<-1\,, \,\,\, -1< q < 0\,, 
\label{eq:ED1-10-IXB-IIIB-add-001} \\
a \Eqn{\sim} a_{\mathrm{s}} \exp \left[ \frac{h_{\mathrm{s}}}{q-1} 
\left( t_{\mathrm{s}} - t 
\right)^{-\left(q-1\right)}
\right]
\quad 
\mathrm{for}\,\,\, 
0<q<1\,, \,\,\, 1<q\,, 
\label{eq:ED1-10-IXB-2.14} \\
a \Eqn{\sim} a_{\mathrm{s}} \frac{h_{\mathrm{s}}}{\left( t_{\mathrm{s}} - t 
\right)^{h_{\mathrm{s}}}}
\quad 
\mathrm{for}\,\,\, q = 1\,,
\label{eq:ED1-10-IXB-IIIB-add-01}
\end{eqnarray}
with $h_{\mathrm{s}} (> 0)$, $H_{\mathrm{s}} (> 0)$ and $a_{\mathrm{s}} (> 0)$ 
being positive constants, $q (\neq 0, \, -1)$ a non-zero constant, 
and $t_{\mathrm{s}}$ the time when the finite-time future singularity 
appears. We only deal with the period $0< t < t_{\mathrm{s}}$ 
because $H$ should be a real number. 
For the expression of $H$ in Eqs.~(\ref{eq:ED1-10-IXB-2.13}) and (\ref{eq:ED1-10-IXB-IIIB-add-001}), the finite-time future singularities occur in the 
case of each range of the value of $q$. 
It follows from $\rho_{\mathrm{DE}} \approx \rho_{\mathrm{eff}} = 3 H^2/\kappa^2$ in Eq.~(\ref{eq:IXA-4.3}), $P_{\mathrm{DE}} \approx P_{\mathrm{eff}} = -\left(2\dot H + 3H^2\right)/\kappa^2$ in Eq.~(\ref{eq:IXA-4.4}), 
and the relation $T=-6H^2$ that in the limit of $t \to t_{\mathrm{s}}$, 
if $H \to \infty$, both $\rho_{\mathrm{DE}}$ 
and $P_{\mathrm{DE}}$ diverge; if $H$ becomes finite but $\dot{H}$ does 
infinite, $P_{\mathrm{DE}}$ diverges, although $\rho_{\mathrm{DE}}$ does not. 
We note that $J_1$ in Eq.~(\ref{eq:IXA-IIB-Add-01}) only depends on $T$, i.e., $H$ 
and $J_2$ in Eq.~(\ref{eq:IXA-IIB-Add-02}) is proportional to $\dot{H}$. 
In Table \ref{tb:table1}, we summarize the conditions to produce 
the finite-time future singularities in the limit of $t \to t_{\mathrm{s}}$.

\begin{table*}[tbp]
\caption{
Conditions to produce 
the finite-time future singularities in the limit of $t \to t_{\mathrm{s}}$. 
}
\begin{center}
\begin{tabular}
{lllllll}
\hline
\hline
Type \quad 
& $q (\neq 0, \, -1)$ \quad 
& $a$  
& $H$  
& $\dot{H}$ 
& $\rho_{\mathrm{DE}}$
& $P_{\mathrm{DE}}$
\\[0mm]
\hline
I
& $q \geq 1$ 
& $a \to \infty$ \quad 
& $H \to \infty$
& $\dot{H} \to \infty$
& $J_1 \neq 0$ \quad 
& $J_1 \neq 0$  
\\[0mm]
&
&
&
&
&
& or $J_2 \neq 0$
\\[0mm]
III
& $0 < q < 1$ 
& $a \to a_{\mathrm{s}}$
& $H \to \infty$
& $\dot{H} \to \infty$
& $J_1 \neq 0$ \quad 
& $J_1 \neq 0$
\\[0mm]
II
& $-1 < q < 0$ 
& $a \to a_{\mathrm{s}}$
& $H \to H_{\mathrm{s}}$ \quad 
& $\dot{H} \to \infty$
& 
& $J_2 \neq 0$
\\[0mm]
IV 
& $q < -1$ 
& $a \to a_{\mathrm{s}}$
& $H \to H_{\mathrm{s}}$
& $\dot{H} \to 0$
& 
& 
\\[0mm]
& ($q \neq \mathrm{integer}$) 
&
& 
& (Higher 
&
&
\\[0mm]
&
&
&
& derivatives
&
&
\\[0mm]
&
&
&
& of $H$ diverge.)
&
&
\\[1mm]
\hline
\hline
\end{tabular}
\end{center}
\label{tb:table1}
\end{table*}

\subsubsection{Reconstruction of an $f(T)$ gravity model}

Next, we reconstruct an $f(T)$ gravity model with realizing 
the finite-time future singularities in the limit of $t \to t_{\mathrm{s}}$. 
{}From Eqs.~(\ref{eq:IXA-4.3}) and (\ref{eq:IXA-4.4}), 
the EoS of dark energy is given by  
\begin{equation} 
w_{\mathrm{DE}} 
= \frac{P_{\mathrm{DE}}}{\rho_{\mathrm{DE}}}
= \frac{-\left[ 
4\left(1 -F -2TF^{\prime} \right) \dot{H} 
+\left( -T -f +2TF \right)
\right]}{-T -f +2TF}\,. 
\label{eq:ED1-10-IXB2-IVB-1} 
\end{equation}
This expression can be rewritten to 
a fluid description explained in Sec.~III A as 
\begin{eqnarray} 
P_{\mathrm{DE}} \Eqn{=} -\rho_{\mathrm{DE}} 
+ \mathcal{J} (H, \dot{H})\,,
\label{eq:ED1-10-IXB2-IVB-2} \\
\mathcal{J} \Eqn{\equiv} 
- \frac{1}{{\kappa}^2} 
\left[ 
2\left(1 -F -2TF^{\prime} \right) \dot{H} \right]\,. 
\label{eq:ED1-10-IXB2-IVB-3}
\end{eqnarray} 
Here, $\mathcal{J}$ corresponds to $-f(\rho)$ in Eq.~(\ref{eq:2.21}). 
The comparison of Eq.~(\ref{eq:ED1-10-IXB2-IVB-2}) with 
$P_{\mathrm{eff}} = - \rho_{\mathrm{eff}} -2\dot{H}/\kappa^2$ 
leads to 
%
$
\dot{H} + \left( \kappa^2/2\right) \mathcal{J} (H, \dot{H}) = 0
$. 
By combining Eq.~(\ref{eq:ED1-10-IXB2-IVB-3}) and this equation, we obtain 
%
$
\dot{H} \left( F +2TF^{\prime} \right) = 0
$. 
This yields the condition $F +2TF^{\prime} = 0$ because $\dot{H} \neq 0$ for 
$H$ in Eqs.~(\ref{eq:ED1-10-IXB-2.13}) and (\ref{eq:ED1-10-IXB-IIIB-add-001}). 

On the other hand, 
Eqs.~(\ref{eq:IXA-4.1}) and (\ref{eq:IXA-4.2}) can be reduced to 
\begin{eqnarray} 
-f +2TF \Eqn{=} 0\,, 
\label{eq:ED1-10-IXB2-IV-BB-add-03} \\ 
F + 2TF^{\prime} \Eqn{=}  0\,, 
\label{eq:ED1-10-IXB2-IV-BB-add-04} 
\end{eqnarray}
where we have also used Eqs.~(\ref{eq:IXA-4.3})--(\ref{eq:IXA-IIB-Add-02}) and 
$\dot{H} \neq 0$ for $H$ in Eqs.~(\ref{eq:ED1-10-IXB-2.13}) and (\ref{eq:ED1-10-IXB-IIIB-add-001}). 
We see that Eq.~(\ref{eq:ED1-10-IXB2-IV-BB-add-04}) is equivalent to 
the above condition and Eq.~(\ref{eq:ED1-10-IXB2-IV-BB-add-03}) corresponds to 
a consistency condition. 
For a power-law model given by 
\begin{equation} 
f(T) = A T^{\alpha}\,, 
\label{eq:ED1-10-IXB2-IVA2-addition-01}
\end{equation} 
with $A (\neq 0)$ and $\alpha (\neq 0)$ being non-zero constants, 
which with $A=1$ and $\alpha = 1$ corresponds to general relativity, 
Eq.~(\ref{eq:ED1-10-IXB2-IV-BB-add-04}) becomes 
%
$
F +2TF^{\prime} 
= A\left(-6\right)^{\alpha -1} \left(2\alpha-1\right) 
H^{2\left(\alpha-1\right)} 
= 0
$, 
and Eq.~(\ref{eq:ED1-10-IXB2-IV-BB-add-03}) reads 
$
-f +2TF = 
A\left(-6\right)^{\alpha} \left(2\alpha-1\right) 
H^{2\alpha} 
= 0
$. 
In the limit of $t \to t_{\mathrm{s}}$, 
both of these equations have to be satisfied asymptotically. 
If $\alpha = 1/2$, these equations are always met. 
Hence, for $q>0$, $\alpha <0$, while for $q<0$, $\alpha =1/2$. 
We remark that these are not sufficient but necessary conditions 
to realize the finite-time future singularities. 
In fact, if $\alpha <0$, 
the Type I singularity occurs faster than the Type III and 
finally the Type I singularity happens because the speed of the 
divergence depends on the absolute value of $q$. 
{}From the same reason, 
if $\alpha =1/2$, the Type IV singularity appears faster than 
the Type II singularity and eventually the Type IV singularity occurs. 
We also note that for 
an exponential model 
$
f(T) = C_{\mathrm{e}} \exp \left( \lambda_{\mathrm{e}} T \right) 
$ with $C_{\mathrm{e}} (\neq 0)$ and $\lambda_{\mathrm{e}} (\neq 0)$ being non-zero constants 
and 
an logarithmic model
$
f(T) = D_{\mathrm{l}} \ln \left( \gamma_{\mathrm{l}} T \right) 
$ with $D_{\mathrm{l}} (\neq 0)$ being a non-zero constant and $\gamma_{\mathrm{l}} (> 0)$ 
being a positive constant, 
both 
Eqs.~(\ref{eq:ED1-10-IXB-2.13}) and (\ref{eq:ED1-10-IXB-IIIB-add-001}) 
cannot be simultaneously met, 
and therefore these models cannot produce 
the finite-time future singularities. 

For ``$w$'' singularity, 
the scale factor is given by~\cite{Dabrowski:2009zzb} 
\begin{eqnarray} 
a (t) \Eqn{=} a_{\mathrm{s}} \left( 1-\frac{3\sigma}{2} 
\left\{ \frac{n-1}{n-\left[ 2/\left(3\sigma\right) \right]}
\right\}^{n-1}\right)^{-1} 
\left\{ 1 - \left[ 1
- \frac{1-2/\left(3\sigma \right)}{n-2/\left(3\sigma \right)} 
\frac{t}{t_{\mathrm{s}}} 
\right]^{n} 
\right\} 
\nonumber \\ 
&& 
{}+ \frac{1-2/\left(3\sigma \right)}{n-2/\left(3\sigma \right)} 
n a_{\mathrm{s}} 
\left( 1-\frac{2}{3\sigma} 
\left\{ \frac{n-\left[ 2/\left(3\sigma\right) \right]}{n-1}
\right\}^{n-1}\right)^{-1} 
\left( \frac{t}{t_{\mathrm{s}}} 
\right)^{2/\left(3\sigma \right)}\,. 
\label{eq:ED1-11-IXB2-FT4-13-Add-IIIB-01} 
\end{eqnarray}
Here, $\sigma$ and $n$ are arbitrary constants. 
When $t \to t_{\mathrm{s}}$, 
both $H$ and $\dot{H}$ becomes zero, whereas 
the effective EoS for the universe 
$w_{\mathrm{eff}} = \left(1/3\right) \left(2q_{\mathrm{dec}} -1\right) 
\to \infty$ with $q_{\mathrm{dec}} \equiv -\ddot{a}a/\dot{a}^2$ 
being the deceleration parameter.  
Thus, in the above limit of $t \to t_{\mathrm{s}}$, 
we find $F +2TF^{\prime} = 0$ because $\dot{H} (t \to t_{\mathrm{s}}) = 0$. 
For a power-law model in Eq.~(\ref{eq:ED1-10-IXB2-IVA2-addition-01})  
with $A \neq 0$ and $\alpha > 1$, 
Eq.~(\ref{eq:ED1-10-IXB2-IV-BB-add-04}) can asymptotically be met 
due to $\dot{H} (t \to t_{\mathrm{s}}) = 0$, 
and hence ``$w$'' singularity can occur. 

\subsubsection{Removing the finite-time future singularities}

We examine the possibility to remove the finite-time future singularities 
by taking a power-law correction term $f_{\mathrm{c}} (T)$, given by 
\begin{equation} 
f_{\mathrm{c}} (T) = B T^\beta\,. 
\label{eq:ED1-10-IXB3-IV-BB-add-05}
\end{equation} 
Here, $B (\neq 0)$ and $\beta (\neq 0)$ are non-zero constants. 
By plugging the total form of $f(T) = A T^{\alpha} + B T^\beta$ 
into Eqs.~(\ref{eq:ED1-10-IXB2-IV-BB-add-03}) and (\ref{eq:ED1-10-IXB2-IV-BB-add-04}), we obtain 
$
-f +2TF = A\left(2\alpha-1\right) T^{\alpha} 
+ B\left(2\beta-1\right) T^{\beta} \neq 0
$ 
and 
$
-F -2TF^{\prime} = -A\alpha\left(2\alpha-1\right) T^{\alpha-1} 
-B\beta\left(2\beta-1\right) T^{\beta-1} \neq 0
$. 
{}From the investigations in Sec.~IX B 2, 
the latter inequality is satisfied when 
the condition that 
for $q > 0$, $\beta > 0$, whereas for $q < 0$, $\beta \neq 1/2$, 
is met. 
Accordingly, 
if $\beta >1$, 
in the limit of $t \to t_{\mathrm{s}}$ 
both (\ref{eq:IXA-4.1}) and (\ref{eq:IXA-4.2}) cannot be met. 
Thus, the finite-time future singularities in $f(T)$ gravity 
can be removed by a power-low correction as $T^\beta$, where $\beta>1$. 
It is remarkable to mention that a $T^2$ term 
can cure all the four types of the finite-time future singularities 
in $f(T)$ gravity, similar to that in $F(R)$ gravity~\cite{Review-Nojiri-Odintsov}. 
In Table~\ref{tb:table2}, we show necessary conditions for 
a power-law $f(T)$ model in Eq.~(\ref{eq:ED1-10-IXB2-IVA2-addition-01}) to prodece the finite-time future singularities and those appearance, and 
necessary conditions for a power-law correction term 
$f_{\mathrm{c}} (T) = B T^\beta$ in Eq.~(\ref{eq:ED1-10-IXB3-IV-BB-add-05}) to 
remove the finite-time future singularities.

\begin{table*}[tbp]
\caption{
Necessary conditions for the appearance of the finite-time future 
singularities on a power-law $f(T)$ model in 
Eq.~(\ref{eq:ED1-10-IXB2-IVA2-addition-01}) and 
those for the removal of the finite-time future singularities on 
a power-law correction term  
$f_{\mathrm{c}} (T) = B T^\beta$ in Eq.~(\ref{eq:ED1-10-IXB3-IV-BB-add-05}). 
}
\begin{center}
\begin{tabular}
{llcll}
\hline
\hline
Type \quad
& $q (\neq 0, \, -1)$
& Final appearance\,\,\,\,\, 
& $f(T) = A T^\alpha$ 
& $f_{\mathrm{c}} (T) = B T^\beta$ 
\\[0mm]
&
& 
& ($A \neq 0$, $\alpha \neq 0$)\,\,\,\,\,
& ($B \neq 0$, $\beta \neq 0$)
\\[0mm]
\hline
I
& $q \geq 1$  
& Occur
& $\alpha < 0$
& $\beta > 1$
\\[0mm]
III 
& $0 < q < 1$ 
& ---
& $\alpha < 0$
& $\beta > 1$
\\[0mm]
II 
& $-1 < q < 0$ 
& ---
& $\alpha = 1/2$
& $\beta \neq 1/2$
\\[0mm]
IV 
& $q < -1$ ($q \neq \mathrm{integer}$)
\,\,\,\,\,
& Occur
& $\alpha = 1/2$
& $\beta \neq 1/2$
\\[1mm]
\hline
\hline
\end{tabular} 
\end{center}
\label{tb:table2}
\end{table*}

We also remark that 
In terms of the ``$w$'' singularity, 
if a power-law correction term in 
Eq.~(\ref{eq:ED1-10-IXB3-IV-BB-add-05}) with $B \neq 0$ and $\beta<0$ 
is taken, the gravitational field equations 
(\ref{eq:IXA-4.1}) and (\ref{eq:IXA-4.2}) cannot be met asymptotically. 
As a consequence, the power-law correction term 
can cure the ``$w$'' singularity.

\subsection{Reconstructed $f(T)$ models performing various cosmologies} 

Furthermore, 
we describe the reconstructed $f(T)$ models in which 
the following various cosmologies are realized: 
(i) inflation, (ii) the $\Lambda$CDM model, (iii) Little Rip scenario and (iv) Pseudo-Rip cosmology. 
We present expressions of $a$, $H$ and $f(T)$ realizing 
the above cosmologies in Table~\ref{tb:table3}. 
Here, $a_{\mathrm{inf}} (> 0)$ and $a_{\mathrm{LR}} (>0)$ are 
positive constants and
$h_{\mathrm{inf}} (> 1)$ is a constant larger than unity. 
We note that the form of $f(T)$ and the conditions for it are 
derived so that the gravitational field equations (\ref{eq:IXA-4.1}) 
and (\ref{eq:IXA-4.2}). 

\begin{table*}[tbp]
\caption{Forms of $H$ and $f(T)$ with realizing 
(i) inflation, (ii) the $\Lambda$CDM model, 
(iii) Little Rip scenario and (iv) Pseudo-Rip cosmology. 
}
\begin{center}
\begin{tabular}
{llll}
\hline
\hline
Cosmology
& $a$
& $H$ 
& $f(T)$ 
\\[0mm]
\hline
(i) Power-law inflation
& $a = a_{\mathrm{inf}} t^{h_{\mathrm{inf}}}$
& $H = h_{\mathrm{inf}}/t$\,,  
& $f(T) = A T^\alpha$\,,  
\\[0mm]
[when $t \to 0$]
& $a_{\mathrm{inf}} > 0$
& $h_{\mathrm{inf}} > 1$
& $\alpha <0$ or $\alpha = 1/2$
\\[0mm]
(ii) $\Lambda$CDM model  
& $a = a_{\Lambda} \exp \left( H_{\Lambda} t \right)$\,,
& $H =\sqrt{\Lambda/3}$ 
& $f(T) = T - 2\Lambda$\,, 
\\[0mm]
or exponential inflation
& $a_{\Lambda} >0$
& $= \mathrm{constant}$
& $\Lambda >0$
\\[0mm]
& 
& $\Lambda >0$
& 
\\[0mm]
(iii) Little Rip scenario
& $a = a_{\mathrm{LR}}$ 
& $H = H_{\mathrm{LR}} \exp \left( \xi t \right)$\,, 
& $f(T) = A T^\alpha$\,, 
\\[0mm]
[when $t \to \infty$]
& $\times \exp \left[ \left(H_{\mathrm{LR}}/\xi\right) 
\exp \left( \xi t \right)\right]$\,, 
& $H_{\mathrm{LR}} > 0$ and $\xi >0$
& $\alpha <0$ or $\alpha = 1/2$
\\[0mm]
& $a_{\mathrm{LR}} > 0$
& 
& 
\\[0mm]
(iv) Pseudo-Rip cosmology\,\,\,\,\,
& $a = a_{\mathrm{PR}} \cosh \left( t/t_0 \right)$\,,
& $H = H_{\mathrm{PR}} \tanh \left(t/t_0\right)$\,,
& $f(T) = A \sqrt{T}$ 
\\[0mm]
& $a_{\mathrm{PR}} > 0$
& $H_{\mathrm{PR}} > 0$
& 
\\[1mm]
\hline
\hline
\end{tabular}
\end{center}
\label{tb:table3}
\end{table*}

As another quantity to show the deviation of a dark energy model from 
the $\Lambda$CDM model, 
in addition to the EoS $w_{\mathrm{DE}}$ of dark energy 
in Eq.~(\ref{eq:2.17}),  
the deceleration parameter $q_{\mathrm{dec}}$ in Eq.~(\ref{eq:ED1-9-Add-IIIE-13}), and the jerk parameter $j$ in Eq.~(\ref{eq:ED1-9-Add-IIIE-14}), 
the snark parameter $s$ is used, which defined as~\cite{Sahni:2002fz} 
\begin{equation} 
s \equiv \frac{j - 1}{3 \left( q_{\mathrm{dec}} -1/2 \right)}\,.
\label{eq:ED1-11-IXC-FT4-8-IVC2-addition-04} 
\end{equation}
For the $\Lambda$CDM model, $w_{\mathrm{DE}} = -1$, 
$q_{\mathrm{dec}} = -1$, $j = 1$ and $s = 0$. 
Hence, 
by examining the deviations of $(w_{\mathrm{DE}}, q_{\mathrm{dec}}, j, s)$ 
from $(-1, -1, 1, 0)$ for the $\Lambda$CDM model 
and using these four parameters as a tool of observational tests, 
we can distinguish a dark energy model from the $\Lambda$CDM model. 
In Table~\ref{tb:table4}, we display the expressions of $w_{\mathrm{DE}}$, 
$q_{\mathrm{dec}}$, $j$ and $s$ at the present time $t_0$ 
for the $\Lambda$CDM model, Little Rip scenario and Pseudo-Rip cosmology~\cite{Bamba:2012vg}. 
Here, $s_0 \equiv s (t=t_0)$. 

Finally, we mention another feature of $f(T)$ gravity. 
It has been discussed that 
in the star collapse, the time-dependent matter instability 
found in $F(R)$ gravity~\cite{Time-dependent-matter-instability}, 
which is related to the well-studied matter instability~\cite{Dolgov:2003px} leading to the appearance of a singularity in the relativistic star formation process~\cite{Rel-stars}, 
can also happen in the framework of $f(T)$ gravity~\cite{Bamba:2012vg}. 

\begin{table*}[tbp]
\caption{Expressions of $w_{\mathrm{DE}}$, 
$q_{\mathrm{dec}}$, $j$ and $s$ at the present time $t=t_0$ 
for the $\Lambda$CDM model, Little Rip scenario and Pseudo-Rip cosmology~\cite{Bamba:2012vg}. 
}
\begin{center}
\begin{tabular}
{lllll}
\hline
\hline
Model \quad
& $w_{\mathrm{DE} (0)}$
& $q_{\mathrm{dec} (0)}$
& $j_0$ 
& $s_0$ 
\\[0mm]
\hline
$\Lambda$CDM 
& $-1$
& $-1$
& $1$
& $0$
\\[0mm]
model
&
&
&
&
\\[0mm]
Little Rip 
& $-1 -\left(2/3\right) \tilde{\chi}$\,, 
& $-1 - \tilde{\chi}$
& $1 +  \chi \left( \tilde{\chi} + 3 \right)$
& $-\left[2\tilde{\chi} \left( \tilde{\chi} + 3 \right)\right]$ 
\\[0mm]
scenario
& $\tilde{\chi} \equiv H_0/\left( H_{\mathrm{LR}} e \right)$  
& 
& 
& $\times \left[3 \left( 2\tilde{\chi} + 3 \right)\right]^{-1}$
\\[0mm]
& $\leq 0.36$\,,
& 
& 
& 
\\[0mm]
& $e = 2.71828$
&
&
&
\\[0mm]
Pseudo-Rip 
& $-1-\left[2\delta/\left(3 \tilde{\mathrm{s}}^2 \right)\right]$\,, 
& $-1 + \left(\delta^2 \tilde{\mathrm{t}}^2 - 1\right)/\left(\delta^2 
\tilde{\mathrm{t}}^2\right)$\,, 
& $1 + \left(1-\delta^3 \tilde{\mathrm{t}}^2\right)/\left(\delta^3 \tilde{\mathrm{t}}^2\right)$\,, 
& $\left[2/\left(3 \delta\right)\right]$  
\\[0mm]
cosmology
&
&
&
& $\times \left(\delta^3 \tilde{\mathrm{t}}^2 -1\right)$
\\[0mm]
&
&
&
& $\times \left(\delta^2 \tilde{\mathrm{t}}^2 +2\right)^{-1}$
\\[0mm]
& $\delta \equiv H_0/H_{\mathrm{PR}}$  
& $\tilde{\mathrm{s}}^2 \equiv \sinh^2 1 = 1.38$
& $\tilde{\mathrm{t}}^2 \equiv \tanh^2 1 = 0.580$
& 
\\[0mm]
& $\leq 0.497$
& 
& 
& 
\\[1mm]
\hline
\hline
\end{tabular} 
\end{center}
\label{tb:table4}
\end{table*}

\subsection{Thermodynamics in $f(T)$ gravity} 

In this section, 
to explore whether 
$f(T)$ gravity is worthy of an alternative gravitational theory to 
general relativity, 
we investigate thermodynamics in $f(T)$ gravity. 
In particular, 
the second law of thermodynamics around the finite-time future singularities 
is studied by applying the procedure proposed 
in Refs.~\cite{Bamba:2011pz, Bamba:2010kf}. 
Black hole thermodynamics~\cite{BCH-B-H} 
suggested the fundamental relation of gravitation to thermodynamics 
(for recent reviews, see, e.g.,~\cite{Rev-T-G-Pad}). 
With the proportionality of the entropy to the horizon 
area, in general relativity 
the Einstein equation was obtained from 
the Clausius relation in thermodynamics~\cite{Jacobson:1995ab}. 
This consideration has been extended to 
more general gravitational 
theories~\cite{Modified-gravity-EOS, Brustein-Hadad-Medved}. 

\subsubsection{First law of thermodynamics} 
 
It is known in~\cite{Bamba:2010kf, Bamba:2011pz, BGT} that 
when the continuity equation of dark component is met as 
$
\dot{\rho}_{\mathrm{DE}}+3H \left( 
\rho_{\mathrm{DE}} + P_{\mathrm{DE}}
\right)  
= 0
$, we can have an equilibrium description of thermodynamics. 
In the flat FLRW universe with the metric 
%
$
d s^2 = h_{\alpha \beta} d x^{\alpha} d x^{\beta}
+\tilde{r}^2 d \Omega^2
$, where 
$\tilde{r}=a(t)r$, $x^0=t$ and $x^1=r$ with the two-dimensional 
metric $h_{\alpha \beta}={\rm diag}(1, -a^2(t))$, 
$d \Omega^2$ is the metric of two-dimensional sphere with unit radius. 
The radius $\tilde{r}_A$ of 
the apparent horizon is described by 
$
\tilde{r}_A= 1/H
$. 
The relation 
$h^{\alpha \beta} \partial_{\alpha} \tilde{r} \partial_{\beta} \tilde{r}=0$ 
leads to the dynamical apparent horizon. 
The time derivative of $\tilde{r}_A= 1/H$ yields 
$
-d\tilde{r}_A/\tilde{r}_A^3 
=\dot{H}H dt
$.  
Combining the Friedmann equation (\ref{eq:IXA-4.1}) with this equation 
presents  
$
\left[1/\left(4\pi G\right)\right] d\tilde{r}_A=\tilde{r}_A^3 H
\left( \rho_{\mathrm{t}}+P_{\mathrm{t}} \right) dt
$ with 
$\rho_{\mathrm{t}} \equiv \rho_{\mathrm{DE}}+
\rho_{\mathrm{M}}$ and $P_{\mathrm{t}} \equiv P_{\mathrm{DE}}+P_{\mathrm{M}}$ 
being the total energy density and pressure of the universe, respectively. 
The Bekenstein-Hawking horizon entropy in general relativity is 
written by  
$
S=\mathcal{A}/\left(4G\right)
$. 
Here, $\mathcal{A}=4\pi \tilde{r}_A^2$ is the area of the apparent 
horizon~\cite{BCH-B-H}. 
Thus, with the horizon entropy as well as the above relation, 
we obtain 
%
$
\left[1/\left(2\pi \tilde{r}_A \right)\right ]dS=
4\pi \tilde{r}_A^3 H 
\left( \rho_{\mathrm{t}}+P_{\mathrm{t}} \right) dt 
$.
%
The Hawking temperature 
%
$
T_{\mathrm{H}} = |\kappa_{\mathrm{sg}}|/\left(2\pi\right)
$ 
is considered to be the associated temperature of the apparent horizon, 
%
and the surface gravity $\kappa_{\mathrm{sg}}$ is given 
by~\cite{Cai:2005ra} 
\begin{equation}
\kappa_{\mathrm{sg}} 
\equiv 
\frac{1}{2\sqrt{-h}} \partial_\alpha 
\left( \sqrt{-h}h^{\alpha\beta} \partial_\beta \tilde{r} \right)  
= -\frac{1}{\tilde{r}_A}
\left( 1-\frac{\dot{\tilde{r}}_A}{2H\tilde{r}_A} \right) 
= -\frac{2\pi G}{3F} \tilde{r}_A 
\left(1 -3w_{\mathrm{t}} \right) \rho_{\mathrm{t}}\,. 
\label{eq:ED1-11-IXD1-FT4-13-VA-add-1}
\end{equation}
Here, $h$ is the determinant of the metric $h_{\alpha\beta}$ 
and $w_{\mathrm{t}} \equiv P_{\mathrm{t}}/\rho_{\mathrm{t}}$ is 
the EoS for the total of energy and matter in the universe. 
{}From Eq.~(\ref{eq:ED1-11-IXD1-FT4-13-VA-add-1}), we see that if 
$w_{\mathrm{t}} \le 1/3$, $\kappa_{\mathrm{sg}} \le 0$. 
The substitution of Eq.~(\ref{eq:ED1-11-IXD1-FT4-13-VA-add-1}) into 
$T_{\mathrm{H}} = |\kappa_{\mathrm{sg}}|/\left(2\pi\right)$ 
yields 
\begin{equation}
T_{\mathrm{H}}=\frac{1}{2\pi \tilde{r}_A}
\left( 1-\frac{\dot{\tilde{r}}_A}{2H\tilde{r}_A} \right)\,.
\label{eq:ED1-11-IXD1-VA-3.14}
\end{equation}
By combining the above relation of the horizon entropy $S$ with Eq.~(\ref{eq:ED1-11-IXD1-VA-3.14}), we find 
%
$
T_{\mathrm{H}} dS = 4\pi \tilde{r}_A^3 H \left(\rho_{\mathrm{t}}+P_{\mathrm{t}} \right) dt 
-2\pi  \tilde{r}_A^2 \left(\rho_{\mathrm{t}}+P_{\mathrm{t}} \right) 
d\tilde{r}_A
$. 
Moreover, 
the Misner-Sharp energy~\cite{Misner-Sharp-energy} is expressed by 
%
$
E=\tilde{r}_A/\left(2G\right) = 
V\rho_{\mathrm{t}}
$, 
%
where $V=4\pi \tilde{r}_A^3/3$ is the volume inside 
the apparent horizon. {}From the second equality, we see that 
$E$ corresponds to the total intrinsic energy. 
With this equation, we obtain 
%
$
dE=-4\pi \tilde{r}_A^3 H \left(\rho_{\mathrm{t}}+P_{\mathrm{t}} \right) dt 
+4\pi \tilde{r}_A^2 \rho_{\mathrm{t}} d\tilde{r}_A
$. 
In addition, the work density~\cite{W-D} is defined by 
%
$
W \equiv
-\left (1/2\right) \left( T^{(\mathrm{M})\alpha\beta}
h_{\alpha\beta} + T^{(\mathrm{DE})\alpha\beta} h_{\alpha\beta} 
\right) 
=
\left (1/2\right) 
\left( \rho_{\mathrm{t}}-P_{\mathrm{t}} \right) 
%
$ 
with $T^{(\mathrm{M})\alpha\beta}$ and $T^{(\mathrm{DE})\alpha\beta}$ being 
the energy-momentum tensor of matter and that of dark components, 
respectively. We plug the work density $W$ into 
the relation on $d E$ derived above, so that we can represent the first law of equilibrium thermodynamics as 
\begin{equation} 
T_{\mathrm{H}} dS=-dE+W dV\,.
\label{eq:ED1-11-IXD1-VA-4.12}
\end{equation} 
Thus, we acquire an equilibrium description of thermodynamics. 
It follows from the gravitational equations (\ref{eq:IXA-4.1}) and (\ref{eq:IXA-4.2}) as well as the above relation on $dS$ that 
$
\dot{S} = 
8\pi^2 H \tilde{r}_A^4 \left(\rho_{\mathrm{t}}+P_{\mathrm{t}}\right) 
= \left( 6\pi/G \right) \left( \dot{T}/T^2 \right) 
= -\left( 2\pi/G \right) \left[ \dot{H}/\left( 3H^3 \right) \right] 
> 0
$. 
Accordingly, for the expanding universe ($H>0$), 
if the null energy condition 
$\rho_{\mathrm{t}}+P_{\mathrm{t}} \ge 0$ in Eq.~(\ref{eq:Add-2-B-1}) 
is satisfied, namely, $\dot{H} \leq 0$, 
$S$ always becomes large.

\subsubsection{Second law of thermodynamics} 

We then explore the second law of thermodynamics. 
The Gibbs equation of all the matter and energy fluid is 
expressed as 
%
$
T_{\mathrm{H}} dS_{\mathrm{t}} = d\left( \rho_{\mathrm{t}} V \right) +P_{\mathrm{t}} dV 
= V d\rho_{\mathrm{t}} + \left( \rho_{\mathrm{t}} +P_{\mathrm{t}} 
\right) dV
$. 
%
We can write the second law of thermodynamics as 
%
%
$
dS_{\mathrm{sum}}/dt
\equiv 
dS/dt + dS_{\mathrm{t}}/dt 
\geq 0
$. 
%
%
Here, $S_{\mathrm{sum}} \equiv S + S_{\mathrm{t}}$ 
with $S_{\mathrm{t}}$ being the entropy of total energy inside the 
horizon. 
If the temperature of the universe 
is the same as that of the apparent horizon~\cite{GWW-JSS}, 
this can be represented as 
\begin{equation}
\frac{dS_{\mathrm{sum}}}{dt} 
= 
-\frac{6\pi}{G}\left( \frac{\dot{T}}{T} \right)^2 \frac{1}{4HT + \dot{T}}\,, 
\label{eq:ED1-11-IXD2-VB-4.A003}
\end{equation}
where we have used $V=4\pi \tilde{r}_A^3/3$, Eqs.~(\ref{eq:IXA-4.2}), 
(\ref{eq:ED1-11-IXD1-VA-3.14}) and the relation on $\dot{S}$ shown in 
the last part of Sec.~IX D 1. 
Consequently, 
the condition that 
$
Y \equiv -\left( 4HT + \dot{T} \right) 
= 12H \left( 2H^2 + \dot{H} \right) 
\geq 0
$~\cite{Bamba:2011pz} is met, the second law of thermodynamics 
can be verified.  
When $t \to t_{\mathrm{s}}$, 
for all of the four types of the finite-time future singularities 
in Table~\ref{tb:table1}, 
the relation $2H^2 + \dot{H} \geq 0$ 
is always satisfied. 
Since $H>0$ for the expanding universe, 
the second law of thermodynamics described by 
Eq.~(\ref{eq:ED1-11-IXD2-VB-4.A003}) can be met 
around the finite-time future singularities 
including in the phantom phase ($\dot{H} >0$), 
although at the exact time of the appearance of singularity 
of $t=t_{\mathrm{s}}$ this classical description of thermodynamics 
could not be applicable.


\section{Testing Dark Energy and Alternative Gravity by cosmography: generalities}

Next, we move to the comparison of the theoretical studies 
on dark energy and modified gravity with the observational data. 
In this section, we introduce the idea and concept of cosmography to 
observationally test dark energy and alternative gravitational theory to 
general relativity. 

The observed accelerated expansion of the cosmic fluid can be faced in several equivalent ways. In other words, both dark energy models and modified
gravity theories seem to be in agreement with data. As a consequence, 
unless higher precision probes of the expansion rate 
and the growth of structure will be available, these two rival 
approaches could not be discriminated. This confusion about the 
theoretical background suggests that a more conservative approach 
to the problem of the cosmic acceleration, relying on as less model 
dependent quantities as possible, is welcome.  A possible solution 
could be to come back to the cosmography~\cite{W72} rather than 
finding out solutions of the Friedmann equations and testing them. 
Being only related to the derivatives of the scale factor, the 
cosmographic parameters make it possible to fit the data on the 
distance\,-\,redshift relation without any {\it a priori} 
assumption on the underlying cosmological model: in this case, the 
only assumption is that the metric is the FLRW one 
(and hence not relying on the solution of cosmological equations). 
Almost eighty years after Hubble's discovery of the expansion of the 
universe, we can now extend, in principle, cosmography well 
beyond the search for the value of the only Hubble constant. 
The SNeIa Hubble diagram extends up to $z = 1.7$ thus invoking the 
need for, at least, a fifth order Taylor expansion of the scale 
factor in order to give a reliable approximation of the 
distance\,-\,redshift relation. As a consequence, it could be, in 
principle, possible to estimate up to five cosmographic parameters, 
although the still too small data set available does not allow to get 
a precise and realistic determination of all of them.

Once these quantities have been determined, one could use them to 
put constraints on the models. In a sense, we can revert to the usual 
approach, consisting with deriving the cosmographic parameters as a 
sort of byproduct of an assumed theory. Here, we follow the other 
way of 
expressing the 
quantities characterizing the model 
as a function of the cosmographic parameters. 
Such a program is particularly suited for the study of alternative theories 
as $F(R)$ or $f(T)$ gravity~\cite{salzano,mariam,hossein} and any equivalent 
description of dynamics by effective scalar fields. 
As it is well known, the mathematical difficulties in analyzing 
the solution of field equations make it 
quite problematic to find out analytical expressions for the scale factor 
and hence predict the values of the cosmographic parameters. 
A key role in $F(R)$ gravity and $f(T)$ gravity is played by the choice 
of the function. 
Under quite general hypotheses, it is possible to 
derive relations between cosmographic parameters and the present time values 
of $F^{(n)}(R) = d^nF/dR^n$ or $f^{(n)}(T) = d^nf/dT^n$, 
with $n = 0, \ldots, 3$, whatever $F(R)$, $f(T)$ or their equivalent 
scalar-field descriptions are. 
 
Once the cosmographic parameters are determined, 
the method allows to investigate the cosmography of alternative 
theories matching with observational data. 

It is worth stressing that the definition of the cosmographic 
parameters only relies on the assumption of the FLRW metric. 
As such, it is however difficult to state a priori to what extent the fifth 
order expansion provides an accurate enough description of the quantities of 
interest. 
Actually, the number of cosmographic parameters to be used depends 
on the problem one is interested in. 

To illustrate the method, one can be concerned only with the SNeIa Hubble 
diagram so that one has to check that the distance modulus $\mu_{\mathrm{cp}}(z)$ obtained using the fifth order expansion of the scale factor is the same 
(within the errors) as the one $\mu_{\mathrm{DE}}(z)$ of the underlying 
physical model. Being such a model of course unknown, one can adopt a 
phenomenological parameterization for the dark energy 
EoS and look at the percentage deviation 
$\Delta \mu/\mu_{\mathrm{DE}}$ as a function of the EoS parameters. 
Note that one can always use a phenomenological dark energy model to get a 
reliable estimate of the scale factor evolution (see, for example,~\cite{unified}). 

Here, we will carry out such an approach using the so called Chevallier-Polarski-Linder (CPL) model~\cite{CPL, Linder03}, introduced below, 
and verified that $\Delta \mu/\mu_{\mathrm{DE}}$ is an increasing function 
of $z$ (as expected), but still remains smaller than $2\%$ up to 
$z \sim 2$ over a wide range of the CPL parameter space. 
On the other hand, halting the Taylor expansion to a lower order may 
introduce significant deviation for $z > 1$ that can potentially 
bias the analysis if the measurement errors are as small as those 
predicted by future observational surveys. 
However, the fifth order expansion is both sufficient to get an accurate 
distance modulus over the redshift range probed by SNeIa and 
necessary to avoid dangerous biases. As shown in~\cite{izzo,ruth}, 
the method can highly be improved by adopting BAO and Gamma Ray Bursts (GRBs) 
as cosmic indicators. 

As stated above, the key rule in cosmography is the Taylor series expansion of 
the scale factor with respect to the cosmic time. To this aim, it is 
convenient to introduce the following functions: 
\begin{eqnarray}
H(t) \Eqn{\equiv} + \frac{1}{a}\frac{da}{dt}\, ,
\\
q(t) \Eqn{\equiv} - \frac{1}{a}\frac{d^{2}a}{dt^{2}}\frac{1}{H^{2}}\,
,
\\
j(t) \Eqn{\equiv} + \frac{1}{a}\frac{d^{3}a}{dt^{3}}\frac{1}{H^{3}}\,
,
\\
s(t) \Eqn{\equiv} + \frac{1}{a}\frac{d^{4}a}{dt^{4}}\frac{1}{H^{4}}\,
,
\\
l(t) \Eqn{\equiv} + \frac{1}{a}\frac{d^{5}a}{dt^{5}}\frac{1}{H^{5}}\,
,
\end{eqnarray}
which are usually referred to as the \textit{Hubble}, 
\textit{deceleration}, \textit{jerk}, \textit{snap}, and 
\textit{lerk} parameters, respectively. 
(Here, for clear understanding, we have again defined 
the Hubble, deceleration (as $q$, which is the same as $q_{\mathrm{dec}}$ in 
Eq.~(\ref{eq:ED1-9-Add-IIIE-13})), jerk and snap parameters.) 
It is then a matter of
algebra to demonstrate the following useful relations\,:
\begin{eqnarray}
\dot{H} \Eqn{=} -H^2 (1 + q) \ , \label{eq: hdot}
\\
\ddot{H} \Eqn{=} H^3 (j + 3q + 2) \ , \label{eq: h2dot}
\\
d^3H/dt^3 \Eqn{=} H^4 \left [ s - 4j - 3q (q + 4) - 6 \right ] \ ,
\label{eq: h3dot}
\\
d^4H/dt^4 \Eqn{=} H^5 \left [ l - 5s + 10 (q + 2) j + 30 (q + 2) q +
24 \right] \ . 
\label{eq: h4dot}
\end{eqnarray}
Equations (\ref{eq: hdot})--(\ref{eq: h4dot}) make it possible
to relate the derivative of the Hubble parameter to the other
cosmographic parameters. The distance\,-\,redshift relation may 
then be obtained, starting from the Taylor expansion of $a(t)$
along the lines described in \cite{Visser,WM04,CV07}.

By these definitions, the series expansion to the 5th order in
time of the scale factor is 
\begin{eqnarray}
a(t) \Eqn{=} a(t_{0}) \left\{ 1 + H_{0} (t-t_{0}) - \frac{q_{0}}{2}
H_{0}^{2} (t-t_{0})^{2} +  \frac{j_{0}}{3!} H_{0}^{3}
(t-t_{0})^{3} + 
\right. \nonumber \\ 
&& 
\left. 
{}+ \frac{s_{0}}{4!} H_{0}^{4} (t-t_{0})^{4} +
\frac{l_{0}}{5!} H_{0}^{5} (t-t_{0})^{5}
+\emph{O}[(t-t_{0})^{6}] \right\} \,, 
\end{eqnarray}
%
%
%
{}from which, we find 
\begin{eqnarray}
\label{eq:a_series}
\frac{a(t)}{a(t_{0})} \Eqn{=} 1 + H_{0} (t-t_{0}) -\frac{q_{0}}{2}
H_{0}^{2} (t-t_{0})^{2} +\frac{j_{0}}{3!} H_{0}^{3} (t-t_{0})^{3} +
\nonumber \\ 
&& 
{}+ \frac{s_{0}}{4!} H_{0}^{4} (t-t_{0})^{4}+ \frac{l_{0}}{5!}
H_{0}^{5} (t-t_{0})^{5} +\emph{O}[(t-t_{0})^{6}]\,.
\end{eqnarray}
%
It is easy to see that Eq.~(\ref{eq:a_series}) is the inverse of
redshift $z$, being the redshift defined by
\begin{equation}
1 + z = \frac{a(t_{0})}{a(t)} \,. 
\nonumber 
\end{equation}
The physical distance travelled by a photon that is emitted at 
time $t_{*}$ and absorbed at the current epoch $t_{0}$ is 
\begin{equation}
D = c \int dt = c (t_{0} - t_{*}) \,, 
\nonumber 
\end{equation}
where $c$ is the speed of light. 
Assuming $t_{*} = t_{0} - \frac{D}{c}$ and inserting it into 
Eq.~(\ref{eq:a_series}), we have
%
\begin{eqnarray}
\hspace{-15mm}
&&
1 + z = \frac{a(t_{0})}{a(t_{0}-\frac{D}{c})} = 
\nonumber \\
\hspace{-15mm}
&=& 
\frac{1}{1 -
\frac{H_{0}}{c}D -
\frac{q_{0}}{2}\left(\frac{H_{0}}{c}\right)^{2}D^{2}  -
\frac{j_{0}}{6}\left(\frac{H_{0}}{c}\right)^{3}D^{3}  +
\frac{s_{0}}{24}\left(\frac{H_{0}}{c}\right)^{4}D^{4} -
\frac{l_{0}}{120}\left(\frac{H_{0}}{c}\right)^{5}D^{5}
+\emph{O}[(\frac{H_{0} D}{c})^{6}]} \,.
\end{eqnarray}
The inverse of this expression becomes 
%
\begin{eqnarray}
\hspace{-10mm}
1 + z \Eqn{=} 1 + \frac{H_{0}}{c}D + \left(1 +
\frac{q_{0}}{2}\right)\left(\frac{H_{0}}{c}\right)^{2}D^{2} +
\left(1 + q_{0} +\frac{
j_{0}}{6}\right)\left(\frac{H_{0}}{c}\right)^{3}D^{3} +
\nonumber \\
\hspace{-10mm}
\Eqn{+} \left(1 +
\frac{3}{2}q_{0} + \frac{q_{0}^{2}}{4} + \frac{j_{0}}{3} -
\frac{s_{0}}{24}\right)\left(\frac{H_{0}}{c}\right)^{4}D^{4} +
\nonumber \\
\hspace{-10mm}
\Eqn{+}
\left( 1 + 2 q_{0} + \frac{3}{4} q_{0}^{2} + \frac{q_{0}
j_{0}}{6} + \frac{j_{0}}{2} - \frac{s}{12} +
l_{0}\right)\left(\frac{H_{0}}{c}\right)^{5}D^{5} +
\emph{O}\left[\left(\frac{H_{0} D}{c}\right)^{6}\right]\,.
\end{eqnarray}
Then, we reverse the series $z(D) \rightarrow D(z)$ to have the 
physical distance $D$ expressed as a function of the redshift $z$
%
\begin{eqnarray}
z(D) \Eqn{=} \mathcal{Z}_{D}^{1} \left(\frac{H_{0} D}{c}\right) +
\mathcal{Z}_{D}^{2} \left(\frac{H_{0} D}{c}\right)^{2} +
\mathcal{Z}_{D}^{3} \left(\frac{H_{0} D}{c}\right)^{3} +
\mathcal{Z}_{D}^{4} \left(\frac{H_{0} D}{c}\right)^{4} + 
\nonumber \\
\Eqn{+}
\mathcal{Z}_{D}^{5} \left(\frac{H_{0} D}{c}\right)^{5} +
\emph{O}\left[\left(\frac{H_{0} D}{c}\right)^{6}\right]
\end{eqnarray}
%
with
%
\begin{eqnarray}
\mathcal{Z}_{D}^{1} \Eqn{=} 1\,, \\
\mathcal{Z}_{D}^{2} \Eqn{=} 1 + \frac{q_{0}}{2}\,, \\
\mathcal{Z}_{D}^{3} \Eqn{=} 1 + q_{0} +\frac{j_{0}}{6}\,, \\
\mathcal{Z}_{D}^{4} \Eqn{=} 1 + \frac{3}{2}q_{0} + \frac{q_{0}^{2}}{4} + \frac{j_{0}}{3} - \frac{s_{0}}{24}\,, \\
\mathcal{Z}_{D}^{5} \Eqn{=} 1 + 2 q_{0} + \frac{3}{4} q_{0}^{2} +
\frac{q_{0} j_{0}}{6} + \frac{j_{0}}{2} - \frac{s}{12} + l_{0}\,.
\end{eqnarray}
%
{}From this, we obtain 
%
\begin{eqnarray}
D(z) &=& \frac{c z}{H_{0}} \left\{ \mathcal{D}_{z}^{0} +
\mathcal{D}_{z}^{1} \ z + \mathcal{D}_{z}^{2} \ z^{2} +
\mathcal{D}_{z}^{3} \ z^{3} + \mathcal{D}_{z}^{4} \ z^{4} +
\emph{O}(z^{5}) \right\}
\end{eqnarray}
with
%
\begin{eqnarray}
\hspace{-10mm}
\mathcal{D}_{z}^{0} \Eqn{=} 1\,, \\
\hspace{-10mm}
\mathcal{D}_{z}^{1} \Eqn{=} - \left(1 +\frac{ q_{0}}{2}\right)\,, \\
\hspace{-10mm}
\mathcal{D}_{z}^{2} \Eqn{=} 1 + q_{0} + \frac{q_{0}^{2}}{2} 
- \frac{j_{0}}{6}\,, \\
\hspace{-10mm}
\mathcal{D}_{z}^{3} \Eqn{=} - \left(1 + \frac{3}{2}q_{0}+
\frac{3}{2}q_{0}^{2} + \frac{5}{8} q_{0}^{3} - \frac{1}{2} j_{0} -
\frac{5}{12} q_{0} j_{0} - \frac{s_{0}}{24}\right)\,, \\
\hspace{-10mm}
\mathcal{D}_{z}^{4} \Eqn{=} 1 + 2 q_{0} + 3 q_{0}^{2} + \frac{5}{2}
q_{0}^{3} + \frac{7}{2} q_{0}^{4} - \frac{5}{3} q_{0}
j_{0} - \frac{7}{8} q_{0}^{2} j_{0} - \frac{1}{8} q_{0} s_{0} - j_{0} +\frac{j_{0}^{2}}{12} - \frac{s_{0}}{6} - \frac{l_{0}}{120}\,. 
\end{eqnarray}
In standard applications, other quantities can result in become useful
\begin{itemize}
    \item the luminosity distance:
    \begin{equation}
    d_{L} = \frac{a(t_{0})}{a(t_{0}-\frac{D}{c})} \: (a(t_{0}) r_{0})\,,
    \end{equation}
    \item the angular-diameter distance:
    \begin{equation}
    d_{A} = \frac{a(t_{0}-\frac{D}{c})}{a(t_{0})} \: (a(t_{0}) r_{0})\,,
    \end{equation}
\end{itemize}
where $r_{0}(D)$ is given by 
%
\begin{equation}\label{eq:r_sin}
r_{0}(D) = \left\{
\begin{array}{lr}
  \sin ( \int_{t_{0}- \frac{D}{c}}^{t_{0}} \frac{c \ \mathrm{d}t}{a(t)} ) &  K = +1; \\
  &  \\
  \int_{t_{0}- \frac{D}{c}}^{t_{0}} \frac{c \ \mathrm{d}t}{a(t)} &  K = 0; \\
  &  \\
  \sinh ( \int_{t_{0}- \frac{D}{c}}^{t_{0}} \frac{c \ \mathrm{d}t}{a(t)} ) &  K = -1.
\end{array} \right.
\end{equation}
If we consider the expansion for short distances, namely, if we insert 
the series expansion of $a(t)$ into $r_{0}(D)$, we find 
%
\begin{eqnarray}
r_{0}(D) \Eqn{=} \int_{t_{0} - \frac{D}{c}}^{t_{0}} \frac{c \
\mathrm{d}t}{a(t)} = \int_{t_{0} - \frac{D}{c}}^{t_{0}} \frac{c \
\mathrm{d}t}{a_{0}} \left\{ 1 + H_{0} (t_{0} - t) + \left(1 +
\frac{q_{0}}{2}\right) H_{0}^{2}(t_{0} - t)^{2} + 
\right. \nonumber \\
&+& \left. 
\left(1 + q_{0}
+\frac{ j_{0}}{6}\right)H_{0}^{3}(t_{0} - t)^{3} 
+ 
\left(1 + \frac{3}{2}q_{0} + \frac{q_{0}^{2}}{4} +
\frac{j_{0}}{3} - \frac{s_{0}}{24}\right)H_{0}^{4}(t_{0} - t)^{4} + 
\right. \nonumber \\
&+& \left. 
\left(1 + 2 q_{0} + \frac{3}{4} q_{0}^{2} + \frac{q_{0}
j_{0}}{6} + \frac{j_{0}}{2} - \frac{s}{12} +
l_{0}\right)H_{0}^{5}(t_{0} - t)^{5} + \emph{O}[(t_{0} - t)^{6}]
\right\} = \nonumber \\ 
\Eqn{=} \frac{D}{a_{0}} \left\{ 1 +
\frac{1}{2} \frac{H_{0} D}{c} + \left[\frac{2 + q_{0}}{6}\right]
\left(\frac{H_{0} D}{c}\right)^{2} + \left[ \frac{6 + 6 q_{0} +
j_{0}}{24} \right] \left(\frac{H_{0} D}{c}\right)^{3} + 
\right. \nonumber \\
&+& \left. 
\left[ \frac{24 + 36 q_{0} + 6 q_{0}^{2} + 8 j_{0} - s_{0}}{120} \right] \left(\frac{H_{0} D}{c}\right)^{4} + 
\right. \nonumber \\
&+& \left. \left[ \frac{12 + 24 q_{0} + 9 q_{0}^{2} + 2 q_{0}
j_{0} + 6 j_{0} - s_{0} + 12 l_{0}}{72} \right] \left(\frac{H_{0}
D}{c}\right)^{5} + 
\right. \nonumber \\
&+& \left. 
\emph{O}\left[\left(\frac{H_{0}
D}{c}\right)^{6}\right] \right\}\,.
\end{eqnarray}
%
To convert from physical distance travelled to $r$ 
coordinate, we have to consider that the Taylor series
expansion of $\sin$-$\sinh$ functions is 
%
\begin{equation}
r_{0}(D) = \left[\int_{t_{0}-\frac{D}{c}}^{t_{0}} \frac{c \
\mathrm{d}t}{a(t)}\right] - \frac{k}{3!}
\left[\int_{t_{0}-\frac{D}{c}}^{t_{0}} \frac{c \
\mathrm{d}t}{a(t)}\right]^{3} + \emph{O}\left(
\left[\int_{t_{0}-\frac{D}{c}}^{t_{0}} \frac{c \
\mathrm{d}t}{a(t)}\right]^{5}  \right)
\end{equation}
so that Eq.(\ref{eq:a_series}) with the spatial curvature $K$ term becomes
%
\begin{eqnarray}
r_{0}(D) \Eqn{=} \frac{D}{a_{0}} \left\{ \mathcal{R}_{D}^{0} +
\mathcal{R}_{D}^{1} \frac{H_{0} D}{c} + \mathcal{R}_{D}^{2}
\left(\frac{H_{0} D}{c}\right)^{2} + \mathcal{R}_{D}^{3}
\left(\frac{H_{0} D}{c}\right)^{3} + \right. \nonumber \\
&+&\left. \mathcal{R}_{D}^{4} \left(\frac{H_{0} D}{c}\right)^{4} +
\mathcal{R}_{D}^{5} \left(\frac{H_{0} D}{c}\right)^{5} +
\emph{O}\left[\left(\frac{H_{0} D}{c}\right)^{6}\right] \right\}
\end{eqnarray}
%
with
%
\begin{eqnarray}
\hspace{-13mm}
\mathcal{R}_{D}^{0} \Eqn{=} 1\,, \\
\hspace{-13mm}
\mathcal{R}_{D}^{1} \Eqn{=} \frac{1}{2} \\
\hspace{-13mm}
\mathcal{R}_{D}^{2} \Eqn{=} \frac{1}{6} \left[2 + q_{0} - \frac{K c^{2}}{H_{0}^{2} a_{0}^{2}}\right]\,,  \\
\hspace{-13mm}
\mathcal{R}_{D}^{3} \Eqn{=} \frac{1}{24} \left[ 6 + 6 q_{0} + j_{0} - 6 \frac{K c^{2}}{H_{0}^{2} a_{0}^{2}}\right]\,, \\
\hspace{-13mm}
\mathcal{R}_{D}^{4} \Eqn{=} \frac{1}{120} \left[ 24 + 36 q_{0} + 6 q_{0}^{2} + 8 j_{0} - s_{0} - \frac{5Kc^{2}(7 + 2 q_{0})}{a_{0}^{2} H_{0}^{2}}\right]\,, \\
\hspace{-13mm}
\mathcal{R}_{D}^{5} \Eqn{=} \frac{1}{144} \left[ 24 + 48 q_{0} + 18
q_{0}^{2} + 4 q_{0} j_{0} + 12 j_{0} - 2 s_{0} + 24 l_{0} -
\frac{3Kc^{2}(15 + 10 q_{0} + j_{0})}{a_{0}^{2} H_{0}^{2}} \right]\,.
\end{eqnarray}
%
Using these definitions for luminosity distance, we acquire
%
\begin{equation}
d_{L}(z) = \frac{c z}{H_{0}} \left\{ \mathcal{D}_{L}^{0} +
\mathcal{D}_{L}^{1} \ z + \mathcal{D}_{L}^{2} \ z^{2} +
\mathcal{D}_{L}^{3} \ z^{3} + \mathcal{D}_{L}^{4} \ z^{4} +
\emph{O}(z^{5}) \right\}
\end{equation}
with
%
\begin{eqnarray}
\hspace{-10mm}
\mathcal{D}_{L}^{0} \Eqn{=} 1\,, \\
\hspace{-10mm}
\mathcal{D}_{L}^{1} \Eqn{=} - \frac{1}{2} \left(-1 + q_{0}\right) \\
\hspace{-10mm}
\mathcal{D}_{L}^{2} \Eqn{=} - \frac{1}{6} \left(1 - q_{0} - 3q_{0}^{2} + j_{0} + \frac{K c^{2}}{H_{0}^{2}a_{0}^{2}}\right)\,, \\
\hspace{-10mm}
\mathcal{D}_{L}^{3} \Eqn{=} \frac{1}{24} \left[2 - 2 q_{0} - 15
q_{0}^{2} - 15 q_{0}^{3} + 5 j_{0} + 10 q_{0} j_{0} + s_{0} +
\frac{2 K c^{2} (1 + 3 q_{0})}{H_{0}^{2} a_{0}^{2}}\right]\,, \\
\hspace{-10mm}
\mathcal{D}_{L}^{4} \Eqn{=} \frac{1}{120} \left[ -6 + 6 q_{0} + 81
q_{0}^{2} + 165 q_{0}^{3} + 105 q_{0}^{4} - 110 q_{0} j_{0} - 105
q_{0}^{2} j_{0} - 15 q_{0} s_{0} + \right. \\
\hspace{-10mm}
&-& \left.  27 j_{0} + 10 j^{2} - 11 s_{0} - l_{0} -
\frac{5Kc^{2}(1 + 8 q_{0} + 9 q_{0}^{2} - 2 j_{0})}{a_{0}^{2}
H_{0}^{2}}\right]\,.
\end{eqnarray}
While, for the angular diameter distance we find 
\begin{equation}
d_{A}(z) = \frac{c z}{H_{0}} \left\{ \mathcal{D}_{A}^{0} +
\mathcal{D}_{A}^{1} \ z + \mathcal{D}_{A}^{2} \ z^{2} +
\mathcal{D}_{A}^{3} \ z^{3} + \mathcal{D}_{A}^{4} \ z^{4} +
\emph{O}(z^{5}) \right\}
\end{equation}
with
%
\begin{eqnarray}
\hspace{-10mm}
\mathcal{D}_{A}^{0} \Eqn{=} 1\,, \\
\hspace{-10mm}
\mathcal{D}_{A}^{1} \Eqn{=} - \frac{1}{2} \left(3 + q_{0}\right)\,, \\
\hspace{-10mm}
\mathcal{D}_{A}^{2} \Eqn{=} \frac{1}{6} \left[11 + 7 q_{0} + 3q_{0}^{2} - j_{0} - \frac{K c^{2}}{H_{0}^{2}a_{0}^{2}}\right]\,, \\
\hspace{-10mm}
\mathcal{D}_{A}^{3} \Eqn{=} - \frac{1}{24} \left[50 + 46 q_{0} + 39
q_{0}^{2} + 15 q_{0}^{3} - 13 j_{0} - 10 q_{0} j_{0} - s_{0}
- \frac{2 K c^{2} (5 + 3 q_{0})}{H_{0}^{2} a_{0}^{2}}\right]\,, \\
\hspace{-10mm}
\mathcal{D}_{A}^{4} \Eqn{=} \frac{1}{120} \left[ 274 + 326 q_{0} + 411
q_{0}^{2} + 315 q_{0}^{3} + 105 q_{0}^{4} - 210 q_{0} j_{0} - 105
q_{0}^{2} j_{0} - 15 q_{0} s_{0} + \right. \\
\hspace{-10mm}
&-& \left. 137 j_{0} + 10 j^{2} - 21 s_{0} - l_{0} -
\frac{5Kc^{2}(17 + 20 q_{0} + 9 q_{0}^{2} - 2 j_{0})}{a_{0}^{2}
H_{0}^{2}}\right]\,. 
\end{eqnarray}
%

We define
$\Omega_{0} = 1 + \frac{K c^{2}}{H_{0}^{2} a_{0}^{2}}$, which can
be considered a purely cosmographic parameter, or $\Omega_{0} = 1
- \Omega_{K}^{(0)} = \Omega_{\mathrm{m}}^{(0)} + \Omega_{\mathrm{r}}^{(0)} + \Omega_{X}^{(0)}$, where $\Omega_{X}^{(0)}$ corresponds to the current fractional 
densities of dark energy, 
if we explore 
the dynamics of the universe. 
With these parameters, we obtain
%
\begin{eqnarray}
\hspace{-10mm}
\mathcal{D}_{L,y}^{0} \Eqn{=} 1\,, \\
\hspace{-10mm}
\mathcal{D}_{L,y}^{1} \Eqn{=} - \frac{1}{2} \left(- 3 + q_{0}\right)\,, \\
\hspace{-10mm}
\mathcal{D}_{L,y}^{2} \Eqn{=} - \frac{1}{6} \left(12 - 5 q_{0} + 3q_{0}^{2} - j_{0} - \Omega_{0} \right)\,, \\
\hspace{-10mm}
\mathcal{D}_{L,y}^{3} \Eqn{=} \frac{1}{24} \left[52 - 20 q_{0} + 21
q_{0}^{2} - 15 q_{0}^{3} - 7 j_{0} + 10 q_{0} j_{0} + s_{0}
- 2 \Omega_{0} (1 + 3 q_{0}) \right]\,, \\
\hspace{-10mm}
\mathcal{D}_{L,y}^{4} \Eqn{=} \frac{1}{120} \left[359 - 184 q_{0} +
186 q_{0}^{2} - 135 q_{0}^{3} + 105 q_{0}^{4} + 90 q_{0} j_{0} -
105 q_{0}^{2} j_{0}
- 15 q_{0} s_{0} + \right. \\
\hspace{-10mm}
&-& \left. 57 j_{0} + 10 j^{2} + 9 s_{0} - l_{0} - 5 \Omega_{0}
(17 - 6 q_{0} + 9 q_{0}^{2} - 2 j_{0}) \right]\,,
\end{eqnarray}
%
and 
%
\begin{eqnarray}
\mathcal{D}_{A,y}^{0} \Eqn{=} 1\,, \\
\mathcal{D}_{A,y}^{1} \Eqn{=} - \frac{1}{2} \left(1 + q_{0}\right)\,, \\
\mathcal{D}_{A,y}^{2} \Eqn{=} - \frac{1}{6} \left(- q_{0} - 3q_{0}^{2} + j_{0} + \Omega_{0} \right)\,, \\
\mathcal{D}_{A,y}^{3} \Eqn{=} - \frac{1}{24} \left(- 2 q_{0} + 3
q_{0}^{2} + 15 q_{0}^{3} - j_{0} - 10 q_{0} j_{0} - s_{0}
+ 2 \Omega_{0} \right)\,, \\
\mathcal{D}_{A,y}^{4} \Eqn{=} - \frac{1}{120} \left( 1 - 6 q_{0} + 9
q_{0}^{2} - 15 q_{0}^{3} - 105 q_{0}^{4} + 10 q_{0} j_{0} + 105
q_{0}^{2} j_{0}+ 15 q_{0} s_{0} + \right. \\
&-& \left. 3 j_{0} - 10 j^{2} + s_{0} + l_{0} + 5 \Omega_{0}
\right)\,.
\end{eqnarray}
%

Previous relations have been derived for any
value of the curvature parameter. 
To illustrate the method, however, 
we can assume a spatially flat universe, using the simplified versions 
for $K = 0$. Now, since we are going to use supernovae data, it 
will be useful to give as well the Taylor series of the expansion 
of the luminosity distance at it enters the modulus distance, 
which is the quantity about which those observational data inform. 
The final expression for the modulus distance based on the Hubble 
free luminosity distance, $\mu(z) = 5 \log_{10} d_{L}(z)$, is 
%
\begin{equation}\label{eq:museries}
\mu(z) = \frac{5}{\log 10} \cdot \left( \log z + \mathcal{M}^{1} z
+ \mathcal{M}^{2} z^2 + \mathcal{M}^{3} z^{3} + \mathcal{M}^{4}
z^{4} \right)\, ,
\end{equation}
with 
%
\begin{eqnarray}
\mathcal{M}^{1} \Eqn{=} - \frac{1}{2} \left( -1 + q_{0} \right) \, ,\\
\mathcal{M}^{2} \Eqn{=} - \frac{1}{24} \left(7 - 10 q_{0} - 9q_{0}^{2}
+ 4j_{0} \right) \, ,\\
\mathcal{M}^{3} \Eqn{=} \frac{1}{24}\left(5 - 9 q_{0} - 16 q_{0}^{2} -
10 q_{0}^{3} + 7 j_{0} + 8 q_{0} j_{0} + s_{0} \right) \, ,\\
\mathcal{M}^{4} \Eqn{=} \frac{1}{2880}\left(-469 + 1004 q_{0} + 2654
q_{0}^{2} + 3300 q_{0}^{3} + 1575 q_{0}^{4} + 200 j_0^{2} -1148 j_{0} +\right. \nonumber \\
&-& \left. 2620 q_{0} j_{0} - 1800 q_{0}^{2} j_{0} - 300 q_{0}
s_{0} - 324 s_{0} - 24 l_{0} \right)\, .
\end{eqnarray}
%

\section{An example: testing $F(R)$ gravity by cosmography}

The cosmographic approach can be used to deal with $F(R)$ gravity~\cite{salzano}. However, similar considerations perfectly hold also for $f(T)$ gravity~\cite{hossein} or any scalar-tensor gravity model. 
In order to construct the cosmographic apparatus, we describe 
the Friedmann equation (\ref{eq:Add-2-01}) in the FLRW space-time 
in $F(R)$ gravity as 
%
\begin{equation}
H^2 = \frac{1}{3} \left( \frac{\rho_{\mathrm{M}}}{F'(R)} + \rho_{\mathrm{curv}}
\right)\,,
\label{eq: hfr}
\end{equation}
where the prime denotes the derivative with respect to $R$, 
the gravitational coupling is taken as 
$\kappa^2 = 1$ 
and $\rho_{\mathrm{curv}}$ is the energy density of an 
{\it effective curvature fluid}: 
\begin{equation}
\rho_{\mathrm{curv}} = \frac{1}{F'(R)} \left[ \frac{1}{2} \left( F(R)  -
R F'(R) \right) - 3 H \dot{R} F''(R) \right] \ . \label{eq:
rhocurv}
\end{equation}
Assuming there is no interaction between the matter and the 
curvature terms (we are in the {\it Jordan frame}), the 
matter continuity equation gives the usual scaling $\rho_{\mathrm{M}} =
\rho_{\mathrm{M}} (t = t_0) a^{-3} = 3 H_0^2 \Omega_{\mathrm{M}} a^{-3}$, 
with $\Omega_{\mathrm{M}}^{(0)}$ the matter density parameter at the present time. 
The continuity equation for $\rho_{\mathrm{curv}}$ then reads
%
\begin{equation}
\dot{\rho}_{\mathrm{curv}} + 3 H (1 + w_{\mathrm{curv}}) \rho_{\mathrm{curv}} = \frac{3 H_0^2 \Omega_{\mathrm{M}}^{(0)} \dot{R} F''(R)}{\left( F'(R) \right)^2} 
a^{-3}
\label{eq: curvcons}
\end{equation}
with
\begin{equation}
w_{\mathrm{curv}} = -1 + \frac{\ddot{R} F''(R) + \dot{R} \left[ \dot{R}
F'''(R) - H F''(R) \right]} {\left[ F(R) - R F'(R) \right]/2 -
3 H \dot{R} F''(R)}\,. 
\label{eq: wcurv}
\end{equation}
the barotropic factor of the curvature fluid. 
It is worth noticing that the curvature fluid quantities 
$\rho_{\mathrm{curv}}$ and $w_{\mathrm{curv}}$ 
only depend on the form of $F(R)$ and its derivatives up to the third order. 
As a consequence, considering only those current values (which 
may naively be obtained by replacing $R$ with $R_0$ everywhere), 
two $F(R)$ theories sharing the same values of $F(R_0)$, 
$F'(R_0)$, $F''(R_0)$, $F'''(R_0)$ will be degenerate from this 
point of view. 
One can argue that this is not strictly true because 
different $F(R)$ theories will lead to different 
expansion rates $H(t)$ and hence different current values of
$R$ and its derivatives. However, it is likely that two $F(R)$ 
functions that exactly match with each other up to the third order 
derivative today will give rise to the same $H(t)$ at least for $t
\simeq t_0$, so that $(R_0, \dot{R}_0, \ddot{R}_0)$ will be almost 
the same. 
Combining Eq.~(\ref{eq: curvcons}) with Eq.~(\ref{eq: hfr}), one 
finally gets the following {\it master equation} for the Hubble
parameter 
%
\begin{equation}
\dot{H} = 
-\frac{1}{2 F'(R)} \left[ 3 H_0^2 \Omega_{\mathrm{M}}^{(0)} a^{-3} 
+ \ddot{R} F''(R)+ 
+ \dot{R} \left( \dot{R} F'''(R) - H F''(R)
\right) \right] \ . 
\label{eq: presingleeq}
\end{equation}
Expressing the scalar curvature $R$ as function of the Hubble parameter as
%
\begin{equation}
R = - 6 \left( \dot{H} + 2 H^2 \right) 
\label{eq: rvsh}
\end{equation}
and inserting the resultant expression into Eq.~(\ref{eq: presingleeq}), 
one ends with a fourth order nonlinear differential equation for the scale 
factor $a(t)$ that cannot easily be solved even 
for the simplest cases (for instance, $F(R) \propto R^n$). 
Moreover, although technically feasible, a numerical solution of 
Eq.~(\ref{eq: presingleeq}) is plagued by the large uncertainties on the 
boundary conditions (i.e., the current 
values of the scale factor and its derivatives up to the third order) that 
have to be set to find out the scale factor.

Motivated by these difficulties, we now approach the problem from 
a different viewpoint. Rather than choosing a parameterized 
expression for $F(R)$ and then numerically solving 
Eq.~(\ref{eq: presingleeq}) for given values of the boundary conditions, 
we try to relate the current 
values of its derivatives to the cosmographic parameters 
$(q_0, j_0, s_0, l_0)$ so that 
constraining them in a model independent way can give us a hint for 
what kind of $F(R)$ theory is able to fit the observed Hubble diagram. 
Note that a similar analysis, but in the 
context of the energy conditions in $F(R)$, has yet been presented 
in~\cite{Bergliaffa}. 
However, in that work, 
an expression for $F(R)$ is given and then 
the snap parameter is computed in order for it to be compared to 
the observed one. 
On the contrary, our analysis does not depend on any assumed functional 
expression for $F(R)$. 

As a preliminary step, it is worth considering again the 
constraint equation (\ref{eq: rvsh}). 
Differentiating with respect to $t$, we easily get the following relations\,: 
\begin{eqnarray}
\dot{R} \Eqn{=} -6 \left( \ddot{H} + 4 H \dot{H} \right)\,, \\ 
\ddot{R} \Eqn{=} -6 \left( d^3H/dt^3 + 4 H \ddot{H} + 4 \dot{H}^2 \right)\,, 
\\ 
d^3R/dt^3 \Eqn{=} -6 \left( d^4H/dt^4 + 4 H d^3H/dt^3 + 12 \dot{H} \ddot{H} \right)\,. 
\label{eq: prederr}
\end{eqnarray}
Evaluating these at the present time and using 
Eqs.~(\ref{eq: hdot})--(\ref{eq: h4dot}), we finally obtain
%
\begin{eqnarray}
R_0 \Eqn{=} -6 H_0^2 (1 - q_0) \ , 
\label{eq: rz} \\
\dot{R}_0 \Eqn{=} -6 H_0^3 \left(j_0 - q_0 - 2\right) \ , 
\label{eq: rdotz} \\
\ddot{R}_0 \Eqn{=} -6 H_0^4 \left( s_0 + q_0^2 + 8 q_0 + 6 \right) \ ,
\label{eq: r2dotz} \\
d^3R_{0}/dt^3 \Eqn{=} -6 H_0^5 \left[ l_0 - s_0 + 2 (q_0 + 4) j_0 - 6
(3q_0 + 8) q_0 - 24 \right] \ , 
\label{eq: r3dotz}
\end{eqnarray}
which will turn out to be useful in the following. 

We come back to the expansion rate and master equations 
(\ref{eq: hfr}) and (\ref{eq: presingleeq}). 
Since they have to hold along the full evolutionary history of the universe, 
they are naively satisfied 
also at the present time. 
Accordingly, we may evaluate them in $t = t_0$ and thus we easily obtain
%
\begin{eqnarray}
H_0^2 \Eqn{=} \frac{H_0^2 \Omega_{\mathrm{M}}^{(0)}}{F'(R_0)} + \frac{F(R_0) - R_0
F'(R_0) - 6 H_0 \dot{R}_0 F''(R_0)}{6 F'(R_0)} \ , 
\label{eq: hfrz} \\ 
- \dot{H}_0 \Eqn{=} \frac{3 H_0^2 \Omega_{\mathrm{M}}^{(0)}}{2 F'(R_0)} +
\frac{\dot{R}_0^2 F'''(R_0) + \left ( \ddot{R}_0 - H_0 \dot{R_0}
\right ) F''(R_0)}{2 F'(R_0)} \ . 
\label{eq: hdotfrz}
\end{eqnarray}
Using Eqs.~(\ref{eq: hdot})--(\ref{eq: h4dot}) and 
(\ref{eq: rz})--(\ref{eq: r3dotz}), we can rearrange 
Eqs.~(\ref{eq: hfrz}) and (\ref{eq: hdotfrz}) as two relations among 
the Hubble constant $H_0$ and the cosmographic parameters $(q_0, j_0, s_0)$, 
on one hand, and the present day values of $F(R)$ and its derivatives up 
to third order. However, two further relations are needed in order
to close the system and determine the four unknown quantities
$F(R_0)$, $F'(R_0)$, $F''(R_0)$, $F'''(R_0)$. A first one may be
easily obtained by noting that, inserting back the physical units,
the rate expansion equation reads
\begin{equation}
H^2 = \frac{8 \pi G}{3 F'(R)} \left(\rho_{\mathrm{m}} + \rho_{\mathrm{curv}} F'(R) \right)\,,
\end{equation}
which clearly shows that, in $F(R)$ gravity, the 
Newton's gravitational 
constant $G$ (restored for the moment) is replaced by an effective (time dependent) $G_{\mathrm{eff}} = G/F'(R)$. On the other hand, it is 
reasonable to assume that the 
value of $G_{\mathrm{eff}}$ at the present time is 
the same as that of the Newton's 
one, so that we can acquire 
the simple constraint 
%
\begin{equation}
G_{\mathrm{eff}}(z = 0) = G \rightarrow F'(R_0) = 1 \ . 
\label{eq: fpz}
\end{equation}
In order to find 
the fourth relation we need to close the system, 
we first differentiate both sides of Eq.~(\ref{eq: presingleeq}) 
with respect to $t$. 
We thus obtain 
%
\begin{eqnarray}
\ddot{H} \Eqn{=} \frac{\dot{R}^2 F'''(R) + \left( \ddot{R} - H
\dot{R} \right) F''(R) + 3 H_0^2 \Omega_{\mathrm{M}}^{(0)} a^{-3}}{2 \left(
\dot{R} F''(R) \right)^{-1} \left( F'(R) \right)^2} -
\frac{\dot{R}^3 F^{(iv)}(R) + \left( 
3 \dot{R} \ddot{R} - H \dot{R}^2 \right) F'''(R)}{2 F'(R)} \nonumber \\
~ & - & \frac{\left( d^3R/dt^3 - H \ddot{R} + \dot{H} \dot{R}
\right) F''(R) - 9 H_0^2 \Omega_{\mathrm{M}}^{(0)} H a^{-3}}{2 F'(R)} \ ,
\label{eq: h2dotfr}
\end{eqnarray}
with $F^{(iv)}(R) = d^4F/dR^4$. 
We now suppose that $F(R)$ may be well approximated by its third order Taylor 
expansion in terms of $\left(R -R_0\right)$, i.e., we set 
%
\begin{equation}
F(R) = F(R_0) + F'(R_0) (R - R_0) +  \frac{1}{2} F''(R_0) (R -
R_0)^2 + \frac{1}{6} F'''(R_0) (R - R_0)^3 \ . 
\label{eq: frtaylor}
\end{equation}
In such an approximation, 
we find $F^{(n)}(R) = d^nF/R^n = 0$ for $n \ge 4$, so that naively $F^{(iv)}(R_0) = 0$. Evaluating 
Eq.~(\ref{eq: h2dotfr}) at the present time, we acquire 
%
\begin{eqnarray}
\ddot{H}_0 \Eqn{=} \frac{\dot{R}_0^2 F'''(R_0) + \left( \ddot{R}_0
- H_0 \dot{R}_0 \right) F''(R_0) + 3 H_0^2 \Omega_{\mathrm{M}}^{(0)}}{2 \left(
\dot{R}_0 F''(R_0) \right)^{-1} \left( F'(R_0) \right)^2} -
\frac{ \left( 3 \dot{R}_0 \ddot{R}_0 - H \dot{R}_0^2 \right)
F'''(R_0)}{2 F'(R_0)} 
\nonumber \\ ~ & - & 
\frac{\left(
d^3R_{0}/dt^3 - H_0 \ddot{R}_0 + \dot{H}_0 \dot{R}_0 \right)
F''(R_0) - 9 H_0^3 \Omega_{\mathrm{M}}^{(0)}}{2 F'(R_0)} \ . 
\label{eq: h2dotfrz}
\end{eqnarray}
Now, we can schematically proceed as follows. 
We evaluate 
Eqs.~(\ref{eq: hdot})--(\ref{eq: h4dot}) at $z = 0$ and plug 
these relations into the left-hand sides of Eqs.~(\ref{eq: hfrz}), 
(\ref{eq: hdotfrz}) and (\ref{eq: h2dotfrz}). 
Then, we insert Eqs.~(\ref{eq: rz})--(\ref{eq: r3dotz}) into 
the right-hand sides of these same equations, so that only the quantities 
related to the cosmographic parameters $(q_0, j_0, s_0, l_0)$ and 
the $F(R)$ term can enter both sides of these relations. 
Finally, we solve them under the constraint (\ref{eq: fpz}) with respect to 
the current 
values of $F(R)$ and its derivatives up to the third order. 
After some algebra, we eventually 
end up with the desired result
%
\begin{eqnarray}
\frac{F(R_0)}{6 H_0^2} \Eqn{=} - \frac{{\cal{P}}_0(q_0, j_0, s_0, l_0)
\Omega_M + {\cal{Q}}_0(q_0, j_0, s_0, l_0)}{{\cal{R}}(q_0, j_0,
s_0, l_0)} \ , 
\label{eq: f0z} \\
F'(R_0) \Eqn{=} 1 \ , 
\label{eq: f1z} \\
\frac{F''(R_0)}{\left ( 6 H_0^2 \right )^{-1}} \Eqn{=} -
\frac{{\cal{P}}_2(q_0, j_0, s_0) \Omega_M + {\cal{Q}}_2(q_0, j_0,
s_0)}{{\cal{R}}(q_0, j_0, s_0, l_0)} \ , 
\label{eq: f2z} \\
\frac{F'''(R_0)}{\left ( 6 H_0^2 \right )^{-2}} \Eqn{=} -
\frac{{\cal{P}}_3(q_0, j_0, s_0, l_0) \Omega_M + {\cal{Q}}_3(q_0,
j_0, s_0, l_0)}{(j_0 - q_0 - 2) {\cal{R}}(q_0, j_0, s_0, l_0)} \ ,
\label{eq: f3z}
\end{eqnarray}
where we have defined 
\begin{eqnarray}
\hspace{-10mm}
{\cal{P}}_0 \Eqn{=} (j_0 - q_0 - 2) l_0 - (3s_0 + 7j_0 + 6q_0^2 +
41q_0 + 22) s_0 + 
\nonumber \\ 
\hspace{-10mm}
\Eqn{-} \left[ (3q_0 + 16) j_0 + 20q_0^2 + 64q_0 + 12
\right] j_0 
-\left( 3q_0^4 + 25q_0^3 + 96q_0^2
+ 72q_0 + 20 \right) \ , 
\label{eq: defp0} \\ 
\hspace{-10mm}
{\cal{Q}}_0 \Eqn{=} (q_0^2 - j_0 q_0 + 2q_0) l_0 +
\left [ 3q_0s_0 + (4q_0 + 6) j_0 + 6q_0^3 + 44q_0^2 + 22q_0 - 12 \right ] s_0 
+ 
\nonumber \\
\hspace{-10mm}
\Eqn{+} \left [ 2j_0^2 + (3q_0^2 + 10q_0 - 6) j_0 + 17q_0^3 +
52q_0^2 + 54q_0 + 36 \right ] j_0 + 
\nonumber \\
\hspace{-10mm}
\Eqn{+} 3q_0^5 + 28q_0^4 + 118q_0^3 + 72q_0^2 - 76q_0 -64\ , 
\label{eq: defq0} \\
\hspace{-10mm}
{\cal{P}}_2 \Eqn{=} 9 s_0 + 6 j_0 + 9q_0^2 + 66q_0 + 42 \ , 
\label{eq: defp2} \\
\hspace{-10mm}
{\cal{Q}}_2 \Eqn{=} - \left \{ 6 (q_0 + 1) s_0 + \left [ 2j_0 - 2 (1
- q_0) \right ] j_0 + 6q_0^3 + 50q_0^2 + 74q_0 + 32 \right \} \ ,
\label{eq: defq2} \\ 
\hspace{-10mm}
{\cal{P}}_3 \Eqn{=} 3 l_0  + 3 s_0 - 9(q_0 + 4) j_0 - (45q_0^2 + 78q_0 +
12) \ , 
\label{eq: defp3} \\
\hspace{-10mm}
{\cal{Q}}_3 \Eqn{=} - \left \{ 2 (1 + q_0) l_0 + 2 (j_0 + q_0) s_0 -
( 2j_0 + 4q_0^2 + 12q_0 + 6 ) j_0 +    
\right . \nonumber \\
\hspace{-10mm}
\Eqn{-} \left . 
(30q_0^3 + 84q_0^2 + 78q_0 +24) \right \} \ , 
\label{eq: defq3} \\
\hspace{-10mm}
{\cal{R}} \Eqn{=} (j_0 - q_0 - 2) l_0 - (3s_0 - 2j_0 + 6q_0^2 +
50q_0 + 40) s_0 + 
\nonumber \\ 
\hspace{-10mm}
\Eqn{+}
\left[ (3q_0 + 10) j_0 + 11q_0^2 + 4q_0 +
18 \right] j_0 - (3q_0^4 +
34q_0^3 + 246q_0 + 104) \ . 
\label{eq: defr}
\end{eqnarray}
Equations (\ref{eq: f0z})--(\ref{eq: defr}) make it possible to 
estimate the current 
values of $F(R)$ and its first three 
derivatives as function of the Hubble constant $H_0$ and the 
cosmographic parameters $(q_0, j_0, s_0, l_0)$ provided a value 
for the matter density parameter $\Omega_{\mathrm{M}}^{(0)}$ is given. 
This is a somewhat problematic point. 
Indeed, while the cosmographic parameters may be estimated in a model 
independent way, the fiducial value for $\Omega_{\mathrm{M}}^{(0)}$ is usually 
the outcome of fitting a given dataset in the framework of an assumed dark 
energy scenario. 
However, it is worth noting that all the different models converge on 
the concordance value $\Omega_{\mathrm{M}}^{(0)} \simeq 0.25$ which is 
also in agreement with astrophysical (model independent) estimates from 
the gas mass fraction in galaxy clusters. 
On the other hand, it has been proposed that $F(R)$ theories may avoid the 
need for dark matter in galaxies and 
galaxy clusters~\cite{noipla,CapCardTro07,Frigerio,sobouti,Mendoza}. 
In such a case, the total matter content of the universe is 
essentially equal to the baryonic one. According to the primordial 
elements abundance and the standard Big bang nucleosynthesis (BBN) scenario, 
we therefore find 
$\Omega_{\mathrm{M}}^{(0)} \simeq \omega_b/h^2$ with $\omega_b = \Omega_b^{(0)} h^2 \simeq 0.0214$~\cite{Kirk} and $h$ the Hubble constant in units of 
$100 {\rm km/s/Mpc}$. Setting $h = 0.72$ in agreement with the 
results of the HST Key project~\cite{Freedman}, we hence obtain 
$\Omega_{\mathrm{M}}^{(0)} = 0.041$ for a universe consisting of baryons only. 
Thus, in the following we will consider both cases when numerical 
estimates are needed. 

It is worth noticing that $H_0$ only plays the role of a scaling 
parameter giving the correct physical dimensions to $F(R)$ and its 
derivatives. As such, it is not surprising that we need four 
cosmographic parameters, namely $(q_0, j_0, s_0, l_0)$, to fix the 
four $F(R)$ related quantities $F(R_0)$, $F'(R_0)$, $F''(R_0)$, 
$F'''(R_0)$. It is also worth stressing that 
Eqs.~(\ref{eq: f0z})--({\ref{eq: f3z}) are linear in the $F(R)$ 
quantities, so that $(q_0, j_0, s_0, l_0)$ can uniquely determine the former 
ones. On the contrary, inverting them to acquire the cosmographic parameters 
as a function of the $F(R)$ ones, we do not obtain linear relations. 
Indeed, the field equations in $F(R)$ theories are nonlinear 
fourth order differential equations in terms of the scale factor $a(t)$, so 
that fixing the derivatives of $F(R)$ up to the third order can make it 
possible to find out a class of solutions, not a single one. 
Each one of these solutions will be characterized by a different set of 
cosmographic parameters. This explains why the inversion of 
Eqs.~(\ref{eq: f0z})--(\ref{eq: defr}) does not give a unique 
result for $(q_0, j_0, s_0, l_0)$. 

As a final comment, we again investigate the underlying assumptions 
leading to the above derived relations. While Eqs.~(\ref{eq: hfrz}) 
and (\ref{eq: hdotfrz}) are exact relations deriving from a 
rigorous application of the field equations, 
Eq.~(\ref{eq: h2dotfrz}) heavily relies on having approximated $F(R)$ with 
its third order Taylor expansion (\ref{eq: frtaylor}). 
If this assumption fails, the system should not be closed because a fifth 
unknown parameter enters the game, namely $F^{(iv)}(R_0)$. 
Actually, replacing $F(R)$ with its Taylor expansion is not 
possible for all the class of $F(R)$ theories. As such, the above 
results only hold in those cases where such an expansion is 
possible. Moreover, by truncating the expansion to the third 
order, we are implicitly assuming that higher order terms are 
negligible over the redshift range probed by the data. That is to 
say, we are assuming that
%
\begin{equation}
F^{(n)}(R_0) (R - R_0)^n << \sum_{m = 0}^{3}{\frac{F^{(m)}(R_0)}{m
!} (R - R_0)^m} \ \ {\rm for} \ n \ge 4 
\label{eq: checkcond}
\end{equation}
over the redshift range probed by the data. Checking the validity 
of this assumption is not possible without explicitly solving the 
field equations, but we can guess an order of magnitude estimate 
considering that, for all the viable models, the background dynamics 
should not differ too much from the $\Lambda$CDM one at least up 
to $z \simeq 2$. Using the expression of $H(z)$ for the 
$\Lambda$CDM model, it is easily to see that $R/R_0$ is a quickly 
increasing function of the redshift, so that, 
in order Eq.~(\ref{eq: checkcond}) should hold, we have to assume that 
$F^{(n)}(R_0) << F'''(R_0)$ for $n \ge 4$. This condition is easier to check 
for many analytical $F(R)$ models. 

Once such a relation is verified, we have to still worry about 
Eq.~(\ref{eq: fpz}) relying on the assumption that the {\it
cosmological} gravitational constant is {\it exactly} the same as 
the {\it local} one. Although reasonable, this requirement is not 
absolutely demonstrated. Actually, the numerical value usually 
adopted for the Newton's constant $G$ 
is obtained from laboratory experiments in settings that can hardly be 
considered homogenous and isotropic. 
Similarly, 
the space-time metric in such conditions 
has nothing to do with the cosmological one, so that strictly speaking, 
matching the two values of $G$ should be an extrapolation. 
Although commonly accepted and quite reasonable, 
the condition $G_{\mathrm{local}} =
G_{\mathrm{cosmo}}$ could (at least, in principle) be violated, so that 
Eq.~(\ref{eq: fpz}) could be reconsidered. Indeed, as we will see, 
the condition $F'(R_0) = 1$ may not be verified for some popular 
$F(R)$ models which has recently been proposed in the literature. 
However, it is reasonable to assume that $G_{\mathrm{eff}} (z = 0) = G 
(1 + \varepsilon)$ with $\varepsilon << 1$. 
When this be the case, we should repeat the derivation of 
Eqs.~(\ref{eq: f0z})--(\ref{eq: f3z}) by 
using the condition $F'(R_0) = (1 + \varepsilon)^{-1}$. 
By executing the 
Taylor expansion of the results in terms of $\varepsilon$ to the first order 
and comparing with the above derived equations, we can estimate the 
error induced by our assumption $\varepsilon = 0$. The resultant 
expressions are too lengthy to be reported and depend on the values of the 
matter density parameter $\Omega_{\mathrm{M}}^{(0)}$, the cosmographic 
parameters $(q_0, j_0, s_0, l_0)$ and $\varepsilon$ in a 
complicated way. 
Nevertheless, we have numerically checked that the error 
induced on $F(R_0)$, $F''(R_0)$, $F'''(R_0)$ are much lower than 
$10\%$ for the value of $\varepsilon$ as high as an unrealistic 
$\varepsilon \sim 0.1$. However, results are 
reliable also for these cases~\cite{salzano}.

\subsection{The CPL model}

A determination of $F(R)$ and its derivatives in terms of the 
cosmographic parameters need for an estimate of these latter from
the data in a model independent way. Unfortunately, even in the
nowadays era of {\it precision cosmology}, such a program is still 
too ambitious to give useful constraints on the $F(R)$ 
derivatives, as we will see later. On the other hand, the 
cosmographic parameters may also be expressed in terms of the dark 
energy density and EoS parameters so that we can work out what are 
the current 
values of $F(R)$ and its derivatives giving the 
same $(q_0, j_0, s_0, l_0)$ of the given dark energy model. 
To this aim, it is convenient to adopt a parameterized expression for 
the dark energy EoS in order to reduce the dependence of the 
results on any underlying theoretical scenario. Following the 
prescription of the Dark Energy Task Force \cite{DETF}, we will 
use the CPL parameterization for the EoS of dark energy by 
setting~\cite{CPL,Linder03}
%
\begin{equation}
w = w_0 + w_a (1 - a) = w_0 + w_a z (1 + z)^{-1}\,, 
\label{eq: cpleos}
\end{equation}
so that, in a flat universe filled by dust matter and dark energy, 
the dimensionless Hubble parameter $E(z) = H/H_0$ 
in Eq.~(\ref{eq:A.3}) reads 
%
\begin{equation}
E^2(z) = \Omega_{\mathrm{M}}^{(0)} (1 + z)^3 + \Omega_X^{(0)} (1 + z)^{3(1 + w_0 + w_a)}
\exp \left(-\frac{3 w_a z}{1 + z}\right) 
\label{eq: ecpl}
\end{equation}
with $\Omega_X^{(0)} = 1 - \Omega_{\mathrm{M}}^{(0)}$ because of the 
assumption that the universe is flat. 
Here and in the following, we omit the inferior ``DE'' of $w_0$ and 
$w_a$. 
In order to determine the cosmographic parameters for such a 
model, we avoid integrating $H(z)$ to get $a(t)$ by noting that 
$d/dt = -(1 + z) H(z) d/dz$. We can use such a relation to 
evaluate $(\dot{H}, \ddot{H}, d^3H/dt^3, d^4H/dt^4)$ and then 
solve Eqs.~(\ref{eq: hdot})--(\ref{eq: h4dot}), evaluated at 
$z = 0$, with respect to the parameters of interest. Some algebra 
finally gives 
%
\begin{eqnarray}
q_0 \Eqn{=} \frac{1}{2} + \frac{3}{2} \left(1 - \Omega_{\mathrm{M}}^{(0)}
\right) w_0 \ , 
\label{eq: qzcpl} \\
j_0 \Eqn{=} 1 + \frac{3}{2} \left(1 - \Omega_{\mathrm{M}}^{(0)}\right) \left[ 3w_0 (1 + w_0) + w_a \right] \ , 
\label{eq: jzcpl} \\
s_0 \Eqn{=} -\frac{7}{2} - \frac{33}{4} \left(1 - \Omega_{\mathrm{M}}^{(0)}\right) w_a -
\frac{9}{4} \left(1 - \Omega_{\mathrm{M}}^{(0)}\right) \left[ 9 + \left(7 - \Omega_{\mathrm{M}}^{(0)}\right) w_a \right] w_0 + 
\nonumber \\
\Eqn{-} \frac{9}{4} \left(1 - \Omega_{\mathrm{M}}^{(0)}\right) 
\left(16 - 3\Omega_{\mathrm{M}}^{(0)}\right) w_0^2 - 
\frac{27}{4} \left(1 - \Omega_{\mathrm{M}}^{(0)}\right) \left(3 - \Omega_{\mathrm{M}}^{(0)}\right) 
w_0^3 \ , 
\label{eq: szcpl} \\ 
l_0 \Eqn{=} \frac{35}{2} + \frac{1 - \Omega_{\mathrm{M}}^{(0)}}{4} 
\left[ 213 + \left(7 - \Omega_{\mathrm{M}}^{(0)}\right) w_a \right] w_a + \frac{1 - 
\Omega_{\mathrm{M}}^{(0)}}{4} \left[ 489 +
9\left(82 - 21 \Omega_{\mathrm{M}}^{(0)}\right) w_a \right] w_0 + 
\nonumber \\ 
\Eqn{+} \frac{9}{2} \left(1 - \Omega_{\mathrm{M}}^{(0)}\right) 
\left[ 67 - 21 \Omega_{\mathrm{M}}^{(0)} + \frac{3}{2} 
\left(23 - 11 \Omega_{\mathrm{M}}^{(0)}\right) w_a \right] w_0^2 + 
\nonumber \\ 
\Eqn{+} 
\frac{27}{4} 
\left(1 - \Omega_{\mathrm{M}}^{(0)}\right) \left(47 - 24 \Omega_{\mathrm{M}}^{(0)}\right) w_0^3 + \frac{81}{2}
\left(1 - \Omega_{\mathrm{M}}^{(0)}\right) \left(3 - 2\Omega_{\mathrm{M}}^{(0)}\right) w_0^4 
\ . 
\label{eq: lzcpl}
\end{eqnarray}
Inserting Eqs.~(\ref{eq: qzcpl})--(\ref{eq: lzcpl}) into 
Eqs.~(\ref{eq: f0z})--(\ref{eq: defr}), we acquire lengthy 
expressions (which we do not report here) giving the current 
values of $F(R)$ and its first three derivatives as a function of
$(\Omega_{\mathrm{M}}^{(0)}, w_0, w_a)$. 
It is worth noting that the $F(R)$ model 
obtained is not dynamically equivalent to the starting CPL one. 
Indeed, the two models have the same cosmographic parameters 
only today. 
For instance, the scale factor is the same 
between the two theories only over the time period during which 
the fifth order Taylor expansion is a good approximation of the 
actual $a(t)$. It is also meaningful to stress 
that such a procedure does not select a unique $F(R)$ model, 
but rather a class of fourth order theories all sharing the same third order 
Taylor expansion of $F(R)$.

\subsection{The $\Lambda$CDM case}

With these caveats in mind, it is significant to first examine 
the $\Lambda$CDM model, which is described by setting $(w_0, w_a) = (-1,
0)$ in the above expressions, and hence we have 
%
\begin{eqnarray}
\begin{cases} 
q_0 = \displaystyle{\frac{1}{2} - \frac{3}{2} 
\left(1-\Omega_{\mathrm{M}}^{(0)}\right)}\,, \\ 
j_0 = \displaystyle{1}\,, \\  
s_0 = \displaystyle{1 - \frac{9}{2} \Omega_{\mathrm{M}}^{(0)}}\,, \\  
l_0 = \displaystyle{1 + 3 \Omega_{\mathrm{M}}^{(0)} + \frac{27}{2} 
\left(\Omega_{\mathrm{M}}^{(0)}\right)^2}\,. 
\end{cases}
\label{eq: cplcdm}
\end{eqnarray}
When inserted into the expressions for the $F(R)$ quantities, 
these relations give the remarkable result\,:
\begin{equation}
F(R_0) = R_0 + 2\Lambda \,, 
\quad
F''(R_0) = F'''(R_0) = 0 \,, 
\label{eq: frlcdm}
\end{equation}
so that we obviously conclude that the only $F(R)$ theory having
exactly the same cosmographic parameters as the $\Lambda$CDM model
is just $F(R) \propto R$, i.e., general relativity (GR). 
It is important to mention 
that such a result comes out as a consequence of the values of $(q_0, j_0)$ 
in the $\Lambda$CDM model. 
Indeed, should we have left $(s_0,
l_0)$ undetermined and only fixed $(q_0, j_0)$ to the values in
(\ref{eq: cplcdm}), we should have got the same result in
(\ref{eq: frlcdm}). Since the $\Lambda$CDM model fits a large 
set of different data well, we do expect that the actual values of 
$(q_0, j_0, s_0, l_0)$ do not differ too much from those in 
$\Lambda$CDM model. Therefore, we plug 
Eqs.~(\ref{eq: f0z})--(\ref{eq: defr}) into the following expressions\,:
%
\begin{equation} 
q_0 = q_0^{\Lambda} {\times} (1 + \varepsilon_q)\,, 
\quad 
j_0 = j_0^{\Lambda} {\times} (1 + \varepsilon_j)\,, 
\quad 
s_0 = s_0^{\Lambda} {\times} (1 + \varepsilon_s)\,, 
\quad 
l_0 = l_0^{\Lambda} {\times} (1 + \varepsilon_l)\,, 
\end{equation}
%
with $(q_0^{\Lambda}, j_0^{\Lambda}, s_0^{\Lambda}, l_0^{\Lambda})$ 
given by Eqs.~(\ref{eq: cplcdm}) and 
$(\varepsilon_q, \varepsilon_j, \varepsilon_s, \varepsilon_l)$ 
quantify the deviations from the values in the $\Lambda$CDM model allowed by 
the data. A numerical estimate of these quantities may be 
obtained, e.g., from a Markov chain analysis, but this is outside 
our aims. Since we are here interested in a theoretical 
examination, we prefer to 
study an idealized situation where 
the four quantities above all share the same value $\varepsilon <<
1$. In such a case, we can easily investigate how much the 
corresponding $F(R)$ deviates from GR 
by exploring 
the two ratios $F''(R_0)/F(R_0)$ and $F'''(R_0)/F(R_0)$. 
Inserting the above expressions for the cosmographic parameters into the exact 
(not reported) formulae for $F(R_0)$, $F''(R_0)$ and $F'''(R_0)$, 
taking those ratios and then expanding to first order in 
$\varepsilon$, we finally find
%
\begin{eqnarray}
\hspace{-10mm}
\eta_{20} \Eqn{\equiv} 
\frac{F''(R_0)}{F(R_0)} 
H_0^4 
= 
\frac{64 - 6 \Omega_{\mathrm{M}}^{(0)} \left(9\Omega_{\mathrm{M}}^{(0)} + 8\right)} {\left[ 3
\left(9\Omega_{\mathrm{M}}^{(0)} + 74\right) \Omega_{\mathrm{M}}^{(0)} - 556 \right] \left(\Omega_{\mathrm{M}}^{(0)}\right)^2 + 16} \
{\times} \ \frac{\varepsilon}{27} \ , 
\label{eq: e20eps} \\
\hspace{-10mm}
\eta_{30} \Eqn{\equiv} 
\frac{F'''(R_0)}{F(R_0)} 
H_0^6 
= 
\frac{6 \left[ \left(81 \Omega_{\mathrm{M}}^{(0)} - 110\right) \Omega_{\mathrm{M}}^{(0)} + 40
\right] \Omega_{\mathrm{M}}^{(0)} + 16} {\left[ 3 \left(9\Omega_{\mathrm{M}}^{(0)} + 74\right) \Omega_{\mathrm{M}}^{(0)} - 556
\right] \left(\Omega_{\mathrm{M}}^{(0)}\right)^2 + 16} \ {\times} \ \frac{\varepsilon}{243
\left(\Omega_{\mathrm{M}}^{(0)}\right)^2} \ , 
\label{eq: e30eps}
\end{eqnarray}
%
which are dimensionless quantities and hence more suitable to estimate 
the order of magnitudes of the different terms. 
Inserting our fiducial values into 
$\Omega_{\mathrm{M}}^{(0)}$, we have 
%
\begin{eqnarray}
\eta_{20} \simeq
\begin{cases}
0.15 \ {\times} \ \varepsilon 
&\mbox{for $\Omega_{\mathrm{M}}^{(0)} = 0.041$\,,} \\
-0.12 \ {\times} \ \varepsilon 
&\mbox{for $\Omega_{\mathrm{M}}^{(0)} = 0.250$\,,} \\
\end{cases} \\
\eta_{30} \simeq 
\begin{cases}
4 \ {\times} \ \varepsilon 
&\mbox{for $\Omega_{\mathrm{M}}^{(0)} = 0.041$\,,} \\
-0.18 \ {\times} \ \varepsilon 
&\mbox{for $\Omega_{\mathrm{M}}^{(0)} = 0.250$\,.} \\
\end{cases} 
\end{eqnarray}
For values of $\varepsilon$ up to $0.1$, the above relations show 
that the second and third derivatives are at most two orders of 
magnitude smaller than the zeroth order term $F(R_0)$. 
Actually, the values of $\eta_{30}$ for a baryon only model (first row) 
seems to argue in favor of a larger importance of the third order 
term. However, we have numerically checked that the above 
relations approximates very well the exact expressions up to 
$\varepsilon \simeq 0.1$ with an accuracy depending on the value 
of $\Omega_{\mathrm{M}}^{(0)}$, being smaller for smaller matter density 
parameters. 
Using the exact expressions for $\eta_{20}$ and $\eta_{30}$, our conclusion on 
the negligible effect of the second 
and third order derivatives are significantly strengthened. 

Such a result holds under the hypotheses that the narrower are the 
constraints on the validity of the $\Lambda$CDM model, the smaller 
are the deviations of the cosmographic parameters from the values in the 
$\Lambda$CDM model. It is possible to show that this indeed the 
case for the CPL parameterization we are considering. 
On the other hand, we have also assumed that the deviations 
$(\varepsilon_q, \varepsilon_j, \varepsilon_s, \varepsilon_l)$ take 
the same values. 
Although such hypothesis is somewhat ad hoc, we argue that 
the main results are not affected by giving it away. 
In fact, 
although different from each other, we can still assume that all 
of them are very small so that Taylor expanding to the first order 
should lead to additional terms into 
Eqs.~(\ref{eq: e20eps})--(\ref{eq: e30eps}) which are likely of the same 
order of magnitude. 
We may thus conclude that, if the observations 
confirm that the values of the cosmographic parameters agree 
within $\sim 10\%$ with those predicted by the $\Lambda$CDM model, 
we must recognize that the deviations of $F(R)$ from the GR
case, $F(R) \propto R$, should be vanishingly small. 

It should be emphasized 
however, that such a conclusion only holds 
for those $F(R)$ models satisfying the constraint 
(\ref{eq: checkcond}). 
It is indeed possible to work out a model having 
$F(R_0) \propto R_0$, $F''(R_0) = F'''(R_0) = 0$, but 
$F^{(n)}(R_0) \ne 0$ for some $n$. For such a (somewhat ad hoc) 
model, Eq.~(\ref{eq: checkcond}) is clearly not satisfied so that 
the cosmographic parameters have to be evaluated from the solution 
of the field equations. 
Accordingly, 
the conclusion above does not hold, so that one cannot exclude that 
the resultant values of $(q_0, j_0, s_0, l_0)$ are within $10\%$ of 
those in the $\Lambda$CDM model.

\subsection{The constant EoS model}

Let us now take into account the condition $w = -1$, but still
retains $w_a = 0$, thus obtaining the so called {\it quiescence} 
models. In such a case, some problems arise because both the terms
$(j_0 - q_0 - 2)$ and ${\cal{R}}$ may vanish for some combinations 
of the two model parameters $(\Omega_{\mathrm{M}}^{(0)}, w_0)$. 
For instance, we 
find that $j_0 - q_0 - 2 = 0$ for $w_0 = (w_1, w_2)$ with
%
\begin{eqnarray}
w_1 \Eqn{=} \frac{1}{1 - \Omega_{\mathrm{M}}^{(0)} + \sqrt{\left(1 - \Omega_{\mathrm{M}}^{(0)}\right) \left(4 -
\Omega_{\mathrm{M}}^{(0)}\right)}} \ , \\ 
w_2 \Eqn{=} - \frac{1}{3} \left[ 1 + \frac{4 - \Omega_{\mathrm{M}}^{(0)}}{\sqrt{\left(1 - \Omega_{\mathrm{M}}^{(0)}\right) \left(4 - \Omega_{\mathrm{M}}^{(0)}\right)}} \right] \ .
\end{eqnarray}
%
On the other hand, the equation ${\cal{R}}(\Omega_{\mathrm{M}}^{(0)}, w_0) = 0$ may have different real roots for $w$ depending on the adopted value
of $\Omega_{\mathrm{M}}^{(0)}$. 
Denoting collectively with ${\bf w}_{null}$ the values of $w_0$ that, 
for a given $\Omega_{\mathrm{M}}^{(0)}$, make 
$(j_0 - q_0 -2) {\cal{R}}(\Omega_{\mathrm{M}}^{(0)}, w_0)$ 
taking the null value, we individuate 
a set of quiescence models whose cosmographic parameters give rise 
to divergent values of $F(R_0)$, $F''(R_0)$ and $F'''(R_0)$. For 
such models, $F(R)$ is clearly not defined, so that we have to 
exclude these cases from further consideration. We only note that 
it is still possible to work out an $F(R)$ theory reproducing the 
same background dynamics of such models, but a different route has 
to be used. 

\begin{figure}
\centering
\resizebox{8.5cm}{!}{\includegraphics{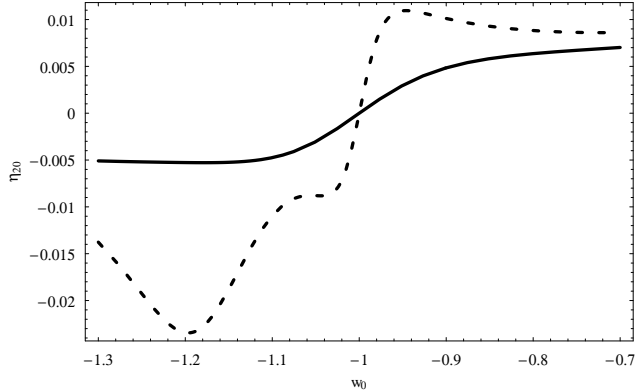}}
\caption{The dimensionless ratio between the current 
values of $F''(R)$ and $F(R)$ as a function of the constant EoS $w_0$ of 
the corresponding quiescence model. Short dashed and solid lines refer 
to models with $\Omega_{\mathrm{M}}^{(0)} = 0.041$ and $0.250$, respectively.}
\label{fig: r20}
\end{figure}

Since both $q_0$ and $j_0$ now deviate from those in 
the $\Lambda$CDM model 
it is not surprising that both $F''(R_0)$ and $F'''(R_0)$ 
take finite non null values. However, it is more interesting to 
study the two quantities $\eta_{20}$ and $\eta_{30}$ defined above 
to investigate the deviations of $F(R)$ from the GR case. These 
are plotted in Figs.~\ref{fig: r20} and \ref{fig: r30} for the 
two fiducial values of $\Omega_{\mathrm{M}}^{(0)}$. 
Note that the range of $w_0$ in these plots have been chosen in order to avoid 
divergences, but the lessons we will draw also hold for 
the other $w_0$ values. 

As a general comment, it is clear that, even in this case, 
$F''(R_0)$ and $F'''(R_0)$ are from two to three orders of 
magnitude smaller than the zeroth order term $F(R_0)$. Such a 
result could yet be guessed from the previous discussion for the 
$\Lambda$CDM case. Actually, relaxing the hypothesis $w_0 = -1$ is 
the same as allowing the values of the cosmographic parameters to deviate from 
those in the $\Lambda$CDM model. Although a direct mapping between the two 
cases cannot be established, it is nonetheless evident that such a 
relation can be argued and hence make the outcome of the above plots 
not fully surprising. 
It is nevertheless worth noting that, while 
in the $\Lambda$CDM case, $\eta_{20}$ and $\eta_{30}$ always have 
opposite signs, this is not the case for quiescence models with $w > -1$. 
Indeed, depending on the value of $\Omega_{\mathrm{M}}^{(0)}$, 
we can have $F(R)$ theories with both $\eta_{20}$ and $\eta_{30}$ positive. 
Moreover, the lower $\Omega_{\mathrm{M}}^{(0)}$ is, the higher the ratios 
$\eta_{20}$ and $\eta_{30}$ are for a given value of $w_0$. This can 
be explained qualitatively noticing that, for a lower 
$\Omega_{\mathrm{M}}^{(0)}$, 
the density parameter of the curvature fluid (playing the role of 
an effective dark energy) must be larger and thus claim for higher 
values of the second and third derivatives.

\begin{figure}
\centering
\resizebox{8.5cm}{!}{\includegraphics{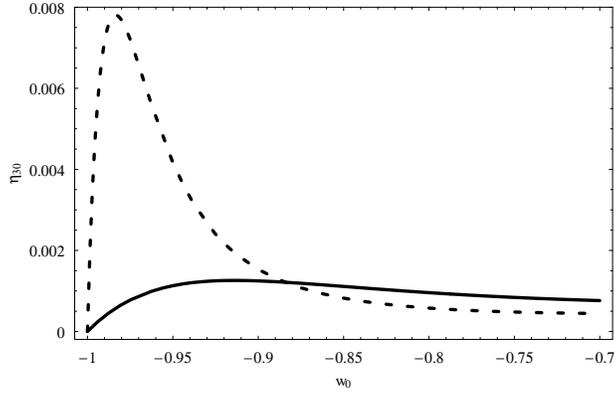}}
\caption{The dimensionless ratio between the current 
values of $F'''(R)$ and $F(R)$ as a function of the constant EoS $w_0$ of 
the corresponding quiescence model. 
Legend is the same as in Fig.~\ref{fig: r20}. 
}
\label{fig: r30}
\end{figure}

\subsection{The general case}

Finally, we study 
evolving dark energy models with $w_a \ne 0$. 
Needless to say, varying three parameters allows to get a wide 
range of models that cannot be discussed in detail. Therefore, we 
only concentrate on evolving dark energy models with $w_0 = -1$ in 
agreement with some most recent analysis. The results on 
$\eta_{20}$ and $\eta_{30}$ are plotted in Figs.~\ref{fig: r20cpl} and 
\ref{fig: r30cpl} where these quantities are displayed as functions of $w_a$. 
Note that we are exploring 
models with positive $w_a$ so that 
$w(z)$ can tend to $w_0 + w_a > w_0$ for $z \rightarrow \infty$ and 
the EoS for dark energy can eventually approach the dust value $w = 0$. 
Actually, this is also the range favored by the data. 
We have, however, excluded values where $\eta_{20}$ or 
$\eta_{30}$ diverge. Considering how they are defined, it is clear 
that these two quantities diverge when $F(R_0) = 0$, so that the 
values of $(w_0, w_a)$ making $(\eta_{20}, \eta_{30})$ diverge 
may be found by solving 
%
\begin{equation}
{\cal{P}}_0(w_0, w_a) \Omega_{\mathrm{M}}^{(0)} + {\cal{Q}}_0(w_0, w_a) = 0\,, 
\end{equation}
%
where ${\cal{P}}_0(w_0, w_a)$ and ${\cal{Q}}_0(w_0, w_a)$ are 
obtained by inserting Eqs.(\ref{eq: qzcpl})--(\ref{eq: lzcpl}) 
into the definitions (\ref{eq: defp0})--(\ref{eq: defq0}). 
For such CPL models, there is no $F(R)$ model having the same 
cosmographic parameters and, at the same time, satisfying all the 
criteria needed for the validity of our procedure. 
In fact, 
if $F(R_0) = 0$, the condition (\ref{eq: checkcond}) is likely to be 
violated so that higher than third order must be included in the 
Taylor expansion of $F(R)$ thus invalidating the derivation of 
Eqs.(\ref{eq: f0z})--(\ref{eq: f3z}). 

Under these caveats, Figs.~\ref{fig: r20cpl} and \ref{fig: r30cpl} 
demonstrate that allowing the EoS for dark energy to evolve 
does not change significantly our conclusions. 
Indeed, the second and third derivatives, although being not null, 
are nevertheless negligible with respect to the zeroth order term, 
and therefore the consequence is 
in favor of a GR\,-\,like $F(R)$ with only very small corrections. 
Such a result is, however, not fully unexpected. {}From 
Eqs.~(\ref{eq: qzcpl}) and (\ref{eq: jzcpl}), we see that, having 
set $w_0 = -1$, the parameter $q_0$ is the same as that in the 
$\Lambda$CDM model, while $j_0$ reads $j_0^{\Lambda} + (3/2)\left(1 -
\Omega_{\mathrm{M}}^{(0)}\right) w_a$. 
As we have stressed above, the Einstein-Hilbert 
Lagrangian $F(R) = R + 2 \Lambda$ is recovered when $(q_0, j_0) =
(q_0^{\Lambda}, j_0^{\Lambda})$ whatever the values of $(s_0, 
l_0)$ are. Introducing a $w_a \ne 0$ makes $(s_0, l_0)$ differ 
from those in the $\Lambda$CDM model, but the first two cosmographic 
parameters are only mildly affected. Such deviations are then 
partially washed out by the complicated way they enter in the 
determination of the values of $F(R)$ at the present time 
and its first three derivatives. 

\begin{figure}
\centering \resizebox{8.5cm}{!}{\includegraphics{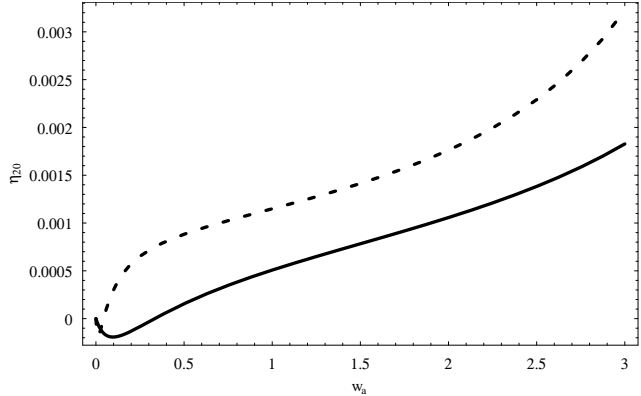}}
\caption{The dimensionless ratio between the present day values of
$F''(R)$ and $F(R)$ as function of the $w_a$ parameter for models
with $w_0 = -1$. 
Legend is the same as in Fig.~\ref{fig: r20}. 
} 
\label{fig: r20cpl}
\end{figure}

\section{Theoretical constraints on the model parameters}

In the preceding 
section, we have worked out a method to 
estimate $F(R_0)$, $F''(R_0)$ and $F'''(R_0)$ resorting to a model 
independent parameterization of the EoS for dark energy. 
However, in the ideal case, the cosmographic parameters are directly 
estimated from the data so that Eqs.~(\ref{eq: f0z})--(\ref{eq: defr}) 
can be used to infer the values of the quantities related to $F(R)$. 
These latter can then be used to put constraints on the parameters 
entering an assumed fourth order theory assigned by an $F(R)$ 
function characterized by a set of parameters 
${\bf p} = (p_1, \ldots, p_n)$ provided that the hypotheses underlying 
the derivation of Eqs.~(\ref{eq: f0z})--(\ref{eq: defr}) are indeed 
satisfied. We show below two interesting cases which clearly 
highlight the potentiality and the limitations of such an 
analysis.

\begin{figure}
\centering \resizebox{8.5cm}{!}{\includegraphics{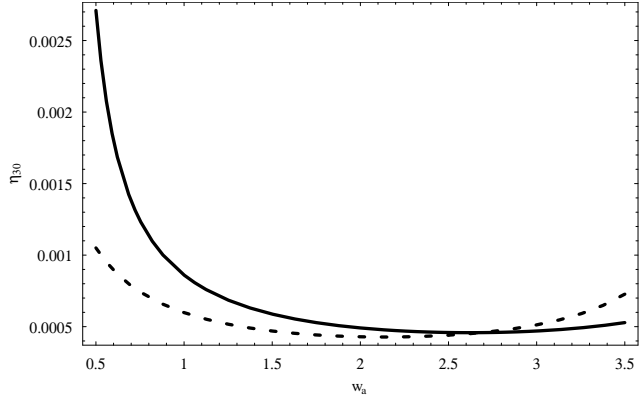}}
\caption{The dimensionless ratio between the present day values of
$F'''(R)$ and $F(R)$ as function of the $w_a$ parameter for models
with $w_0 = -1$. 
Legend is the same as in Fig.~\ref{fig: r20}. 
} 
\label{fig: r30cpl}
\end{figure}

\subsection{Double power law Lagrangian}

As a first interesting example, we take~\cite{Nojiri:2003ft}
%
\begin{equation}
F(R) = R \left(1 + \alpha R^{n} + \beta R^{-m} \right) 
\label{eq: frdpl}
\end{equation}
with $n$ and $m$ two positive real numbers. The following 
expressions are immediately obtained\,:
\begin{eqnarray}
\begin{cases}  
F(R_0) = R_0 \left(1 + \alpha R_0^{n} + \beta R_0^{-m} \right)\,, \\ 
F'(R_0) = 1 + \alpha (n + 1) R_0^n - \beta (m - 1) R_0^{-m}\,, \\ 
F''(R_0) = \alpha n (n + 1) R_0^{n - 1} + \beta m (m - 1) 
R_0^{-(1 + m)}\,, \\ 
F'''(R_0) = \alpha n (n + 1) (n - 1) R_0^{n - 2} - \beta m (m + 1) 
(m - 1) R_0^{-(2 + m)}\,. 
\end{cases}
\label{eq: derdpl}
\end{eqnarray}
Denoting the values of 
$F^{(i)}(R_0)$ determined through Eqs.(\ref{eq: f0z})--(\ref{eq: defr}) 
by $\phi_i$ (with $i = 0, \ldots, 3$), 
we can solve
%
\begin{eqnarray}
\begin{cases}  
F(R_0) = \phi_0\,, \\ 
F'(R_0) = \phi_1\,, \\ 
F''(R_0) = \phi_2\,, \\ 
F'''(R_0) = \phi_3\,, 
\end{cases}
\end{eqnarray}
which is a system of four equations in the four unknowns 
$(\alpha, \beta, n, m)$ that can analytically be solved proceeding as 
follows. First, we solve the first and second equation with 
respect to $(\alpha, \beta)$ and we obtain 
%
\begin{eqnarray}
\begin{cases}  
\alpha = \displaystyle{\frac{1 - m}{n + m} \left( 1 - \frac{\phi_0}{R_0} 
\right) R_0^{-n}}\,, \\ 
\beta = \displaystyle{- \frac{1 + n}{n + m} \left( 1 - \frac{\phi_0}{R_0} 
\right) R_0^{m}}\,, 
\end{cases}
\label{eq: ab12}
\end{eqnarray}
while, solving the third and fourth equations, we get
%
\begin{eqnarray}
\begin{cases}  
\alpha = \displaystyle{\frac{\phi_2 R_0^{1 - n} \left[ 1 + m + (\phi_3/\phi_2) R_0 \right]}{n (n + 1) (n + m)}}\,, \\ 
\beta = \displaystyle{\frac{\phi_2 R_0^{1 + n} \left[ 1 - n + (\phi_3/\phi_2) R_0 \right]}{m (1 - m) (n + m)}}\,. 
\end{cases}
\label{eq: ab34}
\end{eqnarray}
Equating the two solutions, we acquire a system of two equations in
the two unknowns $(n, m)$, given by
%
\begin{eqnarray}
\begin{cases}  
\displaystyle{\frac{n (n + 1) (1 - m) \left( 1 - \phi_0/R_0
\right)}{\phi_2 R_0 \left[ 1 + m + (\phi_3/\phi_2) R_0 \right]}} 
= 1\,, \\ 
\displaystyle{\frac{m (n + 1) (m - 1) \left( 1 - \phi_0/R_0 \right)}{\phi_2 
R_0 \left[ 1 - n + (\phi_3/\phi_2) R_0 \right]}} 
= 1\,. 
\end{cases}
\end{eqnarray}
Solving with respect to $m$, we find two solutions. The first one 
is $m = -n$, which has to be discarded because it makes $(\alpha, \beta)$ 
go to infinity. The only acceptable solution is 
%
\begin{equation}
m = - \left[ 1 - n + (\phi_3/\phi_2) R_0 \right] 
\label{eq: msol}
\end{equation}
which, inserted back into the above system, leads to a second 
order polynomial equation for $n$ with solutions 
%
\begin{equation}
n = \frac{1}{2} \left[1 + \frac{\phi_3}{\phi_2} R_0 {\pm}
\frac{\sqrt{{\cal{N}}(\phi_0, \phi_2, \phi_3)}}{\phi_2 R_0 (1 +
\phi_0/R_0)} \right]\,. 
\label{eq: nsol}
\end{equation}
Here, we have defined
%
\begin{eqnarray}
{\cal{N}}(\phi_0, \phi_2, \phi_3) 
\Eqn{=} \left( R_0^2 \phi_0^2 - 2 R_0^3 \phi_0 + R_0^4 \right) 
\phi_3^2 + 6 \left( R_0 \phi_0^2 - 2 R_0^2 \phi_0 + R_0^3 \right) 
\phi_2 \phi_3 + 
\nonumber \\ 
\Eqn{+} 
9 \left( \phi_0^2 - 2 R_0 \phi_0 + R_0^2 \right)
\phi_2^2 + 4 \left( R_0^2 \phi_0 - R_0^3
\right) \phi_2^3 \ . 
\label{eq: defn}
\end{eqnarray}
Depending on the values of $(q_0, j_0, s_0, l_0)$, 
Eq.~(\ref{eq: nsol}) may lead to one, two or any acceptable solution, i.e. 
real positive values of $n$. This solution has to be inserted back 
into Eq.~(\ref{eq: msol}) to determine $m$ and then into 
Eqs.~(\ref{eq: ab12}) or (\ref{eq: ab34}) to estimate $(\alpha,
\beta)$. If the final values of $(\alpha, \beta, n, m)$ are 
physically viable, we can conclude that the model in 
Eq.~(\ref{eq: frdpl}) is in agreement with the data giving the same 
cosmographic parameters inferred from the data themselves. 
Exploring analytically what is the region of parameter space of 
$(q_0, j_0, s_0, l_0)$ 
which leads to acceptable solutions of $(\alpha, \beta, n, m)$ 
is a daunting task far outside the aim of the present work.

\subsection{The Hu-Sawicki model}

One of the most pressing problems of $F(R)$ theories is the need 
to escape the severe constraints imposed by the Solar System 
tests. A successful model has recently been proposed by Hu and 
Sawicki~\cite{Hu} (HS)\footnote{Note that such a model
does not pass the matter instability test and therefore some viable
generalizations~\cite{Odi,Cogno08,Nodi07,Bamba:2012qi} have been proposed.}
%
\begin{equation}
F(R) = R - R_{\mathrm{c}} \frac{\alpha (R/R_{\mathrm{c}})^n}{1 + \beta (R/R_{\mathrm{c}})^n} \ .
\label{eq: frhs}
\end{equation}
As for the double power law model discussed above, there are four 
parameters which we can be expressed in terms of the cosmographic 
parameters $(q_0, j_0, s_0, l_0)$. 

As a first step, it is trivial to have 
%
\begin{eqnarray}
\begin{cases}   
F(R_0) = \displaystyle{R_0 - R_{\mathrm{c}} \frac{\alpha R_{0 \mathrm{c}}^n}{1 +
\beta R_{0 \mathrm{c}}^n}}\,, \\ 
F'(R_0) = \displaystyle{1 - \frac{\alpha n R_{\mathrm{c}} R_{0 \mathrm{c}}^{n}}{R_0 (1 + \beta R_{0 \mathrm{c}}^n)^2}}\,, \\ 
F''(R_0) = \displaystyle{\frac{\alpha n R_{\mathrm{c}}
R_{0 \mathrm{c}}^n \left [ (1 - n) + \beta (1 + n) R_{0 \mathrm{c}}^n \right ]}{R_0^2
(1 + \beta R_{0 \mathrm{c}}^n)^3}}\,, \\ 
F'''(R_0) = \displaystyle{\frac{\alpha n R_{\mathrm{c}} R_{0 \mathrm{c}}^n (A n^2 + B n + C)}{R_0^3
(1 + \beta R_{0 \mathrm{c}}^n)^4}}\,, 
\end{cases}
\label{eq: derhs}
\end{eqnarray}
with $R_{0 \mathrm{c}} = R_0/R_{\mathrm{c}}$ and 
%
\begin{eqnarray}
\begin{cases}  
A = - \beta^2 R_{0 \mathrm{c}}^{2n} + 4 \beta R_{0 \mathrm{c}}^n - 1\,, \\ 
B = 3 \left(1 - \beta^2 R_{0 \mathrm{c}}^{2n}\right)\,, \\ 
C = -2 \left(1 - \beta R_{0 \mathrm{c}}^{n}\right)^2\,. 
\end{cases}
\end{eqnarray}
Equating Eqs.~(\ref{eq: derhs}) to the four quantities 
$(\phi_0, \phi_1, \phi_2, \phi_3)$ defined as above, we could, in principle, 
solve this system of four equations in four unknowns to obtain 
$(\alpha, \beta, R_c, n)$ in terms of $(\phi_0, \phi_1, \phi_2, \phi_3)$. 
Then, by using Eqs.~(\ref{eq: f0z})--(\ref{eq: defr}) 
we acquire the expressions of $(\alpha, \beta, R_{\mathrm{c}}, n)$ 
as functions of the cosmographic parameters. 
However, setting $\phi_1 = 1$ as required by Eq.~(\ref{eq: f1z}) gives 
the only trivial solution $\alpha n R_{\mathrm{c}} = 0$ so that the HS model reduces 
to the Einstein\,-\,Hilbert Lagrangian $F(R) = R$. 
In order to escape this problem, we can relax the condition $F'(R_0) = 1$ to 
$F'(R_0) = (1 + \varepsilon)^{-1}$. 
As we have discussed in Sec.~XI, 
this is the same as assuming that the current 
effective gravitational constant $G_{\mathrm{eff}, 0} = G/F'(R_0)$ only 
slightly differs from the usual Newton's one, which seems to be a 
quite reasonable assumption. Under this hypothesis, we can 
analytically solve the equations for $(\alpha, \beta, R_{\mathrm{c}}, n)$ in terms of 
$(\phi_0, \varepsilon, \phi_2, \phi_3)$. The actual values of 
$(\phi_0, \phi_2, \phi_3)$ will be no more given by 
Eqs.~(\ref{eq: f0z})--(\ref{eq: f3z}), but we have checked that they 
deviate from those expressions\footnote{Note that the correct expressions 
for $(\phi_0, \phi_2, \phi_3)$ may still formally be written as 
Eqs.~(\ref{eq: f0z})--(\ref{eq: f3z}), but the polynomials 
entering them are now different and also depend on powers of 
$\varepsilon$.} much less than $10\%$ for $\varepsilon$ up to 
$10\%$ well below any realistic expectation. 

With this caveat in mind, we first solve 
\begin{equation}
F(R_0) = \phi_0\,, 
\quad 
F''(R_0) = (1 + \varepsilon)^{-1}\,, 
\end{equation}
to acquire 
%
\begin{eqnarray}
\alpha \Eqn{=} \frac{n (1 + \varepsilon)}{\varepsilon} \left(
\frac{R_0}{R_{\mathrm{c}}} \right)^{1 - n} \left( 1 - \frac{\phi_0}{R_0} \right)^2 \ , \nonumber \\ 
\beta \Eqn{=} \frac{n (1 + \varepsilon)}{\varepsilon} \left(
\frac{R_0}{R_{\mathrm{c}}} \right)^{-n} \left[ 1  - \frac{\phi_0}{R_0}  -
\frac{\varepsilon}{n (1 + \varepsilon)}\right] \ . \nonumber
\end{eqnarray}
Inserting these expressions into the equations in (\ref{eq: derhs}), 
it is easy to check that $R_{\mathrm{c}}$ cancels out, so that we can no more determine 
its value. Such a result is, however, not unexpected. Indeed, 
Eq.~(\ref{eq: frhs}) can trivially be rewritten as
%
\begin{equation}
F(R) = R - \frac{\tilde{\alpha} R^n}{1 + \tilde{\beta} R^n}
\end{equation}
with $\tilde{\alpha} = \alpha R_{\mathrm{c}}^{1 - n}$ and $\tilde{\beta} =
\beta R_{\mathrm{c}}^{-n}$ which are indeed the quantities 
determined by the above expressions for $(\alpha, \beta)$. 
Reversing the discussion, the current values of $F^{(i)}(R)$ 
depend on $(\alpha, \beta, R_{\mathrm{c}})$ only through the two parameters 
$(\tilde{\alpha}, \tilde{\beta})$. 
Accordingly, 
the use of cosmographic parameters is unable to break this degeneracy. 
However, since $R_{\mathrm{c}}$ only plays the role of a scaling parameter, 
we can arbitrarily set its value without loss of generality. 

On the other hand, this degeneracy allows us to get a consistency 
relation to immediately check whether the HS model is viable or
not. Indeed, solving the equation $F''(R_0) = \phi_2$, we find 
%
\begin{equation}
n = \frac{\left(\phi_0/R_0\right) + 
\left[(1 + \varepsilon)/\varepsilon\right](1 - \phi_2 R_0) - 
(1 - \varepsilon)/(1 + \varepsilon)}{1 - \phi_0/R_0}\ ,
\end{equation}
which can then be inserted into the equations $F'''(R_0) = \phi_3$
to obtain a complicated relation among $(\phi_0, \phi_2, \phi_3)$
which we do not report for sake of shortness. 
Solving such a relation with respect to $\phi_3/\phi_0$ and 
executing the Taylor expansion to 
first order in $\varepsilon$, the resultant constraint reads 
%
\begin{equation}
\frac{\phi_3}{\phi_0} \simeq - \frac{1 + \varepsilon}{\varepsilon}
\frac{\phi_2}{R_0} \left[ R_0 \left( \frac{\phi_2}{\phi_0}
\right) + \frac{\varepsilon \phi_0^{-1}}{1 + \varepsilon} \left(
1 - \frac{2 \varepsilon}{1 - \phi_0/R_0} \right) \right] \ .
\end{equation}
If the cosmographic parameters $(q_0, j_0, s_0, l_0)$ are known 
with sufficient accuracy, one could compute the values of 
$(R_0, \phi_0, \phi_2. \phi_3)$ for a given $\varepsilon$ 
(eventually, using the expressions obtained for $\varepsilon = 0$) 
and then check if they satisfied this relation. 
If this is not the case, 
one can immediately give off the HS model also without the need of 
solving the field equations and fitting the data. 
In fact, given the still large errors on the cosmographic parameters, 
such a test only remains in the realm of (quite distant) future applications. 
However, the HS model works for other tests as shown in~\cite{Hu} 
and thus a consistent cosmography analysis has to be combined with them.

\section{Constraints coming from observational data}

Eqs.~(\ref{eq: f0z})--(\ref{eq: defr}) relate the 
values of $F(R)$ and its first three derivatives at the present time 
to the cosmographic parameters $(q_0, j_0, s_0, l_0)$ and the matter
density $\Omega_{\mathrm{M}}^{(0)}$. In principle, therefore, 
a measurement of these latter quantities makes it possible to put constraints 
on $F^{(i)}(R_0)$, with $i = \{0, \ldots, 3\}$, and hence on the 
parameters of a given fourth order theory through the method shown 
in the previous section. 
Actually, the cosmographic parameters are 
affected by errors which obviously propagate onto the $F(R)$ 
quantities. 
Indeed, since the covariance matrix for the cosmographic 
parameters is not diagonal, one has to also take care of this fact 
to estimate the final errors on $F^{(i)}(R_0)$. A similar 
discussion also holds for the errors on the dimensionless ratios 
$\eta_{20}$ and $\eta_{30}$ introduced above. As a general rule, 
indicating with $g(\Omega_{\mathrm{M}}^{(0)}, {\bf p})$ a generic $F(R)$ 
related quantity depending on $\Omega_{\mathrm{M}}^{(0)}$ and the set of 
cosmographic parameters ${\bf p}$, its uncertainty reads 
%
\begin{equation}
\sigma_{g}^2 = \left | \frac{\partial g}{\partial \Omega_{\mathrm{M}}^{(0)}} \right
|^2 \sigma_{M}^2 + \sum_{i = 1}^{i = 4}{ \left | \frac{\partial
g}{\partial p_i} \right |^2 \sigma_{p_i}^2} + \sum_{i \neq j}{2
\frac{\partial g}{\partial p_i} \frac{\partial g}{\partial p_j}
C_{ij}} \label{eq: error}
\end{equation}
where $C_{ij}$ are the elements of the covariance matrix (being
$C_{ii} = \sigma_{p_i}^2$), we have set $(p_1, p_2, p_3, p_4) =
(q_0, j_0, s_0, l_0)$. and assumed that the error $\sigma_M$ on 
$\Omega_{\mathrm{M}}^{(0)}$ is uncorrelated with those on ${\bf p}$. 
Note that this latter assumption strictly holds if the matter density 
parameter is estimated from an astrophysical method 
(such as estimating the total matter in the universe from the evaluated 
halo mass function). 
Alternatively, we will assume that $\Omega_{\mathrm{M}}^{(0)}$ 
is constrained by the CMB radiation related to experiments. 
Since these latter mainly probes the very high redshift universe 
($z \simeq z_{\mathrm{lss}} \simeq 1089$), while the cosmographic parameters 
are concerned with the cosmos at the present time, one can argue that the 
determination of $\Omega_{\mathrm{M}}^{(0)}$ is not affected by the details of 
the model adopted for describing the late universe. 
Indeed, we can reasonably assume that, whatever the candidate for dark energy 
or alternative gravitational theory such as $F(R)$ gravity is, the decoupling 
epoch, i.e., the era when we can observe through the CMB radiation, is 
well approximated by the standard GR with a model comprising only dust matter. 
Hence, 
we will make the simplifying 
(but well motivated) assumption that $\sigma_M$ may be reduced to
very small values and is uncorrelated with the cosmographic 
parameters. 

Under this assumption, the problem of estimating the errors on 
$g(\Omega_{\mathrm{M}}^{(0)}, {\bf p})$ reduces to the analysis of 
the covariance matrix for the cosmographic parameters given the details of the 
data set used as observational constraints. We address this issue by 
computing the Fisher information matrix (see, e.g.,~\cite{Teg97} 
and references therein) defined as 
%
\begin{equation}
F_{ij} = \left \langle \frac{\partial^2 L}{\partial \theta_i
\partial \theta_j} \right \rangle \label{eq: deffij}
\end{equation}
with $L = -2 \ln{{\cal{L}}(\theta_1, \ldots, \theta_n)}$, where 
${\cal{L}}(\theta_1, \ldots, \theta_n)$ is the likelihood of the 
experiment and $(\theta_1, \ldots, \theta_n)$ is the set of parameters 
to be constrained, and $\langle \ldots \rangle$ denotes the 
expectation value. 
In fact, 
the expectation value is computed by 
evaluating the Fisher matrix elements for fiducial values of the 
model parameters $(\theta_1, \ldots, \theta_n)$, while the 
covariance matrix ${\bf C}$ is finally obtained as the inverse of
${\bf F}$. 

A key ingredient in the computation of ${\bf F}$ is the definition 
of the likelihood which depends, of course, of what experimental 
constraint one is using. To this aim, it is worth remembering that
our analysis is based on fifth order Taylor expansion of the scale
factor $a(t)$ so that we can only rely on observational tests
probing quantities that are well described by this truncated
series. Moreover, since we do not assume any particular model, we
can only characterize the background evolution of the Universe,
but not its dynamics which, being related to the evolution of
perturbations, unavoidably need the specification of a physical
model. As a result, the SNeIa Hubble diagram is the ideal
test to constrain the cosmographic parameters. We therefore
defined the likelihood as
%
\begin{equation}
{\cal{L}}(H_0, {\bf p}) \propto 
\exp \left( -\frac{\chi^2(H_0, {\bf p})}{2} 
\right)\,, 
\quad 
\chi^2(H_0, {\bf p}) = \sum_{n =
1}^{{\cal{N}}_{\mathrm{SNeIa}}}{\left[ \frac{\mu_{\mathrm{obs}}(z_i)
- \mu_{\mathrm{th}}(z_n, H_0, {\bf p})}{\sigma_i(z_i)} \right]^2}\ ,
\label{eq: deflike}
\end{equation}
where the distance modulus to redshift $z$ is described as 
%
\begin{equation}
\mu_{\mathrm{th}}(z, H_0, {\bf p}) = 25 + 5 \log_{10} \left(\frac{c}{H_0}\right) + 5 \log_{10} {D_L(z, {\bf p})} \ . \label{eq: defmuth}
\end{equation}
Here, $D_L(z)$ is the Hubble free luminosity distance and it follows from 
Eq.~(\ref{eq:A.2}) that $D_L(z)$ 
in the flat universe is expressed as 
%
\begin{equation}
D_L(z) = (1 + z) \int_{0}^{z}{\frac{dz}{H(z)/H_0}} \ . 
\label{eq: defdlhf}
\end{equation}
Using the fifth order Taylor expansion of the scale factor, we find 
an analytical expression for $D_L(z, {\bf p})$, 
so that in the computation of $F_{ij}$ 
no numerical integration need (this consequence makes the estimate faster). 
As a last ingredient, we need to specify the details of the SNeIa survey 
giving the redshift distribution of the sample and the error on 
each measurement. Following~\cite{Kim}, we adopt\footnote{Note
that, in~\cite{Kim}, the authors assume the data are separated in 
redshift bins so that the error becomes $\sigma^2 = 
\sigma_{sys}^2/{\cal{N}}_{bin} + {\cal{N}}_{bin} (z/z_{max})^2
\sigma_m^2$ with ${\cal{N}}_{bin}$ the number of SNeIa in a bin.
However, we prefer to not bin the data so that ${\cal{N}}_{bin} =
1$.}
%
\begin{equation}
\sigma (z) = \sqrt{\sigma_{\mathrm{sys}}^2 + \left( \frac{z}{z_{\mathrm{max}}}
\right)^2 \sigma_m^2}
\end{equation}
where $z_{\mathrm{max}}$ is the maximum redshift of the survey, 
$\sigma_{\mathrm{sys}}$ is an irreducible scatter in the SNeIa distance 
modulus, 
and $\sigma_m$ is to be assigned depending on the photometric accuracy.

In order to run the Fisher matrix calculation, we have to set a 
fiducial model, 
according to the predictions in the $\Lambda$CDM model 
for the cosmographic parameters. 
For $\Omega_{\mathrm{M}}^{(0)} = 0.3$ 
and the Hubble constant 
$h = 0.72$~\cite{Freedman} in units of $100 {\rm km/s/Mpc}$, 
we acquire 
%
\begin{equation}
(q_0, j_0, s_0, l_0) = (-0.55, 1.0, -0.35, 3.11) \ .
\end{equation}
As a first consistency check, we compute the Fisher matrix for a 
survey mimicking the recent database in~\cite{D07} and thus set 
$({\cal{N}}_{\mathrm{SNeIa}}, \sigma_m) = (192, 0.33)$. After marginalizing 
over $h$ (which, as well known, is fully degenerate with the SNeIa 
absolute magnitude ${\cal{M}}$), we find 
the uncertainties 
\begin{equation}
(\sigma_1, \sigma_2, \sigma_3, \sigma_4) = (0.38, 5.4, 28.1, 74.0)\,, 
\end{equation}
where we are still using the indexing introduced above for the 
cosmographic parameters. These values are reasonably well compared with 
those obtained from a cosmographic fitting of the Gold SNeIa 
dataset~\cite{John04,John05}\footnote{Actually, such estimates have been 
obtained by 
computing the mean and the standard deviation from the 
marginalized likelihoods of the cosmographic parameters. 
Hence, 
the central values do not represent exactly the best fit model, 
while the standard deviations do not give a rigorous description 
of the error because the marginalized likelihoods are manifestly 
non-Gaussian. Nevertheless, we are mainly interested in an order 
of magnitude estimate so that we would not care about such 
statistical details.}
%
\begin{equation}
q_0 = -0.90 {\pm} 0.65 \,, 
\quad 
j_0 = 2.7 {\pm} 6.7 \,,
\quad 
s_0 = 36.5 {\pm} 52.9 \,, 
\quad 
l_0 = 142.7 {\pm} 320 \,.
\end{equation}
Because of the Gaussian assumptions we rely on, 
the Fisher matrix forecasts are known to be lower limits on the determination of a set of parameters, the accuracy to which a given experiment can attain. 
This is indeed the case with the comparison suggesting 
that our predictions are quite optimistic. 
It is worth stressing, however, that the analysis in~\cite{John04,John05} used 
the Gold SNeIa dataset which is poorer in high redshift SNeIa than 
the one in~\cite{D07} 
we are mimicking, so that larger errors on the 
higher order parameters $(s_0, l_0)$ could be expected. 

Rather than computing the errors on $F(R_0)$ and its first three
derivatives, it is more interesting to look at the precision 
attainable on the dimensionless ratios $(\eta_{20}, \eta_{30})$ 
introduced above because they quantify how much deviations from the 
linear order exist. For the fiducial model we are 
considering, both $\eta_{20}$ and $\eta_{30}$ vanish, while, using
the covariance matrix for a survey at the present time and setting 
$\sigma_M/\Omega_{\mathrm{M}}^{(0)} \simeq 10\%$, their uncertainties read
%
\begin{equation}
(\sigma_{20}, \sigma_{30}) = (0.04, 0.04) \ .
\end{equation}
As an application, we can look at Figs.~\ref{fig: r20} and 
\ref{fig: r30} showing how $(\eta_{20}, \eta_{30})$ depend on the 
current value of the EoS $w_0$ for $F(R)$ models sharing the same 
cosmographic parameters of a dark energy model with its constant EoS. 
As it is clear, also considering only the $1 \sigma$ range, the 
full region plotted is allowed by such large constraints on 
$(\eta_{20}, \eta_{30})$. Thus, this means that the full class of 
corresponding $F(R)$ theories is viable. As a consequence, we may 
conclude that the SNeIa data at the present time are unable to 
discriminate between a $\Lambda$ dominated universe and this class 
of fourth order gravity theories.

As a next step, we investigate a SNAP\,-\,like survey~\cite{SNAP} and 
therefore take $({\cal{N}}_{\mathrm{SNeIa}}, \sigma_m) = (2000, 0.02)$. 
We use the same redshift distribution in Table 1 of~\cite{Kim} and 
add 300 nearby SNeIa in the redshift range $(0.03, 0.08)$. The 
Fisher matrix calculation gives the uncertainties on the 
cosmographic parameters 
%
\begin{equation}
(\sigma_1, \sigma_2, \sigma_3, \sigma_4) = 
(0.08, 1.0, 4.8, 13.7)\ .
\end{equation}
The significant improvement of the accuracy in the determination 
of $(q_0, j_0, s_0, l_0)$ is translated into a reduction of the errors 
on $(\eta_{20}, \eta_{30})$, which is now given by 
%
\begin{equation}
(\sigma_{20}, \sigma_{30}) = (0.007, 0.008)\,, 
\end{equation}
where we have supposed 
that, when SNAP data will be available, the matter 
density parameter $\Omega_{\mathrm{M}}^{(0)}$ would be determined with 
a precision $\sigma_M/\Omega_{\mathrm{M}}^{(0)} \sim 1\%$. 
Looking again at Figs.~\ref{fig: r20} and \ref{fig: r30}, it is clear that 
the situation is improved. Indeed, the constraints on $\eta_{20}$ makes 
it possible to narrow the range of allowed models with low matter content 
(the dashed line), while models with typical values of 
$\Omega_{\mathrm{M}}^{(0)}$ are still viable for $w_0$ covering almost the 
full horizontal axis. 
On the other hand, the constraint on $\eta_{30}$ is still too weak, 
so that almost the full region plotted can be allowed. 

Finally, we examine 
an hypothetical future SNeIa survey working 
at the same photometric accuracy as SNAP and with the same 
redshift distribution, but increasing the number of SNeIa up to 
${\cal{N}}_{\mathrm{SNeIa}} = 6 {\times} 10^4$ as expected from, e.g., 
DES~\cite{DES}, PanSTARRS~\cite{PanSTARRS}, SKYMAPPER~\cite{SKY}, 
while still larger numbers may potentially be observed by 
ALPACA~\cite{ALPACA} and LSST~\cite{LSST}. Such a survey can achieve 
%
\begin{equation}
(\sigma_1, \sigma_2, \sigma_3, \sigma_4) = (0.02, 0.2, 0.9, 2.7)
\end{equation}
so that, with $\sigma_M/\Omega_{\mathrm{M}}^{(0)} \sim 0.1\%$, we obtain 
%
\begin{equation}
(\sigma_{20}, \sigma_{30}) = (0.0015, 0.0016) \ .
\end{equation}
Fig.~\ref{fig: r20} shows that, with such a precision on 
$\eta_{20}$, the region of $w_0$ values allowed essentially 
reduces to the value in the $\Lambda$CDM model, while, 
from Fig.~\ref{fig: r30}, 
it is clear that the constraint on $\eta_{30}$ definitively 
excludes models with low matter content further reducing the range 
of values of $w_0$ to quite small deviations from the $w_0 = -1$. We 
can therefore conclude that such a survey will be able to 
discriminate between the concordance $\Lambda$CDM model and all 
the $F(R)$ theories giving the same cosmographic parameters as 
quiescence models other than the $\Lambda$CDM itself. 

A similar discussion may be repeated for $F(R)$ models sharing the 
same values of $(q_0, j_0, s_0, l_0)$ as those in the CPL model 
even if it is less intuitive to grasp the efficacy of the survey being the 
parameter space multivalued. For the same reason, we have not 
explored what is the accuracy on the double power\,-\,law or HS 
models, even if this is technically possible. 
In fact, 
one should first estimate the errors on the current values of $F(R)$ and 
its three time derivatives and then propagate them on the model 
parameters by using the expressions obtained above.

In conclusion, notwithstanding the common claim that we live in the era of 
{\it precision cosmology}, the constraints on $(q_0, j_0, s_0, l_0)$ 
are still too weak to efficiently apply the program we have 
outlined above. 
We have shown how it is possible to 
establish a link between the popular CPL parameterization of the EoS for 
dark energy 
and the derivatives of $F(R)$, 
imposing that they share the same values of the cosmographic parameters. 
This analysis has led to the quite interesting 
conclusion that the only $F(R)$ function, which is able to give the same 
values of $(q_0, j_0, s_0, l_0)$ as those in the $\Lambda$CDM model, is 
indeed $F(R) = R + 2 \Lambda$. 
A similar conclusion holds also in the case of $f(T)$ gravity~\cite{hossein}. 
If future observations will inform us 
that the cosmographic parameters are those of the $\Lambda$CDM model, 
we can therefore rule out all $F(R)$ theories satisfying 
the hypotheses underlying our derivation of 
Eqs.~(\ref{eq: f0z})--(\ref{eq: f3z}). 
Actually, such a result should not be considered as a no way out for higher 
order gravity. 
Indeed, one could still work out a model with null values of $F''(R_0)$ and 
$F'''(R_0)$ as required by the above constraints, but 
non\,-\,vanishing higher order derivatives. One could well argue 
that such a contrived model could be rejected on the basis of the 
Occam razor, but nothing prevents from still taking it into 
account if it turns out to be both in agreement with the data and 
theoretically well founded. 

If new SNeIa surveys will determine the cosmographic parameters 
with good accuracy, acceptable constraints on the two 
dimensionless ratios $\eta_{20} \propto F''(R_0)/F(R_0)$ and 
$\eta_{30} \propto F'''(R_0)/F(R_0)$ could be obtained, 
and thus these quantities 
allow us to discriminate among rival $F(R)$ theories. To 
investigate whether such a program is feasible, we have pursued a 
Fisher matrix based forecasts of the accuracy, which 
future SNeIa surveys can achieve, on the cosmographic parameters and hence on 
$(\eta_{20}, \eta_{30})$. It turns out that a SNAP\,-\,like survey 
can start giving interesting (yet still weak) constraints allowing us 
to reject $F(R)$ models with low matter content, while a 
definitive improvement is achievable with future SNeIa survey 
observing $\sim 10^4$ objects and hence makes it possible to 
discriminate between the $\Lambda$CDM model and a large class of fourth 
order theories. 
It is worth emphasizing, however, that the 
measurement of $\Omega_{\mathrm{M}}^{(0)}$ should come out as the result of 
a model independent probe such as the gas mass fraction in galaxy clusters 
which is, at present, still far from the $1\%$ requested precision. 
On the other hand, one can also rely on the 
$\Omega_{\mathrm{M}}^{(0)}$ 
estimate from the anisotropy and polarization spectra of the CMB radiation 
even if this comes to the price of assuming that the physics at 
recombination is strictly described by GR, so that one has to limit 
its attention to $F(R)$ models reducing to $F(R) \propto R$ during 
that epoch. However, such an assumption is quite common in many 
$F(R)$ models available in literature and therefore it is not a too 
restrictive limitation. 

A further remark is in order concerning what kind of data can be 
used to constrain the cosmographic parameters. The use of the 
fifth order Taylor expansion of the scale factor makes it possible 
to not specify any underlying physical model by relying on the 
minimalist assumption that the universe is described by the flat 
FLRW metric. 
While useful from a theoretical 
perspective, such a generality puts severe limitations to the
dataset one can use. Actually, we can only resort to observational 
tests depending only on the background evolution so that the range 
of astrophysical probes reduces to standard candles (such as SNeIa
and possibly GRBs~\cite{izzo}) and standard rods (such 
as the angular size\,-\,redshift relation for compact 
radiosources). Moreover, pushing the Hubble diagram to $z \sim 2$ 
may rise the question of the impact of gravitational lensing 
amplification on the apparent magnitude of the adopted standard 
candle. The magnification probability distribution function 
depends on the growth of perturbations~\cite{Holz,Holz05,Hui,Friem,Coor}, 
so that one should worry about 
the underlying physical model in order to estimate whether this 
effect biases the estimate of the cosmographic parameters.
However, it has been shown~\cite{R06,Jon,Gunna,Nordin,Sark} that 
the gravitational lensing amplification does not alter 
the measured distance modulus for $z \sim 1$ SNeIa significantly. 
Although such an analysis has been executed for models based on GR, 
we can argue that, whatever the $F(R)$ model is, the growth of 
perturbations finally leads to a distribution of structures along 
the line of sight that is as similar as possible to the observed one 
so that the lensing amplification can be approximately the same. 
We can therefore discuss that the systematic error made by 
neglecting lensing magnification is lower than the statistical 
ones expected by the future SNeIa surveys. 
On the other hand, one can also try reducing this possible bias further 
by using the method of flux averaging~\cite{WangFlux} even if, in such a case, 
our Fisher matrix calculation should be repeated accordingly. 
Furthermore, 
it is significant to note that the constraints on the cosmographic 
parameters may be tightened by imposing some physically motivated 
priors in the parameter space. For instance, we can suppose that 
the Hubble parameter $H(z)$ always stays positive over the full 
range probed by the data or that the transition from past 
deceleration to present acceleration takes place over the range 
probed by the data (so that we can detect it). Such priors should 
be included in the likelihood definition so that the Fisher matrix 
should be recomputed. This is left for a forthcoming work.

Although the data at the present time are still too limited to efficiently 
discriminate among rival dark energy models, we are confident that an 
aggressive strategy aiming at a very precise determination of the 
cosmographic parameters could offer stringent constraints on 
higher order gravity without the need of solving the field 
equations or addressing the complicated problems related to the 
growth of perturbations. Almost 80 years after the pioneering 
distance\,-\,redshift diagram by Hubble, the old cosmographic 
approach appears nowadays as a precious observational tool to 
investigate the new developments of cosmology. 


\section{Conclusion}

In summary, we have presented the review of a number of popular dark energy models, such as the $\Lambda$CDM model, Little Rip and Pseudo-Rip scenarios, the phantom and quintessence cosmologies with the four types (I, II, III and IV) of the finite-time future singularities and non-singular universes filled with dark energy. 

In the first part, we have explained the $\Lambda$CDM model and recent various 
cosmological observations to give the bounds on the late-time acceleration of 
the universe. 
Furthermore, we have investigated a fluid description of the universe in which 
the dark fluid has a general form of the EoS covering the inhomogeneous and 
imperfect EoS. We have explicitly shown that 
all the dark energy cosmologies can be realized by different fluids 
and also considered their properties. 
It has also been demonstrated that at the current stage the cosmological evolutions of all the dark energy universes may be similar to that of the $\Lambda$CDM model, and hence these models are compatible with the cosmological observations. In particular, we have intensively studied the equivalence of different dark energy models. We have described single and multiple scalar field theories, tachyon scalar theory and holographic dark energy, in which the quintessence/phantom cosmology with the current cosmic acceleration can be represented, 
and eventually verified those equivalence to the corresponding fluid descriptions. 

In the second part, as another equivalent class of dark energy models, 
in which dark energy has its geometrical origins, namely, modifications 
of gravitational theories, we have examined $F(R)$ gravity including its extension to $F(R)$ Ho\v{r}ava-Lifshitz gravity and $f(T)$ gravity. 
It has clearly been explored that in these models, 
the $\Lambda$CDM model or the late-time cosmic acceleration with 
the quintessence/phantom behavior can be performed. 

Finally, it is significant to remark that 
there are a number of various dark energy models which we did not discuss in 
this review, such as 
$F(\mathcal{G})$ gravity~\cite{F(G)-gravity}, where 
$\mathcal{G} \equiv R^{2}-4R_{\mu\nu}R^{\mu\nu}+R_{\mu\nu\rho\sigma}R^{\mu\nu\rho\sigma}$ 
with $R_{\mu\nu}$ and $R_{\mu\nu\xi\sigma}$ being the Ricci tensor 
Riemann tensors, respectively, is the Gauss-Bonnet invariant, 
$F(R,\mathcal{G})$ gravity~\cite{Cognola:2006eg}, scalar-Gauss-Bonnet dark energy~\cite{Scalar-Gauss-Bonnet}, 
k-essence dark energy models~\cite{k-essence-DE}, 
ghost condensates scenario~\cite{ArkaniHamed:2003uy} (for its extension to inflation, see~\cite{ArkaniHamed:2003uz}), 
viscous dark energy~\cite{Viscous-DE, RMH-VC}, 
non-minimal derivative dark energy models~\cite{Non-minimal-derivative-DE}, 
G-essence dark energy models~\cite{G-essence-DE}, 
non-local gravity~\cite{Non-local-gravity-DW-NO} produce by quantum effect, 
which is investigated to account for the 
coincidence problem of dark energy and dark matter, 
and galileon dark energy models~\cite{Galileon-DE} (for its application to inflation, called G-inflation, which has recently been proposed, see~\cite{G-inflation}) [as recent reviews on galileon models, see, e.g.,~\cite{Reviews-Galileons}] . 
In particular, galileon gravity has recently been studied very extensively in the literature. The most important feature of the Lagrangian for the galileon scalar field is that the equation of motion derived from the Lagrangian 
is up to the second-order, so that the appearance of an extra degree of 
freedom with the existence of a ghost can be avoided. 
The galileon field originates from a brane bending mode in the 
Dvali-Gabadadze-Porrati (DGP) brane world scenario~\cite{DGP-model}, 
and therefore 
galileon gravity might be regarded as an indirect resolution for the issue of 
a ghost in the self-accelerating branch of the DGP model. 
Since we have no enough space to describe the details of all these models, 
we again mention the important procedure of our approach to show the 
equivalence of dark energy models to represent each cosmology. 
In all of the above models,  
it follows from Eqs.~(\ref{eq:Add-2-01}) and (\ref{eq:Add-2-02}) that 
in the flat FLRW background the gravitational equations can be described as  
$H^2 = \left(\kappa^2/3\right) \rho_\mathrm{DE}$  
and 
$\dot{H} = -\left(\kappa^2/2 \right) 
\left( \rho_\mathrm{DE} + P_\mathrm{DE} \right)$. 
In each model, the difference is only the forms of the energy density $\rho_\mathrm{DE}$ and pressure $P_\mathrm{DE}$ of dark energy. 
Hence, the expression of the Hubble parameter $H$ to describe the concrete 
cosmology, e.g., the $\Lambda$CDM, quintessence and phantom cosmologies, 
can be reconstructed by using these gravitational field equations. 
Similarly, by applying $\rho_\mathrm{DE}$ and $P_\mathrm{DE}$, 
the EoS $w_\mathrm{DE} \equiv \mathrm{DE}/\rho_\mathrm{DE}$ in 
the fluid description in Eq.~(\ref{eq:2.20}) with Eq.~(\ref{eq:2.21}) 
can also be presented. 

Finally, 
it is worth stressing the role of cosmography in this discussion. 
As shown, it is a fundamental tool because it allows, in principle, to 
discriminate among models without a priori assumptions but just laying on 
constraints coming from data. However, the main criticism 
to this approach is related to the extension of the Hubble series, the 
quality and the richness of data samples. In particular, observations
cannot be extended at any redshift and, in most of cases, are not suitable 
to track models up to early epochs. However, the forthcoming observational 
campaigns should ameliorate the situation removing the degeneration 
emerging at low redshifts and allowing a deeper insight of models.

\section*{Acknowledgments}

First of all, all of us would like to thank all the collaborators 
in our works explained in this review: Artyom V. Astashenok, Iver Brevik, Vincenzo F. Cardone, Mariafelicia De Laurentis, Emilio Elizalde, Paul H. Frampton, Chao-Qiang Geng, Zong-Kuan Guo, Shih-Hao Ho, 
Yusaku Ito, Win-Fun Kao, 
Shota Kumekawa, Ruth Lazkoz, Antonio Lopez-Revelles, 
Kevin J. Ludwick, Chung-Chi Lee, Ling-Wei Luo, 
Jiro Matsumoto, 
Ratbay Myrzakulov, Nobuyoshi Ohta, Diego S\'{a}ez-G\'{o}mez, 
Rio Saitou, Vincenzo Salzano, Misao Sasaki, Robert J. Scherrer, 
Lorenzo Sebastiani, Norihito Shirai, Yuta Toyozato, Shinji Tsujikawa, 
Jun'ichi Yokoyama, Artyom V. Yurov and Sergio Zerbini. 
K.B. and S.D.O. would like to acknowledge 
the very kind hospitality as well as support 
at Eurasian National University. 
S.D.O. also appreciates the Japan Society for the Promotion of Science (JSPS) 
Short Term Visitor Program S11135 
and the very warm hospitality at Nagoya University 
where the work has progressed. 
The work is supported in part
by Global COE Program
of Nagoya University (G07) provided by the Ministry of Education, Culture,
Sports, Science \& Technology
(S.N.); 
the JSPS Grant-in-Aid for Scientific Research (S) \# 22224003 and (C) 
\# 23540296 (S.N.); 
%
%
and 
MEC (Spain) project FIS2010-15640 and AGAUR (Catalonia) 2009SGR-994
(S.D.O.).

\appendix
\section{Inertial Force and $w_{\mathrm{DE}}$} 

In this appendix, we check if the occurrence of inertial force in the 
universe with rip (Little rip or the finite-time future singularities) may somehow constrain the $w_{\mathrm{DE}}$. 

As the universe expands, the relative acceleration between two points 
separated by a distance $l$ is given by $l \ddot a/a$. 
If there is a particle with mass $m$ at each of the points, an observer 
at one of the masses will measure an inertial force on the other mass 
of 
\be 
\label{i1} 
F_\mathrm{inert}=m l \ddot a/a = m l \left( \dot H + H^2 \right)\, .
\ee
By using the deceleration parameter 
$q_\mathrm{dec} \equiv -\ddot{a} a^{-1} H^{-2}$ 
in Eq.~(\ref{eq:ED1-9-Add-IIIE-13}), 
we may express the inertial force $F_\mathrm{inert}$ as 
\be
\label{i3}
F_\mathrm{inert}= - m l H^2 q_\mathrm{dec}\, .
\ee
The observational constraint of the value $q_\mathrm{dec}=q_{\mathrm{dec} (0)}$ in the present universe is given by~\cite{Amanullah:2010vv}
\be
\label{i4}
-0.60<q_{\mathrm{dec} (0)}< -0.30\, .
\ee
Here, we mean the present value by the suffix $0$. 
The present value of the Hubble rate $H=H_0$ could be 
\be
\label{i5}
H_0 \sim 70\, \mathrm{km}/\left(\mathrm{s}\cdot\mathrm{Mpc}\right) 
= 2.3 \times 10^{-18}\, \mathrm{s}^{-1}\, .
\ee
We now rewrite the expression in (\ref{i3}) by using 
the gravitational field equations in the FLRW space-time, 
\be
\label{i6}
\frac{3}{\kappa^2}H^2 = \rho\, ,\quad 
- \frac{1}{\kappa^2} \left(2 \dot H + 3 H^2 \right) = P\, ,
\ee
in Eqs.~(\ref{eq:IIIB3-add-005}) and (\ref{eq:IIIB3-add-006}). 
Since the pressures of the usual matters and cold dark matter are 
negligible, 
if we neglect the contribution from the radiation in the present universe, 
we may identify the pressure $P$ with that of the dark energy: 
$P=P_\mathrm{DE}=w_\mathrm{DE} \rho_\mathrm{DE}$. 
Thus, we obtain 
\be
\label{i7}
w_\mathrm{DE} \Omega_\mathrm{DE} 
= - \frac{1}{\kappa^2} \left(2 \dot H_0 + 3 H_0^2 \right) \, , 
\ee
where $\Omega_\mathrm{DE} \equiv \rho_\mathrm{DE}/\rho_\mathrm{crit}^{(0)}$ 
with 
$\rho_\mathrm{crit}^{(0)} \equiv \left(3/\kappa^2\right) H_0^2$ 
in the first relation in Eq.~(\ref{eq:II.04}). 
Then, the inertial force $F_\mathrm{inert}$ has the following form: 
\be
\label{i9}
F_\mathrm{inert}= - \frac{m l H_0^2}{2}\left( 
1 + 3 w_\mathrm{DE} \Omega_\mathrm{DE} \right)\, .
\ee

The galaxy group has a size of $10^{23}\, \mathrm{m}$ and the galaxy is moving 
with the speed of $10^5\,\mathrm{m}/\mathrm{s}$. 
Hence, the acceleration by the 
central force can be estimated to be 
$a_\mathrm{central} \sim 
\left(\left(10^5\right)^2/10^{23}\right)\, \mathrm{m}/\mathrm{s}^2 = 10^{-13}\, \mathrm{m}/\mathrm{s}^2$. 
On the other hand, $l H_0^2 \sim 10^{23}\times \left( 10^{-18} \right)^2 
= 10^{-13}\, \mathrm{m}/\mathrm{s}^2$. 
Therefore, we find $a_\mathrm{central} \sim l H_0^2$, which informs that the 
precise measurement of the sizes of galaxy groups and/or galaxy clusters and 
the rotation speeds of galaxies may give 
a constraint on the value of $w_\mathrm{DE} \Omega_\mathrm{DE}$. 

Conversely, 
provided that 
$w_\mathrm{DE}\sim -1$, $\Omega_\mathrm{DE}\sim 0.73$, Eq.~(\ref{i9}) implies 
\be
\label{i9b}
F_\mathrm{inert} \sim 0.6 m l H_0^2 \sim \left(3.2\time 10^{-36}\,\mathrm{s}^{-2}\right) \times m l \, .
\ee
On the other hand, if we choose $l$ as the size of a galaxy group or galaxy 
cluster and let $v$ a rotational speed of a galaxy in the galaxy group or 
galaxy cluster, the central force by 
the gravity $F_\mathrm{gravity}$ minus the inertial force could be estimated 
as 
\be
\label{i10}
F_\mathrm{cent} \sim \frac{m v^2}{l}\, ,
\ee
Thus, since $F_\mathrm{cent} = F_\mathrm{gravity} - F_\mathrm{inert}$, 
by combining Eqs.~(\ref{i9b}) and (\ref{i10}), we acquire 
\be
\label{i11}
\frac{F_\mathrm{gravity}}{lm}  - \frac{v^2}{l^2} \sim 3.2\time 10^{-36}\,\mathrm{s}^{-2}\, .
\ee
If $F_\mathrm{cent} \sim F_\mathrm{gravity}$ with a difference by a factor, we 
may find 
\be
\label{i12}
\frac{v^2}{l^2} \sim 10^{-36}\,\mathrm{s}^{-2}\, .
\ee


We may also consider the constraint coming from the energy density of galaxy 
clusters. 
By using the gravitational field equations (\ref{i6}), 
the inertial force (\ref{i1}) can be rewritten as 
$
F_\mathrm{inert}=- \left(m l \kappa^2/6 \right) \left(\rho + 3 P \right)
$ 
in the first equality in Eq.~(\ref{eq:ED1-9-Add-IIIE-18}). 
Now, we assume the energy density $\rho_\mathrm{cluster}$ in a galaxy cluster is almost homogeneous. 
In this case, the total mass inside the sphere with a radius $l$ whose center is the center of cluster is given by 
\be
\label{i14}
M=\frac{4\pi}{3}l^3 \rho_\mathrm{cluster}\, .
\ee
Accordingly, 
the Newton gravity which the point particle with mass $m$ suffers is given by
\be
\label{i15}
F_\mathrm{grav.}= G\frac{mM}{l^2} = \frac{4\pi G}{3}m l \rho_\mathrm{cluster}
= \frac{m l \kappa^2}{6} \rho_\mathrm{cluster} \, .
\ee
Therefore, 
if $F_\mathrm{inert}>F_\mathrm{grav.}$, that is, 
if $- \left(\rho + 3 P\right)> \rho_\mathrm{cluster}$, 
the point particle is separated from the cluster. 
We here define 
$w_0\equiv P/\rho$. 
Since $\rho_\mathrm{cluster}=200\rho$, we find the bound for the EoS parameter $w_0$ as $w_0 > -67$. 
As a result, this 
is not so strong constraint.


\end{document}